\documentclass[twoside,english,aps,prl,twocolumns,reprint,showpacs,preprintnumbers,superscriptaddress,nofootinbib,floatfix,10pt]{revtex4-1}
\usepackage[LGR,T1]{fontenc}
\usepackage[latin9]{inputenc}
\usepackage{color}
\usepackage{babel}
\usepackage{float}
\usepackage{textcomp}
\usepackage{bm}
\usepackage{amsmath}
\usepackage{amsthm}
\usepackage{amssymb}
\usepackage{natbib}
\usepackage{graphicx}
\usepackage{esint}
\usepackage{multirow}
\usepackage{hyperref}
\usepackage{longtable} 
\usepackage{subfigure}
\usepackage{multirow} 
\usepackage{array}    
\usepackage{booktabs} 
\usepackage{siunitx}
\hypersetup{pdftitle={SU4Interaction},
 pdfauthor={BNLU},
 unicode=true, bookmarksnumbered=false,bookmarksopen=false, breaklinks=false,pdfborder={0 0 1},colorlinks=true, allcolors=blue}

\makeatletter

\ProvideTextCommand{\~}{LGR}[1]{\char126#1}

\usepackage{amsfonts,latexsym}
\usepackage{times}
\usepackage{slashed}

\makeatother

\usepackage{subfigure}

\begin{document}

\title{Sign-Problem-Free Nuclear Quantum Monte Carlo Simulation}

\author{Zhong-Wang Niu}
\affiliation{Graduate School of China Academy of Engineering Physics, Beijing 100193,
China}

\author{{Bing-Nan Lu}\href{bnlv@gscaep.ac.cn}{$^*$}}
\affiliation{Graduate School of China Academy of Engineering Physics, Beijing 100193,
China}

\date{\today}

\begin{abstract}

Quantum Monte Carlo (QMC) methods offer exact solutions for quantum many-body systems but face severe limitations in fermionic systems like atomic nuclei due to the sign problem.
While sign-problem-free QMC algorithms exist and  provide valuable insights across disciplines, they have been restricted to simple models with limited quantitative predictive power.
Here we overcome this barrier by developing a novel lattice nuclear force that is rigorously sign-problem-free for even-even nuclei.
This interaction achieves a standard deviation of $\sigma = 2.932$ MeV from experimental binding energies for 76 even-even nuclei ($N,Z \leq 28$), matching state-of-the-art phenomenological mean-field models. 
Key innovations include the first sign-problem-free implementation of spin-orbit coupling for shell evolutions and an efficient QMC-optimized framework for global parameter fitting.
Using this approach, we compute binding energies from $^4$He to $^{132}$Sn with unprecedented one-thousandth level numerical precision, reproduce symmetric nuclear matter saturation, and reveal novel spin-orbit-driven clustering in light nuclei. 
This work transforms sign-problem-free QMC into a scalable and predictive nuclear structure tool,
while establishing a high-fidelity, non-perturbative foundation for \textit{ab initio} calculations of heavy nuclei.

\end{abstract}

\maketitle


\paragraph{Introduction}

Understanding the emergence of the nuclear structure from fundamental nucleon-nucleon interactions is a central challenge in modern nuclear physics.
While exact diagonalization approaches are limited by the exponential growth of Hilbert space dimensionality, 
quantum Monte Carlo (QMC) methods offer a promising solution by stochastically sampling the vast Hilbert space, reducing computational complexity to polynomial scaling with nucleon number. 
Recent years have seen numerous QMC applications to nuclear many-body problems (see Refs.~\cite{PhysRept278-1, PPNP47-319, PPNP63-117, Lahde2019nuclear, RMP87-1067,    AnnRevNuclPartSci69-279, RMP94-031003} for reviews).

Despite significant successes, most QMC applications are fundamentally constrained by the fermionic sign problem~\cite{PRL94-170201}. 
This arises when integrating highly oscillatory functions, where cancellations between positive and negative amplitudes lead to exponentially large statistical noise scaling with particle number.
Although mitigation techniques exist, 
\textit{e.g.}, fixed-node~\cite{JChemPhys77-349, JChemPhys77-5593, IntlRevPhysChem14-85}/constrained path approximations~\cite{ PRL74-3652, PRB55-7464, PRC62-014001}, the Lefschetz thimble method~\cite{PRD86-074506, PRD88-051501R}, 
and perturbative expansions~\cite{PRL128-242501, PRR5-L042021, EPJA61-85,  PRC111-015801},
their implementations present significant numerical challenges and often introduce systematic biases.

Notably, for certain symmetry-protected fermionic systems, QMC calculations are immune to the sign problem. 
In these rare cases, QMC delivers unbiased results with high numerical precision and exceptional scalability. 
For example, in lattice quantum chromodynamics (QCD), the isospin symmetry between the up and down quarks implies that the fermionic determinant factorizes into a perfect square, guaranteeing positive amplitudes for any gauge-field configuration~\cite{ProgTheorPhys110-615}. It was further shown that a broader class of interactions without a factorizable determinant can also be sign-problem-free, where positivity is protected by the time-reversal symmetry of the fermion matrix~\cite{PRB71-155115, PRB91-241117}.
See Ref.~\cite{AnnuRevCondMattPhys10-337} and references therein for  sign-problem-free QMC applied to phenomenological condensed matter models.

In nuclear physics, sign-problem-free QMC algorithms are highly desirable but remain underutilized, particularly for nuclear \textit{ab initio} calculations.
Such calculations typically employ complex, high-fidelity nuclear forces derived from fundamental symmetries, which inevitably induce sign problems~\cite{PRL111-032501, PRL113-192501, PRC98-044002, Nature630-59, PRL132-232502}. 
Crucially, any non-zero sign problem, even negligible in light nuclei, grows exponentially with particle number, preventing direct QMC simulations of heavy nuclei. 
Remedies like perturbation theory exist~\cite{PRL128-242501, PRR5-L042021} but still require sign-problem-free interactions as the non-perturbative foundation. 
Consequently, developing sign-problem-free QMC methods would profoundly deepen our first-principles understanding of heavy nuclei. 
Currently, a prominent sign-problem-free nuclear force is the spin-isospin independent Wigner-SU(4) interaction~\cite{PLB797-134863, PhysRev51-106, PRL117-132501, PRL127-062501}, which successfully reproduces light-nuclei binding energies and charge radii, describes emergent geometry in $^{12}$C~\cite{NatComm14-2777}, clustering in hot nuclear matter~\cite{PRL125-192502, PLB850-138463}, Beryllium isotope spectra~\cite{PRL134-162503}, and $\alpha$-particle monopole transitions~\cite{PRL132-062501}.
These significant achievements strongly motivate the search for more accurate sign-problem-free interactions.
The core challenge, however, is that the time-reversal principle enabling sign-problem-free simulations inherently forbids essential interactions such as the tensor force and most spin/isospin-dependent terms~\cite{PRL95-232502, PPNP76-76, PRL89-182501, PRL98-132501}, severely constraining the accessible model space.

In this Letter, we tackle this challenge by constructing a novel minimal nuclear force that is simple enough to avoid the Monte Carlo sign problem, yet sufficiently sophisticated to reproduce key experimental data.
To satisfy these competing requirements, we introduce a sign-problem-free spin-orbit interaction in addition to the Wigner-SU(4) force to capture shell evolution, and optimize its parameters against selected nuclear binding energies using an efficient derivative-based method. 
While conceptually inspired by phenomenological mean-field force construction~\cite{RMP75-121, PPNP57-470, JPG46-013001}, crucially, our approach yields exact solutions to the Hamiltonian.

\paragraph{Method}

We formulate the QMC algorithm within the framework of the \textit{ab initio} approach of nuclear lattice effective field theory (NLEFT), where the nuclear force models are discretized and numerically solved on a three-dimensional cubic lattice using auxiliary field Monte Carlo methods. 
Recent advances in NLEFT have yielded significant achievements across multiple domains, \textit{e.g.}, nuclear ground states~\cite{EPJA31-105,PRL104-142501,EPJA45-335,PLB732-110} and excited states~\cite{PRL112-102501}, intrinsic density distributions and clustering phenomena~\cite{PRL106-192501,PRL109-252501,PRL110-112502,PRL119-222505,arXiv2411.17462}, nucleus-nucleus scattering dynamics~\cite{PRC86-034003,Nature528-111}, and hypernucleus~\cite{PRL115-185301,EPJA56-24,EPJA60-215,PRD111-036002}.

The minimal nuclear force employed here comprises both two- and three-body interactions supplemented by a spin-orbit coupling term.
The Hamiltonian is defined on a $L^3$ cubic lattice with integer coordinates $\bm{n}=(n_x, n_y, n_z)$,
\begin{eqnarray}
    H = \sum_{\bm{n}} \left[ -\frac{\Psi^\dagger \nabla^2 \Psi}{2M} + :\frac{C_2}{2} \overline{\rho}^2 + \frac{C_3}{6} \overline{\rho}^3 + C_{s} \overline{\rho}\ \overline{\rho}_{s}: \right],
    \label{eq:Hamiltonian}
\end{eqnarray}
where the summation spans all lattice sites and the colons mean the normal ordering.
Here $\Psi(\bm{n})$ and $\Psi^\dagger(\bm{n})$ denote annihilation and creation operators, $M=938.92$~MeV is the nucleon mass, $\nabla^2$ is the Laplace operator implemented via fast Fourier transform (FFT), and $C_2$, $C_3$, $C_{s}$ are coupling constants.
We fix the lattice spacing to $a=1.32$~fm  and use natural units $\hbar = c = 1$ throughout.
We use symbols with overline to denote smeared quantities,
\begin{equation}
    \overline{\sigma}(\bm{n}) = \sigma(\bm{n}) + s_{\rm L}\sum_{|\bm{n}^\prime - \bm{n}| = 1} \sigma(\bm{n}^\prime),\qquad \sigma = \rho, \rho_s,
    \label{eq:localsmearing}
\end{equation}
with $s_{\rm L}$ a real number controlling local smearing. 
The densities $\rho$ and $\rho_s$ are defined as
\begin{eqnarray}
    \rho &=& \overline{\Psi}^\dagger \overline{\Psi}, \nonumber \\
    \rho_s &=& \frac{i}{4}\sum_{ijk} \epsilon_{ijk} \nabla_i \left[\overline{\Psi}^\dagger(\overrightarrow{\nabla_j} - \overleftarrow{\nabla_j})  \sigma_k \overline{\Psi}\right],
    \label{eq:densities}
\end{eqnarray}
where $\epsilon_{ijk}$ is the Levi-Civita symbol, $\nabla$ represents the spatial gradients implemented with the FFT, and $\sigma_k$ is the spin Pauli matrices.
The smeared operators $\overline{\Psi}$ and $\overline{\Psi}^\dagger$ are defined as,
\begin{equation}
    \overline{\sigma}(\bm{n}) = \sigma(\bm{n}) + s_{\rm NL}\sum_{|\bm{n}^\prime - \bm{n}| = 1} \sigma(\bm{n}^\prime),\qquad \sigma = \Psi, \Psi^\dagger
    \label{eq:nonlocalsmearing}
\end{equation} 
with $s_{\rm NL}$ a real number controlling non-local smearing.
Implicit spin/isospin indices in $\Psi(\bm{n})$, $\Psi(\bm{n})^\dagger$, and $\sigma_k$ are summed.
Both local (Eq.~(\ref{eq:localsmearing})) and nonlocal (Eq.~(\ref{eq:nonlocalsmearing})) smearing govern the effective interaction range.
The relative strength between $s_{\rm L}$ and $s_{\rm NL}$ determines the locality, an essential element for nuclear binding. 
In particular, the heavy nuclei dissociate into small clusters ($A\leq 4$) at the non-local limit ($s_{\rm L}=0$) and exhibit severe overbinding at the local limit ($s_{\rm NL}=0$)~\cite{PRL117-132501}.
The three-body force ($C_3$) significantly improves the light nuclei structure~\cite{PLB797-134863}.
The newly introduced spin-orbit term ($C_s$) generates requisite shell structure evolution.
For $C_s=0$ we return to the Wigner-SU(4) interaction thoroughly studied in Ref.~\cite{PLB797-134863}.
We also incorporate a perturbative Coulomb force for protons~\cite{SM}.



We solve the Hamiltonian Eq.~(\ref{eq:Hamiltonian}) using the imaginary time projection method,
\begin{equation}
   |\Phi\rangle \propto \lim_{\tau\rightarrow\infty} \exp(-\tau H) |\Phi_T\rangle,
\end{equation}
where $|\Phi\rangle$ is the ground state and $|\Phi\rangle_T$ is a trial wave function with non-zero overlap. 
For sufficiently large $\tau$, excited state contributions are exponentially suppressed.
Expectation values are computed through numerical extrapolation
\begin{equation}
   \langle O \rangle = \lim_{\tau \rightarrow \infty} \frac{\langle \Phi_T| e^{-\tau H/2} O e^{-\tau H/2} | \Phi_T\rangle} 
   {\langle \Phi_T| e^{-\tau H} |\Phi_T\rangle}.
   \label{eq:expecation}
\end{equation}
In NLEFT we implement Eq.~(\ref{eq:expecation}) via auxiliary field transformation. 
For the two-body forces,
\begin{eqnarray}
e^{-a_{t}H} & = & \int\mathcal{D}c\exp\left[-\sum_{\bm{n}}\frac{c(\bm{n})^{2}}{2}+\frac{a_{t}}{2M}\Psi^{\dagger}\nabla^{2}\Psi\right.\nonumber \\
 &  & \left.+\sqrt{-a_{t}C_{2}}\sum_{\bm{n}}c(\bm{n})\left(\overline{\rho}(\bm{n})+\frac{C_{s}}{2C_{2}}\overline{\rho}_{s}(\bm{n})\right)\right],
 \label{eq:transferMatrix}
\end{eqnarray}
where $a_t$ is the temporal step and $c(\bm{n})$ is a real auxiliary field. 
The terms of order $O(C_s^2)$ generated from this transformation are negligible 
(Transformation involving 3NF follows \cite{SM}).
The $c$ fields at each imaginary time step are collectively integrated using the importance sampling Monte Carlo.
Without loss of generality, we consider even-even nuclei and take the trial wave function $|\Phi_T\rangle$ as a Slater determinant of nucleon wave functions paired by time-reversal symmetry $\mathcal{T}=i\sigma_y\mathcal{K}$, where $\mathcal{K}$ is the complex conjugate.
These nucleons interact with the $c$-field and propagate independently under Eq.~(\ref{eq:transferMatrix}). 
As the single-particle Hamiltonian in Eq.~(\ref{eq:transferMatrix}) commute with $\mathcal{T}$, the nucleons remain paired during the imaginary-time evolution.
The resulting determinant of the fermionic correlation matrix is positively definite, ensuring completely sign-problem-free simulations (See \cite{SM} for proof).
For odd-even and odd-odd nuclei, the nucleons can not be completely paired up and a mild sign problem remains.


The interaction in Eq.~(\ref{eq:Hamiltonian}) contains five parameters $C_2$, $C_3$, $C_s$, $s_{\rm L}$, and $s_{\rm NL}$. 
We found that the mass prediction is not sensitive to $s_{\rm NL}$ and fix $s_{\rm NL}=0.45$ throughout.
The remaining four parameters are optimized against experimental binding energies of $^4$He, $^{16}$O, $^{24}$Mg, $^{28}$Si, $^{32}$S and $^{40}$Ca.
Here we select three doubly-magic nuclei and three open-shell nuclei to constrain both the bulk properties and the shell evolution.
The optimization minimize the loss function
\begin{equation}
\chi^{2}=\sum_{A}\left[\frac{E(A)-E_{{\rm exp}}(A)}{\Delta(A)}\right]^{2},
\label{eq:lossfunction}
\end{equation}
with weights $\Delta(A)=1$~MeV for $^{4}$He and $3$~MeV for other nuclei.
To apply the steepest descent method, we need to calculate the partial derivatives of $\chi^2$ against the fitting parameters, which in turn depend on the derivatives of the binding energies $E(A)$.
The latter was computed using the Feynman-Hellmann theorem,
\begin{equation}
\partial E(A)/\partial x=\langle\Phi|H(x+\delta)-H(x)|\Phi\rangle/\delta,
\label{eq:FHthreorem}
\end{equation}
where $x$ is one of the fitting parameters and $\delta$ is a small number chosen for convenience.
The expectation values in Eq.~(\ref{eq:FHthreorem}) are measured concurrently with other observables in  Monte Carlo simulations. 
Crucially, our sign-problem-free QMC enables precise, unbiased estimation of all derivatives simultaneously, facilitating rapid convergence with adaptive learning rate.
The optimization with a reasonable initial guess typically converges within ten iterations.
Among multiple local minima, we identify the globally optimal values $C_2 = -4.410\times10^{-7}$~MeV$^{-2}$, $C_3 = 1.561\times10^{-15}$~MeV$^{-5}$, $ C_{s} = 8.590\times10^{-12}$~MeV$^{-4}$ and $s_{\rm L} = 8.082\times10^{-2}$.
In what follows we denote this group of parameters as \texttt{LAT-OPT1}.
Our strategy, distinct from derivative-free methods used in recent chiral force optimizations against finite nuclei~\cite{PRL110-192502, PLB761-87}, exhibits an  efficiency independent of parameter space dimensionality due to simultaneous derivative measurement. 




\begin{figure}[htbp]
    \centering
    \includegraphics[width=\columnwidth]{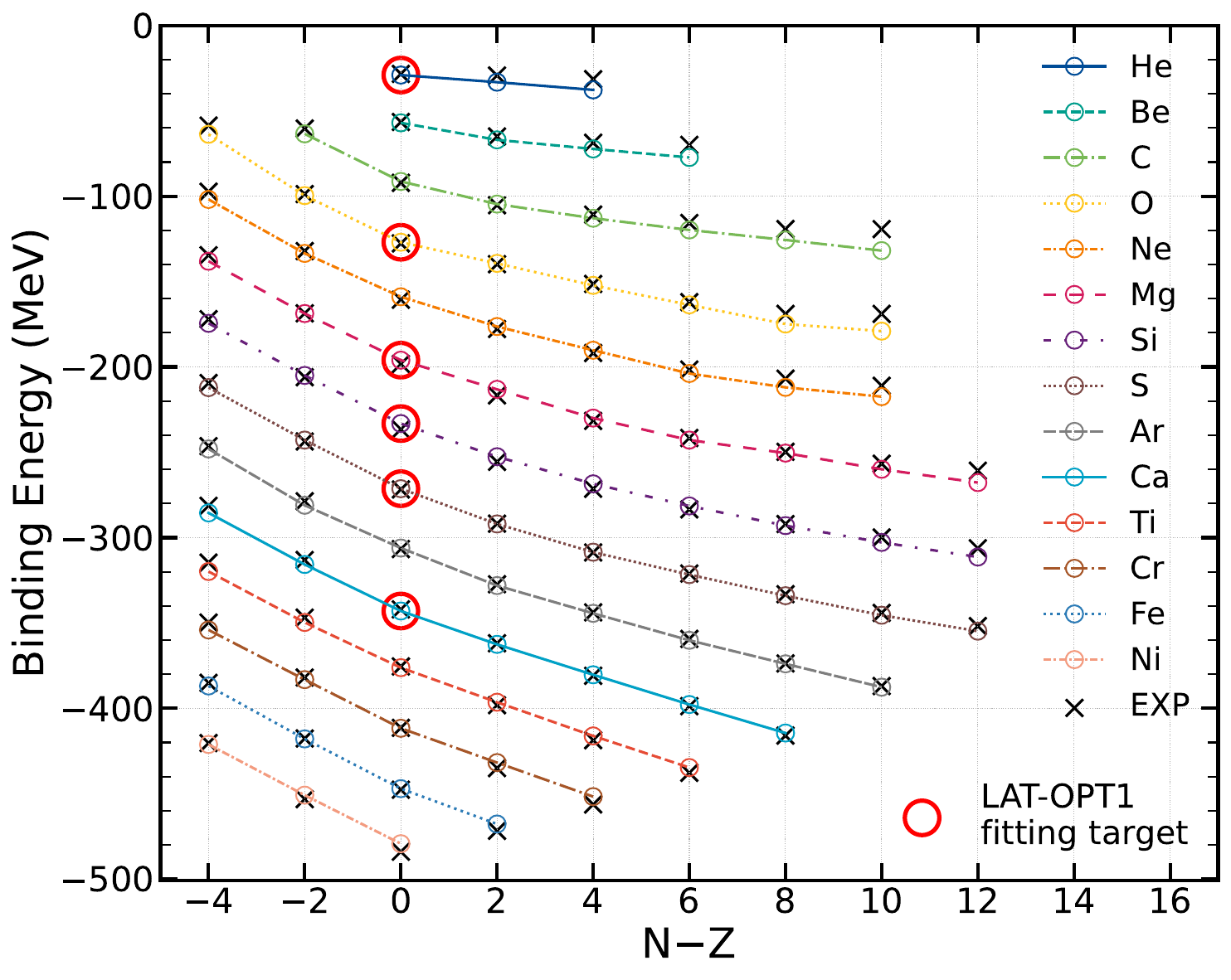}
    \renewcommand{\figurename}{\textbf{Fig.}}
    \caption{Ground state energies from \texttt{LAT-OPT1} (Errors smaller than the symbols).
    Red circles enclose the target nuclei for calibrating the interaction.
    Crosses denote experimental values~\cite{exp}.
    Isotopic chains are connected to guide the eyes.
    See~\cite{SM} for numerical results.}
    \label{fig:binding_energies}
\end{figure}

\paragraph{Results}

Next we present numerical results calculated using the parameter set \texttt{LAT-OPT1}.
Unless otherwise specified, all calculations employ a $L=11$ cubic lattice with periodic boundary conditions. 
The temporal step is set to $a_t=(1000$~MeV$)^{-1}$.
We perform QMC calculations using $50$, $100$, $\cdots$, $350$ temporal slices and extrapolate the results to infinite imaginary time.
The reported errors are combined QMC statistical uncertainties and extrapolation errors.


In Fig.~\ref{fig:binding_energies} we show binding energies for $76$ even-even nuclei with $N,Z\leq 28$.
The results show excellent agreement with experimental values, yielding a standard error of $\sigma=2.932$~MeV.
For comparison, state-of-the-art density functional theories typically give $\sigma \approx 2$-$3$ MeV in this mass region~\cite{PRC85-024304,mean_field_mass_tables,PRC104-054312}.
Note that our model achieves this level of accuracy with only four parameters trained on six nuclei.
This remarkable generalization capability can be attributed to the incorporation of full quantum correlations, with shell effect, pairing, clustering, deformation, collective rotation, and shape fluctuation all treated on an equal footing. 
Systematic overbinding at large neutron numbers indicates missing components in our nuclear force, such as the pion-exchange potentials dictated by the chiral dynamics~\cite{RMP81-1773, PR503-1, frontierinphysics8-98, arXiv:2402.14032}.


\begin{figure}[htbp]
  \centering
  \includegraphics[width=0.9\linewidth]{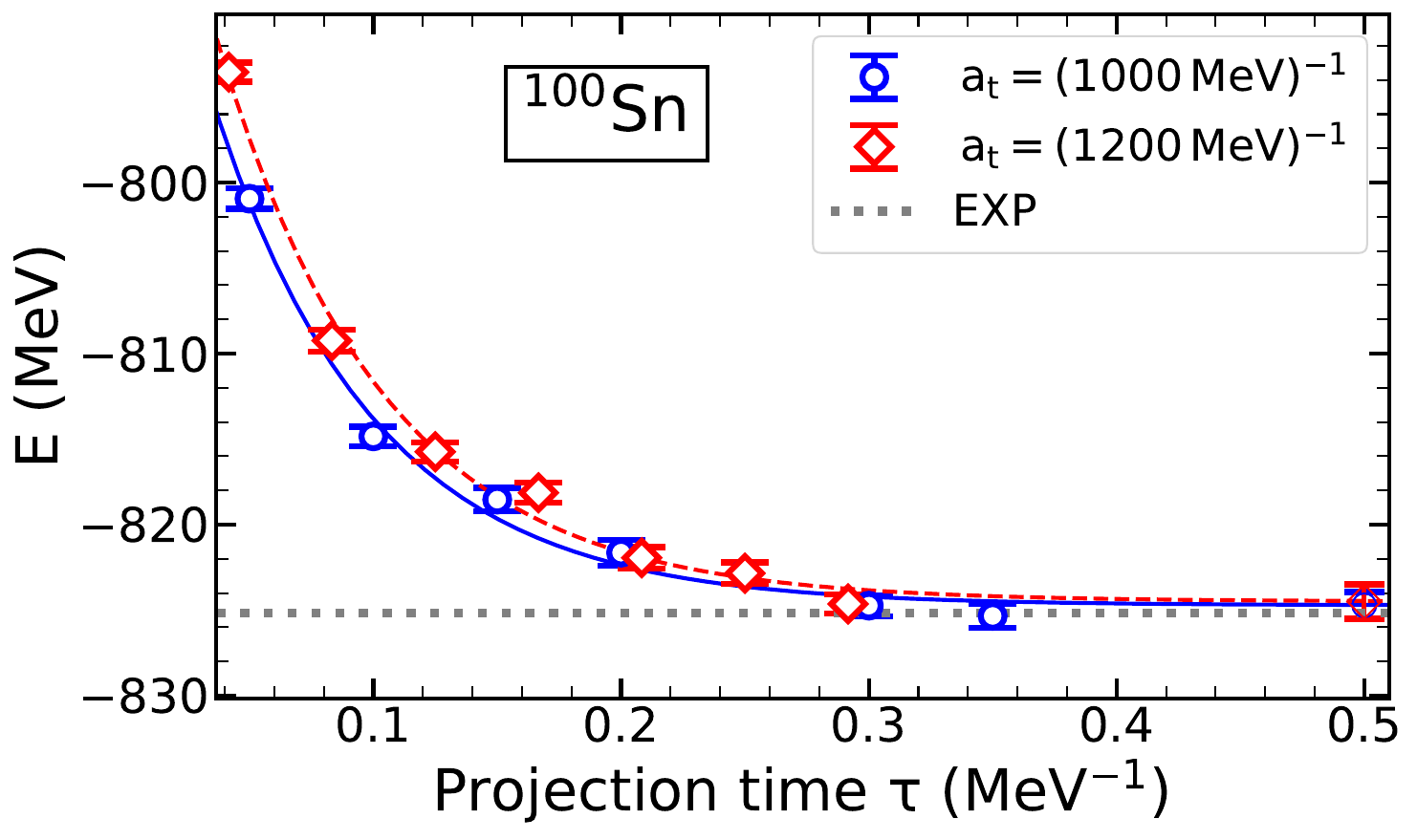}
  \caption{Circles and diamonds denote the $^{100}$Sn binding energies calculated with the temporal step $a_t=(1000$~MeV$)^{-1}$ and $(1200$~MeV$)^{-1}$, respectively.
  Lines represent the exponential fits $E(\tau)=E(\infty)+Ce^{- \tau \Delta}$. 
  Rightmost symbols show extrapolated values.
  Dotted line denotes experimental value~\cite{exp}.}
  \label{fig:Sn100_tcutoff}
\end{figure}

While existing nuclear QMC approaches using high-fidelity forces remain confined to light/medium-mass systems and require error mitigation, our sign-problem-free method can be directly used to investigate heavy nuclei. 
In Fig.~\ref{fig:Sn100_tcutoff} we plot the QMC results for $^{100}$Sn, extrapolated to infinite imaginary time. 
We find a combined QMC statistical and extrapolation error below $1$~MeV.
We also probe dependencies on temporal step size and box volume, finding only negligible variations.
Moreover, The extrapolated binding energy reproduces the experiment within $1$~MeV.
These unprecedented one-thousandth level numerical precision and remarkable predicative capability demonstrate sign-problem-free QMC as an accurate and scalable approach for heavy nuclei.

Tab.~\ref{tab:gs-energies} lists calculated binding energies for representative nuclei.
The relative errors against the experiments are below 3\% for $^4$He and $^{132}$Sn, and below 1\% for other nuclei, demonstrating high accuracy across the nuclear chart.
Furthermore, to quantify the impact of the spin-orbit coupling term proportional to $C_s$, we also list its partial contributions $E_{\rm sl}$.
This term accounts for 5-10\% of the binding energies on average and systematically fluctuates with the number of nucleons.
Notably, the fraction $E_{\rm sl}/E_{\rm bind}$ is always below 5\% for nuclei with traditional harmonic oscillator magic numbers 2, 8, 20, 40 such as $^4$He, $^{16}$O, $^{40}$Ca and $^{80}$Zr, but significantly enhanced for nuclei with new  magic numbers 28, 50, 82 such as $^{56}$Ni, $^{100}$Sn and $^{132}$Sn. 
This extra binding reflects the enhanced stability of these nuclei obtained from shell gaps generated by spin-orbit splitting, directly validating the Mayer-Jensen shell model paradigm within our framework without employing a single-particle mean field.


\begin{table}[htbp]
\centering
\small
\setlength{\tabcolsep}{4pt}
\caption{Calculated binding energies $E_{\rm bind}$, expectation values of the spin-orbit coupling term $E_{\rm sl}$, their ratio and experimental binding energies~\cite{exp} (EXP) for selected nuclei.}
\label{tab:gs-energies}
\begin{tabular}{lcccc}
\toprule
Nucleus  & $E_{\mathrm{bind}}$ (MeV) & $E_{\mathrm{sl}}$ (MeV) & $E_{\mathrm{sl}}/E_{\mathrm{bind}}$ & EXP (MeV) \\
\midrule
$^{4}$He     & \num{-28.8}       & \num{-0.3}     & 0.010 & \num{-28.3} \\
$^{12}$C      & \num{-91.3(1)}       & \num{-13.3}    & 0.146 & \num{-92.2} \\
$^{14}$C      & \num{-104.6(1)}       & \num{-12.7}    & 0.121 & \num{-105.3} \\
$^{16}$O     & \num{-126.9(2)}      & \num{-5.6}     & 0.044 & \num{-127.6} \\
$^{40}$Ca    & \num{-343.0(2)}      & \num{-13.6}    & 0.040 & \num{-342.1} \\
$^{48}$Ca     & \num{-414.5(3)}      & \num{-42.3}    & 0.102 & \num{-416.0} \\
$^{56}$Ni     & \num{-479.3(6)}      & \num{-74.6}    & 0.156 & \num{-484.0} \\
$^{80}$Zr     & \num{-672.1(8)}      & \num{-23.3}    & 0.035 & \num{-669.2} \\
$^{90}$Zr     & \num{-782.1(5)}      & \num{-64.8}    & 0.083 & \num{-783.9} \\
$^{100}$Sn    & \num{-824.7(8)}      & \num{-103.0}   & 0.125 & \num{-825.2} \\
$^{132}$Sn    & \num{-1134.2(27)}    & \num{-110.9}   & 0.098 & \num{-1102.8} \\
\bottomrule
\end{tabular}
\end{table}


Our approach solve the quantum many-body problem non-perturbatively,  enabling studies of strong-correlation phenomena like nuclear clustering that emerges from the intricate interplay among shell structure, quantum fluctuations and continuum effects~\cite{PhysRept432-43, RMP90-035004}.
Fig.~\ref{fig:spinorbit_optimization} (a) shows the loss function $\chi^2$ in Eq.~(\ref{eq:lossfunction}) as a function of the spin-orbit coupling constant $C_s$, with other parameters fixed to \texttt{LAT-OPT1} values.
A pronounced minimum occurs precisely at the \texttt{LAT-OPT1} value $C_s^{\rm opt}$
(vertical line), confirming that our optimization robustly identifies the physical spin-orbit strength stringently constrained by target nuclei.
Fig.~\ref{fig:spinorbit_optimization} (b) displays binding energies for $^4$He and light $\alpha$-conjugate nuclei ($^8$Be, $^{12}$C and $^{16}$O) relative to $n$-$\alpha$ breakup thresholds.
The $^4$He binding is independent of $C_s$, while $^8$Be persists unbound against two-$\alpha$ decay.
The 3-$\alpha$ ($^{12}$C) and 4-$\alpha$ ($^{16}$O) systems are more sensitive to spin-orbit coupling.
In the SU(4)-symmetric limit ($C_s\to 0)$, $^{16}$O is stable, but $^{12}$C is unbound against  $3\alpha$ dissociation. 
Conversely, at the other extreme of strong spin-orbit coupling ($C_s \gg C_s^{\rm opt}$), the 3-$\alpha$ and 4-$\alpha$ binding energies coincide, indicating that $^{12}$C becomes deeply bound, while $^{16}$O turns unstable against $^{12}$C$+\alpha$ breakup. 
Remarkably, only near the $\chi^2$-optimized value $C_s^{\rm opt}$ do we reproduce the experimental binding hierarchy.
While numerical calculations demonstrate that inter-cluster effective interactions are sensitive to the strength and locality of the underlying fundamental interactions~\cite{PRL117-132501, PRL118-232502, PRC103-024318}, here we further establish a direct link between spin-orbit coupling and effective $\alpha$-cluster interactions.
Likewise, other key components for clusterization like the tensor force can be studied similarly~\cite{NatComm13-2234, Phil.Trans.R.Soc.A382-20230123, PLB858-139036, arXiv2506.16947}.
Such investigations pave the way towards a unified mechanism governing both shell and clustering phenomena in light nuclei.







\begin{figure}[htbp]
  \centering
  \includegraphics[width=0.9\linewidth]{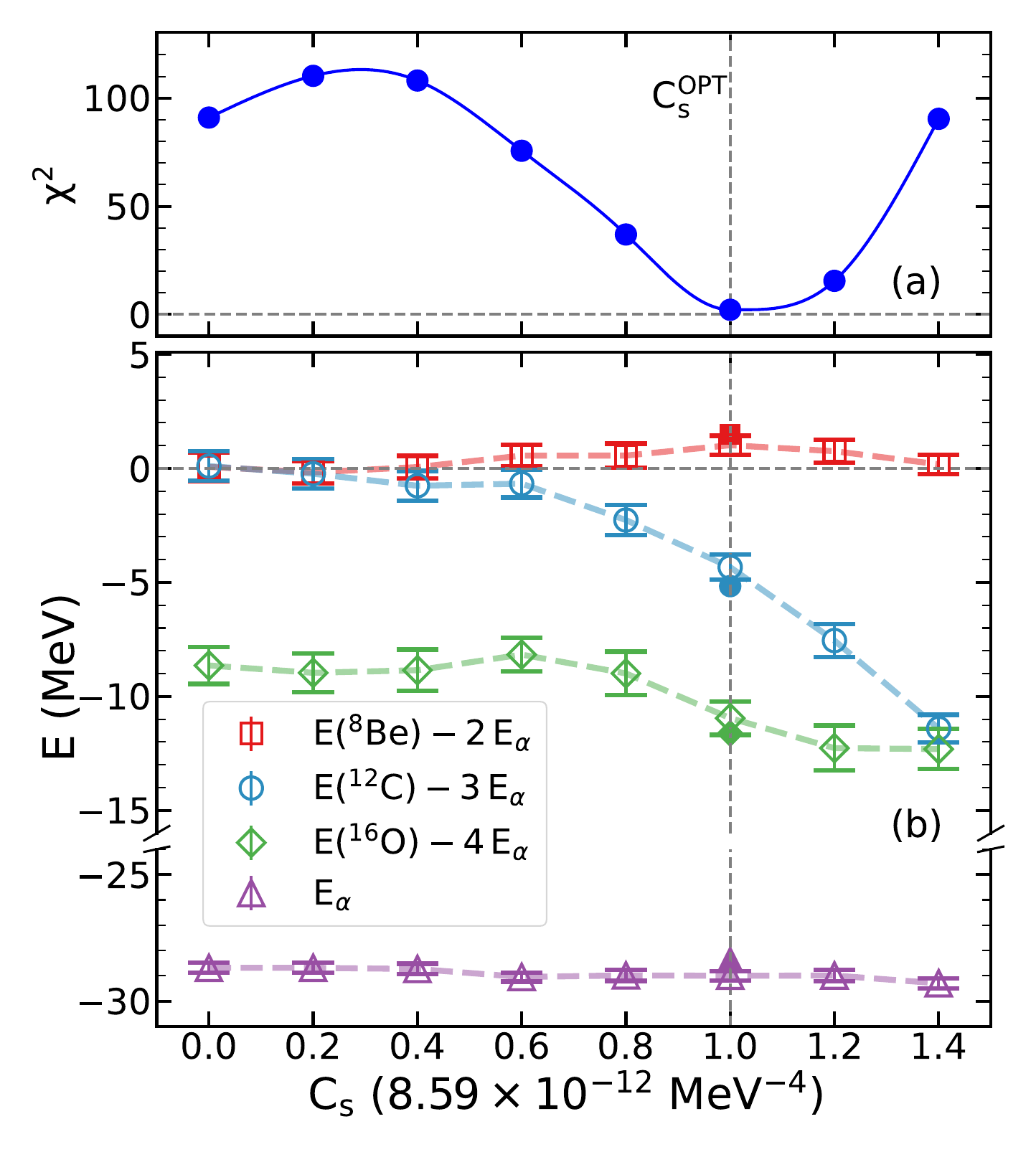}
  \caption{(a) Loss function (Eq.~(\ref{eq:lossfunction})) versus spin-orbit coupling $C_s$. 
(b) Binding energies relative to $n$-$\alpha$ thresholds: $^{8}$Be (squares), $^{12}$C (circles), $^{16}$O (diamonds). 
Full symbols: experimental values \cite{exp}; 
solid/open triangles: calculated/experimental $^4$He energies. 
Vertical line: optimal $C_s^{\rm opt}$ minimizing $\chi^2$.
   } 
  \label{fig:spinorbit_optimization}
\end{figure}

Lastly, Fig.~\ref{fig:EoS_of_SNM} shows the predicted equation of state for symmetric nuclear matter.
Comparison to the Wigner-SU(4) interaction~\cite{PLB797-134863} reveals significant improvement. 
We identify a saturation point at $(\rho_0, E/A)=(0.175$~fm$^{-3}$, $-15.4$~MeV$)$, consistent with empirical values.


\begin{figure}[htbp]
  \centering
  \includegraphics[width=0.8\linewidth]{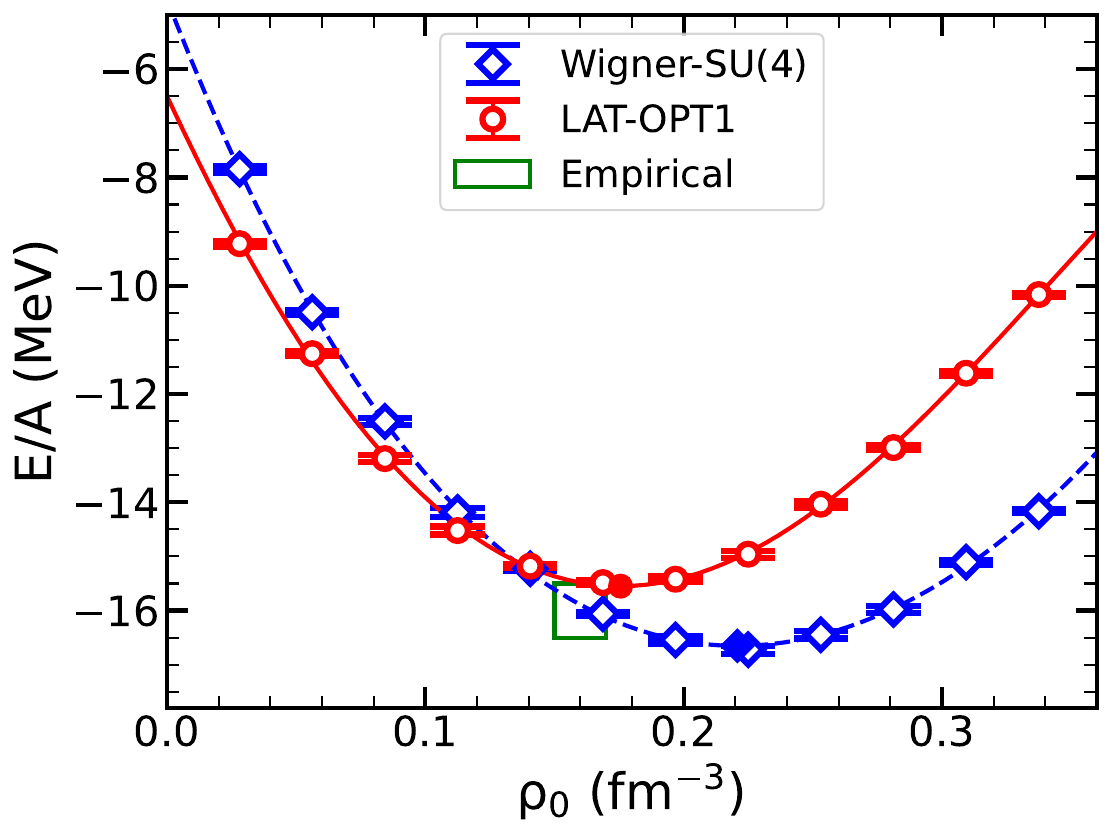}
  \caption{Equation of state for symmetric nuclear matter calculated with \texttt{LAT-OPT1} (circles) and Wigner-SU(4)~\cite{PLB797-134863} (diamonds) interactions.
  Full symbols denote corresponding saturation points.
  Green rectangle mark the empirical saturation point $(0.16\pm0.01$~fm$^{-3}$, $-16.0\pm1.0$~MeV$)$~\cite{Bethe1971}.}
  \label{fig:EoS_of_SNM}
\end{figure}

\paragraph{Summary and perspective}
This work presents the first exactly solvable quantum many-body model for atomic nuclei that quantitatively reproduces experimental data across light to heavy nuclei. 
Exactly solvable quantum many-body problems are rare benchmarks essential for evaluating advanced algorithms, including those based on artificial neural networks~\cite{RMP94-031003} or quantum computing~\cite{CPB30-020306, EPJConf296-01025}. 
Here, exact solvability implies achieving the desired precision within polynomial time. 
For modeling nuclear structure, the proposed sign-problem-free algorithm offers orders of magnitude acceleration compared to calculations using realistic interactions for heavy nuclei.
The approach extends directly to investigating excited-state spectra~\cite{PRL106-192501}, density correlations~\cite{PRL119-222505}, and finite-temperature nuclear properties~\cite{PRL125-192502} using advanced lattice QMC algorithms. 
Furthermore, it establishes a robust non-perturbative foundation for \textit{ab initio} calculations with complex chiral nuclear forces, enabling perturbative incorporation of missing physics order-by-order~\cite{PRL128-242501, Nature630-59}.

Finally, this work establishes a scalable, systematically improvable computational paradigm that bridges complex \textit{ab initio} frameworks and simple phenomenological models.
Recent \textit{ab initio} studies have uncovered inconsistencies among few-body systems, many-body nuclei, and nuclear matter predictions~\cite{PLB736-119, PRC104-064312, PRC104-054001, FewBodySyst64-77}, prompting the development of minimal interactions to resolve these discrepancies~\cite{PLB797-134863, PRL121-072701, PRL133-142501}. 
Our framework represents such a minimal interaction approach, potentially enhancing the explainability of modern \textit{ab initio} nuclear theory by distilling the essential elements of nuclear binding. 
This advances the quest for fundamental nuclear interactions and will likely inspire future research across nuclear physics, quantum many-body methods, and related disciplines.



\paragraph{Acknowledgement}
We thank members of the Nuclear Lattice Effective Field Theory Collaboration for insightful discussions, and Dean Lee for critical reading of the manuscript.
This work has been supported by NSAF No. U2330401 and National Natural Science Foundation of China with Grant No. 12275259.







\bigskip
\textbf{*} \href{bnlv@gscaep.ac.cn}{bnlv@gscaep.ac.cn}

\title{Supplemental Materials}

\author{Zhong-Wang Niu}
\affiliation{Graduate School of China Academy of Engineering Physics, Beijing 100193,
China}

\author{Bing-Nan Lu}
\affiliation{Graduate School of China Academy of Engineering Physics, Beijing 100193,
China}

\date{\today}

\maketitle

\newpage
\appendix

\onecolumngrid

\section{Supplemental Material}

This supplement presents the implementation details of the three-body force, the proof of positivity for our approach, and the treatment of Coulomb interactions. While these elements are documented elsewhere in the literature, they are included here for completeness and self-sufficiency. We further elaborate on: (i) the optimization algorithm employed, (ii) computational complexity analysis, and (iii) finite volume effects. Numerical results and original data are provided in comprehensive tables and composite figures, including nucleon-nucleon phase shifts, odd-even staggering in oxygen isotopes, the pure neutron matter equation of state, and charge density distributions.


\subsection{Auxiliary field transformation for three-body force}

For non-perturbative simulations of many-body forces, we employ a discrete auxiliary field that simultaneously simulates two-, three-, and four-body forces without sign oscillations.
This transformation has been used in Ref.~\cite{PLB797-134863*}:
\begin{align}
: \exp \left( - \frac { 1 } { 2 } C_2 a _ { t } \rho ^ { 2 } - \frac { 1 } { 6 } C _ { 3 } a _ { t } \rho ^ { 3 } - \frac { 1 } { 2 4 } C _ { 4 } a _ { t } \rho ^ { 4 } \right) : = \sum _ { k = 1 } ^ { N } \omega _ { k } : \exp \left( \sqrt { - C_2 a _ { t } } \phi _ { k } \rho \right): ,
\label{eq:discrete_aux_f}
\end{align}
where $\rho$ is a general one-body density operator, and $C_{2}$, $C_{3}$, $C_{4}$ are the two-, three-, and four-body coefficients, respectively.
We solve for the real numbers $\omega _ { k }$ and $\phi _ { k }$.
In this work, we consider only attractive two-body interactions with $C_2 < 0$. To avoid the sign problem, we further require $\omega _ { k } > 0$ for all $k$.

To determine $\phi _ { k }$ and $\omega _ { k }$, we expand Eq.~(\ref{eq:discrete_aux_f}) up to $\mathcal { O } \left( \rho^ { 4 } \right)$ and compare both sides order by order.
For nuclear forces where three- and four-body interactions are typically much weaker than the two-body interaction, we use the ansatz with $N = 3$:
\begin{align}\label{d}
\omega _ { 1 } = \frac { 1 } { \phi _ { 1 } \left( \phi _ { 1 } - \phi _ { 3 } \right) } , \quad \omega _ { 2 } = 1 + \frac { 1 } { \phi _ { 1 } \phi _ { 3 } } , \quad \omega _ { 3 } = \frac { 1 } { \phi _ { 3 } \left( \phi _ { 3 } - \phi _ { 1 } \right) },
\end{align}
where $\phi _ { 2 } = 0$, and $\phi _ { 1 }$ and $\phi _ { 3 }$ are roots of the quadratic equation:
\begin{align}
\phi ^ { 2 } + \frac { C _ { 3 } } { \sqrt { - C_2 ^ { 3 } a _ { t } } } \phi - \frac { C _ { 3 } ^ { 2 } } { C_2 ^ { 3 } a _ { t } } + \frac { C _ { 4 } } { C_2 ^ { 2 } a _ { t } } - 3 = 0 .
\end{align}
Using Vieta's formulas relating polynomial coefficients to sums and products of roots, we verify that Eq.~(\ref{d}) satisfies Eq.~(\ref{eq:discrete_aux_f}) up to $\mathcal { O } \left( \rho ^ { 4 } \right)$. For a pure two-body interaction ($C_{3,4} = 0$), the solution simplifies to $\phi _ { 1 } = - \phi _ { 3 } = \sqrt { 3 }$, $\phi _ { 2 } = 0$, $\omega _ { 1 } = \omega _ { 3 } = 1 / 6$, $\omega _ { 2 } = 2 / 3$. The formalism in Eq.~(\ref{d}) efficiently simulates many-body forces.

In practical calculations, we take $C_4 = 0$ and substitute
\begin{equation}
  \rho = \overline{\rho} + \frac{C_s}{2C_2}\overline{\rho}_s
\end{equation}
into Eq.~(\ref{eq:discrete_aux_f}).
The coefficients $\omega_k$ and $\phi_k$ are determined by $C_2$, $C_3$ and $a_t$ at the outset.
Rather than updating continuous auxiliary fields, we randomly select indices $1\leq k \leq 3$ per lattice site and apply the Metropolis algorithm to accept or reject the new index field.
After sufficient warm-up, thermal equilibrium is achieved, and observables are measured by inserting operators at the central time step.
Energy is obtained by directly measuring the Hamiltonian expectation value.
Note that Eq.~(\ref{eq:discrete_aux_f}) generates additional interactions of order $\mathcal{O}(C_s^2 / C_2)$ or $\mathcal{O}(C_3 C_s / C_2)$,
which are highly suppressed and negligible.

\subsection{Proof for the positivity of the amplitude}

The sufficient condition for a positive definite correlation matrix was given in Ref.~\cite{PRB71-155115*}.
Here we present a step-by-step derivation specific to NLEFT calculations.
The expectation value of any operator $O$ is expressed as
\begin{eqnarray}
\langle O \rangle &=& \lim_{\tau \rightarrow \infty} \frac{\langle \Phi_T| e^{-\tau H/2} O e^{-\tau H/2} | \Phi_T\rangle} 
   {\langle \Phi_T| e^{-\tau H} |\Phi_T\rangle} = 
    \lim_{L_t \rightarrow \infty} \frac{\langle \Phi_T| M^{L_t / 2}  O M^{L_t / 2} | \Phi_T\rangle} 
   {\langle \Phi_T| M^{L_t} |\Phi_T\rangle} 
    \nonumber \\ 
   &=& 
   \lim_{L_t \rightarrow \infty} \frac{
   \int\mathcal{D}c \exp(-\sum_{i=1}^{L_t} c_i^2 / 2)\langle \Phi_T | M(c_{L_t}) \cdots M(c_{L_t/2+1}) O M(c_{L_t/2}) \cdots M(c_1) |\Phi_T \rangle
   }{
   \int\mathcal{D}c \exp(-\sum_{i=1}^{L_t} c_i^2 / 2) \langle \Phi_T | M(c_{L_t}) \cdots M(c_1) |\Phi_T \rangle
   },
   \label{eq:path_integral}
\end{eqnarray}
where we have neglected the three-body force and applied the traditional Hubbard-Stratonovich auxiliary field transformation and $\tau = L_t \ a_t$.
The proof remains essentially unchanged if we include the three-body force using the discrete transformation.
Here $M = {:}\exp(-a_t H){:}$ is the transfer matrix, $c_i(\bm{n})$ is the auxiliary field at the $i$-th time slice,
and
\begin{equation}
 M(c)  = \exp\left[ \sum_{\bm{n}, \bm{n}^\prime} a_t \frac{\Psi^\dagger(\bm{n}) \nabla^2_{\bm{n} \bm{n}^\prime} \Psi(\bm{n}^\prime)}{2M} + \sum_{\bm{n}} \sqrt{-a_{t}C_{2}}c(\bm{n})\left(\overline{\rho}(\bm{n})+\frac{C_{s}}{2C_{2}}\overline{\rho}_{s}(\bm{n})\right)\right]
 \label{eq:transformed_transfer_matrix}
\end{equation}
is the transformed transfer matrix. 
The summations over $\bm{n}$, $\bm{n}^\prime$ run over all $L^3$ lattice sites,  
and $\nabla^2_{\bm{n} \bm{n}^\prime}$ represents the discretized Laplacian operator implemented using FFT.

The transfer matrix in Eq.~(\ref{eq:transformed_transfer_matrix}) is bilinear in the fermion creation and annihilation operators.
When acting on a single-particle wave function, it corresponds to a one-body imaginary-time evolution operator of dimension $L^3 \times L^3$.
For products of single-particle wave functions, it induces no many-body correlations and evolves each particle independently:
\begin{equation}
 M(c) \left( |\psi_{1}\rangle \wedge  |\psi_{2}\rangle \wedge \cdots \wedge  |\psi_{A}\rangle \right) =  M(c) |\psi_{1}\rangle \wedge  M(c) |\psi_{2}\rangle \wedge \cdots \wedge  M(c) |\psi_{A}\rangle,
 \label{eq:Single_particle_evolve}
\end{equation}
where $\psi_{1, \ldots, A}$ are a set of single-particle wave functions, and $\wedge$ denotes the antisymmetrized product, \textit{i.e.}, the Slater determinant.

Without loss of generality, we assume that $|\Phi_T\rangle$ is a Slater determinant composed of single-particle wave functions $\phi_1, \phi_2, \ldots, \phi_A$.
Repeated application of Eq.~(\ref{eq:Single_particle_evolve}) allows us to rewrite the denominator of Eq.~(\ref{eq:path_integral}) as
\begin{equation}
    \mathcal{Z} 
    = \int\mathcal{D}c \exp(-\sum_{i=1}^{L_t} c_i^2 / 2)
    \left[ \bigwedge_{i=1}^A \langle \phi_i| \right] 
    \left[ \bigwedge_{i=1}^A \overline{M}(c) |\phi_i\rangle \right] = \int\mathcal{D}c \exp(-\sum_{i=1}^{L_t} c_i^2 / 2) \det(Z(c)), 
\end{equation}
where $\overline{M}(c) = M(c_{L_t}) \cdots M(c_1)$ is the product of single-particle evolution operators, and $Z$ is an $A\times A$ fermionic correlation matrix:
\begin{align}
Z(c) &= 
\bigl[\,\langle \phi_i \mid \overline{M}(c) \mid \phi_j \rangle\bigr]_{i,j=1}^A
=
\begin{pmatrix}
\langle\phi_1\mid \overline{M}(c) \mid\phi_1\rangle & \langle\phi_1\mid \overline{M}(c)\mid\phi_2\rangle & \cdots & \langle\phi_1\mid \overline{M}(c)\mid\phi_A\rangle\\
\langle\phi_2\mid \overline{M}(c)\mid\phi_1\rangle & \langle\phi_2\mid \overline{M}(c)\mid\phi_2\rangle & \cdots & \langle\phi_2\mid \overline{M}(c)\mid\phi_A\rangle\\
\vdots & \vdots & \ddots & \vdots\\
\langle\phi_A\mid \overline{M}(c)\mid\phi_1\rangle & \langle\phi_A\mid \overline{M}(c)\mid\phi_2\rangle & \cdots & \langle\phi_A\mid \overline{M}(c)\mid\phi_A\rangle
\end{pmatrix}
.
\end{align}

In NLEFT calculations, we generate an ensemble of auxiliary-field configurations $\{c\}$ with the probability distribution
\begin{equation}
    P(c) \propto \exp(-\sum_{i=1}^{L_t} c_i^2 / 2) \left| \det(Z(c)) \right|
    \label{eq:probability_distribution}
\end{equation}
and measure observables as arithmetic averages.
The sign problem occurs when the determinant in Eq.~(\ref{eq:probability_distribution}) is not positive definite.
In this case, the averaged sign
\begin{equation}
  \langle e^{i\theta} \rangle = \int \mathcal{D}c P(c) \frac{\det(Z(c))}{\left| \det(Z(c)) \right|} 
 \end{equation}
quantifies the severity of the sign problem.
For $\langle e^{i\theta} \rangle = 1$, there is no sign problem, while $\langle e^{i\theta} \rangle \approx 0$ indicates a severe sign problem.

The single nucleon wave functions include spin and isospin indices and can be written as complex vectors of dimension $4\times L^3$. 
Correspondingly, the transfer matrix in Eq.~(\ref{eq:transformed_transfer_matrix}) becomes a $4L^3 \times 4L^3$ matrix for any given $c$-field configuration.
Since the interaction used in this work has no isospin-mixing term, the determinant factorizes:
\begin{equation}
   \det(Z(c)) = \det(Z_P(c)) \det(Z_N(c)),
\end{equation}
where $Z_P$ and $Z_N$ are the correlation matrices for protons and neutrons, respectively.
In what follows, we omit the isospin degrees of freedom; all conclusions hold for $Z(c)$, $Z_P(c)$, and $Z_N(c)$.

For even-even nuclei, we prepare the single nucleon wave functions as paired by the time-reversal operation:
\begin{equation}
   |\phi_{A/2 + k}\rangle = \mathcal{T} |\phi_k\rangle,\qquad k = 1, 2, \ldots, A/2,
   \label{eq:tsymmetry}
\end{equation}
where $\mathcal{T} = i\sigma_y \mathcal{K}$ is the time-reversal operator and $\mathcal{K}$ denotes complex conjugation.
The matrix $\overline{M}(c)$ commutes with $\mathcal{T}$, as the spin Pauli matrices and the imaginary factor $i$ in $\rho_s$ both contribute sign changes that cancel under time reversal.
Note that we focus on the $\mathcal{T}$-symmetry of the decomposed transfer matrix in Eq.~(\ref{eq:transformed_transfer_matrix}), not the original interactions, which correspond to squares of the decomposed interactions and are always time-reversal even.
For example, while the Coulomb force is time-reversal invariant, its auxiliary field transformation explicitly involves an imaginary phase and cannot be simulated without a sign problem.

Using the properties of $\mathcal{T}$ and Eq.~(\ref{eq:tsymmetry}), we obtain the relations:
\begin{eqnarray}
   \langle \phi_{A/2 + i} | \overline{M}(c) | \phi_{A/2 + j} \rangle &=& 
   \langle \mathcal{T} \phi_{i} | \overline{M}(c) | \mathcal{T} \phi_{j} \rangle =
   \langle \phi_{i} | \overline{M}(c) | \mathcal{T}^\dagger \mathcal{T} \phi_{j} \rangle^* =
   \langle \phi_{i} | \overline{M}(c) | \phi_{j} \rangle^*, \\\nonumber
   \langle \phi_{A/2 + i} | \overline{M}(c) | \phi_{j} \rangle &=&  
   \langle \mathcal{T} \phi_{i} | \overline{M}(c) |  \phi_{j} \rangle =
   \langle \phi_{i} | \overline{M}(c) | \mathcal{T}^\dagger \phi_{j} \rangle^* =
   -\langle \phi_{i} | \overline{M}(c) | \phi_{A/2 + j} \rangle^*.
\end{eqnarray}
The correlation matrix then has the structure:
\begin{equation}
  Z \;=\;
  \begin{pmatrix}
    U & -V^*\\[4pt]
    V &  U^*
  \end{pmatrix}
,
\label{eq:Zcorr}
\end{equation}
where $U$ and $V$ are $A/2 \times A/2$ complex matrices.

We define a spin-flipping matrix:
\begin{equation}
\Sigma
\;=\;
i\sigma_{y} \otimes I_{A/2}
\;=\;
\begin{pmatrix}
0 & I_{A/2}\\[6pt]
-I_{A/2} & 0
\end{pmatrix}
,
\end{equation}
where $I_{A/2}$ is the $A/2 \times A/2$ identity matrix.
Direct verification shows that
\begin{equation}\label{eq:core}
  Z\,\Sigma
  \;=\;
  \Sigma\,Z^*.
\end{equation}
For any eigenvalue $\lambda\in\mathbb{C}$ with corresponding eigenvector $v$ satisfying $Z\,v = \lambda\,v$, we have
\begin{equation}
  Z\,(\Sigma\,v^*)
  = \Sigma\,Z^*\,v^*
  = \Sigma\,(\lambda^*\,v^*)
  = \lambda^*\,(\Sigma\,v^*),
\end{equation}
implying that $\lambda^*$ is also an eigenvalue with eigenvector $w=\Sigma v^*$.
Thus, complex eigenvalues of $Z$ always appear in conjugate pairs.
If $\lambda$ is real, we find
\begin{equation}
\langle v \mid w \rangle
= v^\dagger\,\Sigma\,v^*
= (v^T\,\Sigma\,v)^*
= 0,
\end{equation}
indicating that $v$ and $w$ are orthogonal eigenvectors corresponding to the same real eigenvalue.
In this case, the real eigenvalue appears twice in the diagonalized form of $Z$.
Consequently, the determinant $\det(Z)$, being the product of all eigenvalues, is always nonnegative.
We note that for odd-even or odd-odd nuclei the correlation matrix can not be written in the form of Eq.~(\ref{eq:Zcorr}) and the resulting determinant could be negative or complex.

\subsection{Dependence on the lattice spacing}

In this work, we adopt a fixed lattice spacing of $a = 1.32$~fm, corresponding to a momentum cutoff $\Lambda = \pi / a \approx 471$~MeV. 
This value is prevalent in recent NLEFT investigations~\cite{PLB797-134863*, PRL128-242501*, Nature630-59*, PRL132-232502*, PRL134-162503*}.
We note that most continuum chiral EFT calculations employ a smooth cutoff $\Lambda$ between $450$ and $550$~MeV~\cite{NPA747-362*, PRC68-041001*}, 
yielding similar cutoff dependencies. 
This cutoff scale also exposes a hidden spin-isospin exchange symmetry originating from QCD~\cite{PRL127-062501*}.

So far, most NLEFT calculations have been performed at a single lattice spacing.
Systematic multi-cutoff studies have been limited to nucleon-nucleon scattering~\cite{EPJA53-83*, PRC98-044002*, PRC112-014009*} and few-body systems ($A \leq 4$)~\cite{PLB747-511*, arXiv:2509.02953*}. 
In this work, we expect that our optimization algorithm also works for other lattice spacings in a reasonable interval and the optimized nuclear force exhibits similar performance, with residual cutoff dependencies that can be estimated within the EFT framework.
In typical nuclear EFTs, cutoff-dependencies scale as $O((Q / \Lambda)^{\nu + 1})$, where $Q$ is the characteristic momentum and $\nu$ is the order of the EFT interaction~\cite{RMP92-025004*}. 
Our interaction incorporates an improved leading-order ($O(Q^0)$) term with partial absorption of next-to-leading-order ($O(Q^2)$) effects, including finite-range smearing and spin-orbit terms, and next-to-next-to-leading-order ($O(Q^3)$) three-body forces. 
We thus make a rough estimate $\nu \approx 2$. 
In nuclei, $Q \sim \sqrt{ME} \approx 100$~MeV, with $M$ the nucleon mass and $E$ the typical single-nucleon separation energy.
For sufficiently large $\Lambda$ (\textit{e.g.}, $\Lambda \geq 300$~MeV, corresponding to $a \leq 2$~fm), cutoff dependencies are suppressed to the order of $O((100/300)^3) \approx O(0.037)$. 
These residual lattice effects can be numerically verified and further reduced by including higher-order interactions, which we leave for future work.



\subsection{Galilean invariance}

The interactions introduced in this work explicitly break Galilean invariance, which originates from the essential nonlocality required to reproduce experimental data. 
In the center-of-mass frame, our local and non-local smearing effectively mimic the $q^2$ and $k^2$ terms at next-to-leading order in the chiral expansion, while the spin-orbit term reduces to the standard contact coupling $\frac{i}{2}(\bm{q}\times\bm{k})\cdot(\bm{\sigma}_1 + \bm{\sigma}_2)$. 
To our knowledge, no existing nuclear force simultaneously satisfies Galilean invariance, avoids the QMC sign problem, and reproduces experimental binding energies.

\subsection{Coulomb force}


A static Coulomb force $V_{\mathrm{cou}}$ is included. 
The form is the same as that used in Ref.~\cite{EPJA61-85*} with a different cutoff.
We compute the first-order perturbative energy of $V_{\mathrm{cou}}$.
In momentum space, we have
\begin{equation}
V_{\mathrm{cou}} = \frac{\alpha}{q^2}
\exp\!\Bigl(-\frac{q^2}{2\Lambda_{\mathrm{cou}}^2}\Bigr),
\end{equation}
where $\alpha = 1/137$ is the fine structure constant,
$\Lambda_{\mathrm{cou}} = 180\,\mathrm{MeV}$ is a momentum cutoff that removes the singularity at $r=0$, and $\bm{q} = \bm{p}^\prime - \bm{p}$ is the momentum transfer.
In coordinate space, the potential is expressed using the error function:
\begin{equation}
 V_{\mathrm{cou}}(r)
= \alpha\,\frac{\mathrm{erf}(\Lambda_{\mathrm{cou}}\,r/2)}{r}.
\end{equation}
The asymptotic behavior satisfies
$$  V_{\mathrm{cou}}(0) = \frac{\alpha\,\Lambda_{\mathrm{cou}}}{\sqrt{\pi}} \quad\text{and}\quad  V_{\mathrm{cou}}(r) \to \frac{\alpha}{r} \quad\text{as}\;\Lambda_{\mathrm{cou}}\to\infty. $$
The potential $V_{\mathrm{cou}}$ is smooth for all $r$ and can be implemented on the lattice using FFT.

\subsection{Gradient Descent Algorithm for parameter optimization}

Gradient descent is an optimization algorithm that leverages gradient information to iteratively update parameters. 
The method used in this work is summarized as follows:

\begin{enumerate}
    \item \textbf{Initialization}: Select a reasonable initial parameter set $\theta_0$.
    
    \item \textbf{Parameter rescaling}: Rescale the parameters to approximately normalize the partial derivatives. 
    During early iterations:
    $$
    \theta_i \rightarrow \theta_i\left| \frac{\partial \chi^2}{\partial \theta_i} \right|_{t=0},
    $$
    yielding partial derivatives of magnitude $\sim 1$. 
    Near convergence when iterations slow, apply fixed scaling:
    $$
    \theta_i \rightarrow \theta_i \times k_i,
    $$
    with scaling factors: $
    k_{s_L} = 1.853 \times 10^5,  k_{C_2} = 3.946 \times 10^{11}, k_{C_3} = 1.004 \times 10^{18},$ and $ k_{C_{s}} = 1.260 \times 10^{15}.$
    This technique significantly accelerates convergence.
    
\item \textbf{Gradient computation}: For each iteration, the partial derivatives of the loss function $\partial \chi^2 / \partial \theta_i$ must be computed. Fig.~\ref{sL_test} compares two methods for computing these partial derivatives.

One straightforward approach is to directly calculate the energies using two sets of parameters differing by small variations $\theta_i$ and $\theta_i + \delta\theta_i$, and then approximate the derivative via finite difference. However, these two calculations are independent and their combined statistical errors are typically similar in magnitude to the difference being measured. Consequently, the resulting derivative estimate often suffers from large statistical uncertainty.

Alternatively, we can apply the Feynman-Hellmann theorem (FHT). This allows us to directly measure the derivatives as observables within a \textit{single} simulation run. Crucially, this approach significantly reduces the statistical errors. The key reason for this improvement is that the FHT method inherently captures the correlation between the energies at the slightly different parameter values within the same ensemble. Furthermore, this method enables the simultaneous measurement of derivatives with respect to \textit{any number} of parameters.

This precise and efficient estimation of the derivatives forms the essential foundation of our optimization algorithm. The algorithm's capability to handle derivatives effectively makes it suitable for application to a wider range of target nuclei and larger sets of adjustable parameters.

    \begin{figure}[htbp]
    \centering
    \includegraphics[width=0.5\columnwidth]{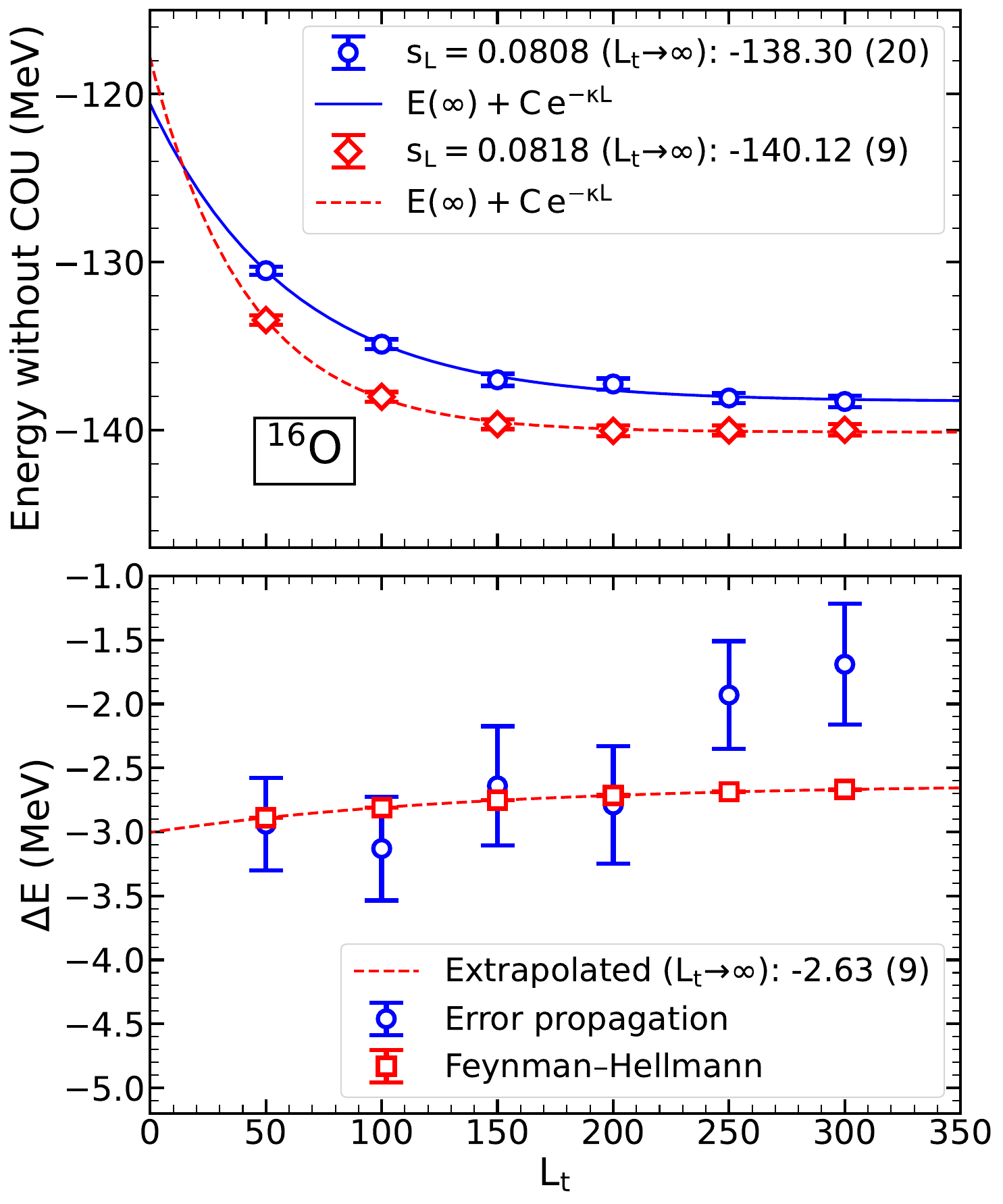}
    \caption{\textbf{Comparison of methods for computing derivatives.}
    (Upper panel) Ground state energies of $^{16}$O calculated with $s_{\rm L}$= 0.0808 (blue circles) and 0.0818 (red diamonds), respectively.
    Other parameters take the \texttt{LAT-OPT1} values.
    Errorbars denote independent statistical errors in each calculation.
    (Lower panel) Blue circles represent the difference between the two groups of points in the upper panel, with combined errors estimated with the error propagation formula $e=\sqrt{e_1^2 + e_2^2}$. Red squares denote the results from the Feynman-Hellmann theorem, with the statistical errors smaller than the symbols.}    
    \label{sL_test}
\end{figure}

    \item \textbf{Parameter update}: 
    $$
    \theta_{t+1} = \theta_t - \alpha \frac{\partial \chi^2(\theta_t)}{\partial \theta},
    $$
    with adaptive learning rate:
    $$
    \alpha =
    \begin{cases} 
    \alpha \times 2 & \text{if } \chi^2 \text{ decreases} \\ 
    \alpha \times 0.2 & \text{otherwise}
    \end{cases}
.
    $$     This strategy increases step size in smooth regions and decreases it to prevent instability or overshooting.  
    
        \begin{figure}[htbp]
    \centering
    \includegraphics[width=0.4\textwidth]{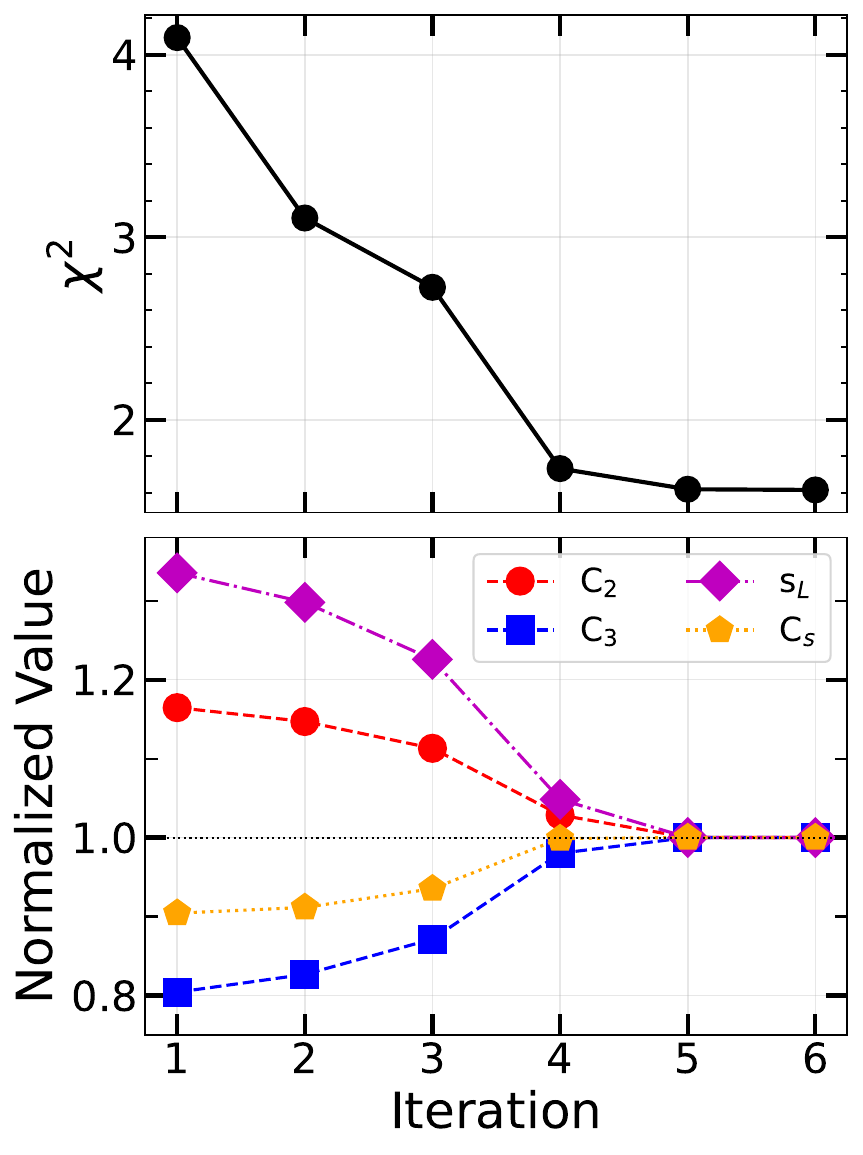}
    \caption{\textbf{optimization of loss function.} (Upper) Loss function $\chi^2$ versus iteration steps. (Lower) Dimensionless parameters normalized to converged values. Deviations of scaled $s_{\rm L}$ and $C_2$ are exaggerated for visibility. We divide each parameter by its value at the last iteration in order to present the convergence of all parameters within the same figure in an illustrative manner.}
    \label{parameters}
\end{figure}
    
    \item \textbf{Convergence check}: Terminate when:     $$
    |\chi^2(\theta_{t+1}) - \chi^2(\theta_t)| < \epsilon.
    $$
\end{enumerate}

Fig.~\ref{parameters} (Upper) demonstrates the monotonic decrease of $\chi^2$ during optimization, confirming the method's efficacy. 
Fig.~\ref{parameters} (Lower) schematically illustrates parameter convergence to final values.

\subsection{Computational complexity for heavy nuclei}

The sign-problem-free QMC method is particularly suited for simulating heavy nuclei, as its computational time scales polynomially with nucleon number, whereas the sign problem inevitably introduces exponential scaling. 
We numerically demonstrate this advantage below.

\begin{figure}[hbtp]
  \centering
  \includegraphics[width=0.75\textwidth]{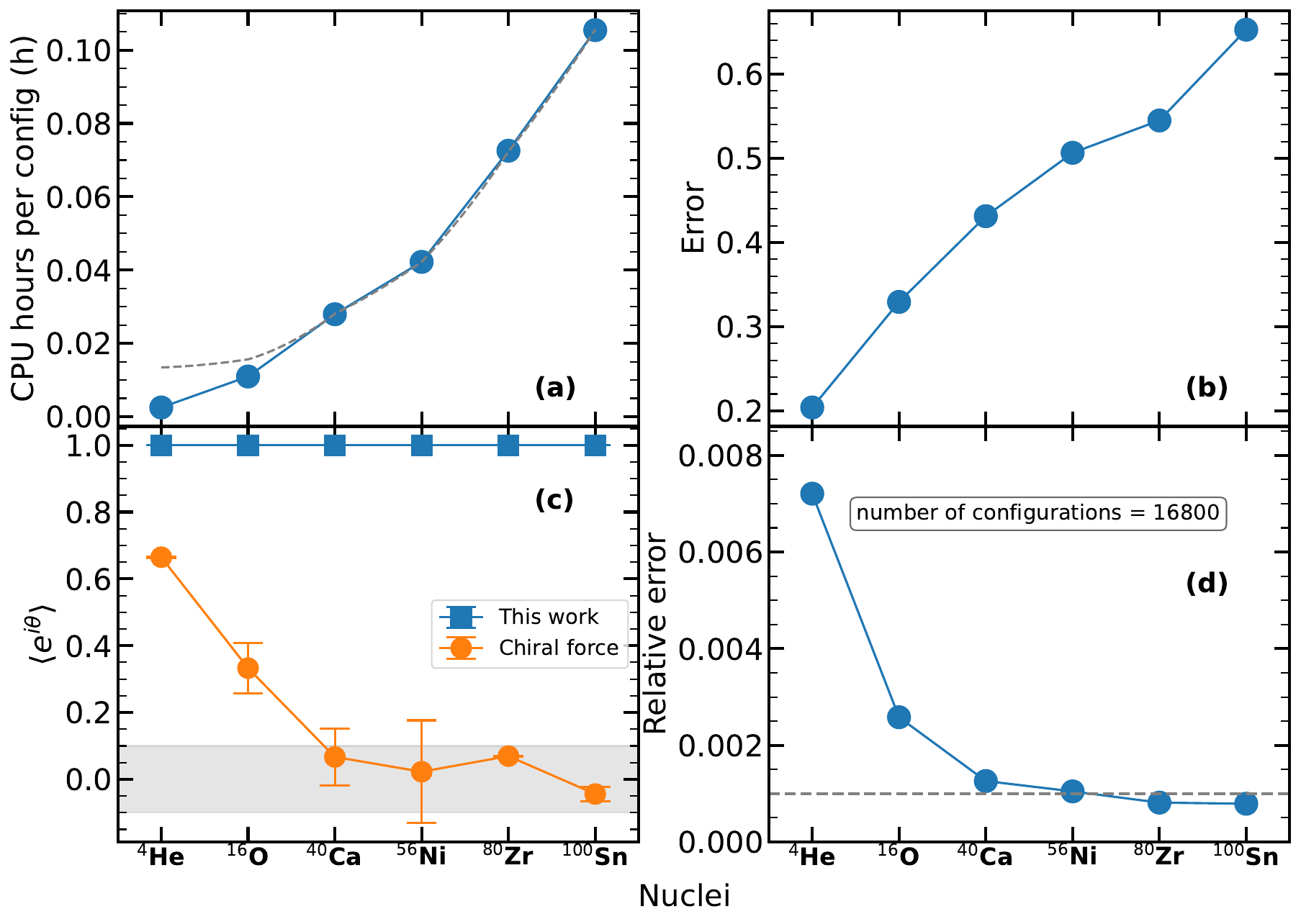}
  \caption{
    \textbf{Computational performance.} 
    (a) Average CPU hours per configuration; 
    (b) Absolute binding energy statistical uncertainties (16,800 configurations); 
    (c) Average sign $\langle e^{i\theta}\rangle$ for \texttt{LAT-OPT1} (orange squares) versus N$^2$LO chiral force (blue circles, sign-problematic);
    (d) Relative binding energy statistical uncertainties.}
  \label{fig:composite}
\end{figure}

Fig.~\ref{fig:composite}(a) depicts the average CPU hours per auxiliary-field configuration, measured on a consistent supercomputing platform. 
The dashed parabolic curve serves as a visual guide. 
Using fixed parameters ($L=11$, $L_t=200$), we confirm polynomial time complexity and observe approximately linear scaling for $A\geq 40$.

Fig.~\ref{fig:composite}(b) displays statistical uncertainties in binding energies across 16,800 configurations. 
Uncertainties increase gradually with mass number but remain below 1~MeV for nuclei up to $A\sim 100$. 
Relative QMC statistical errors (Fig.~\ref{fig:composite}(d)) exhibit rapid convergence in heavy nuclei, stabilizing near one-thousandth for $A\geq 40$.

Combining these observations, we achieve this accuracy level with linearly increasing computational resources. 
This scalability is highly encouraging for QMC studies of heavy nuclei and warrants further investigation. 
Although larger simulation boxes are required for $A > 100$ to suppress finite-volume effects, high-precision calculations remain feasible.

Finally, Fig.~\ref{fig:composite}(c) compares the average sign $\langle e^{i\theta}\rangle$ between our sign-problem-free interaction and a non-perturbative N$^2$LO chiral force. 
The chiral interaction, featuring complex spin/isospin dependencies, exhibits a sign problem where $\langle e^{i\theta}\rangle$ decreases rapidly with mass number. 
Conversely, our approach maintains $\langle e^{i\theta}\rangle = 1$ for all systems.



\subsection{Finite Volume Effects}

\begin{figure}[htbp]
  \centering
  \includegraphics[width=0.8\textwidth]{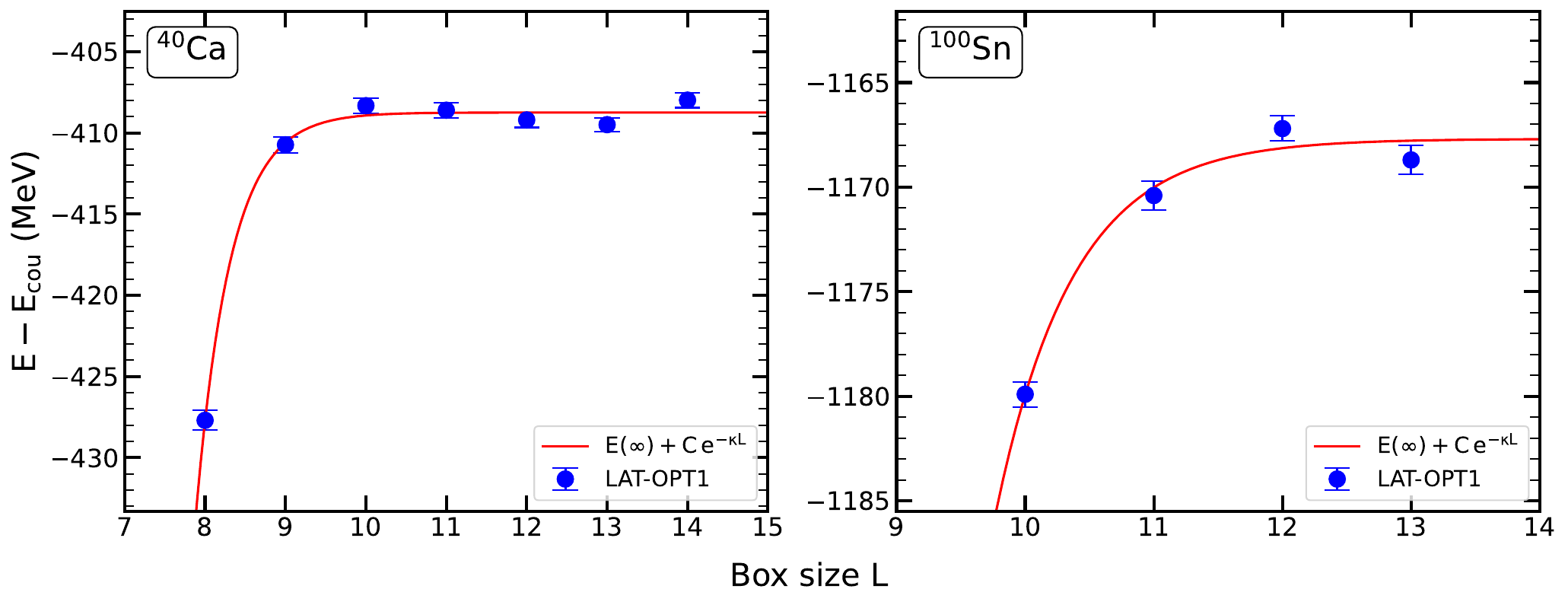}
  \caption{\textbf{Finite-volume convergence of binding energies.} 
    Binding energies (Coulomb excluded) for $^{40}\mathrm{Ca}$ ($L_t=400$, left) and $^{100}\mathrm{Sn}$ ($L_t=200$, right) versus lattice size $L$. 
    Blue symbols: Monte Carlo results with statistical uncertainties (error bars); 
    red curves: exponential fits $E(\infty)+Ce^{-kL}$. 
    Plateau formation at $L\gtrsim10$ confirms negligible finite-volume effects.}
  \label{fig:FVE}
\end{figure}

All calculations in this work employ a lattice box of size $L = 11$, which yields negligible finite-volume effects for binding-energy computations of most nuclei. This is illustrated in Fig.~\ref{fig:FVE}, showing convergence of binding energies (excluding Coulomb interactions) for $^{40}\mathrm{Ca}$ ($L_t = 400$) and $^{100}\mathrm{Sn}$ ($L_t = 200$) as box size increases. The distinct plateau beyond $L = 10$ confirms effective convergence of finite-volume effects.

It is interesting to discuss the possible application of our method to even heavier nuclei, such as $^{208}$Pb or actinides.
Based on the approximately linear scaling observed in Fig.~\ref{fig:composite}, we estimate that computations for nuclei with $A \approx 200$ will require computational time roughly twice that for $^{100}$Sn. This remains computationally feasible with current supercomputing resources.

However, heavier nuclei exhibit larger radii, necessitating a larger simulation box size to mitigate finite-volume effects. The required box size should scale proportionally to the nuclear radius $R$, which itself scales as $A^{1/3}$. Given that our algorithm scales as $O(L^3)$ with box size, this volume increase introduces an additional factor of $A$ into the overall scaling. Consequently, we estimate the scaling for precision calculations of heavy nuclei to be $O(A^2)$. This implies that simulating $^{208}$Pb would require approximately four times the computational resources as $^{100}$Sn.

For such simulations, addressing challenges like efficient memory management and re-optimization of the effective interaction will be essential prerequisites, along with GPU acceleration to facilitate the calculations.


\subsection{Nucleon-nucleon scattering phase shifts}

Nuclear \textit{ab initio} calculations aim to solve nuclear structure from bare nuclear forces constrained solely by nucleon-nucleon phase shifts and few-body data. 
However, reconciling theoretical predictions with experimental data remains challenging. 
Consequently, modern \textit{ab initio} approaches often employ interactions constrained by both NN phase shifts and nuclear properties.

We therefore investigate the inverse problem: whether nuclear forces determined solely by finite-nuclei observables can reproduce NN phase shifts. 
This approach is theoretically feasible since nuclear masses and charge radii can be measured with higher precision than scattering cross sections. 
Conversely, these observables should encode complete information about nuclear forces. 
Solving such inverse many-body problems requires high-precision nuclear solvers. 
While numerous mean-field and density functional interactions are calibrated to finite nuclei, these methods are based on variational principles and not exact. 
The sign-problem-free QMC method provides the essential foundation for this problem, offering both an unbiased high-precision solver and an efficient optimizer for nuclear masses.

\begin{figure}[htbp]
  \centering
  \includegraphics[width=0.5\textwidth]{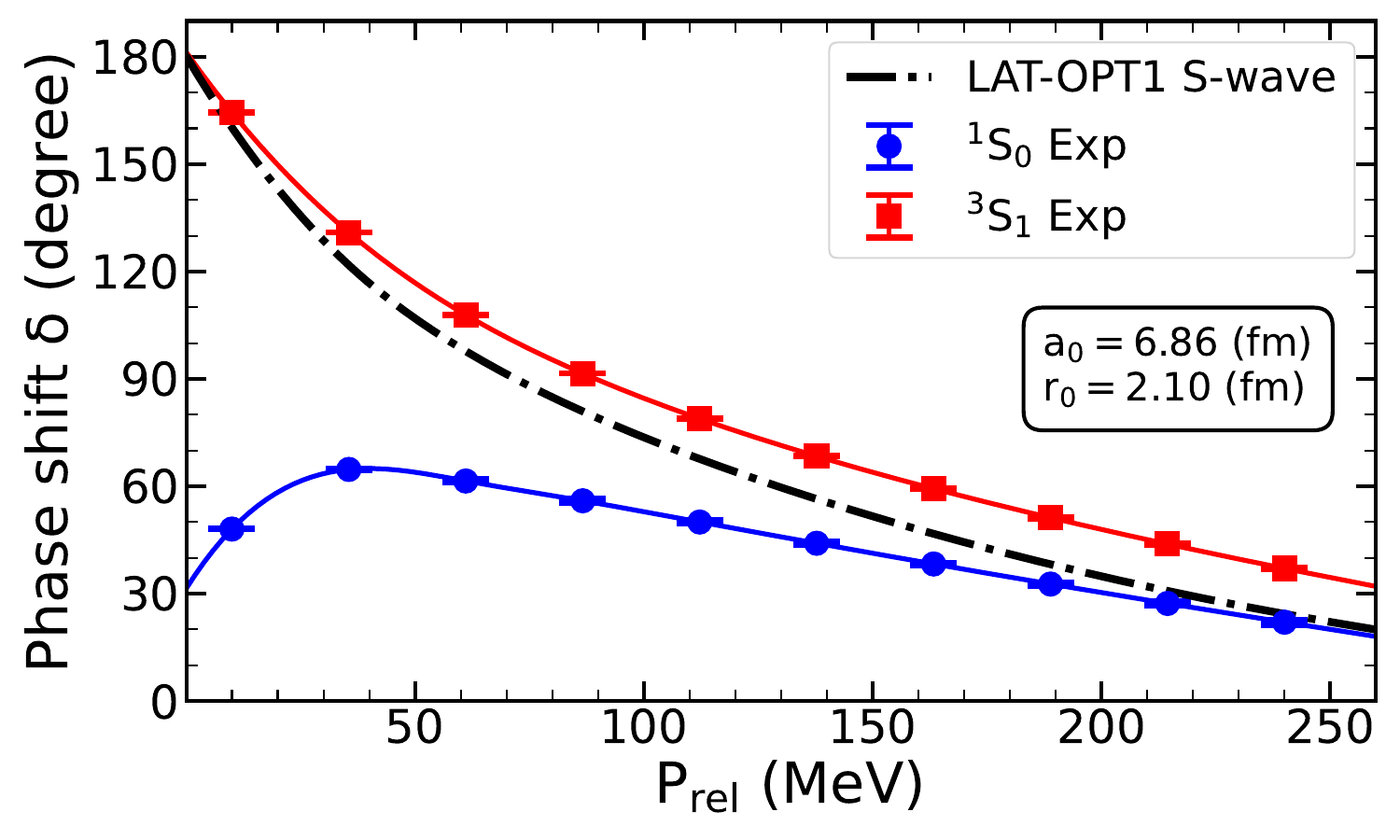}
  \caption{\textbf{S-wave nucleon-nucleon phase shifts.} 
    Empirical phase shifts $\delta$ for ${}^1S_0$ (blue circles) and ${}^3S_1$ (red squares) channels versus relative momentum $P_{\rm rel}$ \cite{PhysRevC.48.792*}. 
    Black dash-dot curve: \texttt{LAT-OPT1} predictions. 
    Inset: Extracted scattering length $a_0$ and effective range $r_0$. 
    \texttt{LAT-OPT1} accurately captures the empirical trend up to $P_{\rm rel}\approx250\,\mathrm{MeV}$.}
  \label{fig:PhaseShifts}
\end{figure}

Fig.~\ref{fig:PhaseShifts} shows that our \texttt{LAT-OPT1} predictions for NN S-wave phase shifts fall between empirical ${}^1S_0$ and ${}^3S_1$ values across the momentum range up to $250\,\mathrm{MeV}$, demonstrating reasonable reproduction of S-wave interaction strength. 
Fitting the phase shifts to the effective-range expansion,
$$ k\cot\delta = -\frac{1}{a_0} + \frac{1}{2}\,r_0\,k^2 + \cdots, $$
yields a scattering length $a_0 = 6.86\,\mathrm{fm}$ and effective range $r_0 = 2.10\,\mathrm{fm}$, both within the range of $^1S_0$ and $^3S_1$ experimental values. 
This represents a leading-order phase shift prediction from nuclear structure data.

While this simple interaction cannot reproduce S-wave splitting, S-D mixing, or P-wave phase shifts, including additional spin/isospin-dependent operators and pion-exchange potentials should improve description accuracy. 
Such extensions would however require sign-problem mitigation techniques.


\subsection{Oxygen isotopes and pure neutron matter}

\begin{figure}[htbp]
  \centering
  \includegraphics[width=0.6\textwidth]{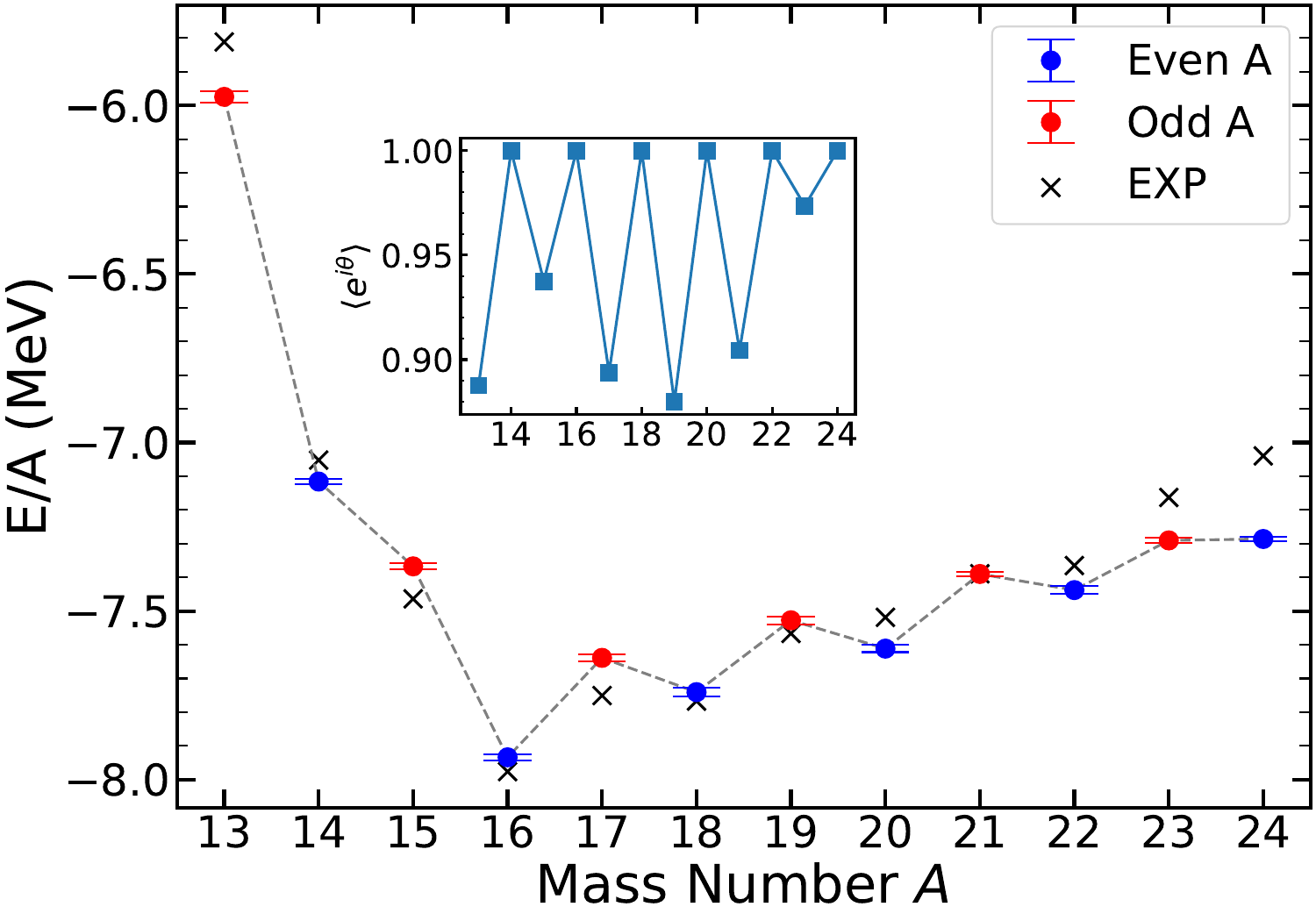}
  \caption{\textbf{Binding energies in oxygen isotopes.} 
    Binding energy per nucleon $E/A$ versus mass number $A$. 
    Blue circles: even-$A$; red circles: odd-$A$; black crosses: experimental values. 
    Inset: Average phase $\langle e^{i\theta} \rangle$ ($=1$: sign-problem-free).}
  \label{fig:Oxygen}
\end{figure}

Fig.~\ref{fig:Oxygen} presents binding energies per nucleon across the oxygen isotopic chain for both even-$A$ and odd-$A$ nuclei. 
The inset displays corresponding average phases. 
Calculations for odd nucleon numbers introduce a mild sign oscillation due to slight time-reversal symmetry breaking, yet yield statistical uncertainties comparable to even-$A$ nuclei, indicating no significant efficiency degradation. 
The computed energies reproduce odd-even staggering with slightly overestimated pairing gaps. 
Discrepancies at large neutron numbers may indicate missing interaction components (e.g., tensor forces).


\begin{figure}[htbp]
  \centering
  \includegraphics[width=0.5\textwidth]{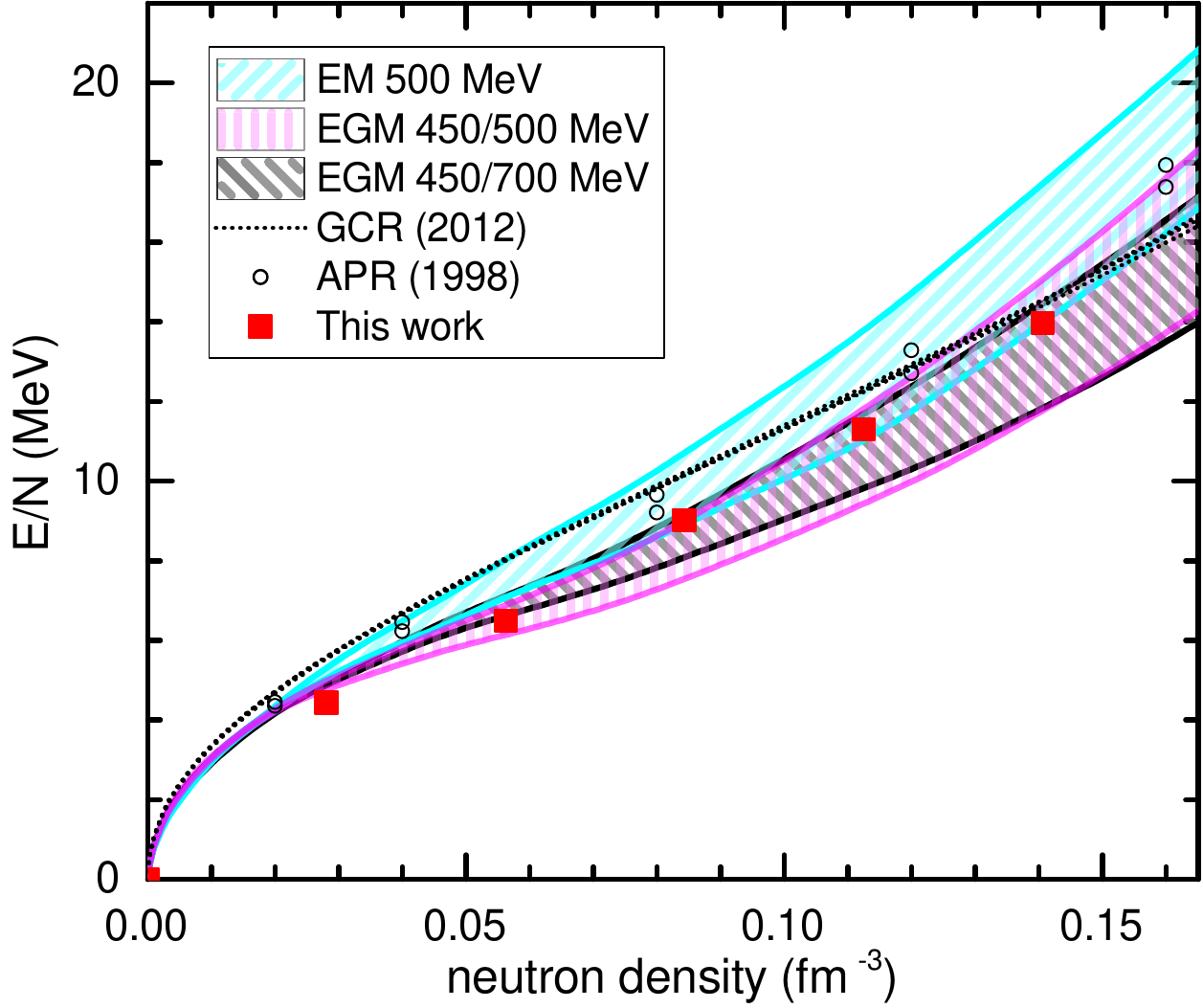}
  \caption{\textbf{Pure neutron matter equation of state.} 
  Energy per particle versus neutron density for \texttt{LAT-OPT1} interaction (red squares). 
  Comparisons include: N$^3$LO chiral interactions (EM~500 MeV, EGM~450/500 MeV, EGM~450/700 MeV) \cite{Tews2013*}; variational APR \cite{Akmal1998*}; Auxiliary Field Diffusion MC (GCR) \cite{Gandolfi2012*}.}
  \label{fig:PNS}
\end{figure}

Fig.~\ref{fig:PNS} compares pure neutron matter equations of state predicted by different interaction models. 
The \texttt{LAT-OPT1} parameterization shows excellent agreement with benchmark results up to saturation density. 
As demonstrated in the main text, this model approximately reproduces empirical saturation properties of symmetric nuclear matter. 
These results confirm that infinite nuclear matter properties can be reliably extracted from light nuclei fitting ($A \leq 40$), suggesting the essential role of many-body correlations.

\subsection{Sign problem for odd-even and odd-odd nuclei}

We have examined the performance of \texttt{LAT-OPT1} for various odd-even and odd-odd heavy nuclei.
Table~\ref{tab:avg_phase} presents the average sign $\langle e^{i\theta}\rangle$ for representative nuclei computed with $L_t = 350$ using identical computational settings as for even-even nuclei.
The selected nuclei include three odd-even systems ($^{41}$Ca, $^{57}$Ni, $^{85}$Zr) and one odd-odd nucleus ($^{82}$Nb).  
For all odd-even nuclei, the average sign exceeds $0.9$, indicating that the QMC statistical errors are comparable to those of neighboring even-even nuclei.
For $^{82}\mathrm{Nb}$, where $\langle e^{i\theta}\rangle \approx 0.75$, the statistical uncertainty is merely $\sim 0.1\,\mathrm{MeV}$ larger than in adjacent even-even nuclei.
We therefore conclude that our method enables precise extraction of ground-state energies for all nuclei with atomic mass numbers $A \leq 100$.
\begin{table}[htbp]
\caption{Average sign $\langle e^{i\theta} \rangle$ for selected odd-even and odd-odd nuclei.}
\begin{tabular}{c | c c c c}
\hline\hline
& $^{41}$Ca & $^{57}$Ni & $^{82}$Nb & $^{85}$Zr \\
\hline
$\langle e^{i\theta} \rangle$ & 0.915  & 0.920  & 0.750 & 0.927 \\
\hline\hline
\end{tabular}
\label{tab:avg_phase}
\end{table}

\subsection{Charge density distributions and charge radii}

\begin{figure}[htbp]
    \centering 
    \includegraphics[width=1.0\textwidth]{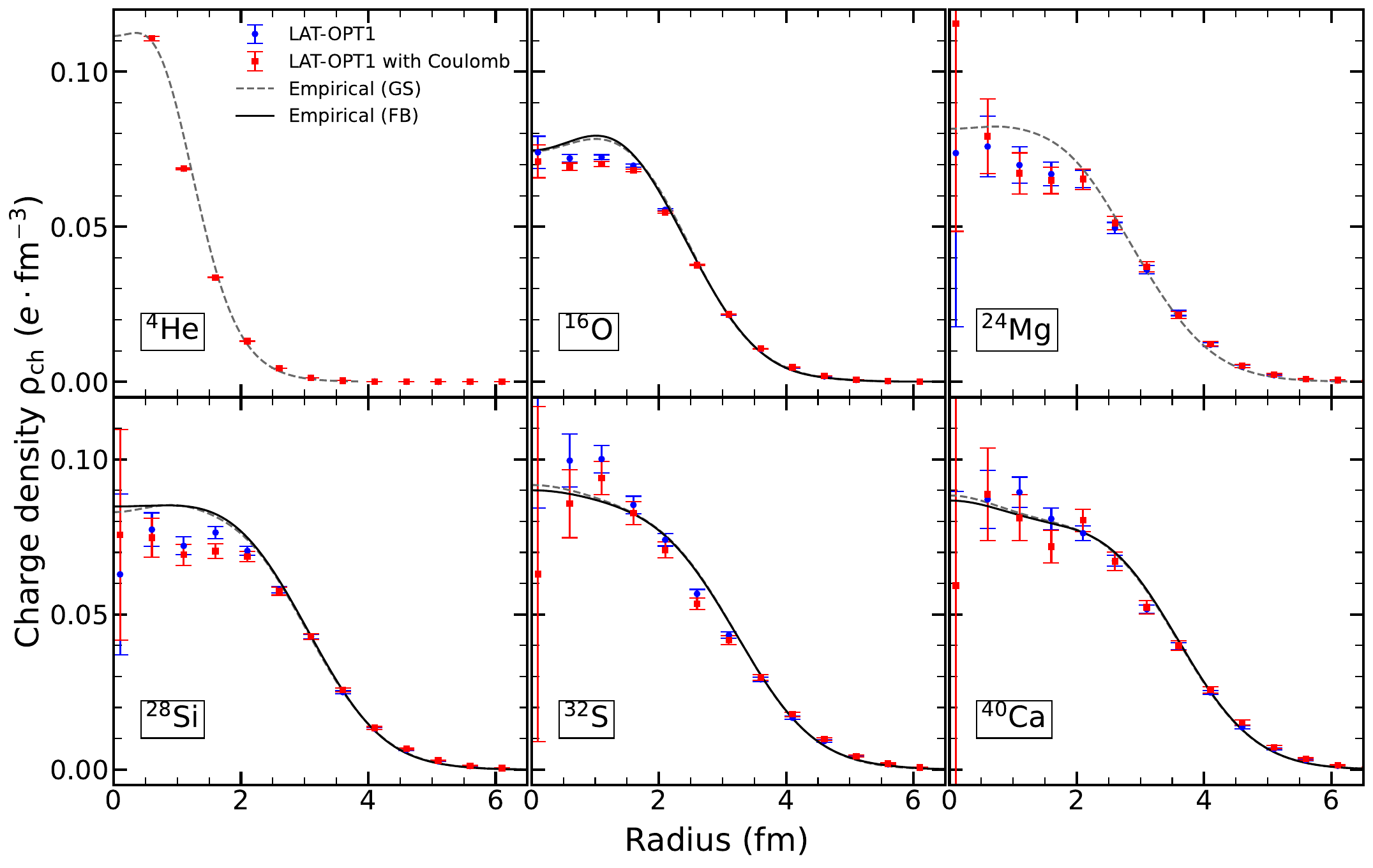}
    \caption{ \textbf{Charge density distributions.} Calculated charge density distributions for selected nuclei without (blue circles) and with (red squares) 
    perturbative Coulomb corrections compared with the empirical results~\cite{DeVries1987*}.}
    \label{fig:charge_radii}
\end{figure}

Charge density distributions (Fig.~\ref{fig:charge_radii}) and corresponding charge radii (Tab.~\ref{tab:rsq}) for the six target nuclei were calculated using the pinhole algorithm \cite{Elhatisari2017*}. 
This method induces sign problems and is typically restricted to light nuclei ($A \leq 40$). 
Calculations incorporated the proton size $R_p = 0.84\,\mathrm{fm}$ via convolution. 

Results with and without perturbative Coulomb inclusion demonstrate only minor corrections for these light nuclei. 
The computed charge densities approximately reproduce empirical profiles, though central regions exhibit large statistical uncertainties and slight underestimation compared to experimental data. 
Charge radii are overestimated by $\sim 5$-$7\%$. 
Inclusion of charge radii as optimization constraints should improve these results.


\subsection{Comparison with phenomenological models}

When evaluated as a nuclear mass model, \texttt{LAT-OPT1} predictions compare favorably with established approaches: 
macroscopic-microscopic (mac-mic) models, relativistic mean field (RMF) theory, and Skyrme density functional theory. 
These models demonstrate strong performance in describing nuclear structures, with parameters constrained by global fitting to binding energies, charge radii, and nuclear matter saturation properties across training sets typically spanning $^{16}$O to $^{208}$Pb.

We compare three published parameterizations: the FRDM~\cite{Moller2016*} (mac-mic), PC-PK1~\cite{Zhang2022*} (RMF), and UNEDF1~\cite{PRC85-024304*} (Skyrme). 
Standard deviations from experimental masses are computed for even-even nuclei with $N, Z \leq 28$ (Tab.~\ref{t1}). 
Phenomenological models show $\sim 2$-$3$ MeV RMS deviations in this region, improving significantly for heavier nuclei where mean-field contributions dominate. 
\texttt{LAT-OPT1} achieves comparable precision ($\sim 3$ MeV RMS) despite fewer parameters, while mac-mic models yield smaller deviations with more adjustable parameters.

Tab.~\ref{t1} also includes results from the Wigner-SU(4) interaction in Ref.~\cite{PLB797-134863*}, demonstrating significant improvement through: 
(1) systematic finite-volume effect elimination via larger boxes; 
(2) comprehensive imaginary-time extrapolation replacing fixed-$\tau$ calculations. 
The inclusion of spin-orbit coupling and efficient optimization further reduced deviations from $\sim 10$ MeV to $3$ MeV. 
Future extensions to larger training sets may further improve accuracy.

Complete numerical results (Tab.~\ref{tab:gse}) compare \texttt{LAT-OPT1} with mean-field calculations, listing Coulomb contributions separately. 
We also present the original data as separate figures for each nuclei.
We show the results without the Coulomb energy, which can be restored by looking up Tab.~\ref{tab:gse}.


\begin{table}[htbp]
\caption{\label{tab:rsq}
\textbf{Charge radii.} RMS charge radii $R$ (fm) for selected nuclei computed with \texttt{LAT-OPT1} versus experimental values \cite{exp*}.}
\begin{ruledtabular}
\begin{tabular}{lcccccc}
Nucleus & $^{4}$He & $^{16}$O & $^{24}$Mg & $^{28}$Si & $^{32}$S & $^{40}$Ca \\
\hline
\texttt{LAT-OPT1}   & 1.797(3)   & 2.872(6)   & 3.235(24)  & 3.319(25)  & 3.409(30)  & 3.537(23) \\
Experiment          & 1.6755(28) & 2.6991(52) & 3.0570(16) & 3.1224(24) & 3.2611(18) & 3.4776(19) \\
Relative error [\%] & 7.3        & 6.4        & 5.8        & 6.3        & 4.5        & 1.7       \\
\end{tabular}
\end{ruledtabular}
\end{table}

\setlength{\tabcolsep}{12pt} 
\renewcommand{\arraystretch}{1.5}

\begin{table}[ht]

\caption{Standard deviations $\sigma$ of ground-state binding energies for even-even nuclei ($N, Z \leq 28$) across mass models. 
Here, \emph{Parameters} is adjustable parameter count, \emph{Nuclei} is nuclei count for $\sigma$, and $\sigma_{\rm all}$ is  standard deviation over full nuclide chart.}

\begin{ruledtabular}
\begin{tabular}{l l c c c c}
Model & Parameterization & Parameters & $\sigma$ (MeV) & Nuclei & $\sigma_{\rm all}$ (MeV) \\
\hline
Macroscopic-microscopic & FRDM~\cite{Moller2016*} & $>30$ & 1.15 & 69 & 0.56 \\
Relativistic mean field & PC-PK1~\cite{Zhang2022*} & 11 & 2.25 & 60 & 1.52 \\
Skyrme DFT & UNEDF1~\cite{PRC85-024304*} & 12 & 3.43 & 75 & 1.91 \\
Lattice EFT & Wigner-SU(4)~\cite{PLB797-134863*} & 4 & 10.21 & 55 & --- \\
Lattice EFT & \texttt{LAT-OPT1} & 5 & 2.93 & 76 & --- \\
\end{tabular}
\end{ruledtabular}
\label{t1}
\end{table}


\begin{longtable}{cccccc>{\centering\arraybackslash}p{0.15\textwidth}}
\caption{Ground-state binding energies of even-even nuclei calculated with  UNEDF1~\cite{PRC85-024304*}, PC-PK1~\cite{Zhang2022*} and \texttt{LAT-OPT1}, alongside experimental values (EXP) and perturbative Coulomb corrections in \texttt{LAT-OPT1} calculations (COU). The \texttt{LAT-OPT1} errors are combined statistical and imaginary-time-extrapolation errors. All energies are in MeV.}\label{tab:gse}\\
\toprule
\textbf{$A$} & \textbf{$N$} & \textbf{UNEDF1} & \textbf{PC-PK1} & \textbf{\texttt{LAT-OPT1}} & \textbf{EXP} & \textbf{COU (MeV)} \\
\midrule
\endfirsthead

\textbf{$A$} & \textbf{$N$} & \textbf{UNEDF1} & \textbf{PC-PK1} & \textbf{\texttt{LAT-OPT1}} & \textbf{EXP} & \textbf{COU (MeV)} \\
\midrule
\endhead

\midrule
\endfoot

\multicolumn{4}{l}{\textbf{Z = 2 (He)}} \\
\midrule
4  & 2  & $\ensuremath{-}$18.70       &               & $\ensuremath{-}$28.81       & $\ensuremath{-}$28.30       & 0.543 \\
6  & 4  & $\ensuremath{-}$23.24       &               & $\ensuremath{-}$33.17(6)    & $\ensuremath{-}$29.27       & 0.527 \\
8  & 6  & $\ensuremath{-}$24.61       &               & $\ensuremath{-}$37.68(10)   & $\ensuremath{-}$31.39       & 0.523 \\
\midrule

\multicolumn{4}{l}{\textbf{Z = 4 (Be)}} \\
\midrule
8  & 4  & $\ensuremath{-}$43.42       &               & $\ensuremath{-}$56.98(20)   & $\ensuremath{-}$56.50       & 2.450 \\
10 & 6  & $\ensuremath{-}$58.52       &               & $\ensuremath{-}$66.96(16)   & $\ensuremath{-}$64.97       & 2.553 \\
12 & 8  & $\ensuremath{-}$64.89       &               & $\ensuremath{-}$72.29(30)   & $\ensuremath{-}$68.64       & 2.560 \\
14 & 10 &                              &               & $\ensuremath{-}$77.28(29)   & $\ensuremath{-}$69.916      & 2.482 \\
\midrule

\multicolumn{4}{l}{\textbf{Z = 6 (C)}} \\
\midrule
10 & 4  & $\ensuremath{-}$54.07       &               & $\ensuremath{-}$63.45(16)   & $\ensuremath{-}$60.32       & 6.056 \\
12 & 6  & $\ensuremath{-}$83.11       &               & $\ensuremath{-}$91.33(12)   & $\ensuremath{-}$92.16       & 6.180 \\
14 & 8  & $\ensuremath{-}$99.80       &               & $\ensuremath{-}$104.60(10)  & $\ensuremath{-}$105.28      & 6.297 \\
16 & 10 & $\ensuremath{-}$105.23      &               & $\ensuremath{-}$112.88(30)  & $\ensuremath{-}$110.75      & 6.175 \\
18 & 12 & $\ensuremath{-}$109.27      &               & $\ensuremath{-}$119.75(51)  & $\ensuremath{-}$115.67      & 6.038 \\
20 & 14 & $\ensuremath{-}$112.72      &               & $\ensuremath{-}$125.61(47)  & $\ensuremath{-}$119.22      & 6.047 \\
22 & 16 & $\ensuremath{-}$114.57      &               & $\ensuremath{-}$131.86(55)  & $\ensuremath{-}$119.26      & 6.001 \\
\midrule

\multicolumn{4}{l}{\textbf{Z = 8 (O)}} \\
\midrule
12 & 4  & $\ensuremath{-}$55.10       & $\ensuremath{-}$59.70      & $\ensuremath{-}$63.70(12)   & $\ensuremath{-}$58.58       & 10.875 \\
14 & 6  & $\ensuremath{-}$93.98       & $\ensuremath{-}$101.16     & $\ensuremath{-}$99.62(10)   & $\ensuremath{-}$98.73       & 11.278 \\
16 & 8  & $\ensuremath{-}$121.71      & $\ensuremath{-}$128.03     & $\ensuremath{-}$126.94(15)  & $\ensuremath{-}$127.62      & 11.318 \\
18 & 10 & $\ensuremath{-}$136.12      & $\ensuremath{-}$141.84     & $\ensuremath{-}$139.33(22)  & $\ensuremath{-}$139.81      & 11.221 \\
20 & 12 & $\ensuremath{-}$147.62      & $\ensuremath{-}$152.98     & $\ensuremath{-}$152.22(21)  & $\ensuremath{-}$151.37      & 11.142 \\
22 & 14 & $\ensuremath{-}$157.22      & $\ensuremath{-}$161.39     & $\ensuremath{-}$163.63(26)  & $\ensuremath{-}$162.03      & 11.099 \\
24 & 16 & $\ensuremath{-}$164.00      & $\ensuremath{-}$165.98     & $\ensuremath{-}$174.87(30)  & $\ensuremath{-}$168.95      & 11.051 \\
26 & 18 & $\ensuremath{-}$165.45      & $\ensuremath{-}$167.41     & $\ensuremath{-}$179.02(34)  & $\ensuremath{-}$168.93      & 11.873 \\
\midrule

\multicolumn{4}{l}{\textbf{Z = 10 (Ne)}} \\
\midrule
16 & 6  &                              & $\ensuremath{-}$100.15     & $\ensuremath{-}$101.86(16)  & $\ensuremath{-}$97.33       & 17.054 \\
18 & 8  & $\ensuremath{-}$129.33      & $\ensuremath{-}$134.37     & $\ensuremath{-}$133.48(40)  & $\ensuremath{-}$132.14      & 17.190 \\
20 & 10 & $\ensuremath{-}$152.93      & $\ensuremath{-}$160.81     & $\ensuremath{-}$158.90(52)  & $\ensuremath{-}$160.64      & 17.123 \\
22 & 12 & $\ensuremath{-}$172.32      & $\ensuremath{-}$177.97     & $\ensuremath{-}$176.33(25)  & $\ensuremath{-}$177.77      & 17.011 \\
24 & 14 & $\ensuremath{-}$186.86      & $\ensuremath{-}$192.61     & $\ensuremath{-}$190.22(23)  & $\ensuremath{-}$191.84      & 17.112 \\
26 & 16 & $\ensuremath{-}$198.32      & $\ensuremath{-}$202.25     & $\ensuremath{-}$203.86(16)  & $\ensuremath{-}$201.55      & 17.052 \\
28 & 18 & $\ensuremath{-}$204.30      & $\ensuremath{-}$208.55     & $\ensuremath{-}$211.98(36)  & $\ensuremath{-}$206.87      & 16.867 \\
30 & 20 & $\ensuremath{-}$207.95      & $\ensuremath{-}$212.44     & $\ensuremath{-}$217.41(29)  & $\ensuremath{-}$211.04      & 16.653 \\
\midrule

\multicolumn{4}{l}{\textbf{Z = 12 (Mg)}} \\
\midrule
20 & 8  & $\ensuremath{-}$132.57      & $\ensuremath{-}$135.54     & $\ensuremath{-}$138.08(20)  & $\ensuremath{-}$134.56      & 24.448 \\
22 & 10 & $\ensuremath{-}$163.81      & $\ensuremath{-}$169.09     & $\ensuremath{-}$168.72(22)  & $\ensuremath{-}$168.58      & 24.334 \\
24 & 12 & $\ensuremath{-}$190.53      & $\ensuremath{-}$206.16     & $\ensuremath{-}$196.03(32)  & $\ensuremath{-}$198.26      & 24.424 \\
26 & 14 & $\ensuremath{-}$211.19      & $\ensuremath{-}$220.52     & $\ensuremath{-}$213.13(32)  & $\ensuremath{-}$216.68      & 24.378 \\
28 & 16 & $\ensuremath{-}$227.37      & $\ensuremath{-}$231.59     & $\ensuremath{-}$229.96(27)  & $\ensuremath{-}$231.63      & 24.412 \\
30 & 18 & $\ensuremath{-}$238.48      & $\ensuremath{-}$243.12     & $\ensuremath{-}$242.85(18)  & $\ensuremath{-}$241.63      & 24.193 \\
32 & 20 & $\ensuremath{-}$246.65      & $\ensuremath{-}$252.21     & $\ensuremath{-}$250.49(21)  & $\ensuremath{-}$249.72      & 23.985 \\
34 & 22 & $\ensuremath{-}$251.27      & $\ensuremath{-}$260.61     & $\ensuremath{-}$259.95(36)  & $\ensuremath{-}$256.71      & 23.616 \\
36 & 24 & $\ensuremath{-}$255.81      & $\ensuremath{-}$266.85     & $\ensuremath{-}$267.76(33)  & $\ensuremath{-}$260.80      & 23.399 \\
\midrule

\multicolumn{4}{l}{\textbf{Z = 14 (Si)}} \\
\midrule
24 & 10 & $\ensuremath{-}$168.73      & $\ensuremath{-}$172.61     & $\ensuremath{-}$174.45(41)  & $\ensuremath{-}$172.01      & 32.929 \\
26 & 12 & $\ensuremath{-}$201.15      & $\ensuremath{-}$207.64     & $\ensuremath{-}$204.98(50)  & $\ensuremath{-}$206.04      & 32.941 \\
28 & 14 & $\ensuremath{-}$229.40      & $\ensuremath{-}$239.37     & $\ensuremath{-}$233.18(35)  & $\ensuremath{-}$236.54      & 33.227 \\
30 & 16 & $\ensuremath{-}$251.31      & $\ensuremath{-}$253.02     & $\ensuremath{-}$252.60(21)  & $\ensuremath{-}$255.62      & 33.015 \\
32 & 18 & $\ensuremath{-}$267.55      & $\ensuremath{-}$271.61     & $\ensuremath{-}$268.49(36)  & $\ensuremath{-}$271.41      & 32.872 \\
34 & 20 & $\ensuremath{-}$280.56      & $\ensuremath{-}$284.74     & $\ensuremath{-}$281.45(27)  & $\ensuremath{-}$283.46      & 32.687 \\
36 & 22 & $\ensuremath{-}$288.46      & $\ensuremath{-}$292.93     & $\ensuremath{-}$292.97(39)  & $\ensuremath{-}$292.05      & 32.318 \\
38 & 24 & $\ensuremath{-}$295.95      & $\ensuremath{-}$303.93     & $\ensuremath{-}$302.81(39)  & $\ensuremath{-}$299.93      & 31.932 \\
40 & 26 & $\ensuremath{-}$301.71      & $\ensuremath{-}$311.49     & $\ensuremath{-}$311.39(76)  & $\ensuremath{-}$306.23      & 31.677 \\
\midrule

\multicolumn{4}{l}{\textbf{Z = 16 (S)}} \\
\midrule
28 & 12 & $\ensuremath{-}$206.45      & $\ensuremath{-}$209.32     & $\ensuremath{-}$212.12(28)  & $\ensuremath{-}$209.41      & 42.472 \\
30 & 14 & $\ensuremath{-}$240.24      & $\ensuremath{-}$241.38     & $\ensuremath{-}$242.86(21)  & $\ensuremath{-}$243.68      & 42.741 \\
32 & 16 & $\ensuremath{-}$267.57      & $\ensuremath{-}$269.35     & $\ensuremath{-}$271.34(32)  & $\ensuremath{-}$271.78      & 42.930 \\
34 & 18 & $\ensuremath{-}$288.88      & $\ensuremath{-}$289.29     & $\ensuremath{-}$291.99(11)  & $\ensuremath{-}$291.84      & 42.443 \\
36 & 20 & $\ensuremath{-}$306.68      & $\ensuremath{-}$308.37     & $\ensuremath{-}$308.54(20)  & $\ensuremath{-}$308.71      & 42.294 \\
38 & 22 & $\ensuremath{-}$318.83      & $\ensuremath{-}$322.56     & $\ensuremath{-}$321.67(26)  & $\ensuremath{-}$321.05      & 41.941 \\
40 & 24 & $\ensuremath{-}$330.10      & $\ensuremath{-}$335.55     & $\ensuremath{-}$334.04(77)  & $\ensuremath{-}$333.17      & 41.665 \\
42 & 26 & $\ensuremath{-}$339.61      & $\ensuremath{-}$346.39     & $\ensuremath{-}$345.48(43)  & $\ensuremath{-}$344.12      & 41.420 \\
44 & 28 & $\ensuremath{-}$347.49      & $\ensuremath{-}$354.35     & $\ensuremath{-}$354.82(54)  & $\ensuremath{-}$351.82      & 41.284 \\
\midrule

\multicolumn{4}{l}{\textbf{Z = 18 (Ar)}} \\
\midrule
32 & 14 & $\ensuremath{-}$244.70      & $\ensuremath{-}$246.95     & $\ensuremath{-}$248.05(29)  & $\ensuremath{-}$246.40      & 53.414 \\
34 & 16 & $\ensuremath{-}$276.88      & $\ensuremath{-}$276.40     & $\ensuremath{-}$281.05(16)  & $\ensuremath{-}$278.72      & 53.397 \\
36 & 18 & $\ensuremath{-}$303.51      & $\ensuremath{-}$306.28     & $\ensuremath{-}$306.11(40)  & $\ensuremath{-}$306.72      & 53.459 \\
38 & 20 & $\ensuremath{-}$326.49      & $\ensuremath{-}$327.13     & $\ensuremath{-}$328.09(14)  & $\ensuremath{-}$327.34      & 53.073 \\
40 & 22 & $\ensuremath{-}$343.13      & $\ensuremath{-}$343.28     & $\ensuremath{-}$344.37(28)  & $\ensuremath{-}$343.81      & 52.717 \\
42 & 24 & $\ensuremath{-}$357.79      & $\ensuremath{-}$360.56     & $\ensuremath{-}$360.18(32)  & $\ensuremath{-}$359.34      & 52.351 \\
44 & 26 & $\ensuremath{-}$371.15      & $\ensuremath{-}$374.98     & $\ensuremath{-}$373.95(46)  & $\ensuremath{-}$373.73      & 52.170 \\
46 & 28 & $\ensuremath{-}$382.74      & $\ensuremath{-}$387.54     & $\ensuremath{-}$387.62(32)  & $\ensuremath{-}$386.97      & 52.050 \\
\midrule

\multicolumn{4}{l}{\textbf{Z = 20 (Ca)}} \\
\midrule
36 & 16 & $\ensuremath{-}$281.41      & $\ensuremath{-}$281.45     & $\ensuremath{-}$285.52(20)  & $\ensuremath{-}$281.37      & 65.306 \\
38 & 18 & $\ensuremath{-}$313.11      & $\ensuremath{-}$313.03     & $\ensuremath{-}$315.74(10)  & $\ensuremath{-}$313.12      & 65.166 \\
40 & 20 & $\ensuremath{-}$341.11      & $\ensuremath{-}$343.06     & $\ensuremath{-}$343.00(18)  & $\ensuremath{-}$342.05      & 65.436 \\
42 & 22 & $\ensuremath{-}$362.24      & $\ensuremath{-}$363.51     & $\ensuremath{-}$362.55(21)  & $\ensuremath{-}$361.90      & 64.697 \\
44 & 24 & $\ensuremath{-}$381.13      & $\ensuremath{-}$382.12     & $\ensuremath{-}$380.36(33)  & $\ensuremath{-}$380.96      & 64.434 \\
46 & 26 & $\ensuremath{-}$398.49      & $\ensuremath{-}$399.38     & $\ensuremath{-}$397.78(12)  & $\ensuremath{-}$398.77      & 64.181 \\
48 & 28 & $\ensuremath{-}$413.75      & $\ensuremath{-}$415.50     & $\ensuremath{-}$414.49(29)  & $\ensuremath{-}$416.00      & 64.110 \\
\midrule

\multicolumn{4}{l}{\textbf{Z = 22 (Ti)}} \\
\midrule
40 & 18 & $\ensuremath{-}$315.78      & $\ensuremath{-}$314.46     & $\ensuremath{-}$319.86(38)  & $\ensuremath{-}$314.63      & 77.408 \\
42 & 20 & $\ensuremath{-}$348.27      & $\ensuremath{-}$348.76     & $\ensuremath{-}$349.70(17)  & $\ensuremath{-}$346.89      & 77.381 \\
44 & 22 & $\ensuremath{-}$373.46      & $\ensuremath{-}$373.09     & $\ensuremath{-}$376.20(45)  & $\ensuremath{-}$375.47      & 77.218 \\
46 & 24 & $\ensuremath{-}$396.43      & $\ensuremath{-}$398.46     & $\ensuremath{-}$396.45(44)  & $\ensuremath{-}$398.20      & 76.715 \\
48 & 26 & $\ensuremath{-}$417.47      & $\ensuremath{-}$419.46     & $\ensuremath{-}$416.18(28)  & $\ensuremath{-}$418.70      & 76.576 \\
50 & 28 & $\ensuremath{-}$436.29      & $\ensuremath{-}$436.72     & $\ensuremath{-}$434.70(46)  & $\ensuremath{-}$437.79      & 76.446 \\
\midrule

\multicolumn{4}{l}{\textbf{Z = 24 (Cr)}} \\
\midrule
44 & 20 & $\ensuremath{-}$351.82      & $\ensuremath{-}$351.51     & $\ensuremath{-}$354.10(46)  & $\ensuremath{-}$349.78      & 90.962 \\
46 & 22 & $\ensuremath{-}$381.05      & $\ensuremath{-}$382.51     & $\ensuremath{-}$383.20(76)  & $\ensuremath{-}$381.98      & 90.616 \\
48 & 24 & $\ensuremath{-}$409.68      & $\ensuremath{-}$412.29     & $\ensuremath{-}$411.70(59)  & $\ensuremath{-}$411.47      & 90.573 \\
50 & 26 & $\ensuremath{-}$434.00      & $\ensuremath{-}$435.67     & $\ensuremath{-}$431.81(20)  & $\ensuremath{-}$435.05      & 90.115 \\
52 & 28 & $\ensuremath{-}$455.12      & $\ensuremath{-}$454.84     & $\ensuremath{-}$451.75(32)  & $\ensuremath{-}$456.35      & 90.032 \\
\midrule

\multicolumn{4}{l}{\textbf{Z = 26 (Fe)}} \\
\midrule
48 & 22 & $\ensuremath{-}$385.80      & $\ensuremath{-}$386.47     & $\ensuremath{-}$386.93(27)  & $\ensuremath{-}$385.09      & 105.36 \\
50 & 24 & $\ensuremath{-}$417.55      & $\ensuremath{-}$418.55     & $\ensuremath{-}$417.94(71)  & $\ensuremath{-}$417.70      & 105.05 \\
52 & 26 & $\ensuremath{-}$445.60      & $\ensuremath{-}$447.92     & $\ensuremath{-}$447.08(78)  & $\ensuremath{-}$447.70      & 105.25 \\
54 & 28 & $\ensuremath{-}$470.35      & $\ensuremath{-}$470.35     & $\ensuremath{-}$467.81(40)  & $\ensuremath{-}$471.76      & 104.86 \\
\midrule

\multicolumn{4}{l}{\textbf{Z = 28 (Ni)}} \\
\midrule
52 & 24 & $\ensuremath{-}$421.10      & $\ensuremath{-}$419.22     & $\ensuremath{-}$421.19(45)  & $\ensuremath{-}$420.36      & 120.98 \\
54 & 26 & $\ensuremath{-}$452.63      & $\ensuremath{-}$451.87     & $\ensuremath{-}$450.88(32)  & $\ensuremath{-}$453.22      & 120.80 \\
56 & 28 & $\ensuremath{-}$481.59      & $\ensuremath{-}$483.65     & $\ensuremath{-}$479.33(60)  & $\ensuremath{-}$484.00      & 121.35 \\
\midrule

\multicolumn{4}{l}{\textbf{Z = 40 (Zr)}} \\
\midrule
80 & 40 & $\ensuremath{-}$669.13      & $\ensuremath{-}$666.35    & $\ensuremath{-}$672.1(8)  & $\ensuremath{-}$669.20  &  233.90 \\
90 & 50 & $\ensuremath{-}$786.02      & $\ensuremath{-}$783.47    & $\ensuremath{-}$782.1(5)  & $\ensuremath{-}$783.90  &  230.52 \\
\midrule

\multicolumn{4}{l}{\textbf{Z = 50 (Sn)}} \\
\midrule
100 & 50 & $\ensuremath{-}$828.83      & $\ensuremath{-}$827.97    & $\ensuremath{-}$824.7(8)  & $\ensuremath{-}$825.16  & 348.76 \\
120 & 70 & $\ensuremath{-}$1020.24      & $\ensuremath{-}$1019.77    & $\ensuremath{-}$1033.65(11)  & $\ensuremath{-}$1020.48  & 339.35 \\
132 & 82 & $\ensuremath{-}$1101.53    & $\ensuremath{-}$1104.47    & $\ensuremath{-}$1134.2(27)  & $\ensuremath{-}$1102.84     & 333.14 \\
\fontsize{12pt}{14pt}\selectfont
\end{longtable}


\begin{figure}[htbp]
    \centering
    \begin{minipage}{0.46\textwidth}
        \centering
        \includegraphics[width=\textwidth]{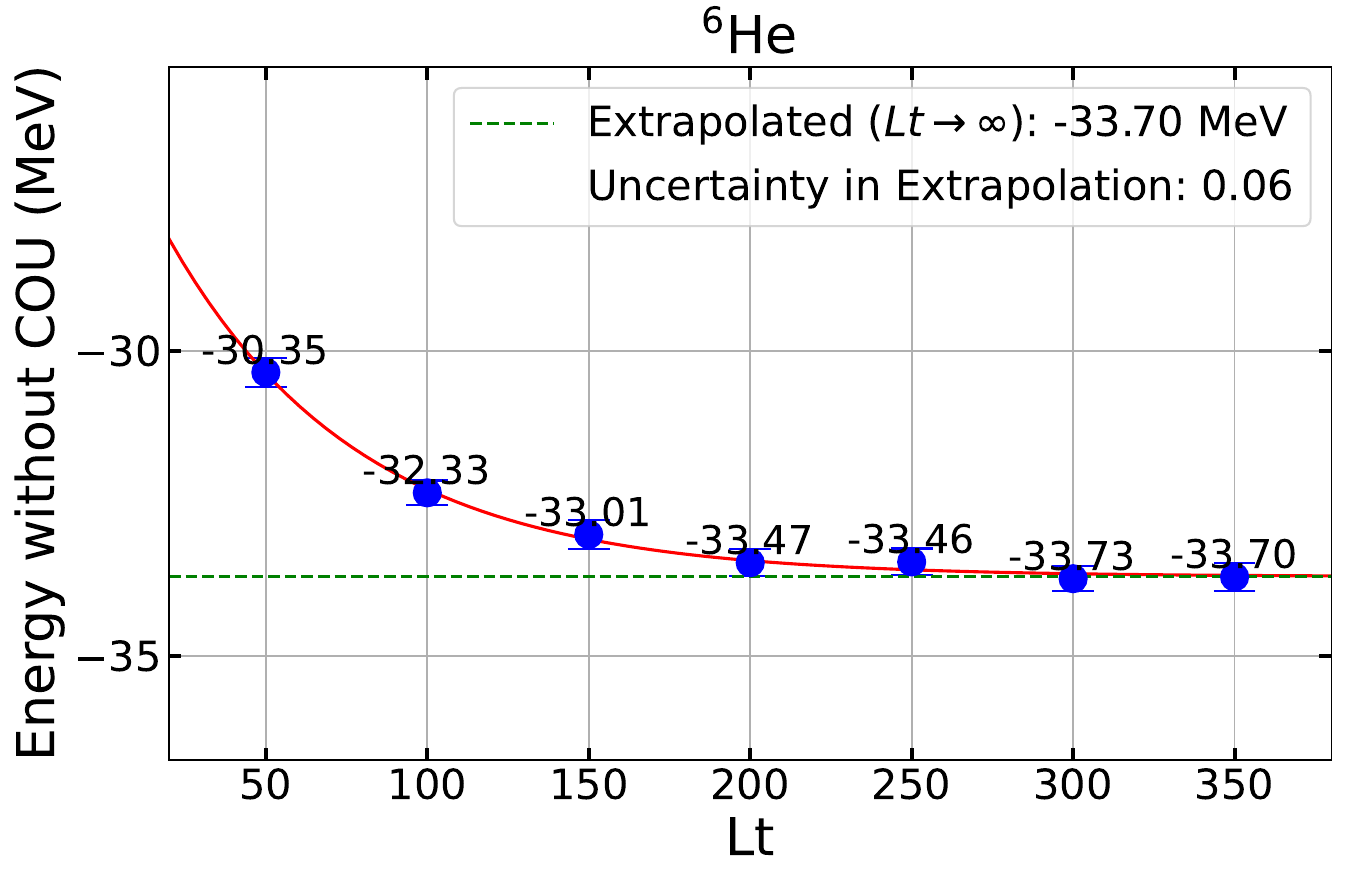}
    \end{minipage}%
    \begin{minipage}{0.46\textwidth}
        \centering
        \includegraphics[width=\textwidth]{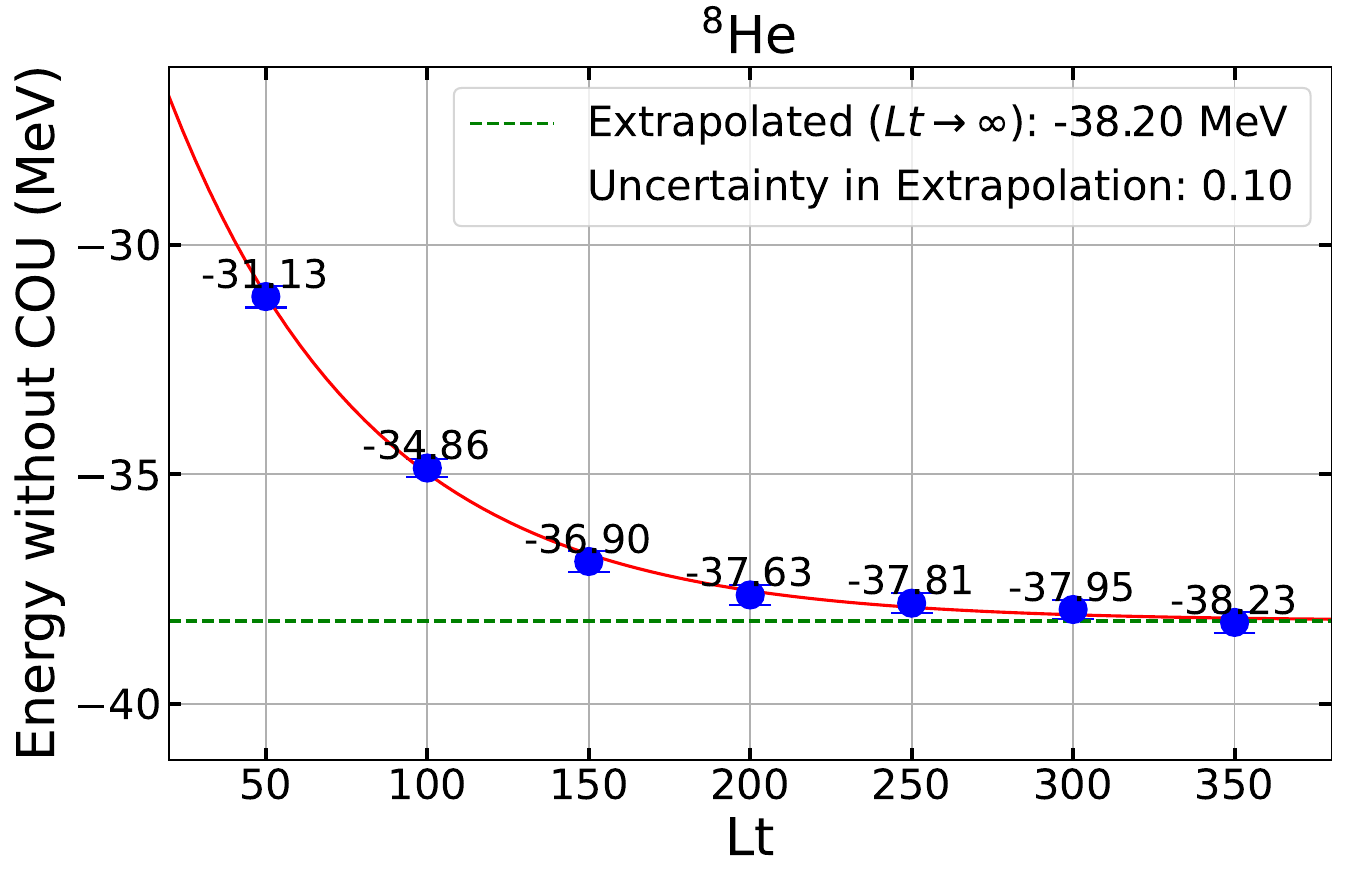}
    \end{minipage}%

    \vspace{0.5cm}
    
    \begin{minipage}{0.46\textwidth}
        \centering
        \includegraphics[width=\textwidth]{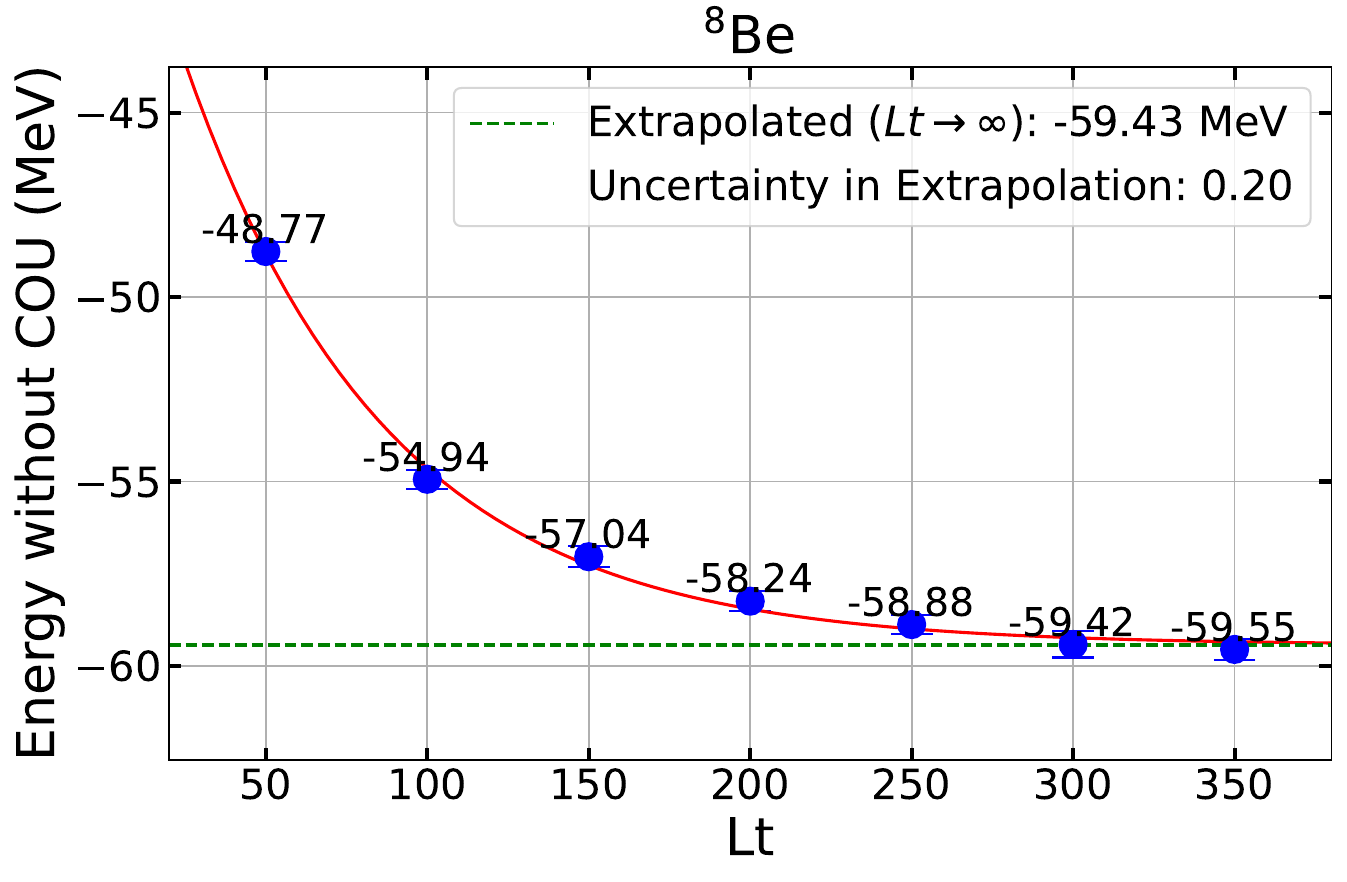}
    \end{minipage}%
    \begin{minipage}{0.46\textwidth}
        \centering
        \includegraphics[width=\textwidth]{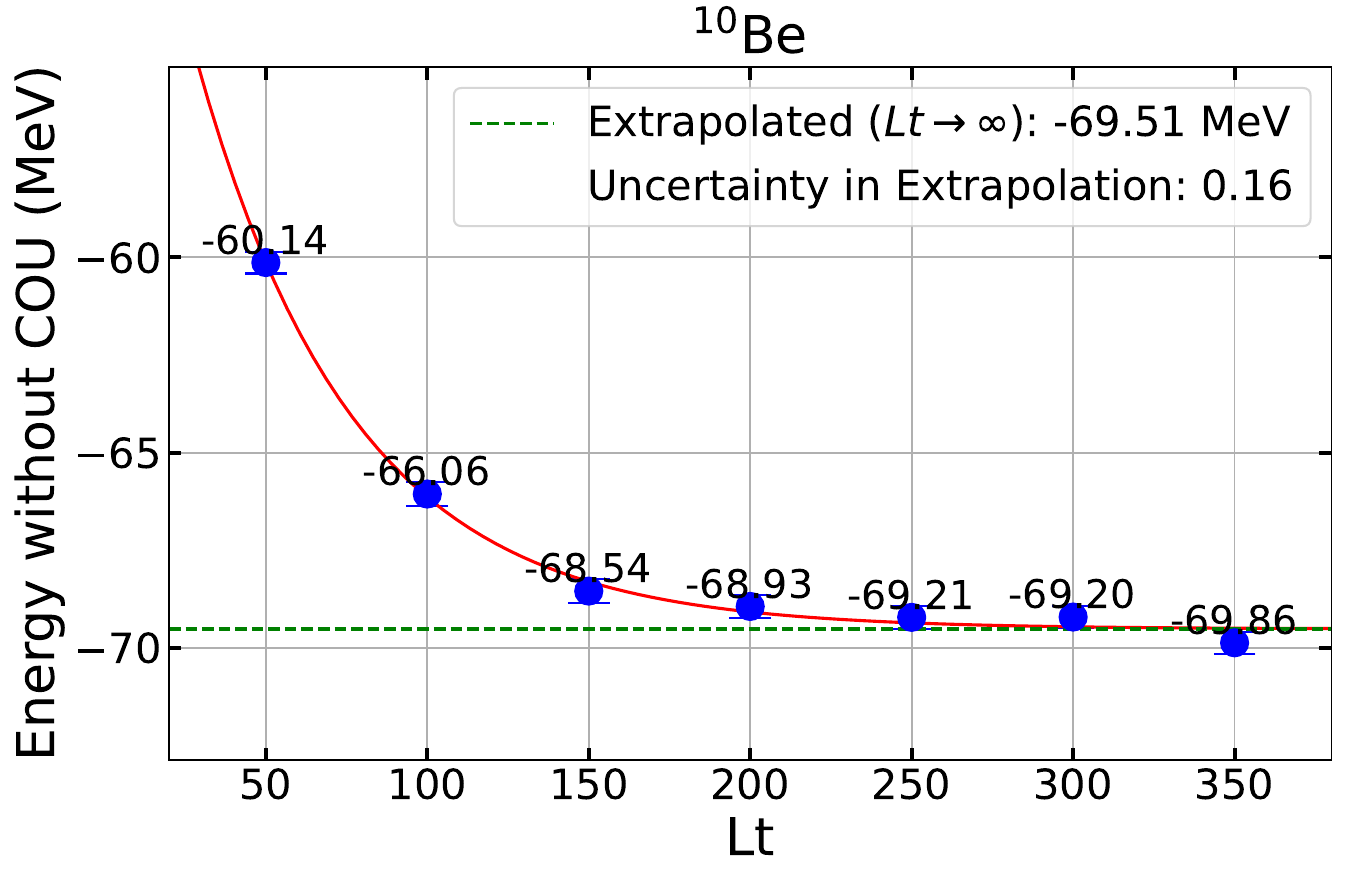}
    \end{minipage}%

    \vspace{0.5cm}
    \begin{minipage}{0.46\textwidth}
        \centering
        \includegraphics[width=\textwidth]{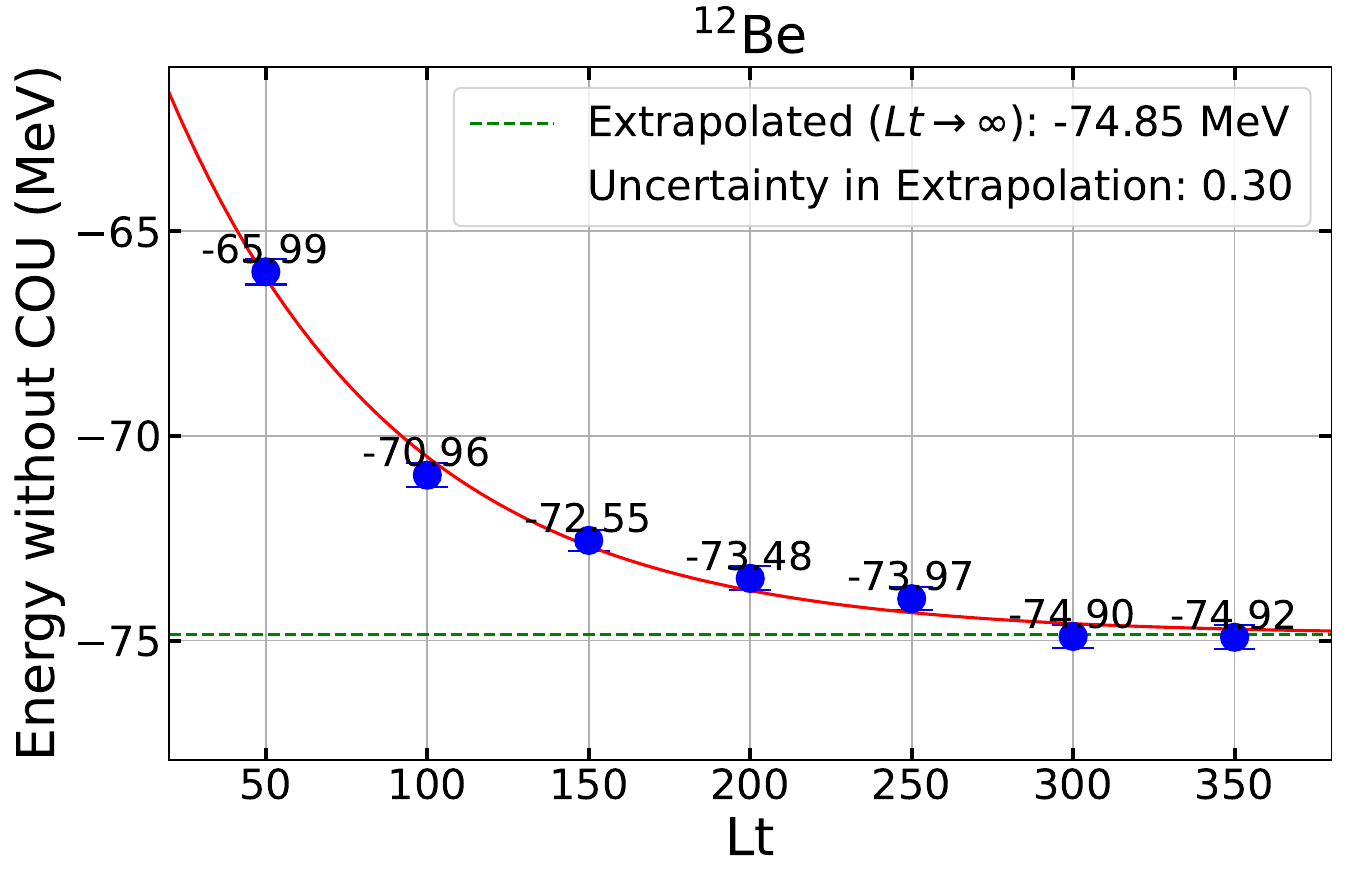}
    \end{minipage}%
    \begin{minipage}{0.46\textwidth}
        \centering
        \includegraphics[width=\textwidth]{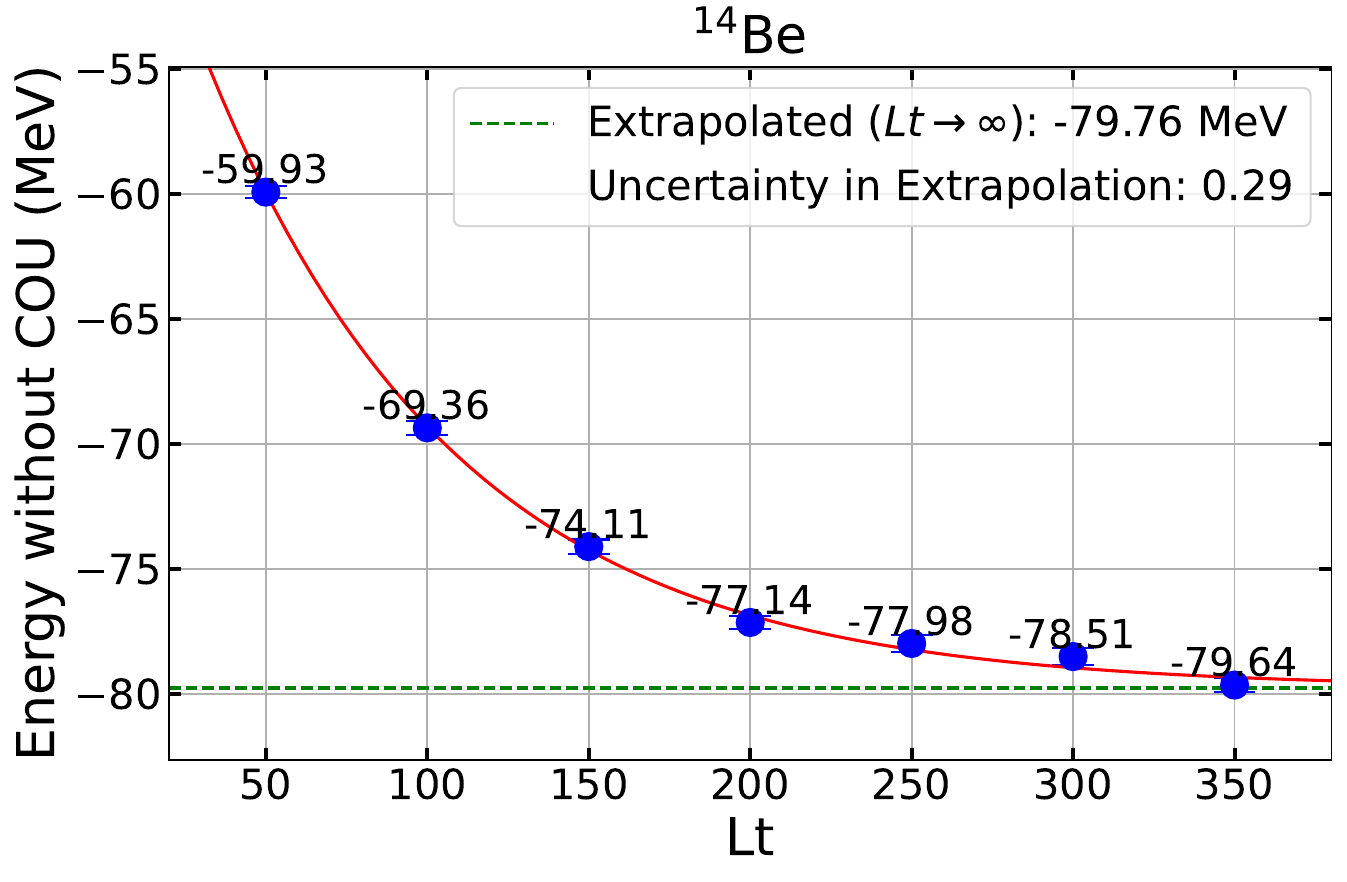}
    \end{minipage}%

    \vspace{0.5cm}
    
    \begin{minipage}{0.46\textwidth}
        \centering
        \includegraphics[width=\textwidth]{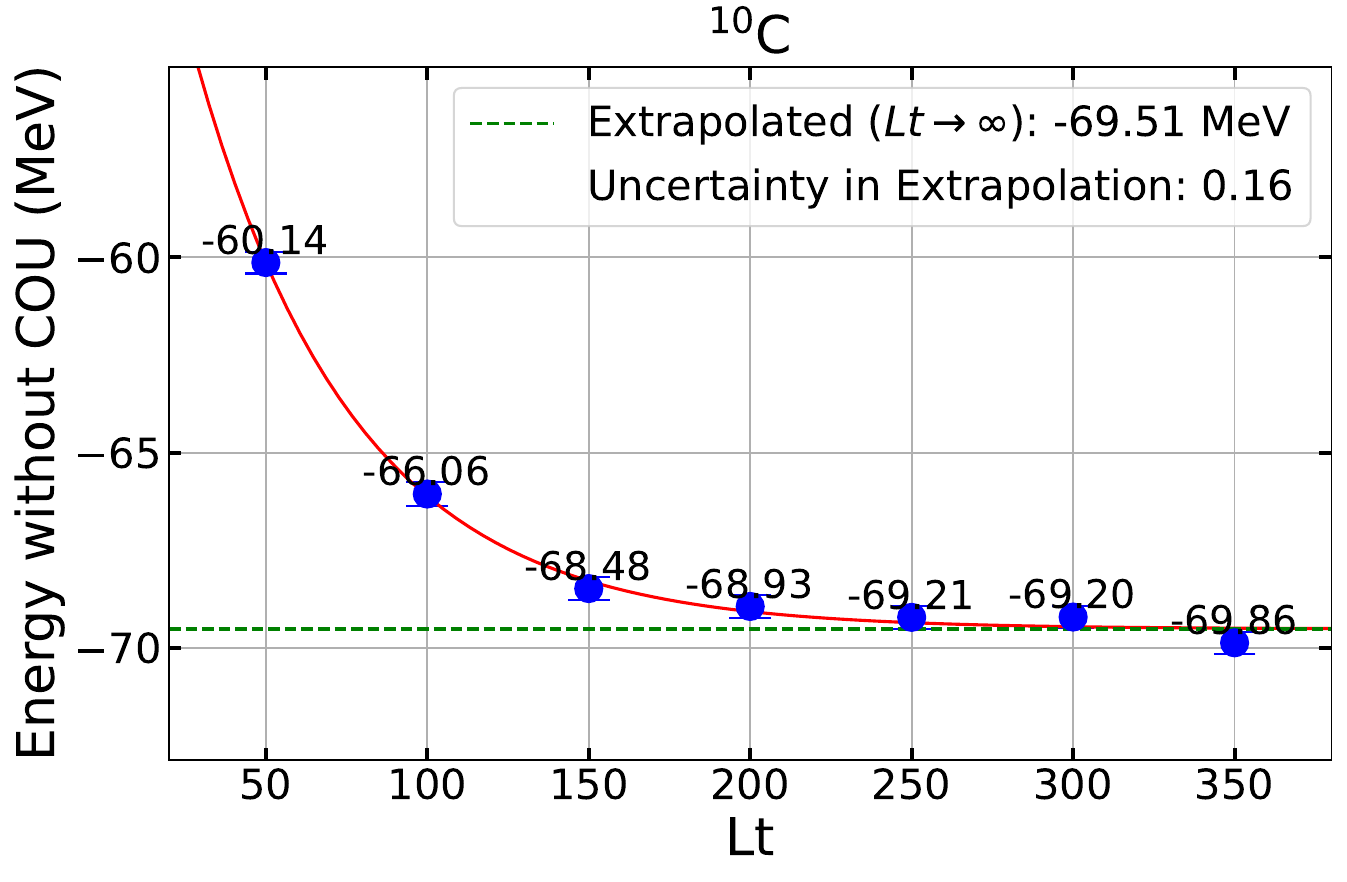}
    \end{minipage}%
    \begin{minipage}{0.46\textwidth}
        \centering
        \includegraphics[width=\textwidth]{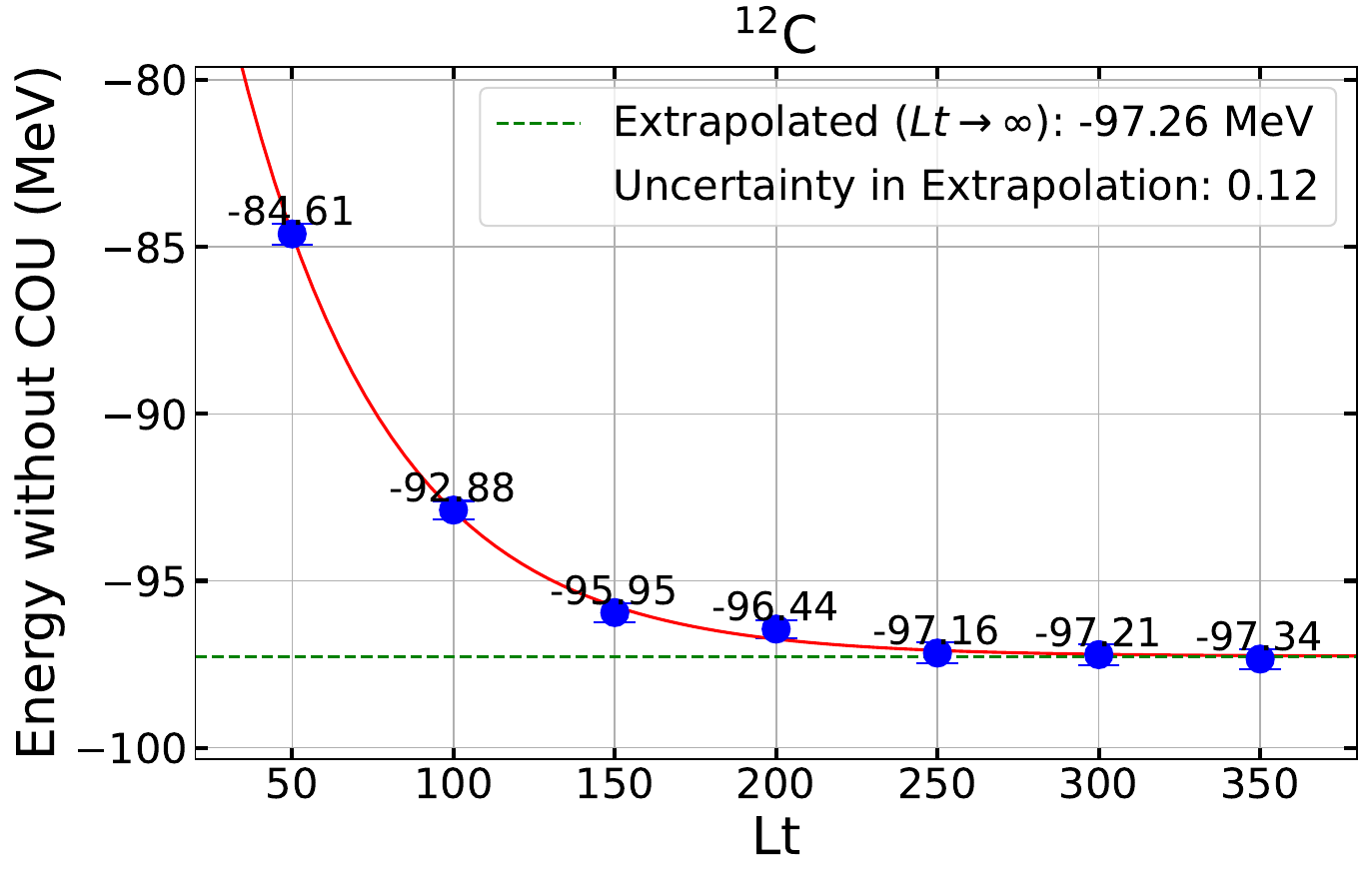}
    \end{minipage}%

\end{figure}

\begin{figure}[H]
    \vspace{0.5cm}
    \begin{minipage}{0.46\textwidth}
        \centering
        \includegraphics[width=\textwidth]{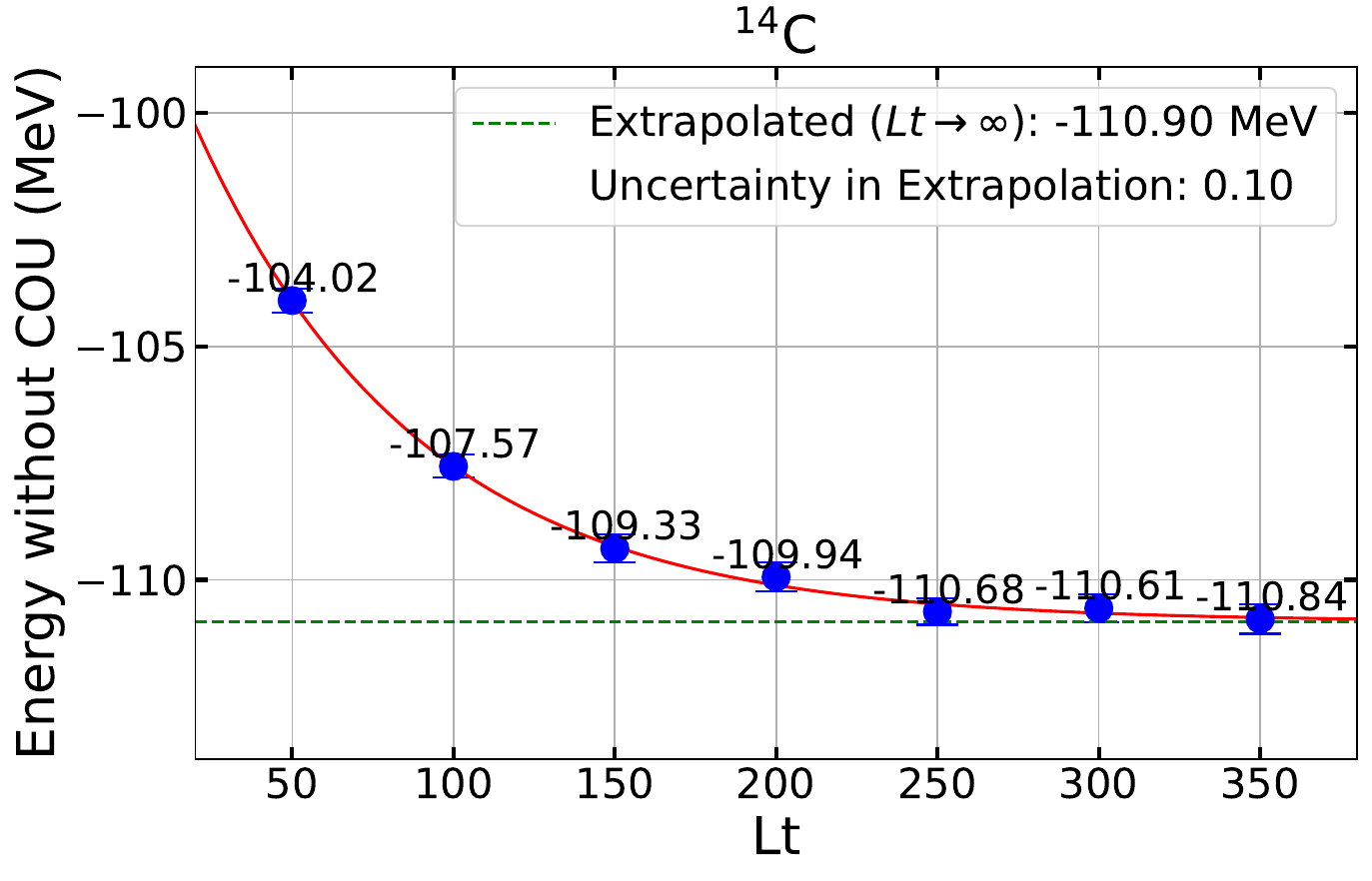}
    \end{minipage}%
    \begin{minipage}{0.46\textwidth}
        \centering
        \includegraphics[width=\textwidth]{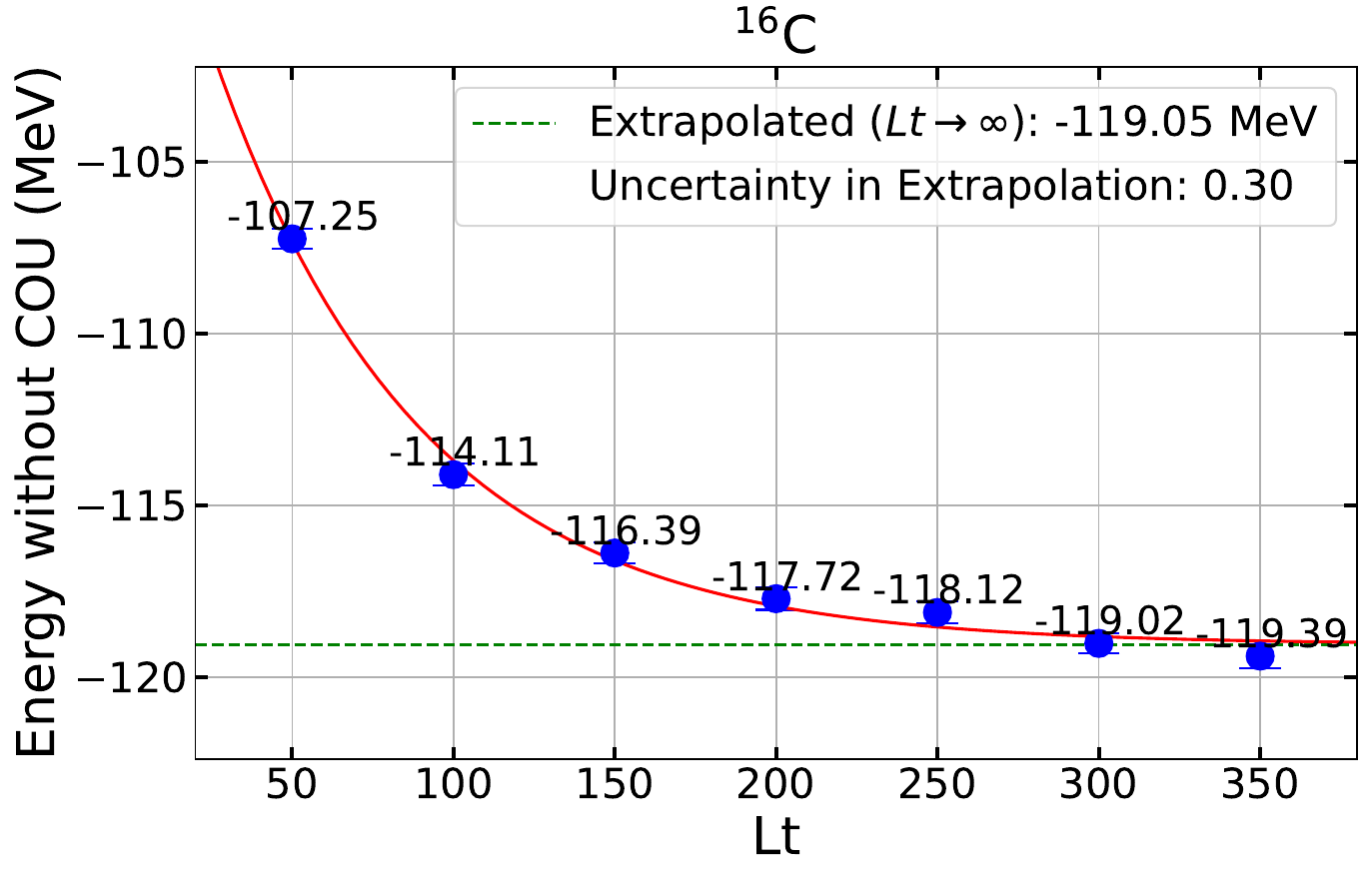}
    \end{minipage}%

    \newpage
    \vspace{0.5cm}
    
    \begin{minipage}{0.46\textwidth}
        \centering
        \includegraphics[width=\textwidth]{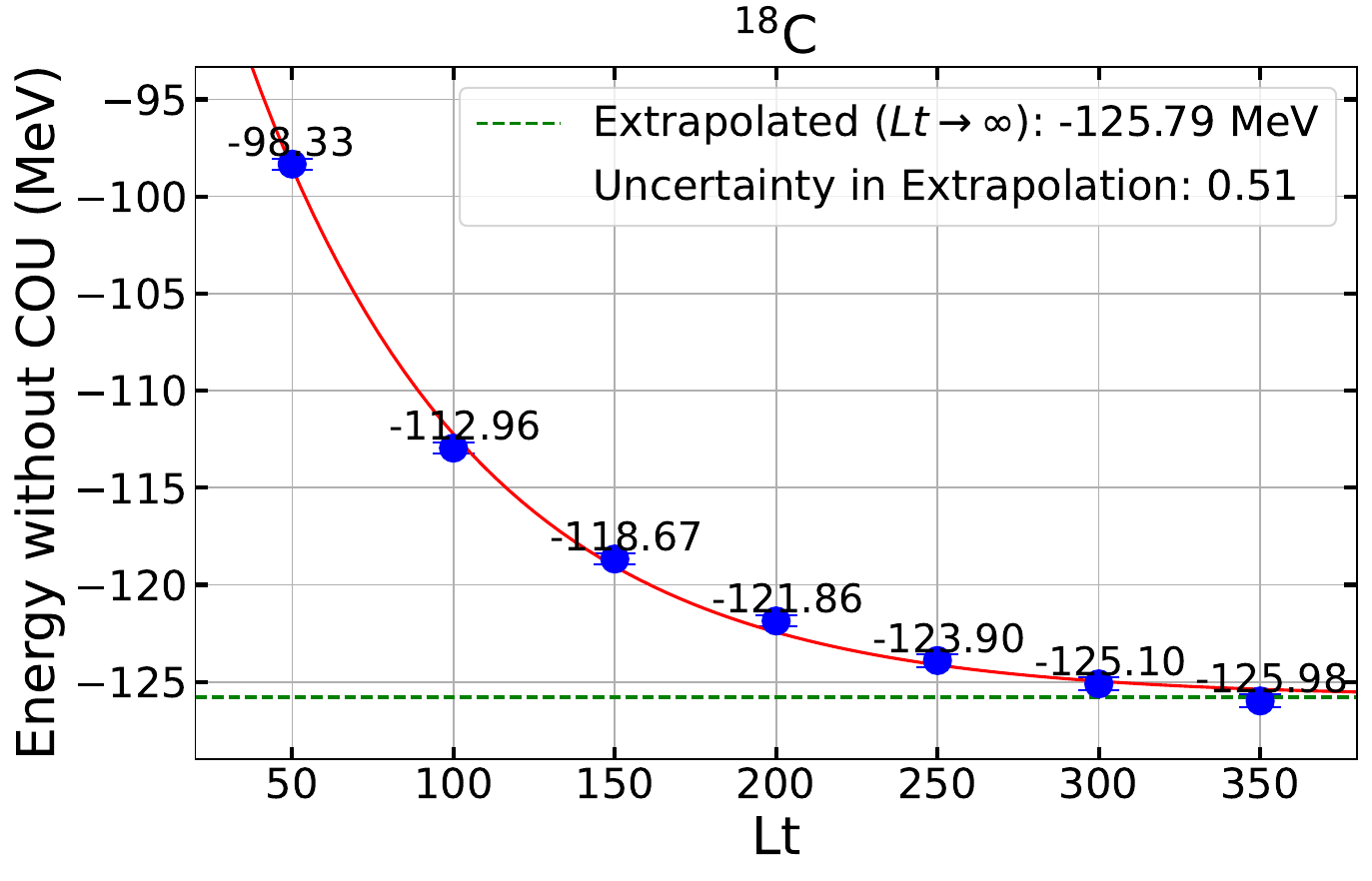}
    \end{minipage}%
    \begin{minipage}{0.46\textwidth}
        \centering
        \includegraphics[width=\textwidth]{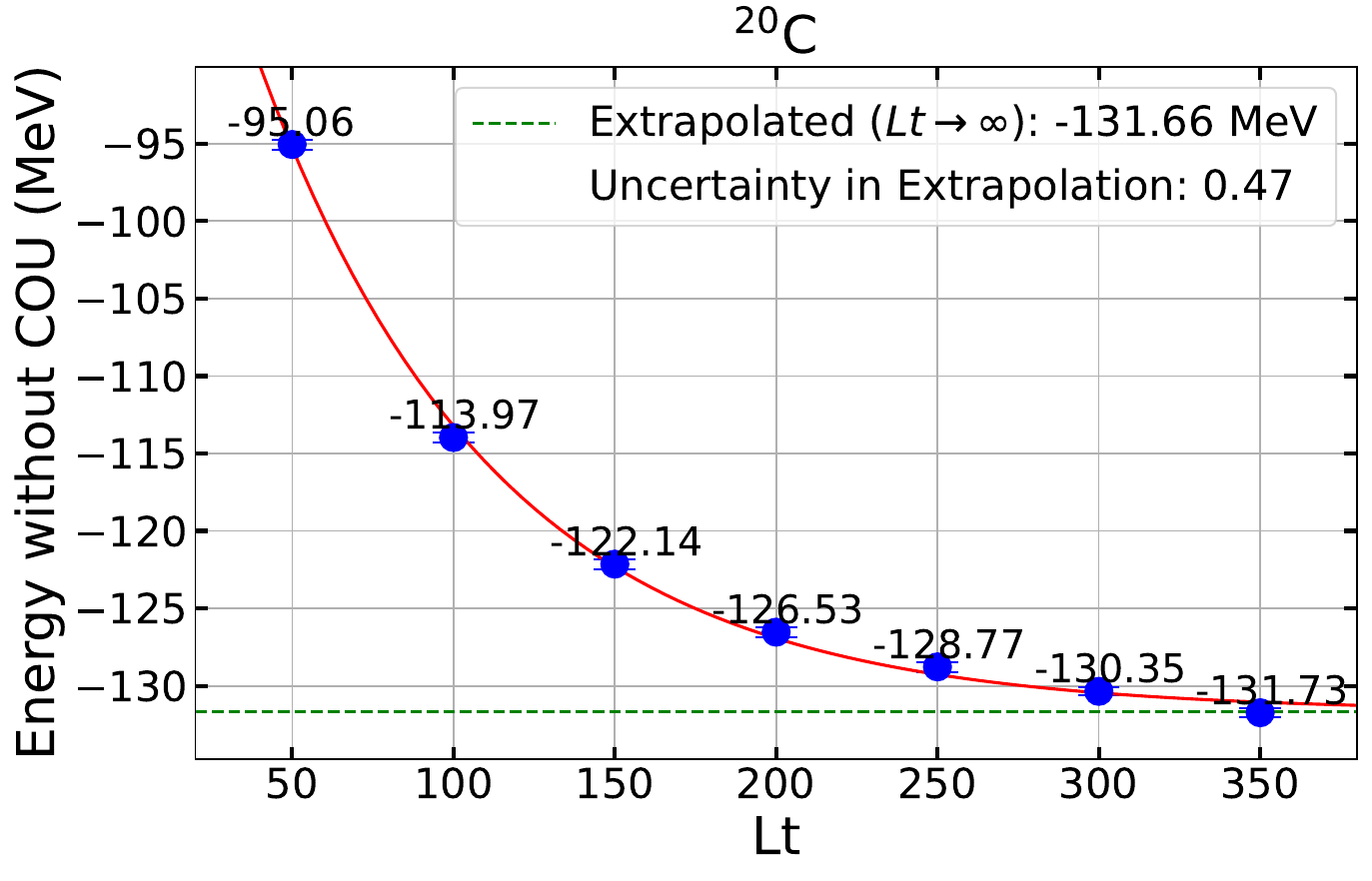}
    \end{minipage}%

    \vspace{0.5cm}
    \begin{minipage}{0.46\textwidth}
        \centering
        \includegraphics[width=\textwidth]{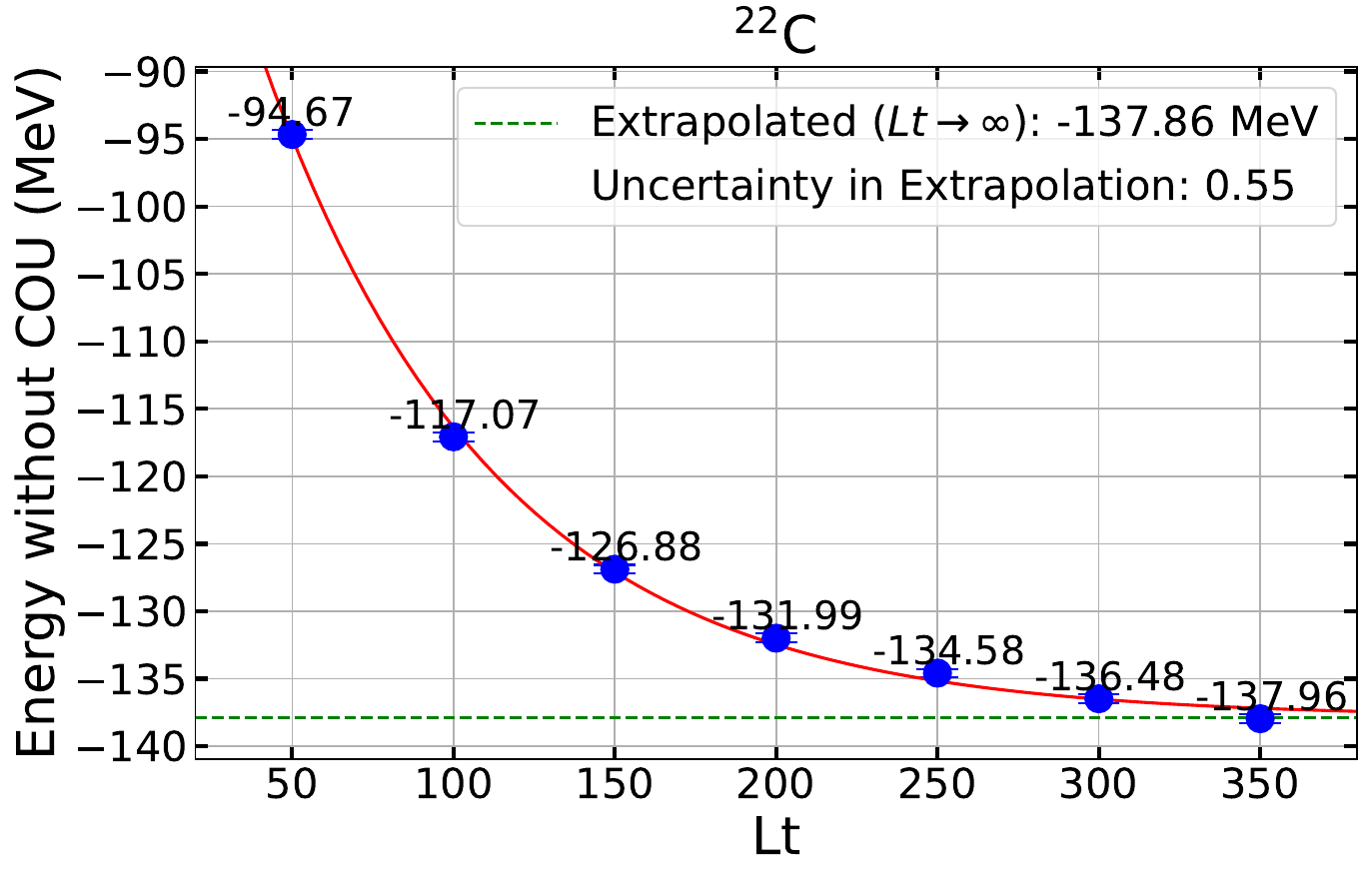}
    \end{minipage}%
    \begin{minipage}{0.46\textwidth}
        \centering
        \includegraphics[width=\textwidth]{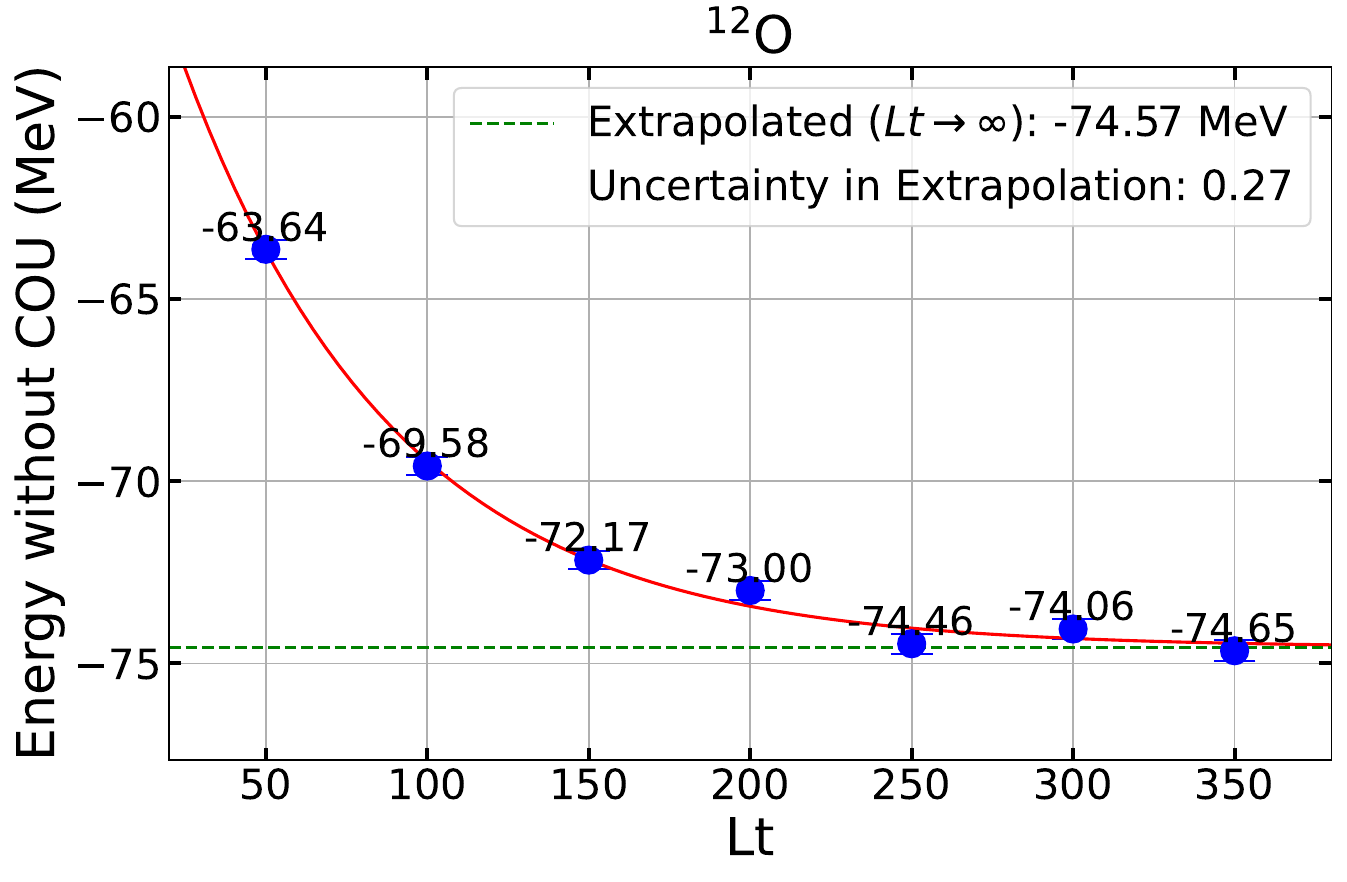}
    \end{minipage}%

    \vspace{0.5cm}
    
    \begin{minipage}{0.46\textwidth}
        \centering
        \includegraphics[width=\textwidth]{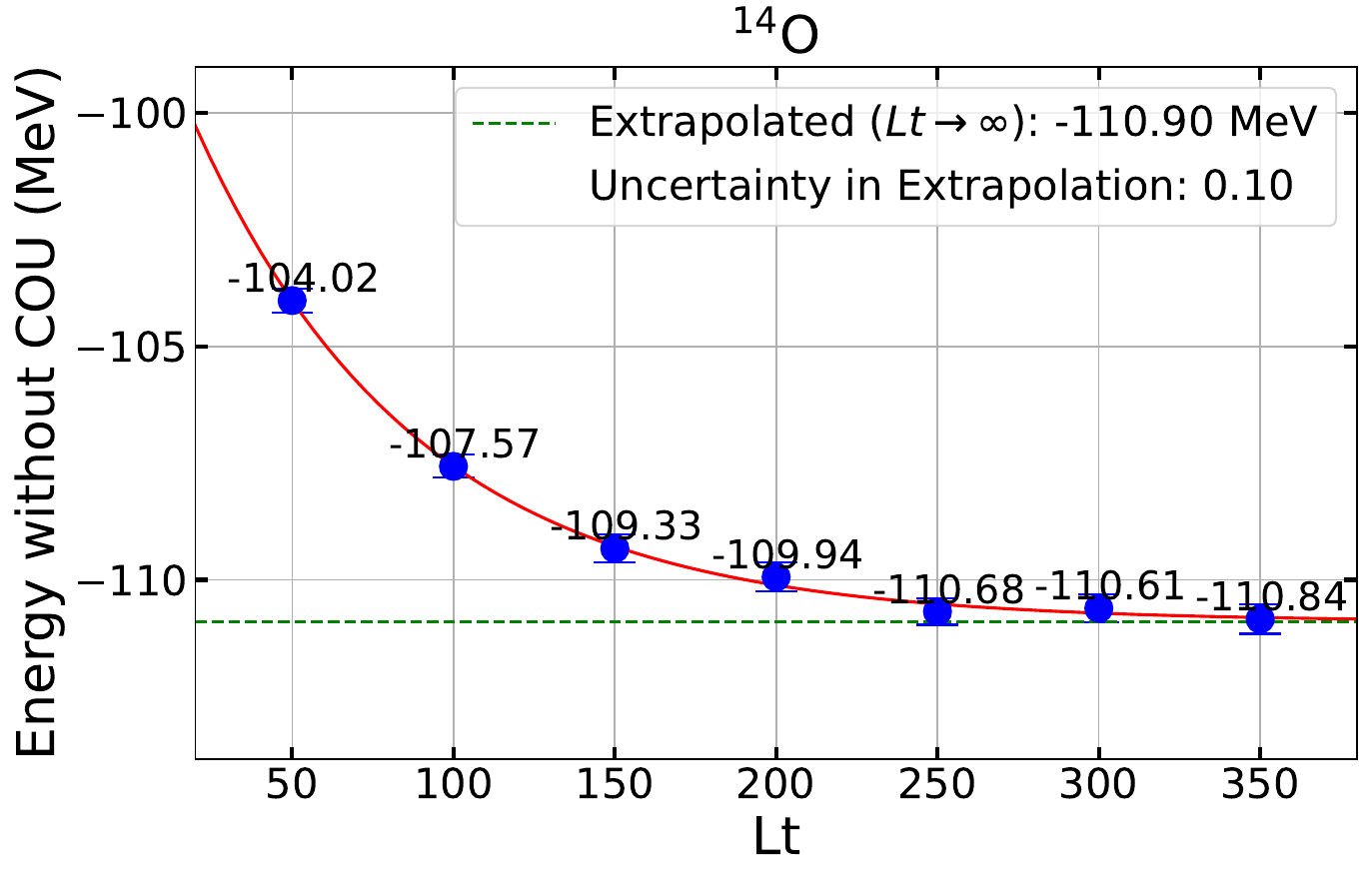}
    \end{minipage}%
    \begin{minipage}{0.46\textwidth}
        \centering
        \includegraphics[width=\textwidth]{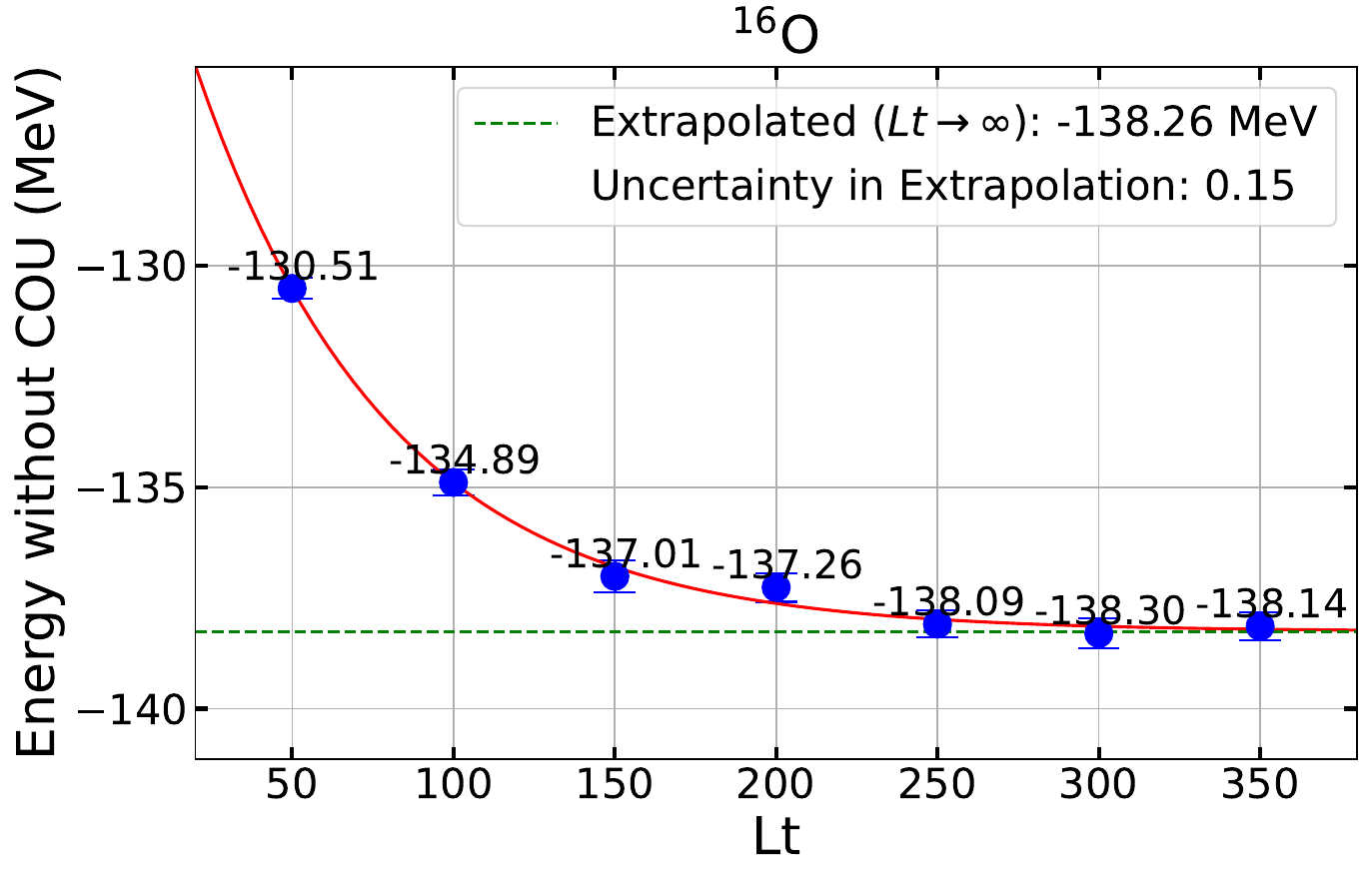}
    \end{minipage}%

\end{figure}

\begin{figure}[H]
    \vspace{0.5cm}
    \begin{minipage}{0.46\textwidth}
        \centering
        \includegraphics[width=\textwidth]{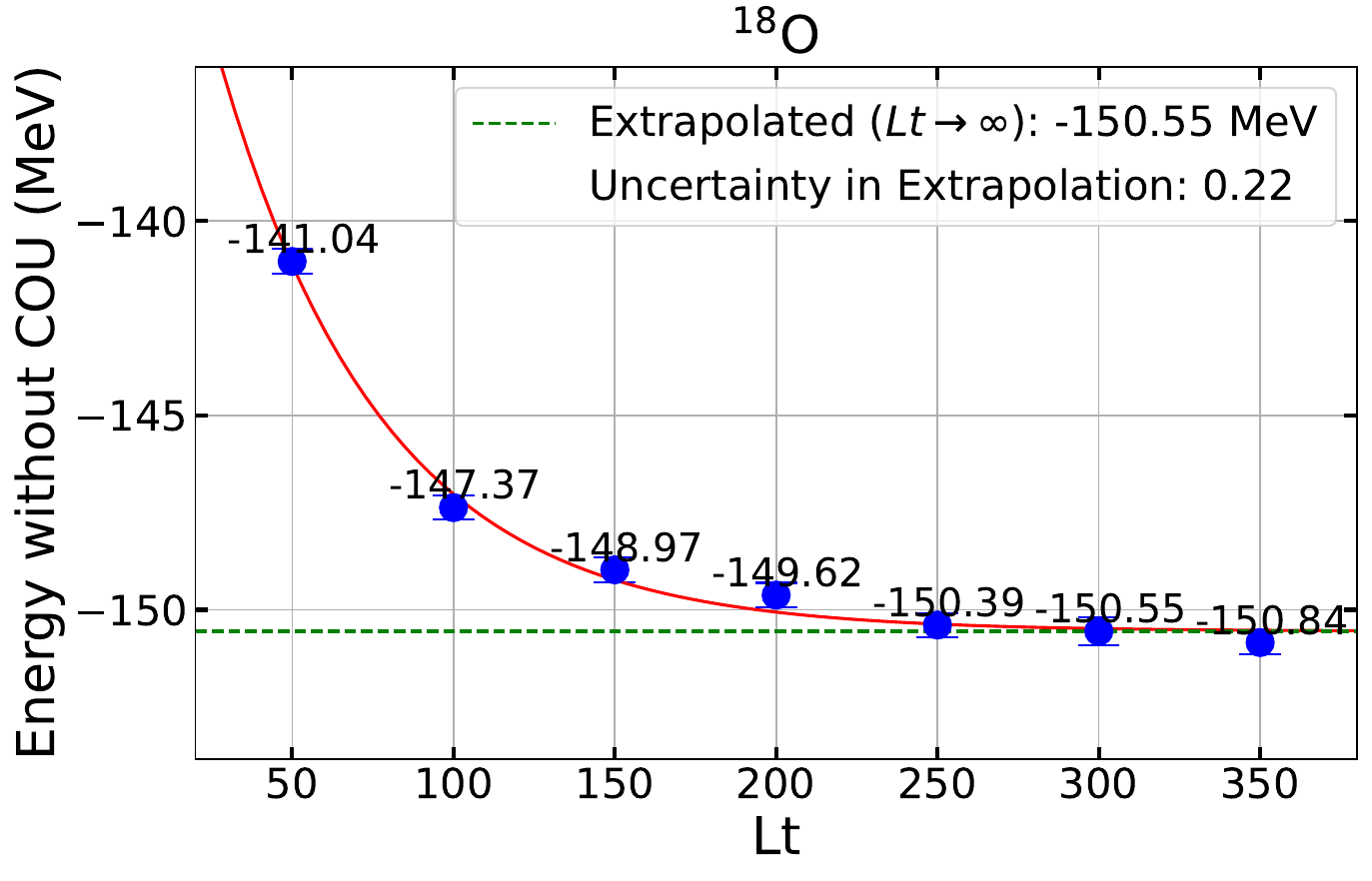}
    \end{minipage}%
    \begin{minipage}{0.46\textwidth}
        \centering
        \includegraphics[width=\textwidth]{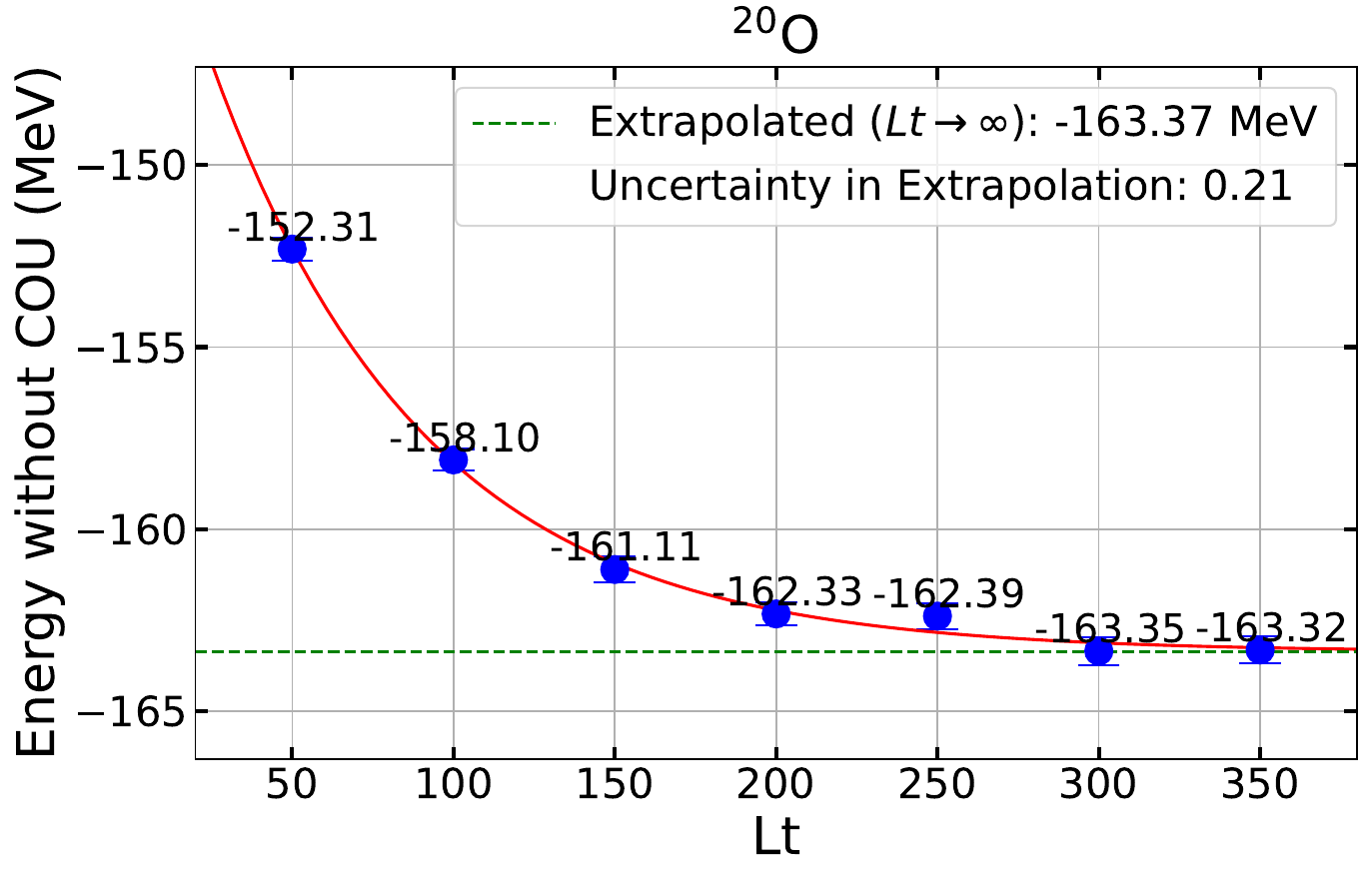}
    \end{minipage}%

    \vspace{0.5cm}
    
    \begin{minipage}{0.46\textwidth}
        \centering
        \includegraphics[width=\textwidth]{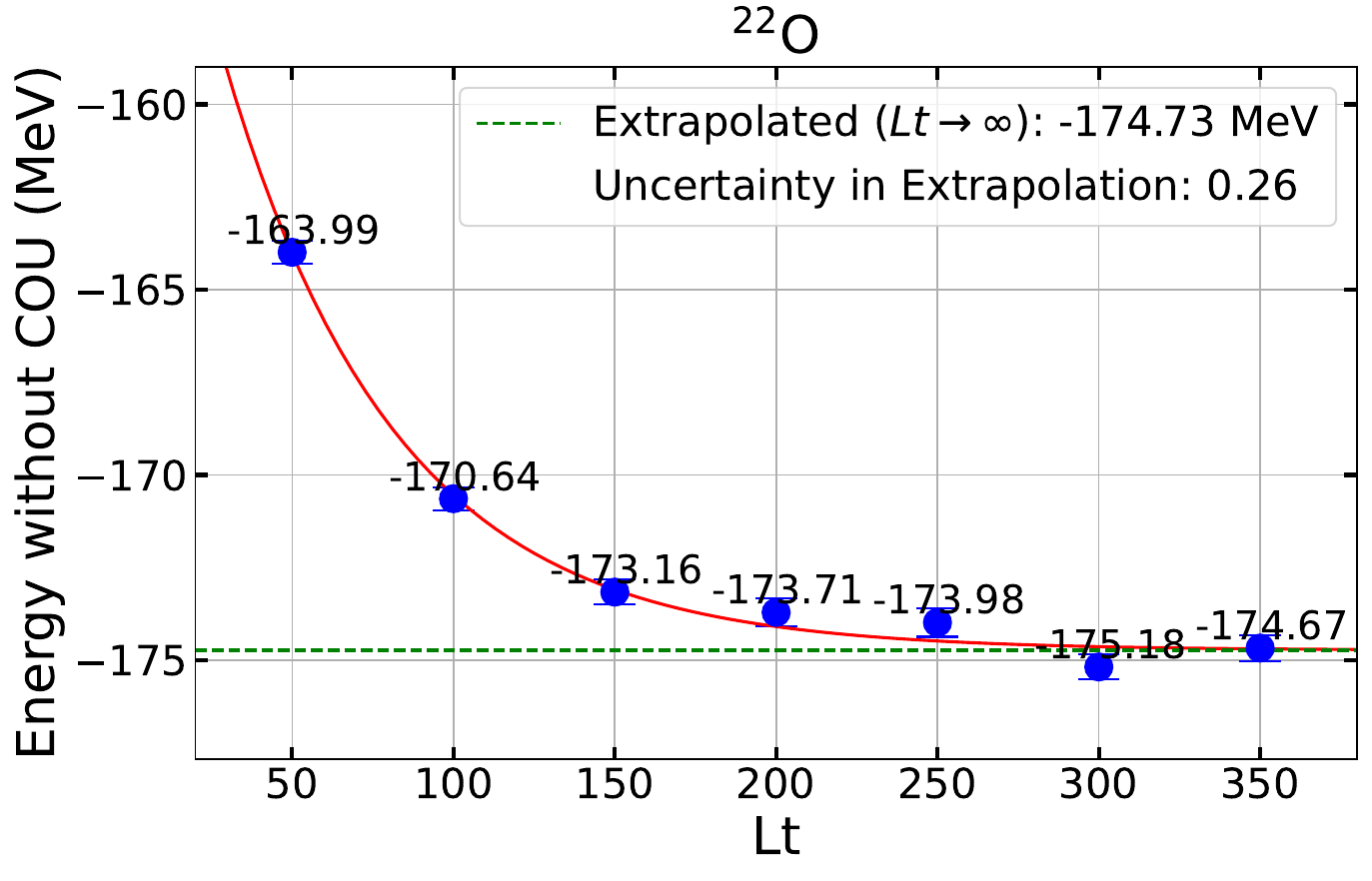}
    \end{minipage}%
    \begin{minipage}{0.46\textwidth}
        \centering
        \includegraphics[width=\textwidth]{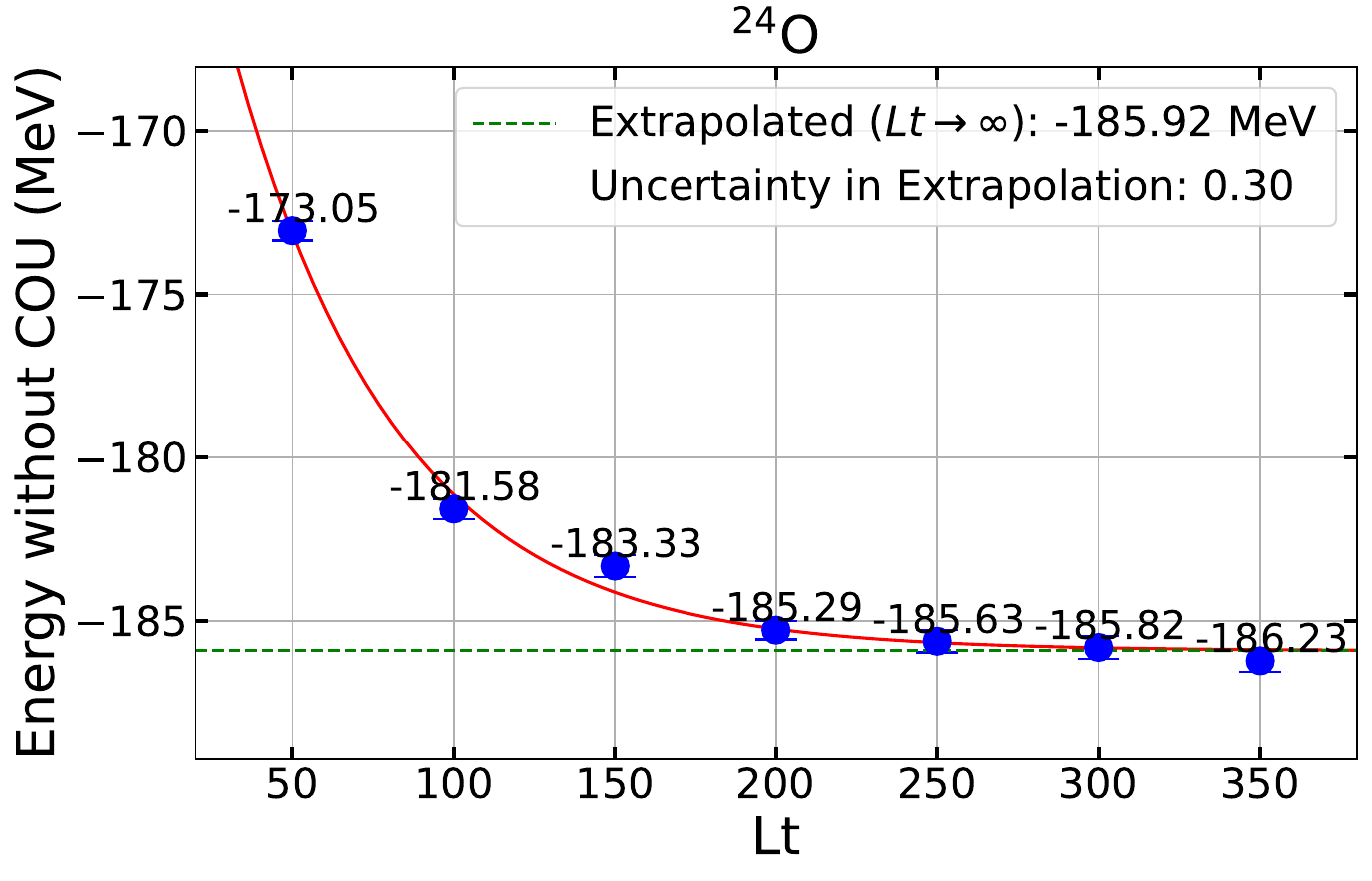}
    \end{minipage}%

    \newpage
    \vspace{0.5cm}
    \begin{minipage}{0.46\textwidth}
        \centering
        \includegraphics[width=\textwidth]{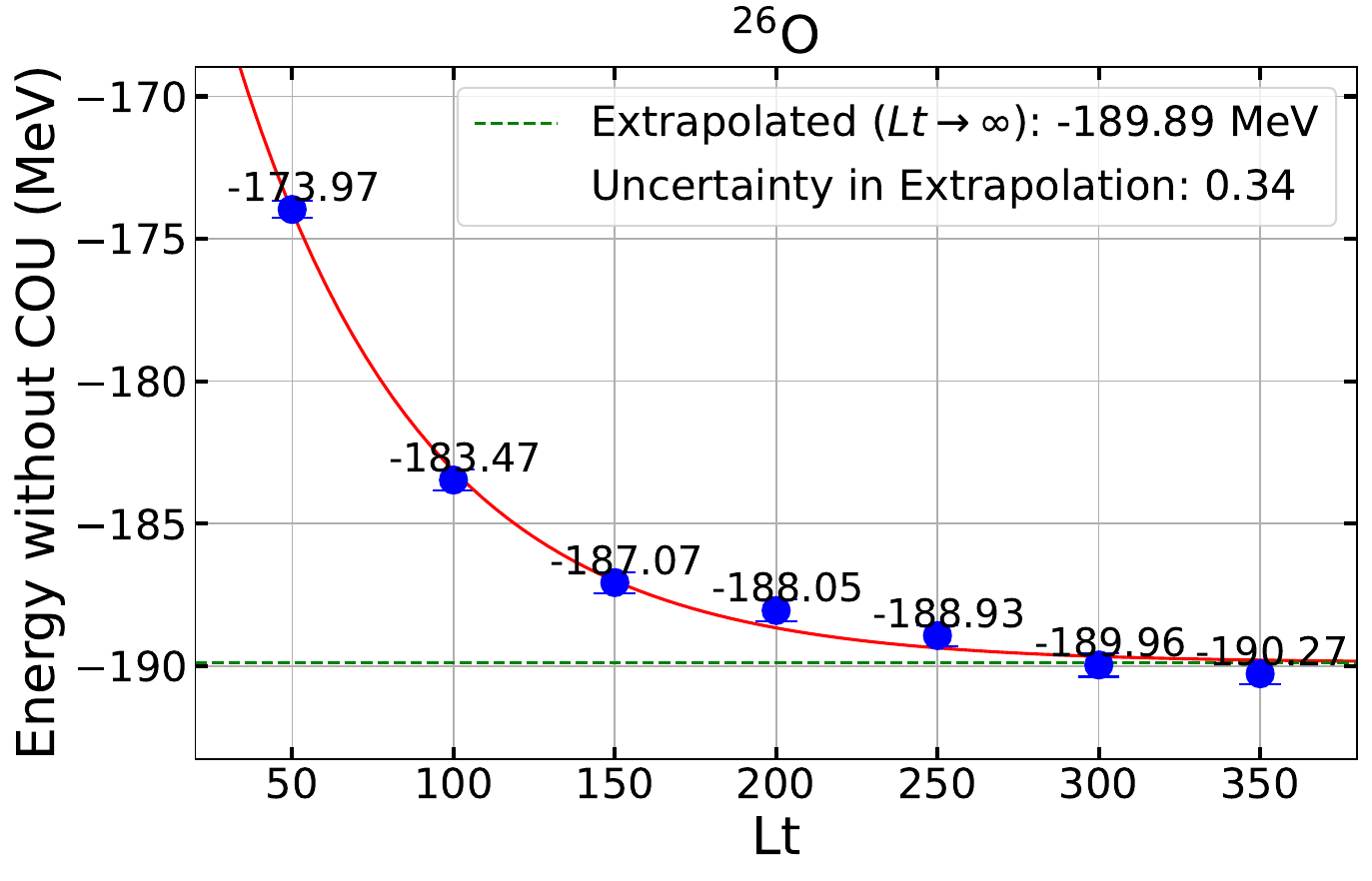}
    \end{minipage}%
    \begin{minipage}{0.46\textwidth}
        \centering
        \includegraphics[width=\textwidth]{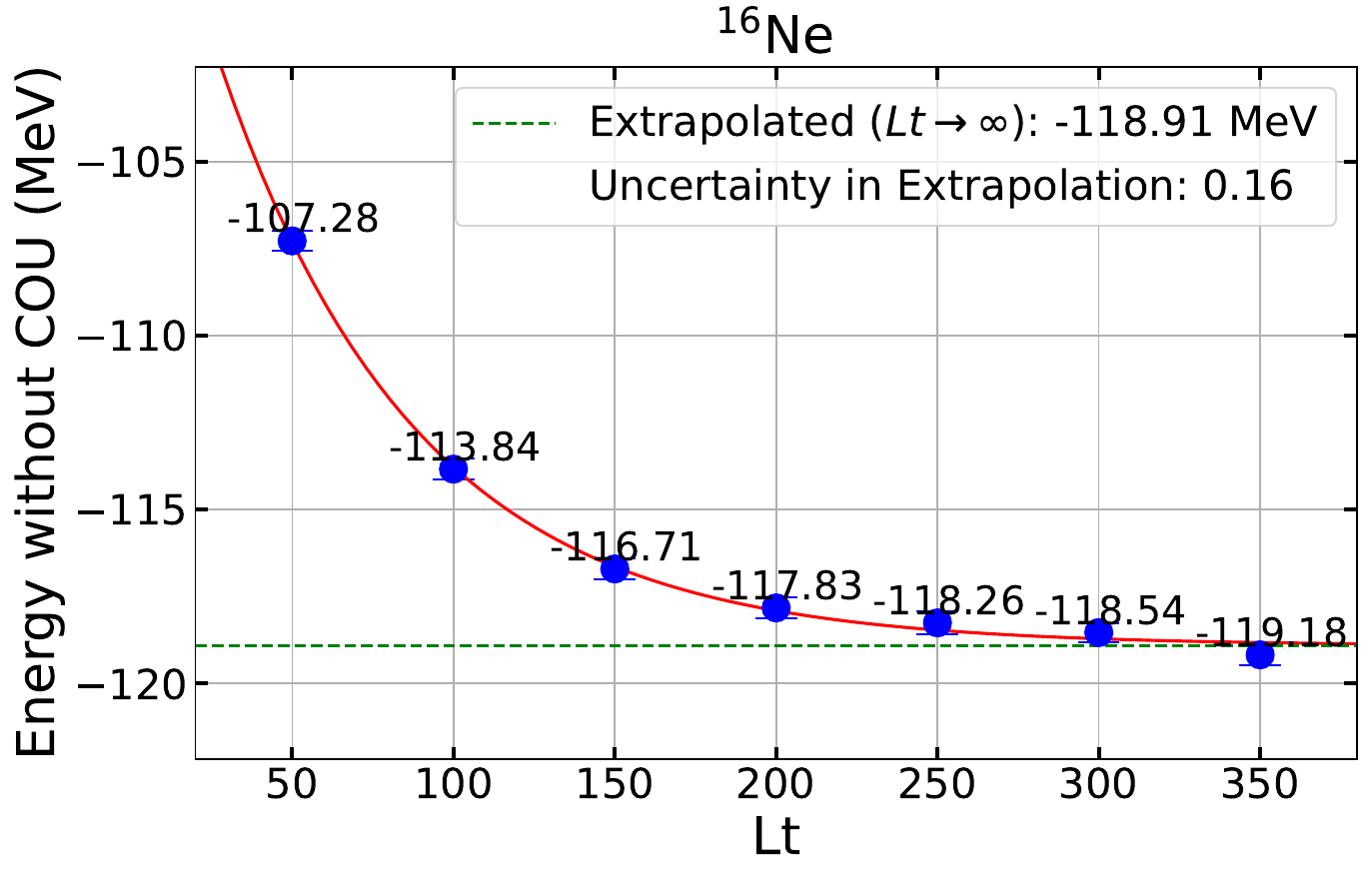}
    \end{minipage}%

    \vspace{0.5cm}
    
    \begin{minipage}{0.46\textwidth}
        \centering
        \includegraphics[width=\textwidth]{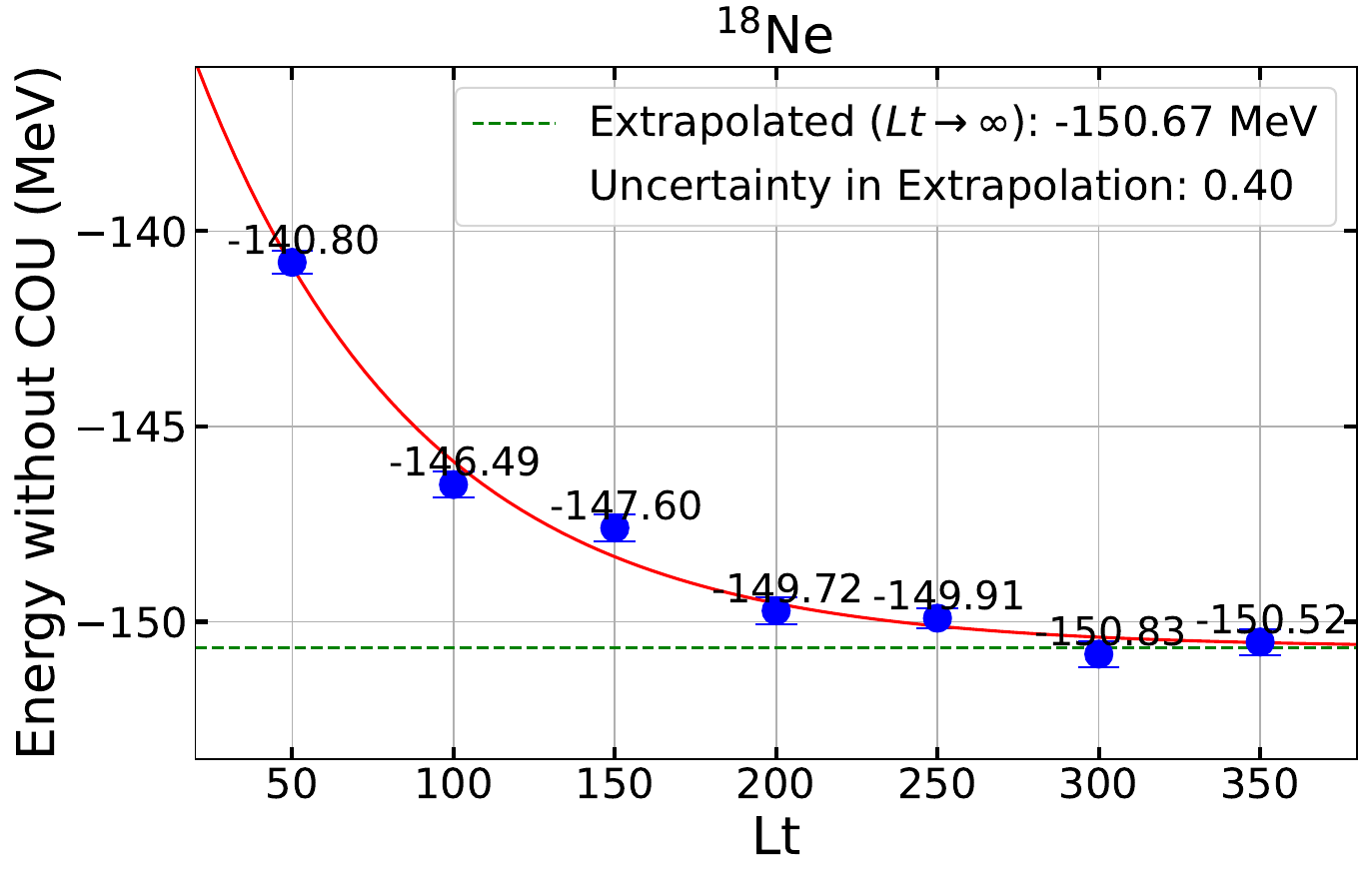}
    \end{minipage}%
    \begin{minipage}{0.46\textwidth}
        \centering
        \includegraphics[width=\textwidth]{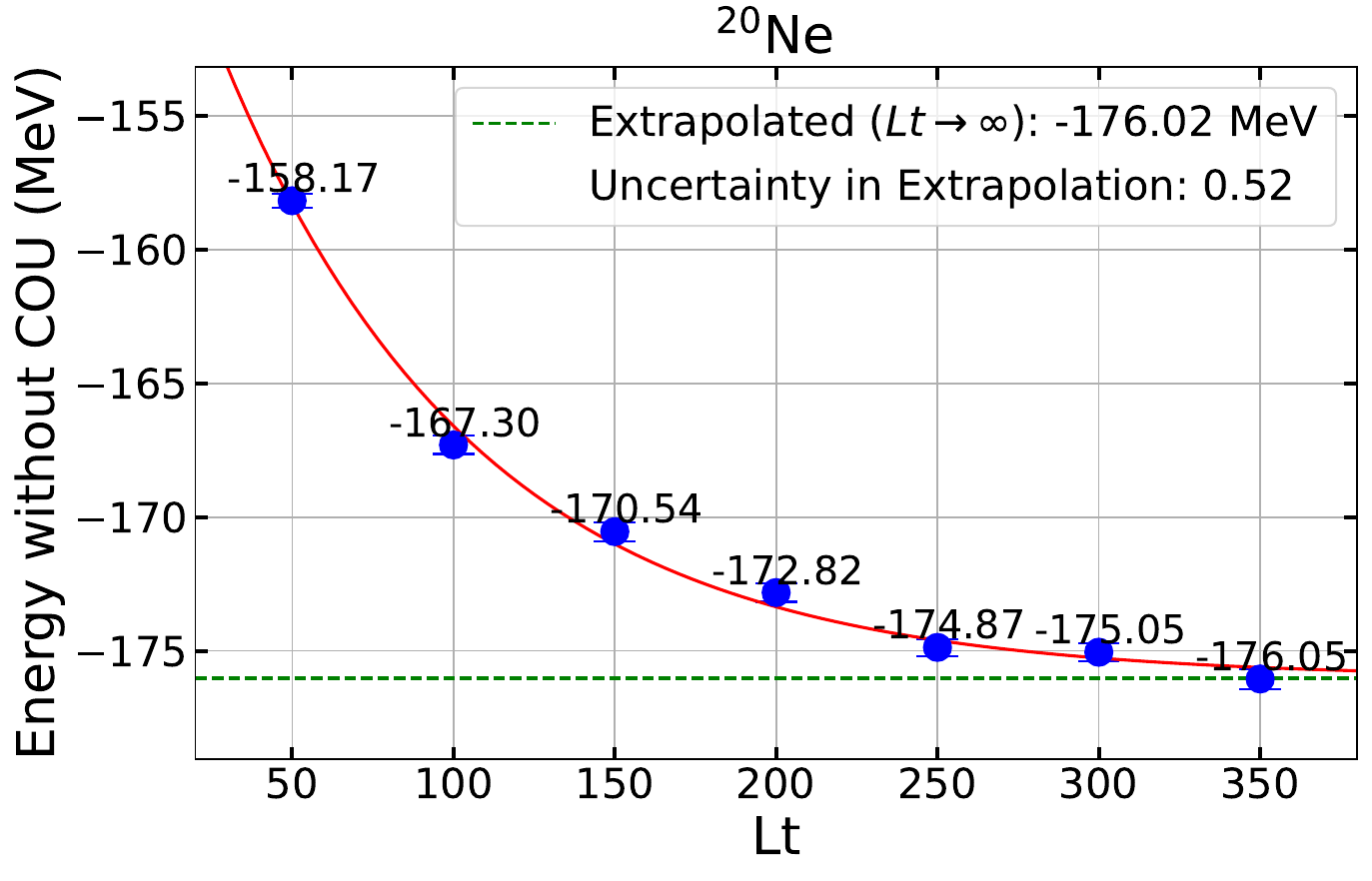}
    \end{minipage}%
\end{figure}

\begin{figure}[H]

    \vspace{0.5cm}
    \begin{minipage}{0.46\textwidth}
        \centering
        \includegraphics[width=\textwidth]{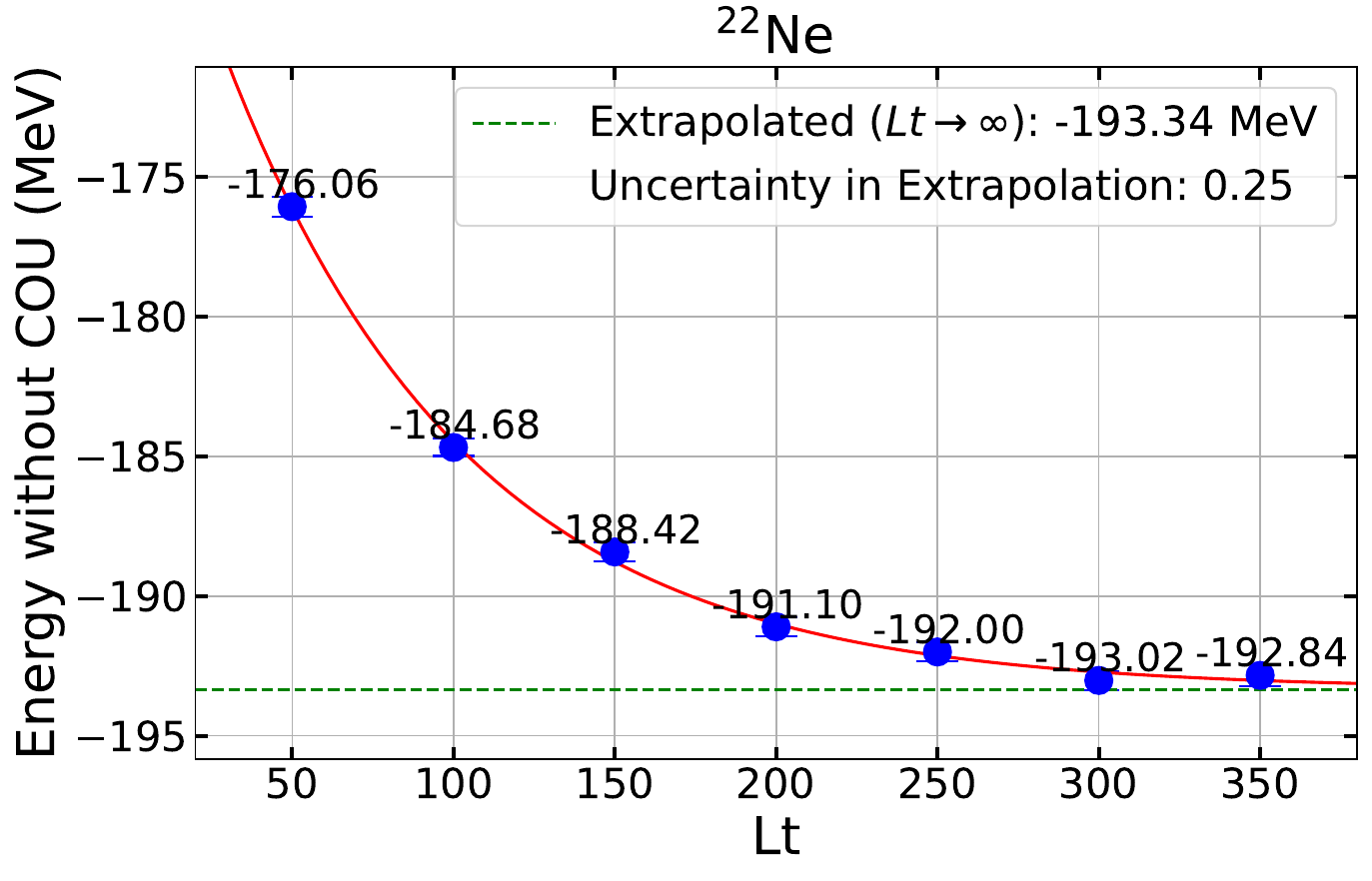}
    \end{minipage}%
    \begin{minipage}{0.46\textwidth}
        \centering
        \includegraphics[width=\textwidth]{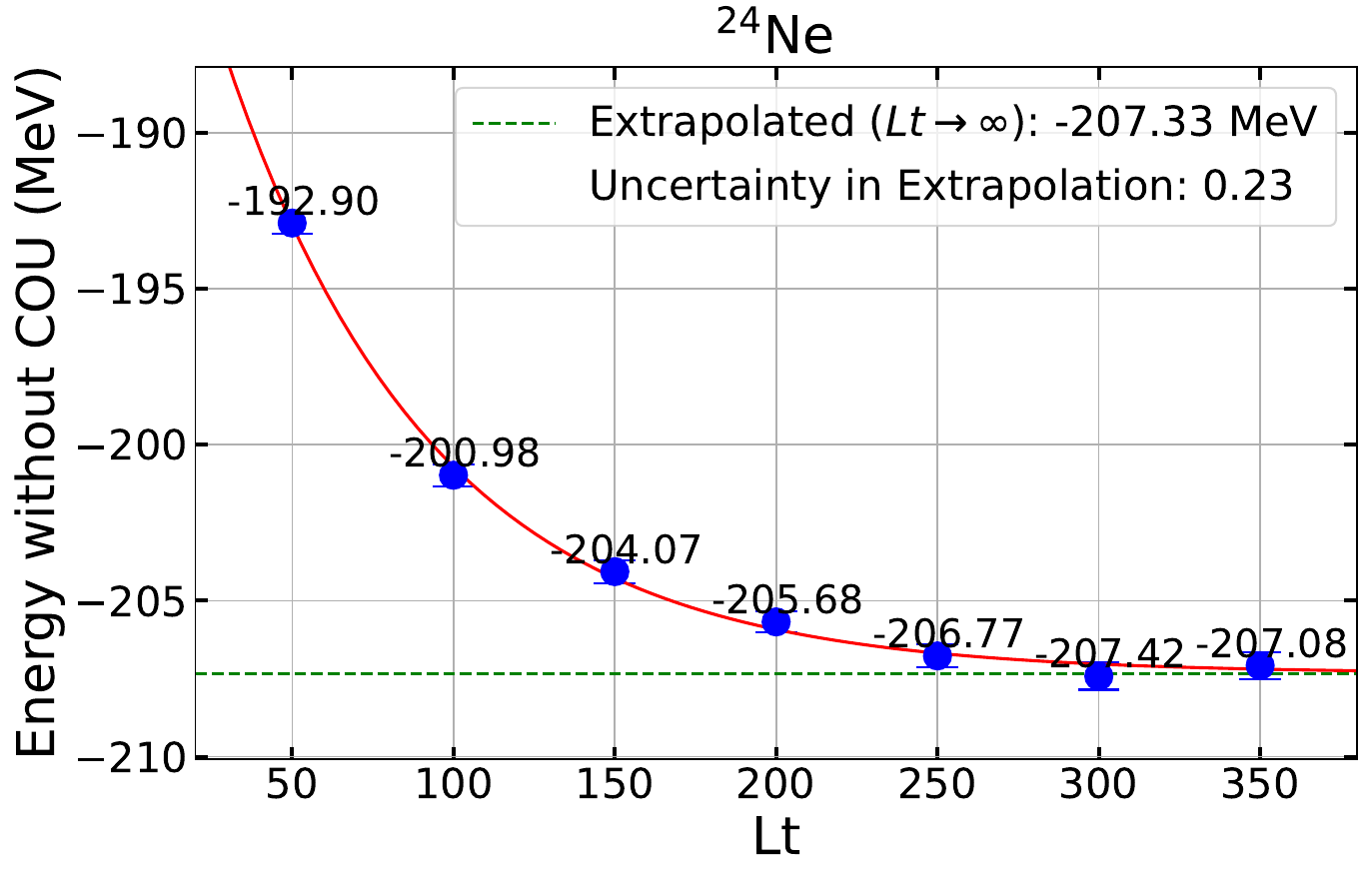}
    \end{minipage}%

    \vspace{0.5cm}
    
    \begin{minipage}{0.46\textwidth}
        \centering
        \includegraphics[width=\textwidth]{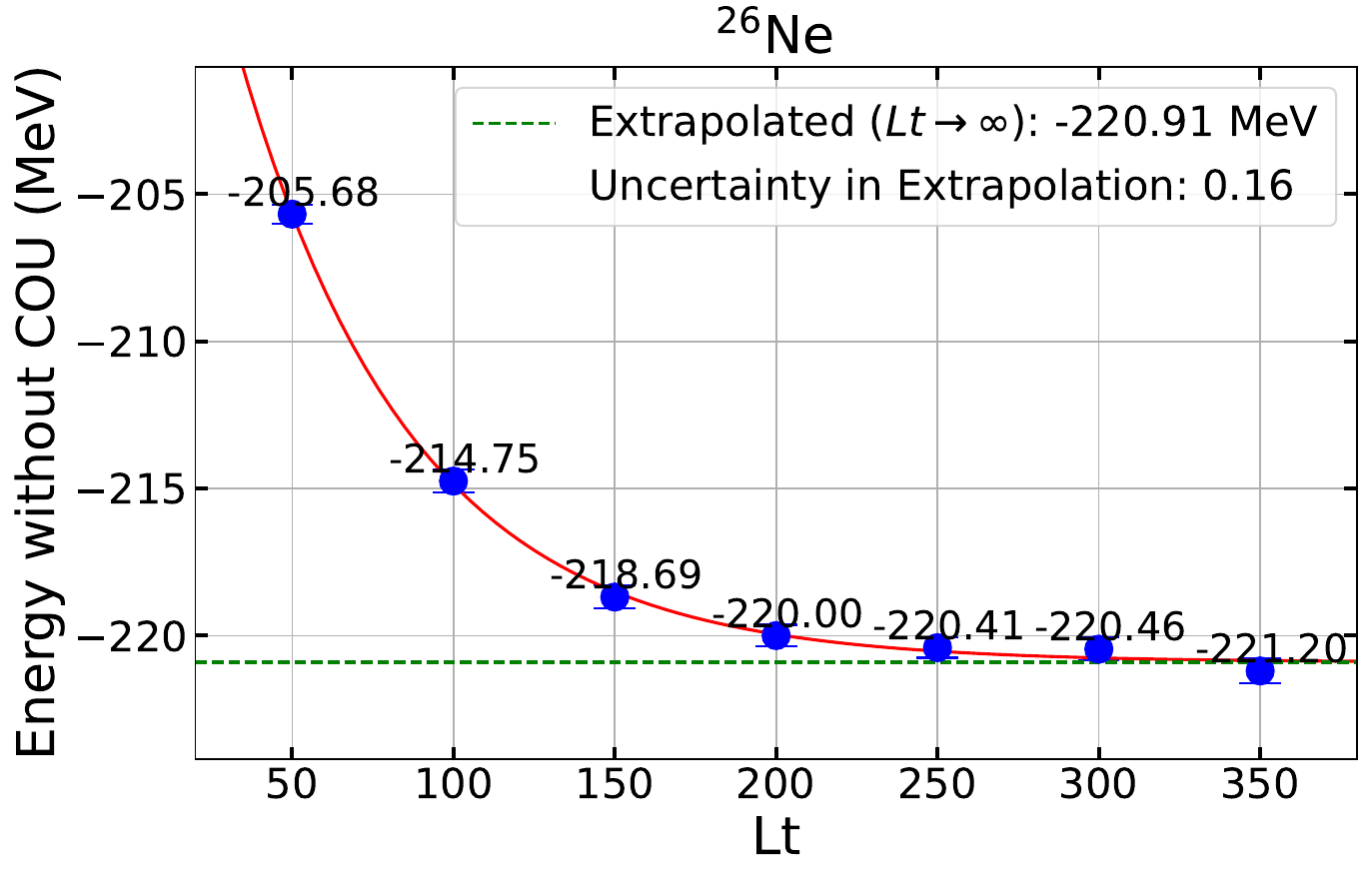}
    \end{minipage}%
    \begin{minipage}{0.46\textwidth}
        \centering
        \includegraphics[width=\textwidth]{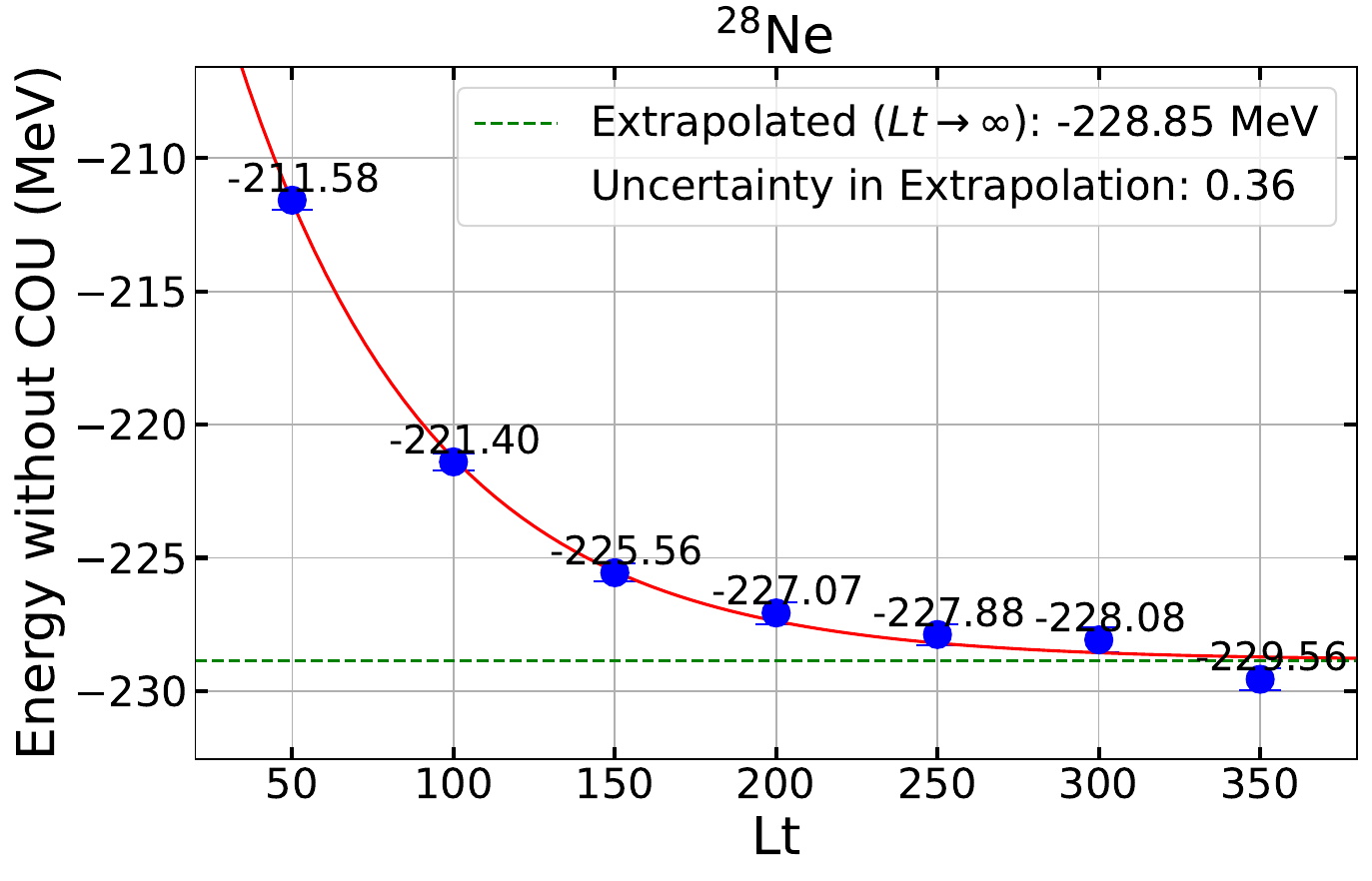}
    \end{minipage}%

    \newpage
    \vspace{0.5cm}
    \begin{minipage}{0.46\textwidth}
        \centering
        \includegraphics[width=\textwidth]{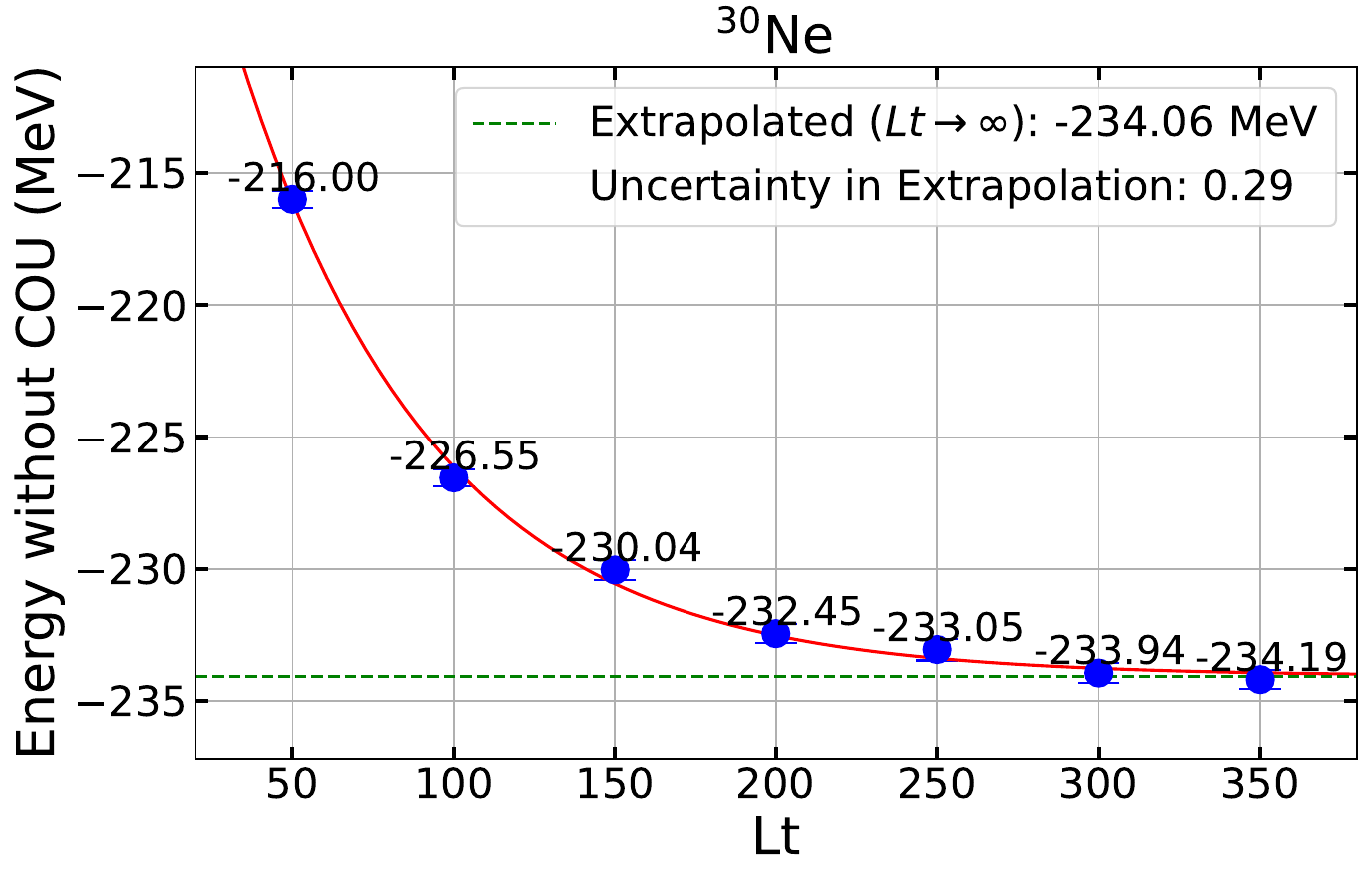}
    \end{minipage}%
    \begin{minipage}{0.46\textwidth}
        \centering
        \includegraphics[width=\textwidth]{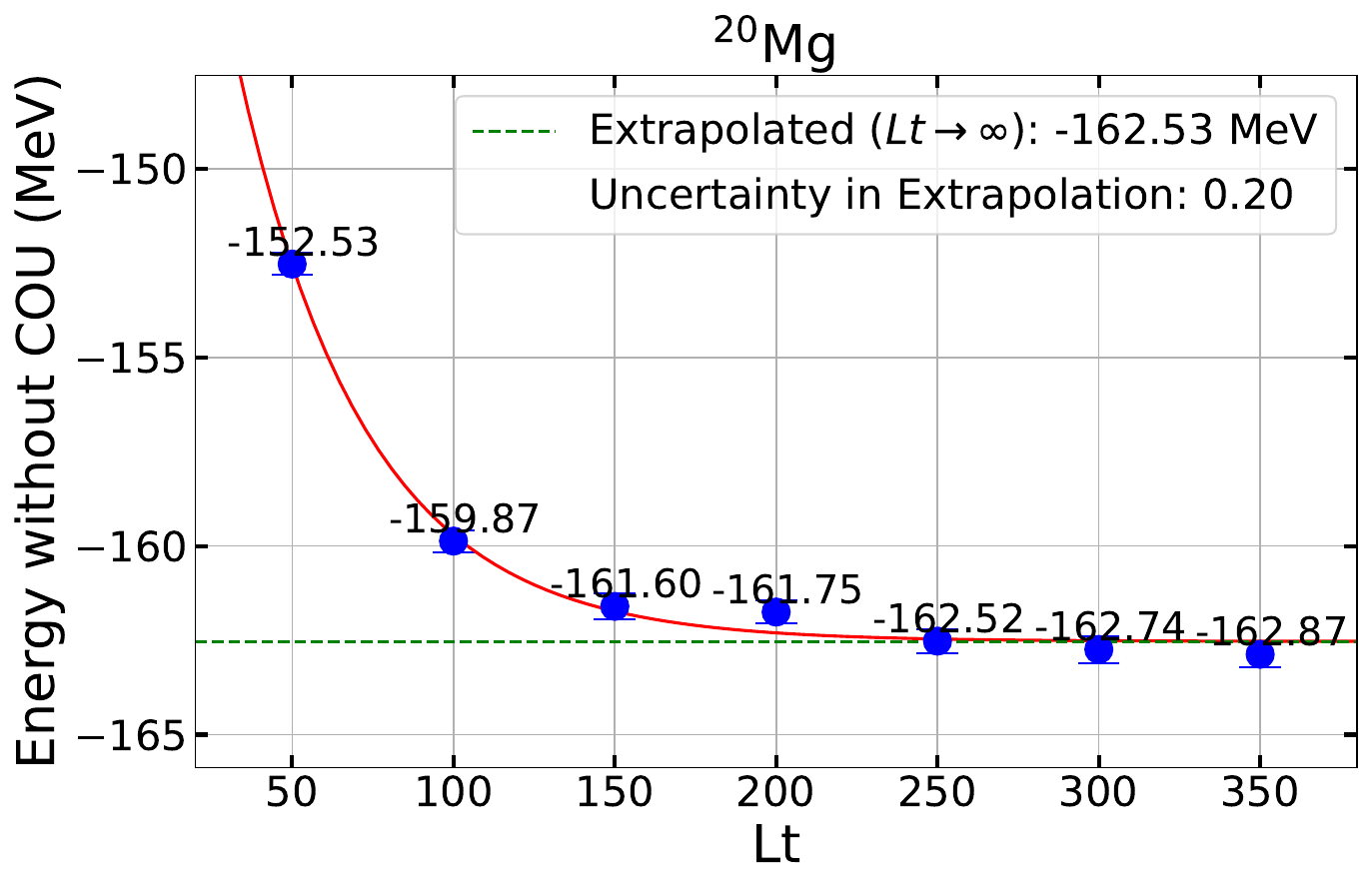}
    \end{minipage}%

    \vspace{0.5cm}
    
    \begin{minipage}{0.46\textwidth}
        \centering
        \includegraphics[width=\textwidth]{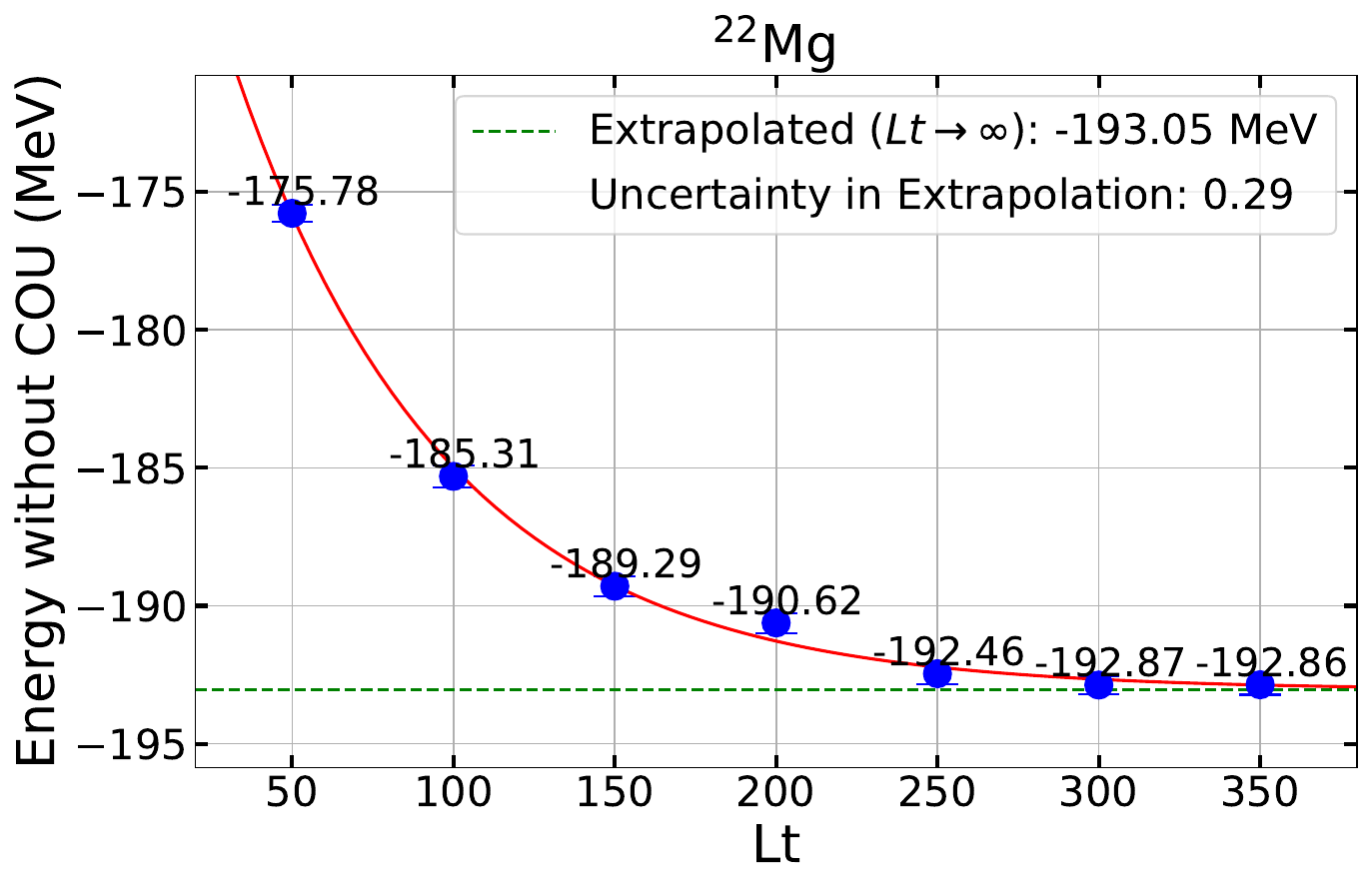}
    \end{minipage}%
    \begin{minipage}{0.46\textwidth}
        \centering
        \includegraphics[width=\textwidth]{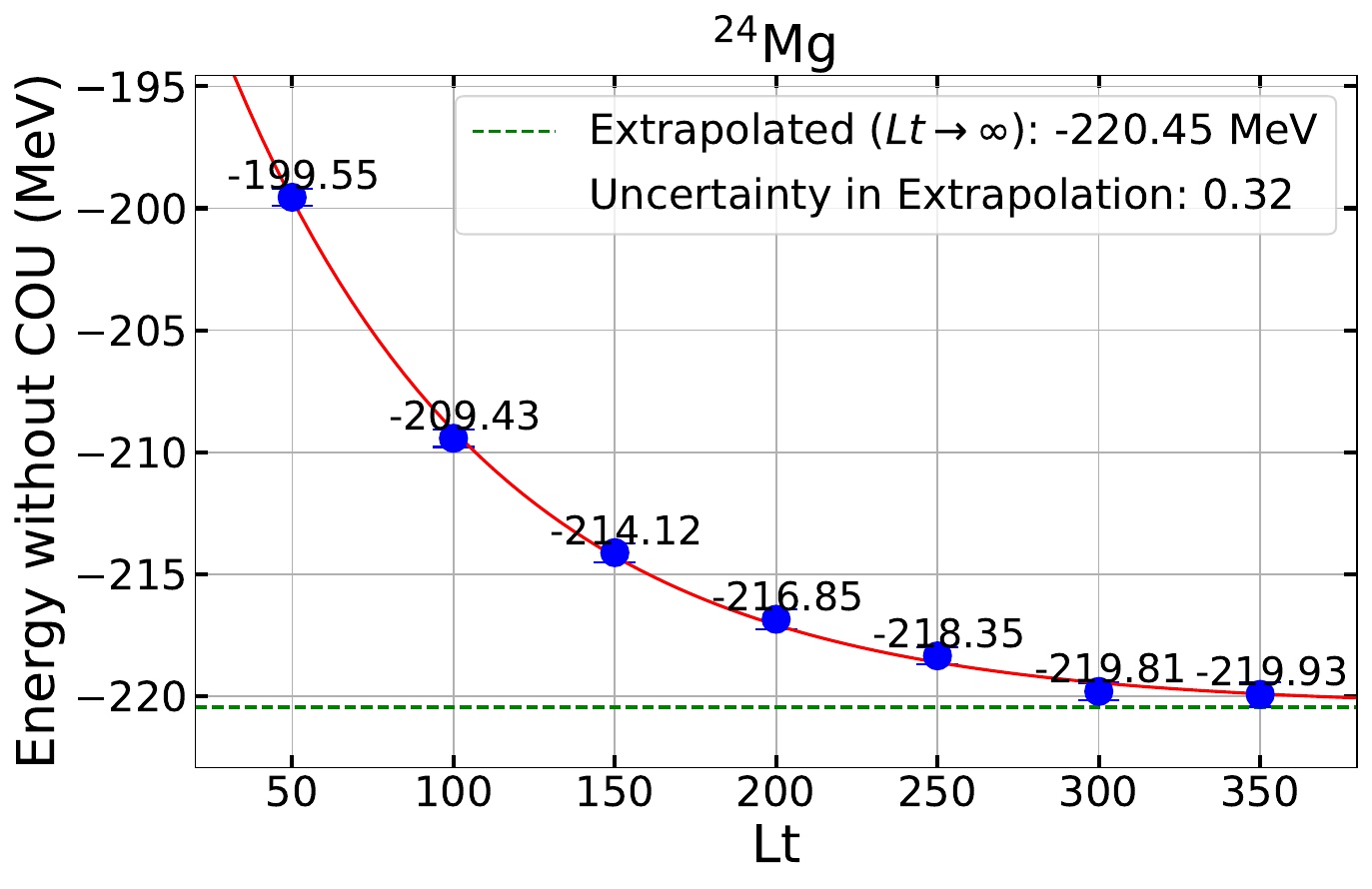}
    \end{minipage}%

\end{figure}

\begin{figure}[H]
    \vspace{0.5cm}

    \begin{minipage}{0.46\textwidth}
        \centering
        \includegraphics[width=\textwidth]{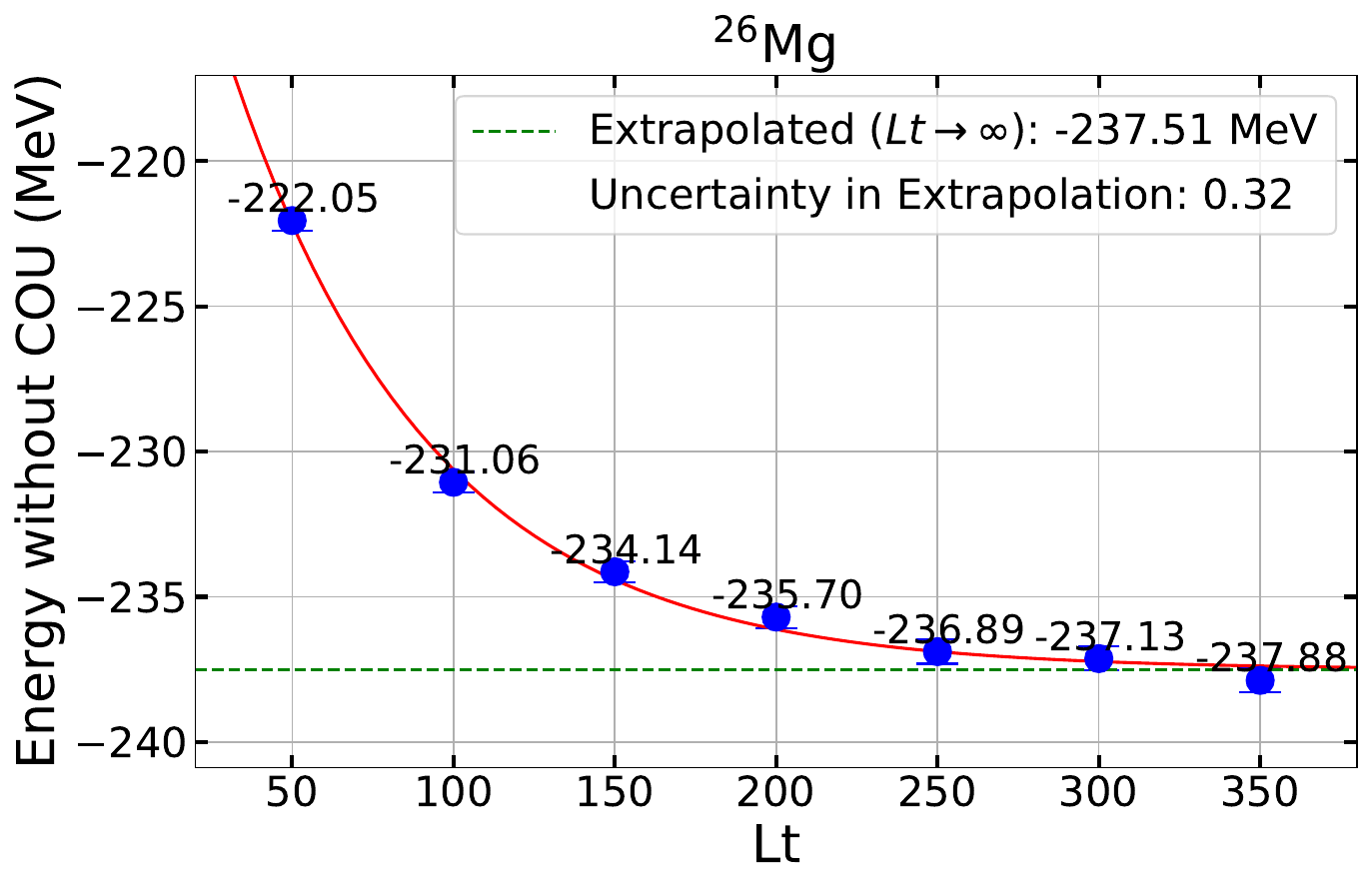}
    \end{minipage}%
    \begin{minipage}{0.46\textwidth}
        \centering
        \includegraphics[width=\textwidth]{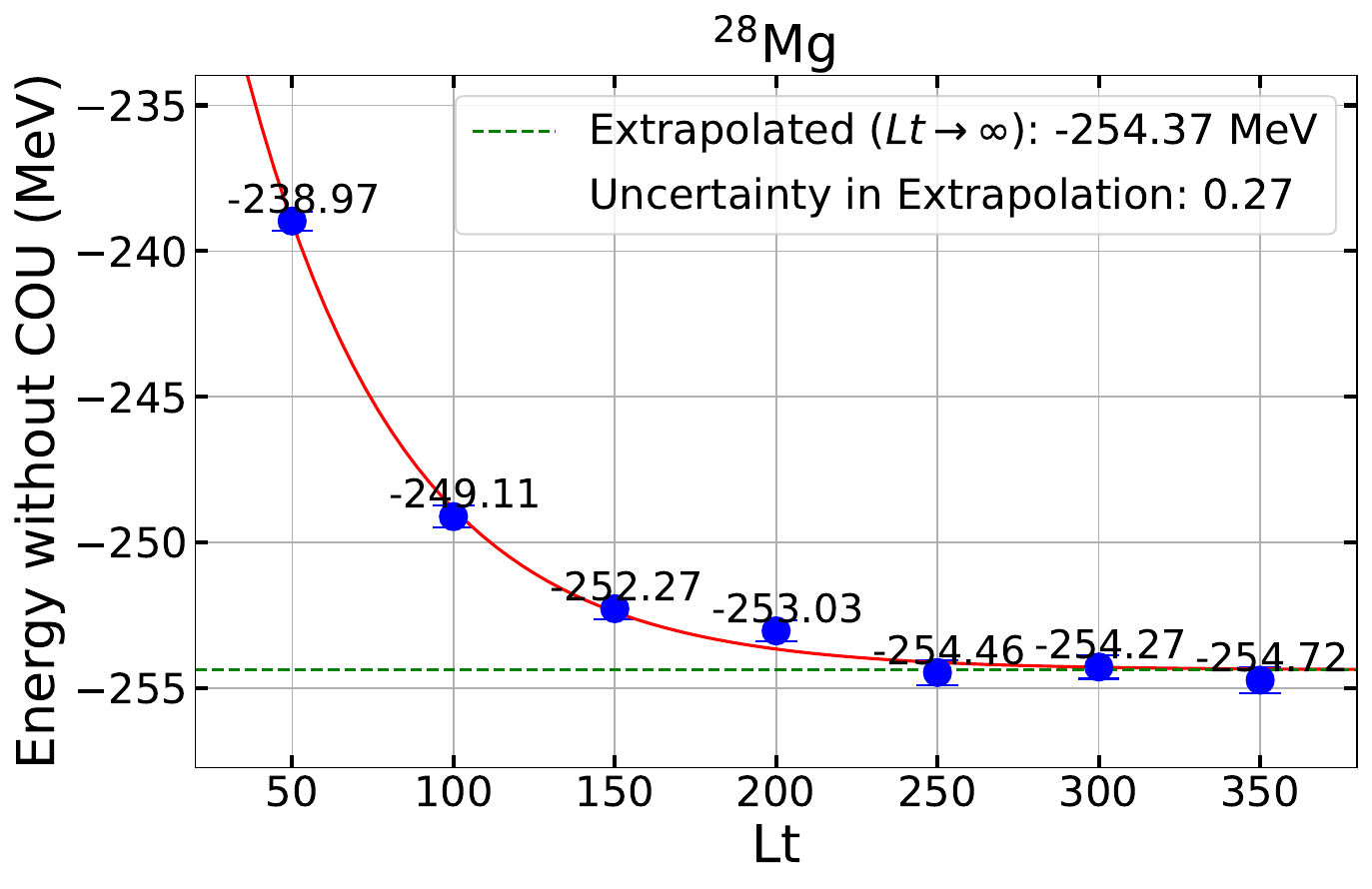}
    \end{minipage}%

    \vspace{0.5cm}
    
    \begin{minipage}{0.46\textwidth}
        \centering
        \includegraphics[width=\textwidth]{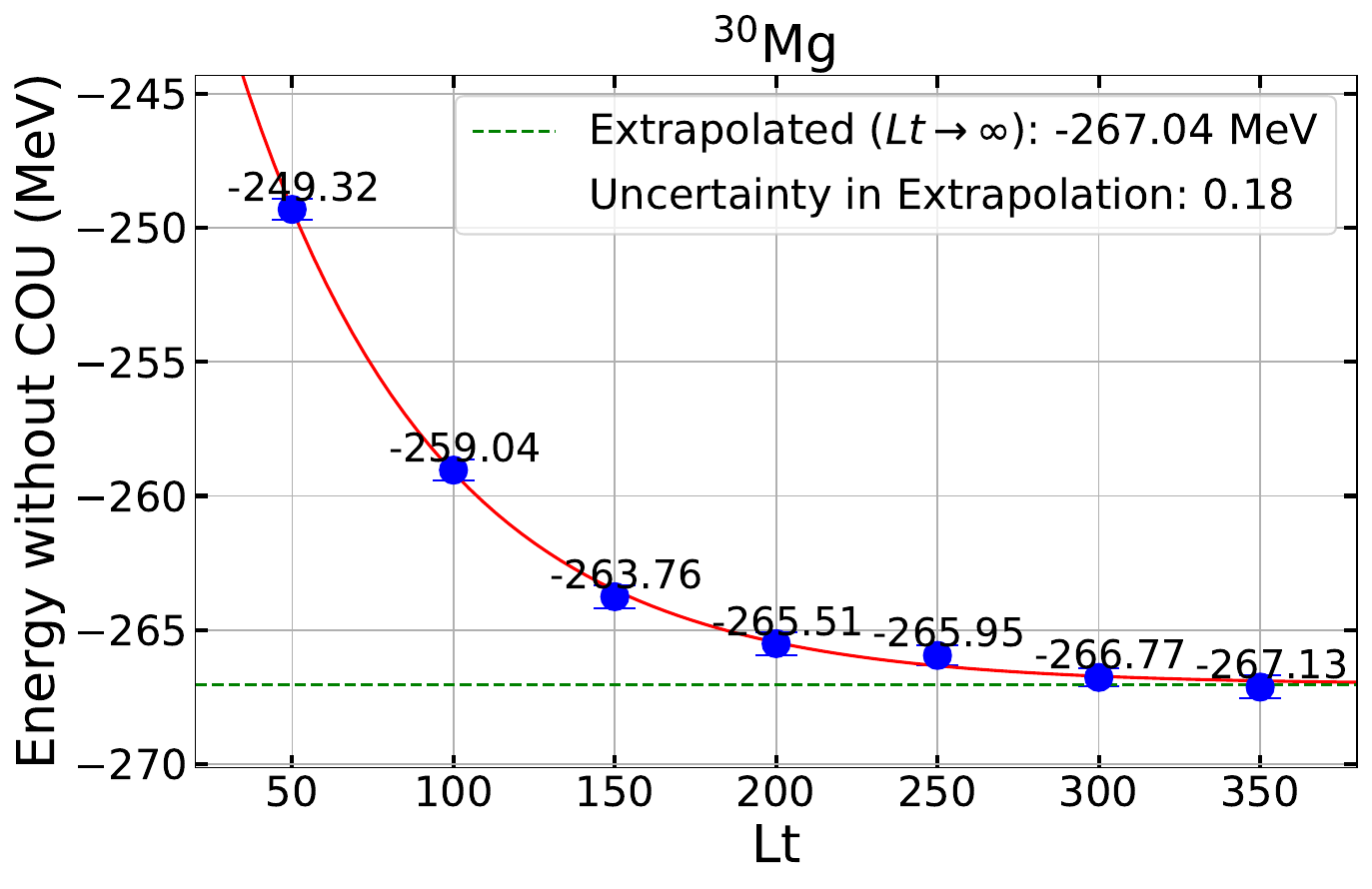}
    \end{minipage}%
    \begin{minipage}{0.46\textwidth}
        \centering
        \includegraphics[width=\textwidth]{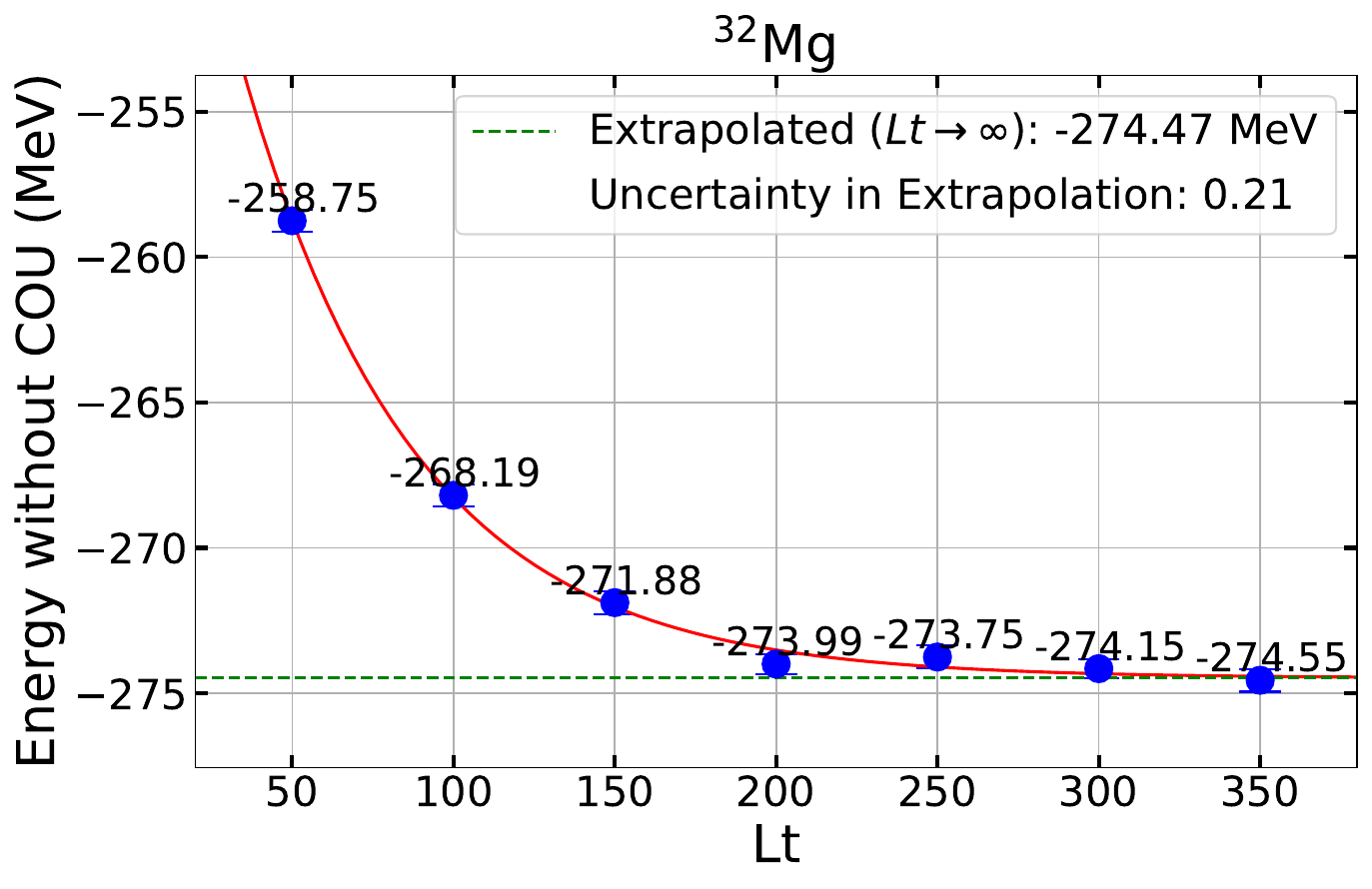}
    \end{minipage}%

    \vspace{0.5cm}

    \begin{minipage}{0.46\textwidth}
        \centering
        \includegraphics[width=\textwidth]{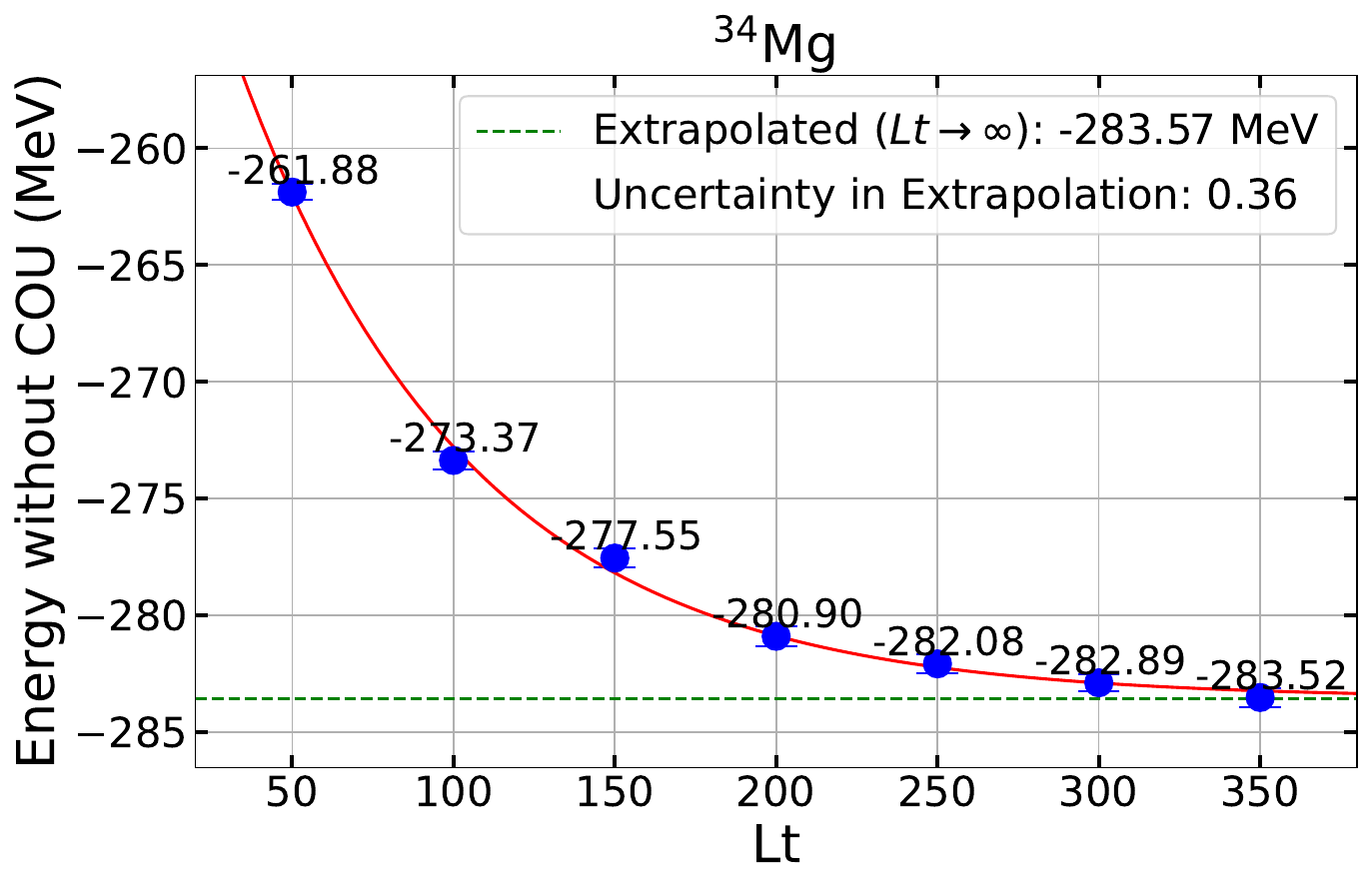}
    \end{minipage}%
    \begin{minipage}{0.46\textwidth}
        \centering
        \includegraphics[width=\textwidth]{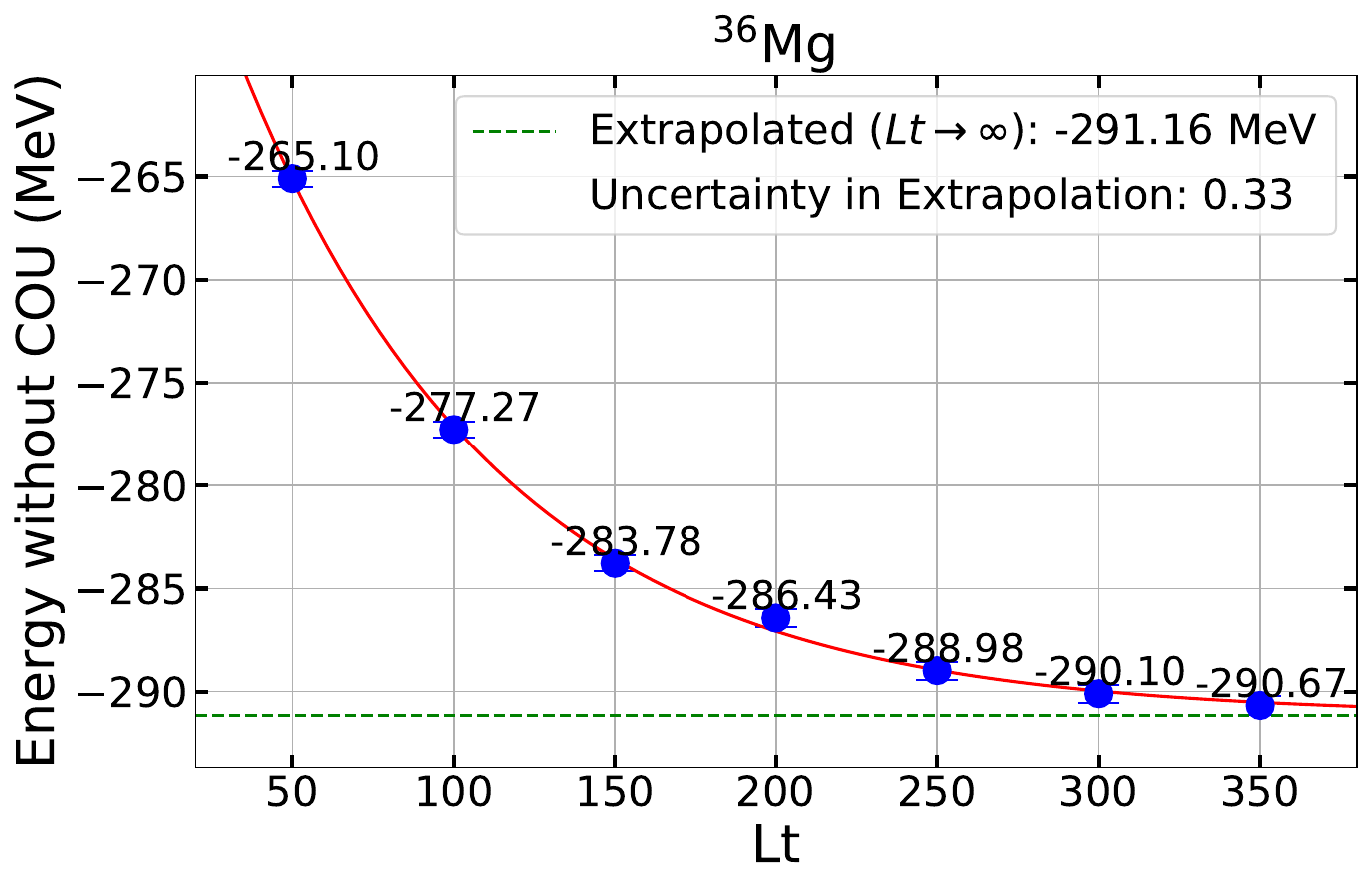}
    \end{minipage}%

    \newpage
    \vspace{0.5cm}
    
    \begin{minipage}{0.46\textwidth}
        \centering
        \includegraphics[width=\textwidth]{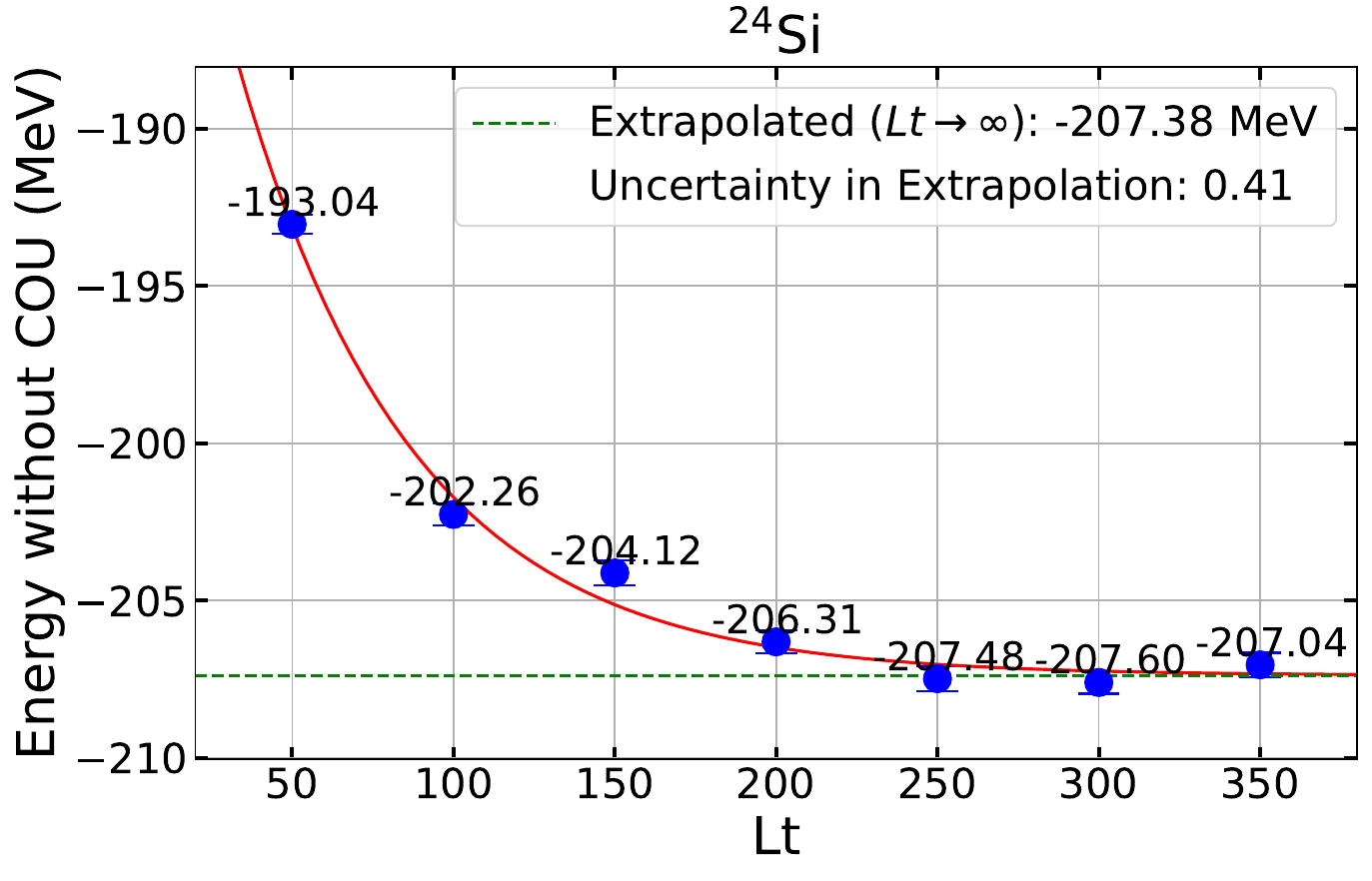}
    \end{minipage}%
    \begin{minipage}{0.46\textwidth}
        \centering
        \includegraphics[width=\textwidth]{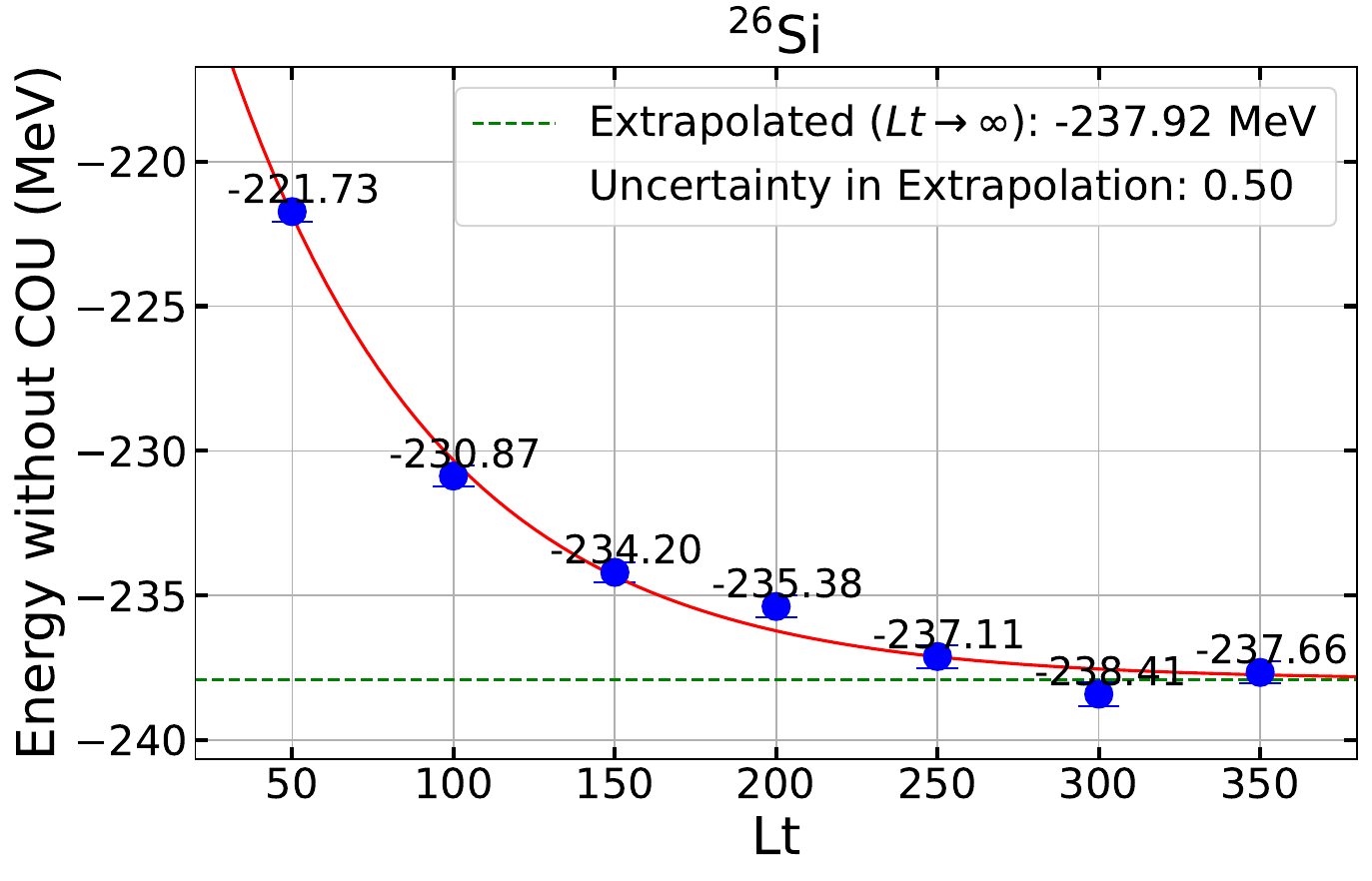}
    \end{minipage}%

\end{figure}

\begin{figure}[H]
    \vspace{0.5cm}

    \begin{minipage}{0.46\textwidth}
        \centering
        \includegraphics[width=\textwidth]{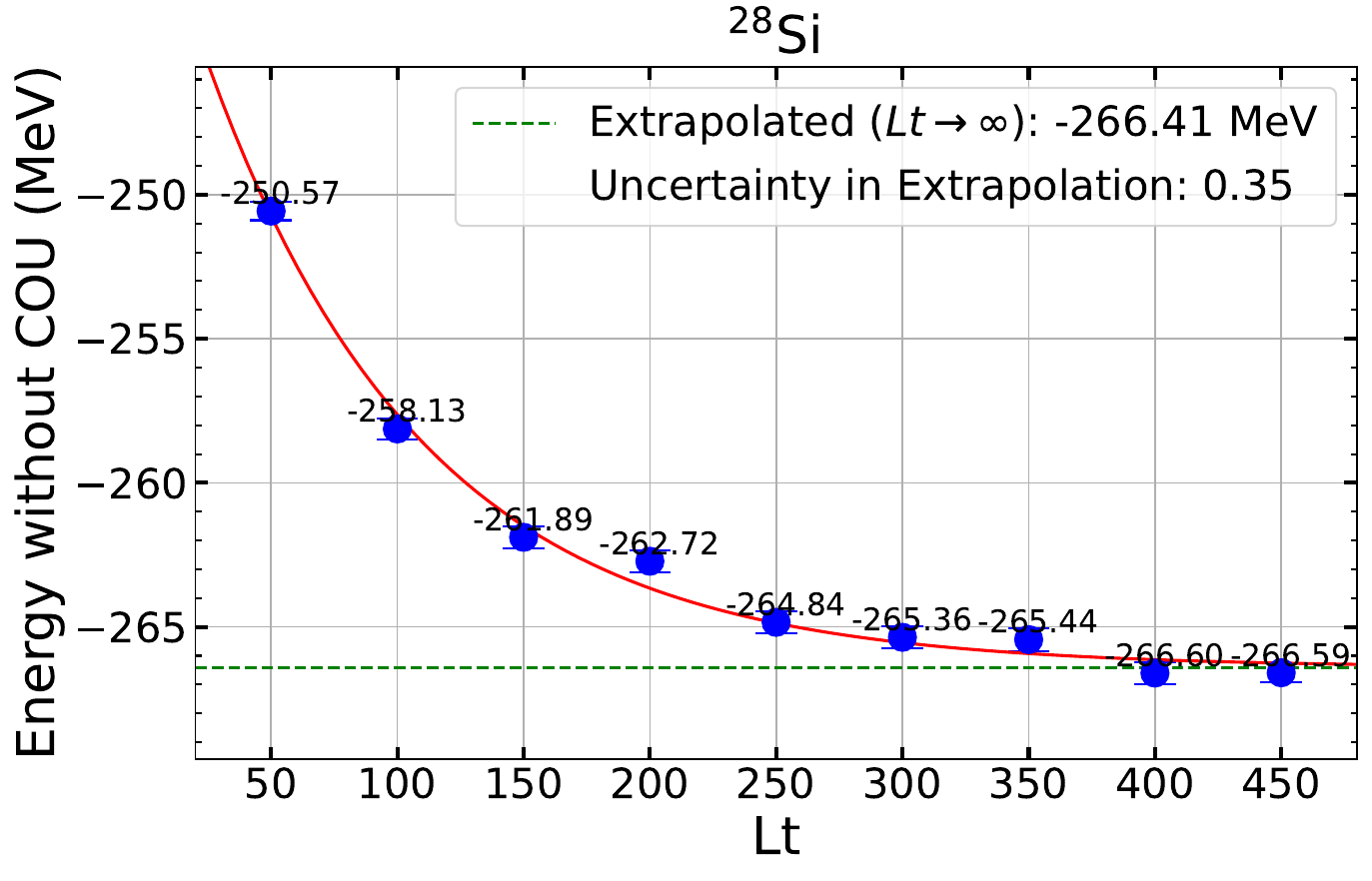}
    \end{minipage}%
    \begin{minipage}{0.46\textwidth}
        \centering
        \includegraphics[width=\textwidth]{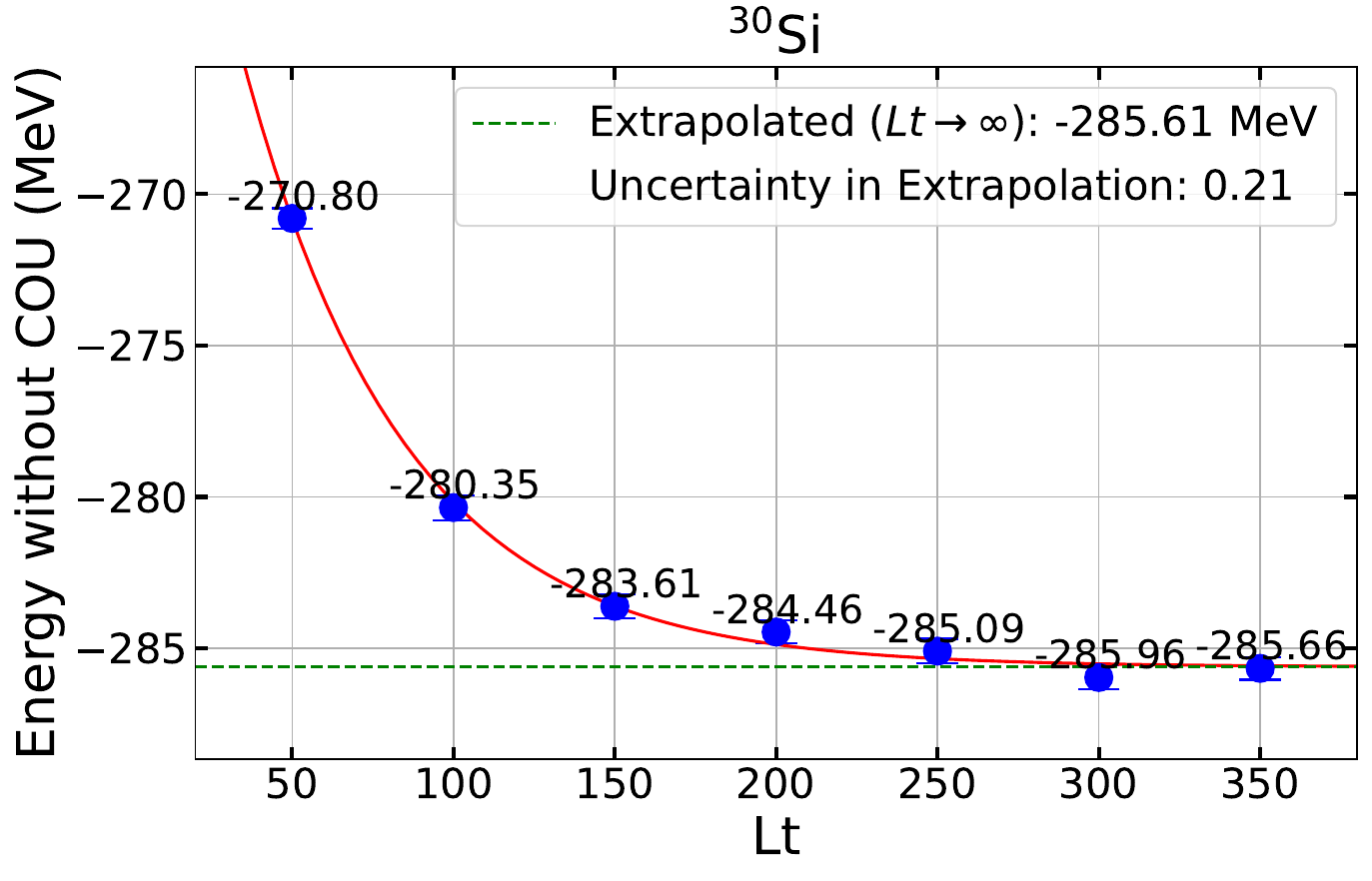}
    \end{minipage}%

    \vspace{0.5cm}
    
    \begin{minipage}{0.46\textwidth}
        \centering
        \includegraphics[width=\textwidth]{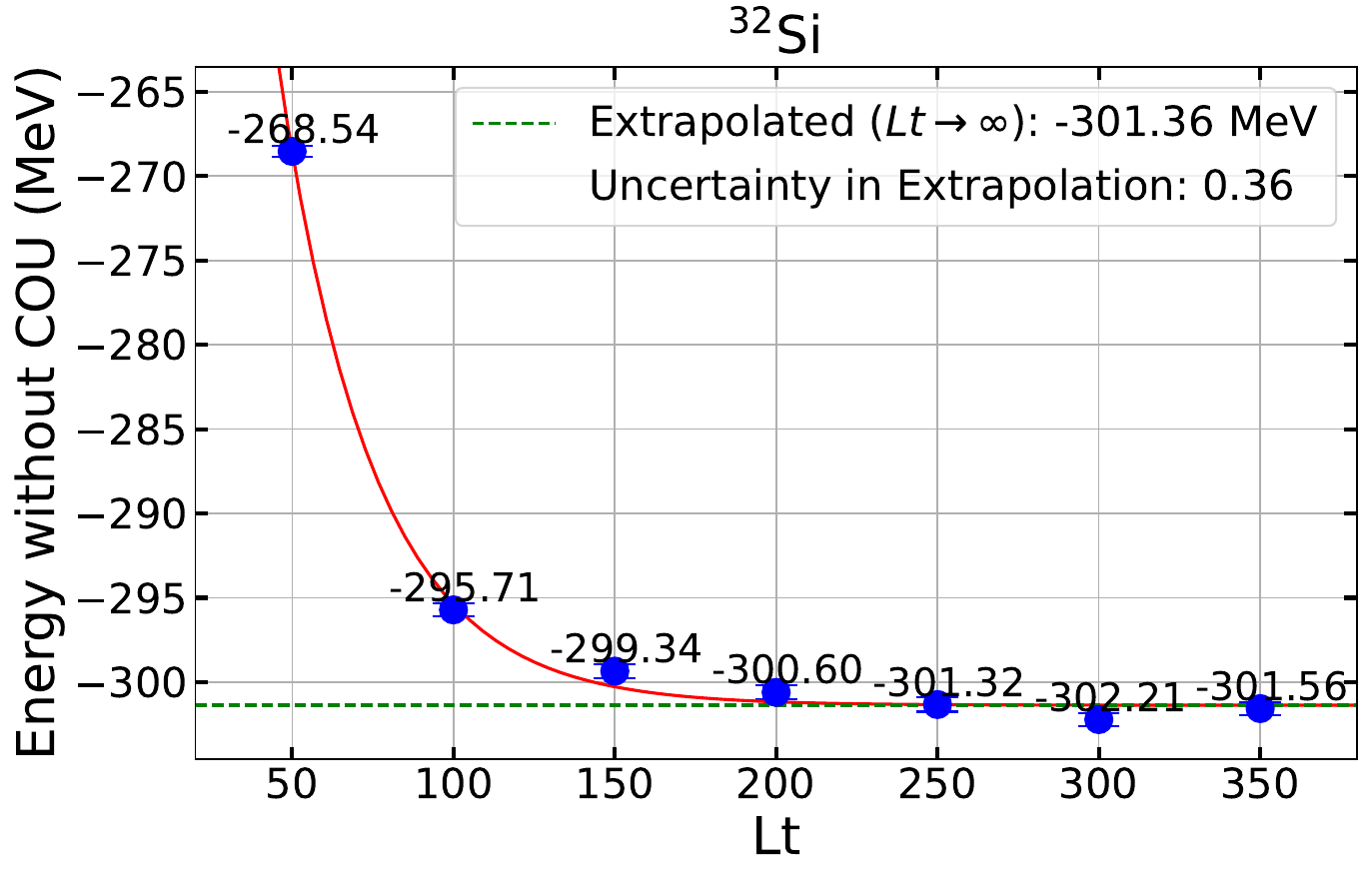}
    \end{minipage}%
    \begin{minipage}{0.46\textwidth}
        \centering
        \includegraphics[width=\textwidth]{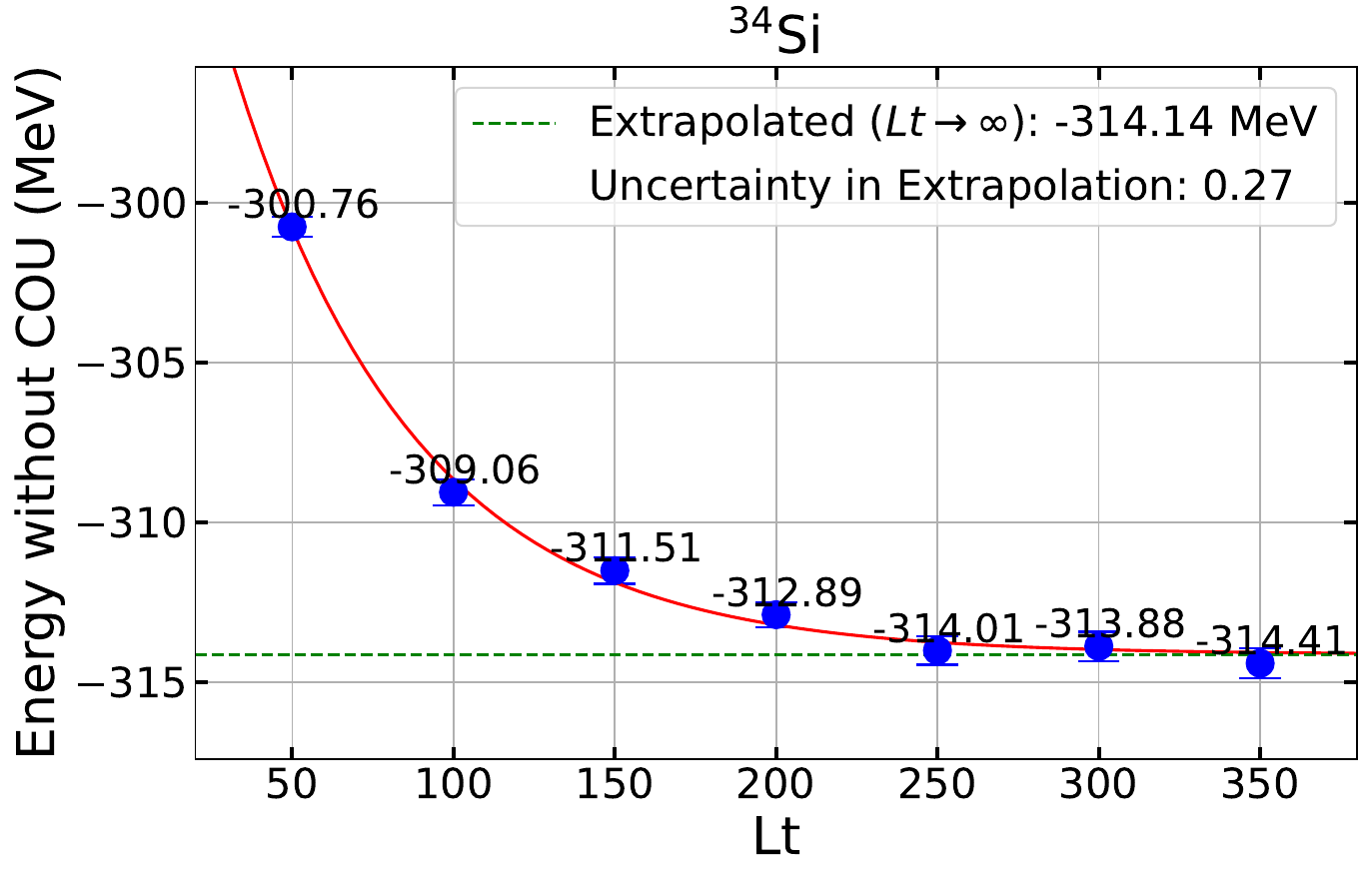}
    \end{minipage}%

    \vspace{0.5cm}

    \begin{minipage}{0.46\textwidth}
        \centering
        \includegraphics[width=\textwidth]{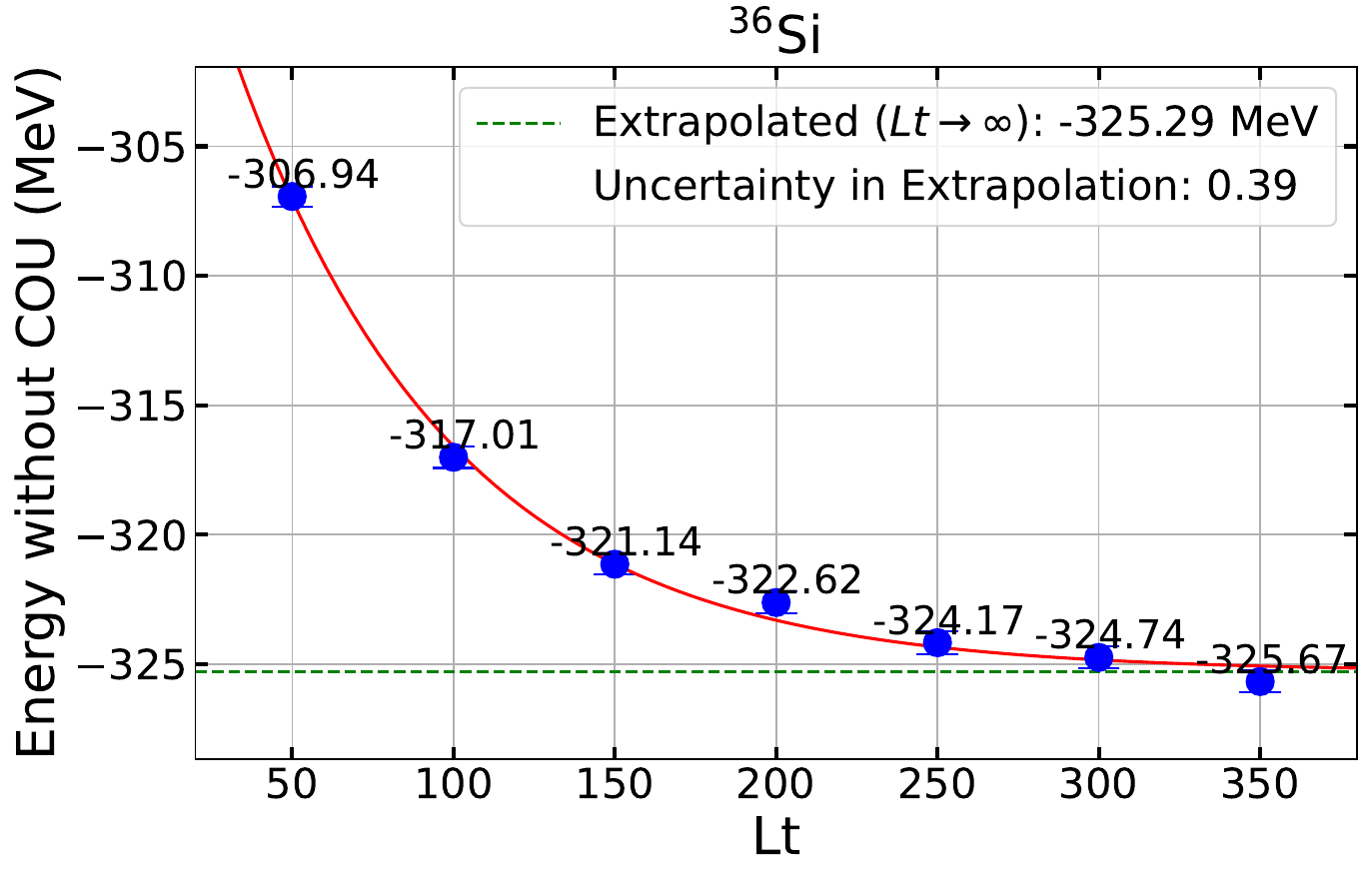}
    \end{minipage}%
    \begin{minipage}{0.46\textwidth}
        \centering
        \includegraphics[width=\textwidth]{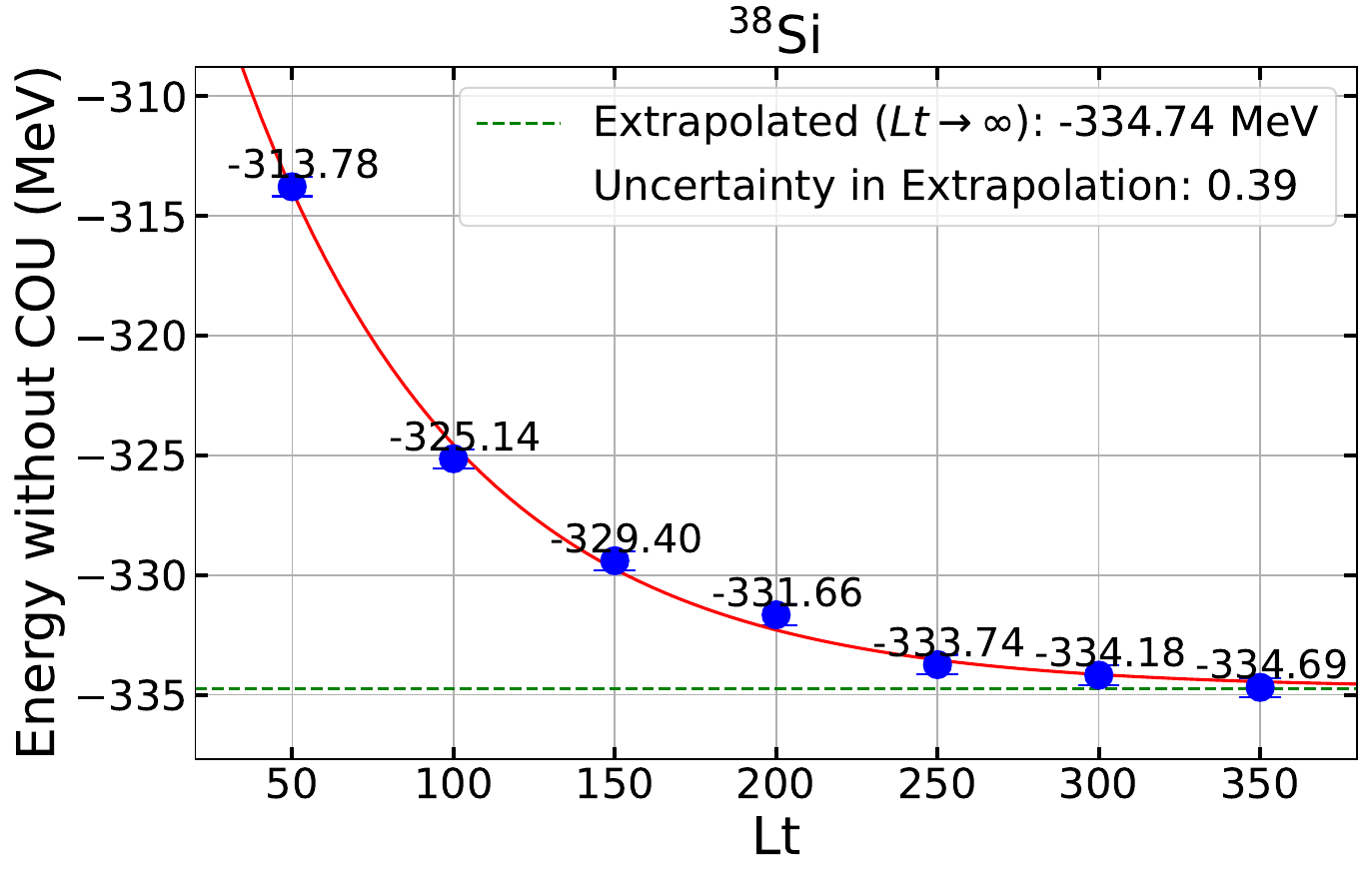}
    \end{minipage}%

    \vspace{0.5cm}
    
    \begin{minipage}{0.46\textwidth}
        \centering
        \includegraphics[width=\textwidth]{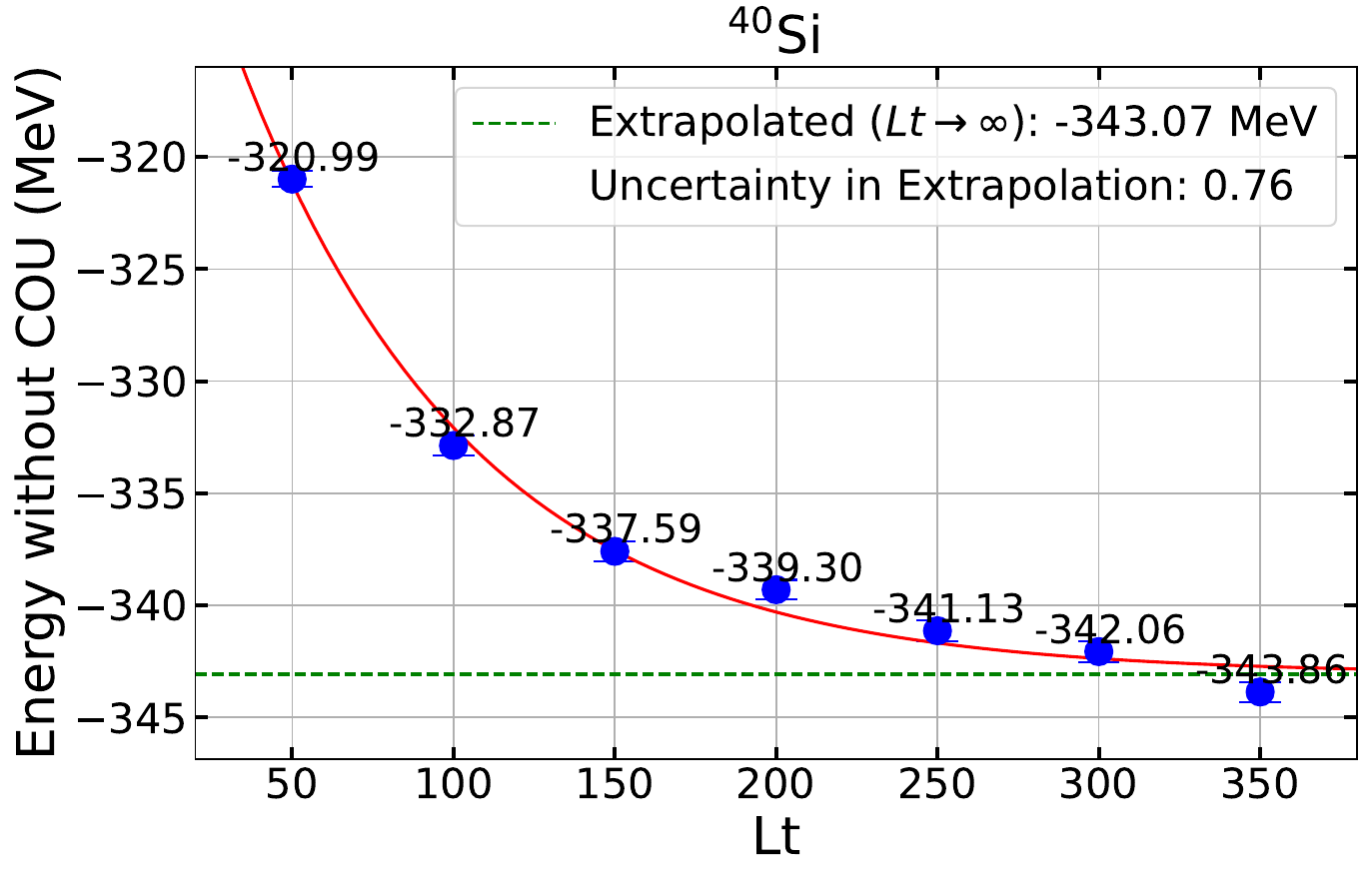}
    \end{minipage}%
    \begin{minipage}{0.46\textwidth}
        \centering
        \includegraphics[width=\textwidth]{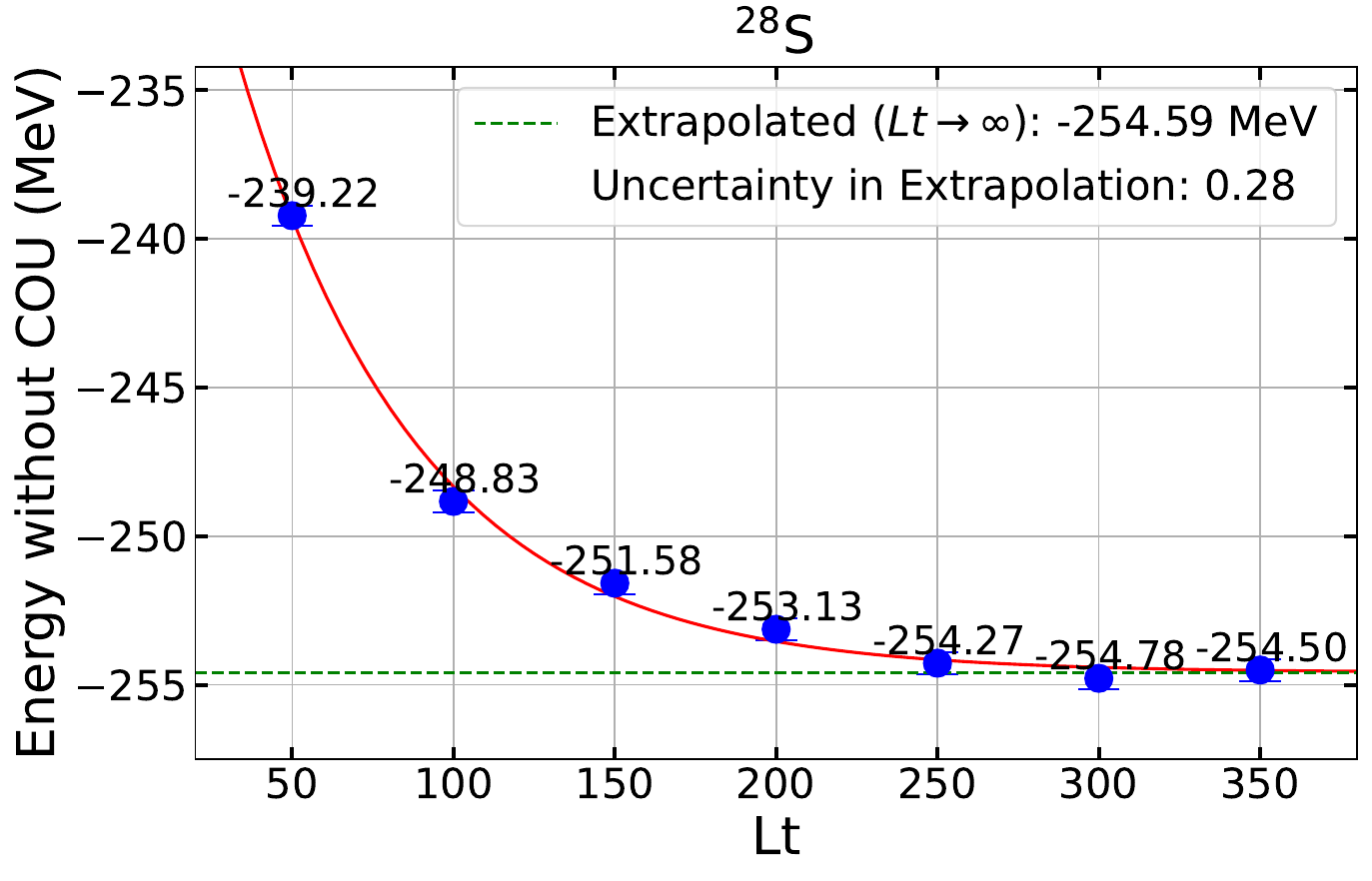}
    \end{minipage}%

\end{figure}

\begin{figure}[H]
    \vspace{0.5cm}

    \begin{minipage}{0.46\textwidth}
        \centering
        \includegraphics[width=\textwidth]{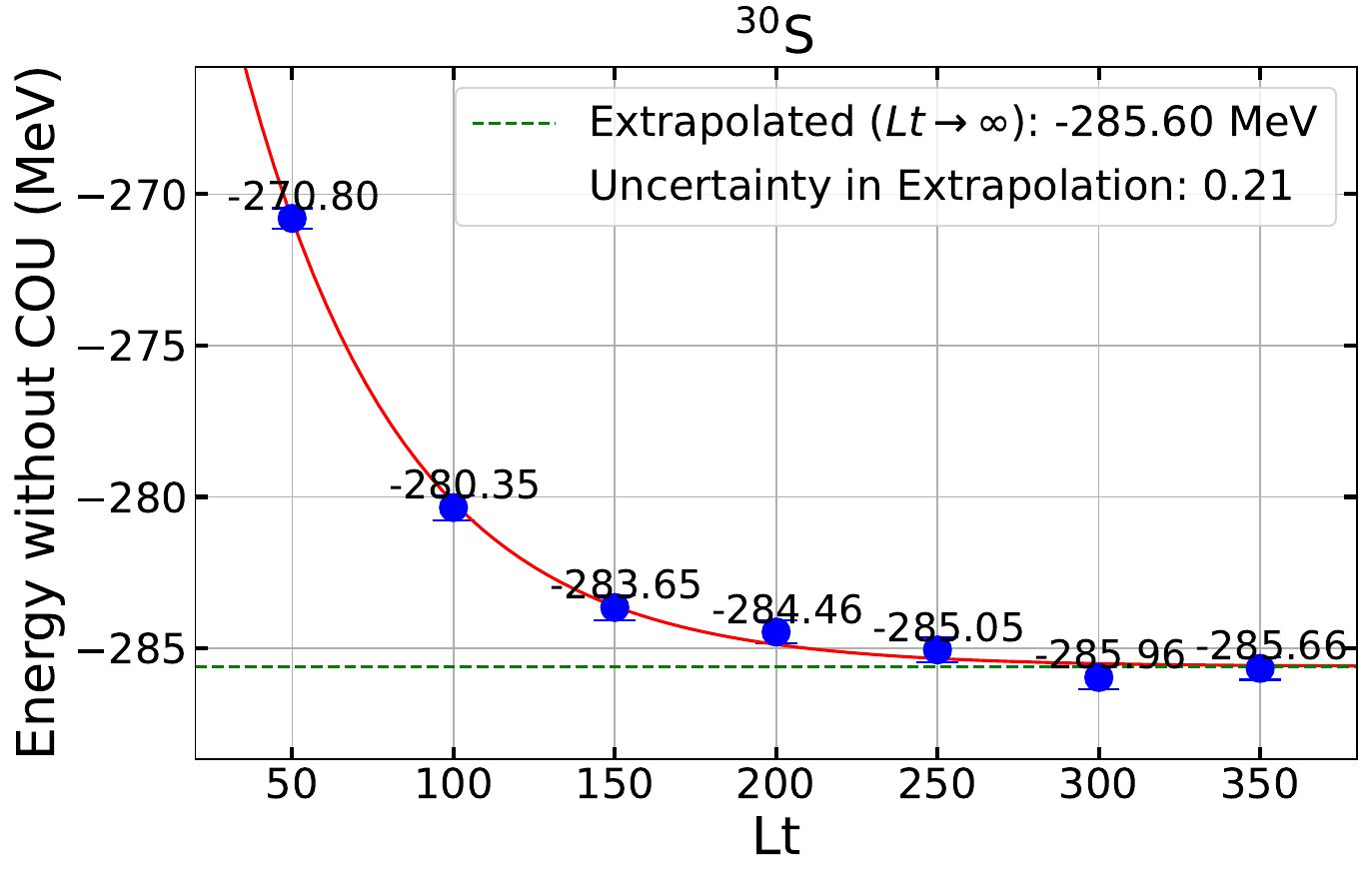}
    \end{minipage}%
    \begin{minipage}{0.46\textwidth}
        \centering
        \includegraphics[width=\textwidth]{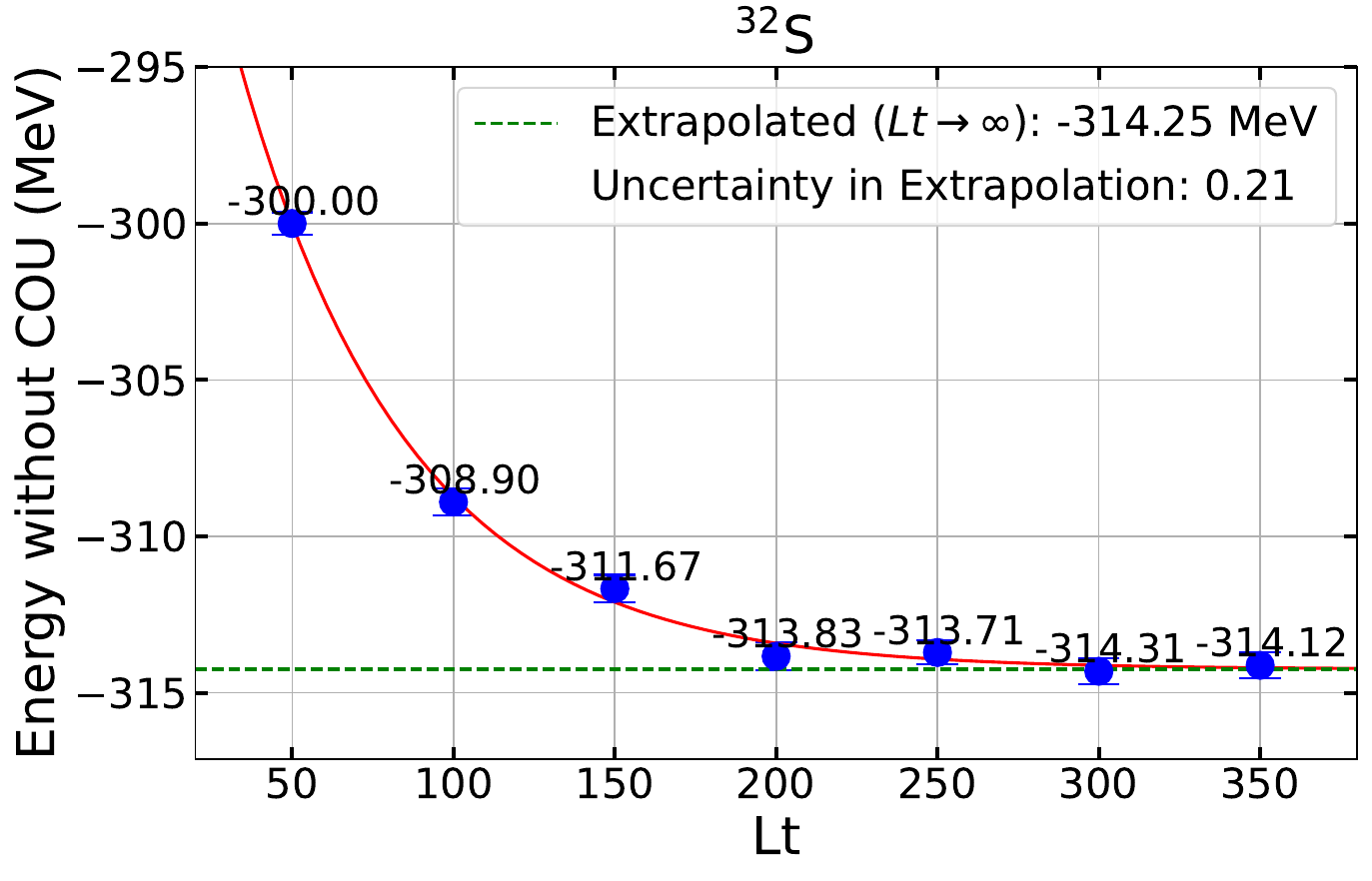}
    \end{minipage}%

    \vspace{0.5cm}
    
    \begin{minipage}{0.46\textwidth}
        \centering
        \includegraphics[width=\textwidth]{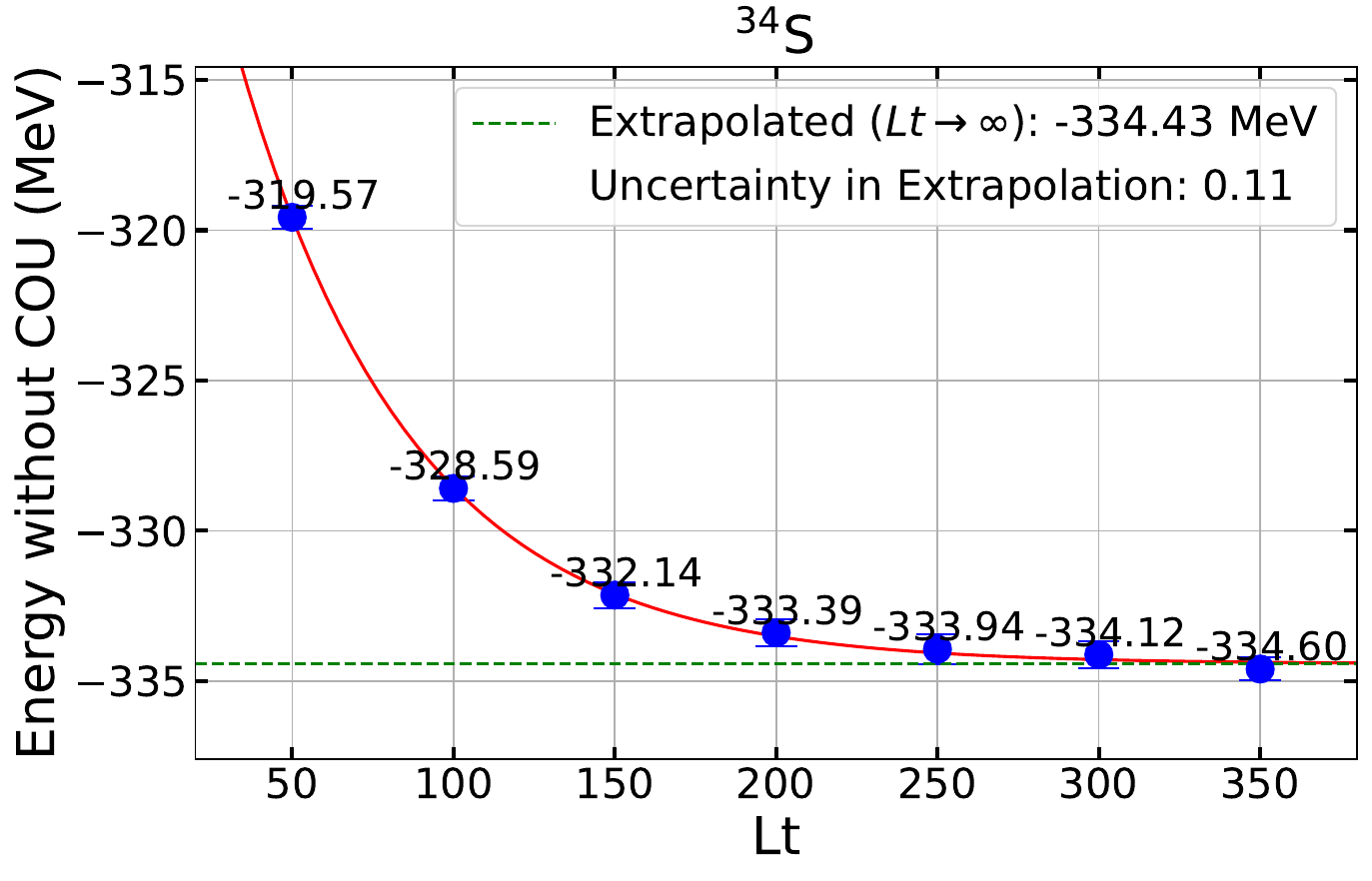}
    \end{minipage}%
    \begin{minipage}{0.46\textwidth}
        \centering
        \includegraphics[width=\textwidth]{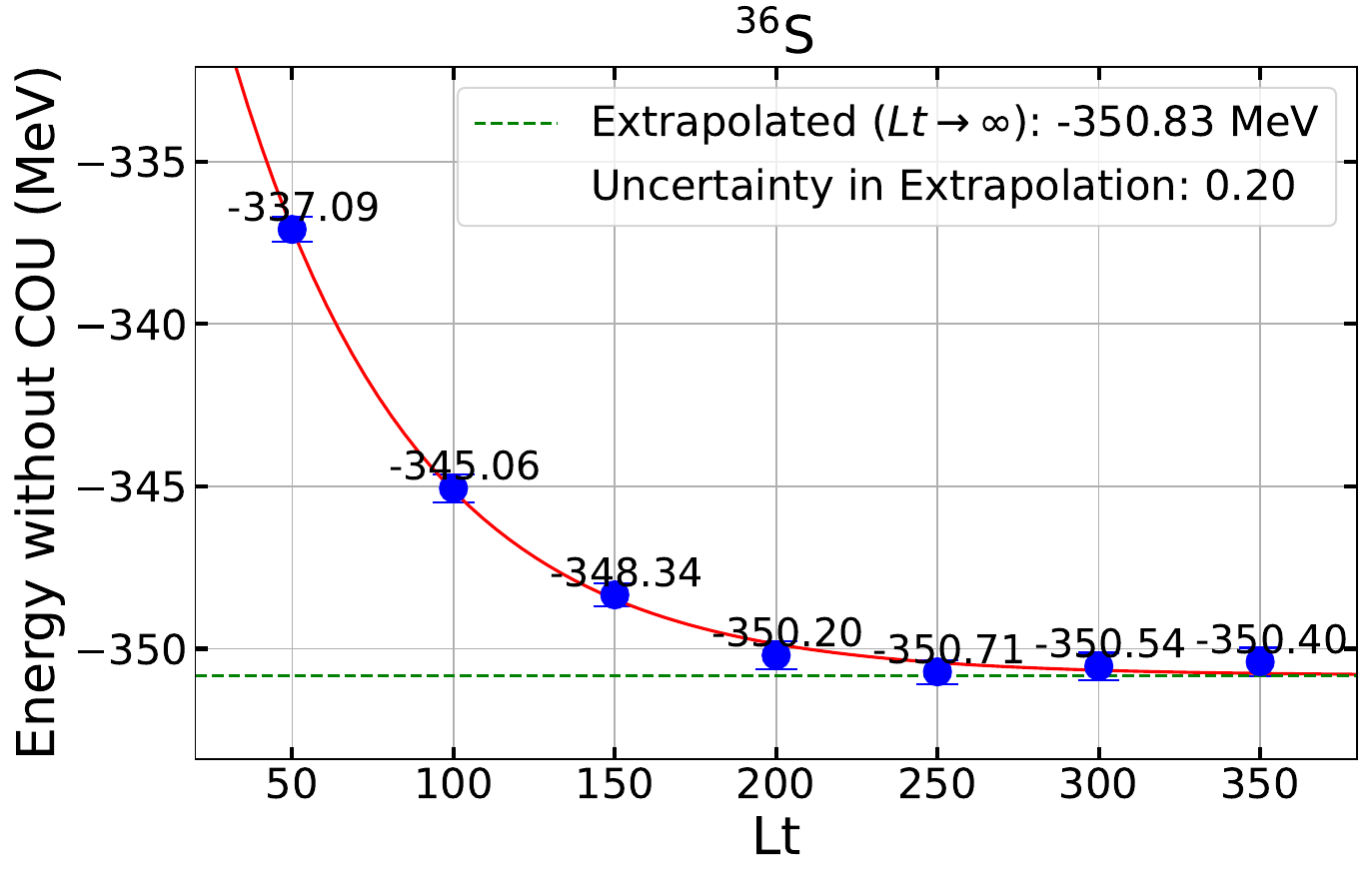}
    \end{minipage}%

    \vspace{0.5cm}

    \begin{minipage}{0.46\textwidth}
        \centering
        \includegraphics[width=\textwidth]{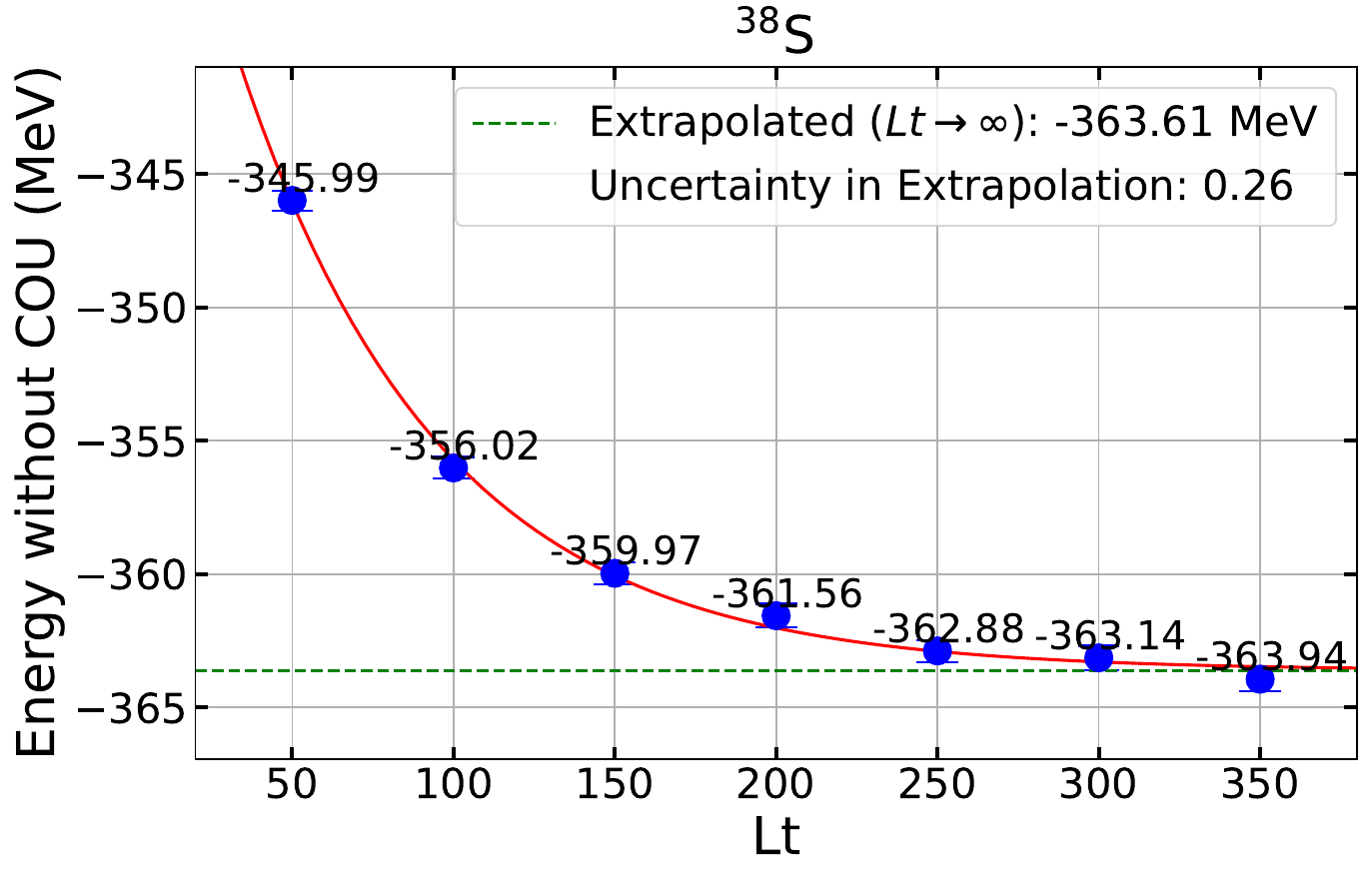}
    \end{minipage}%
    \begin{minipage}{0.46\textwidth}
        \centering
        \includegraphics[width=\textwidth]{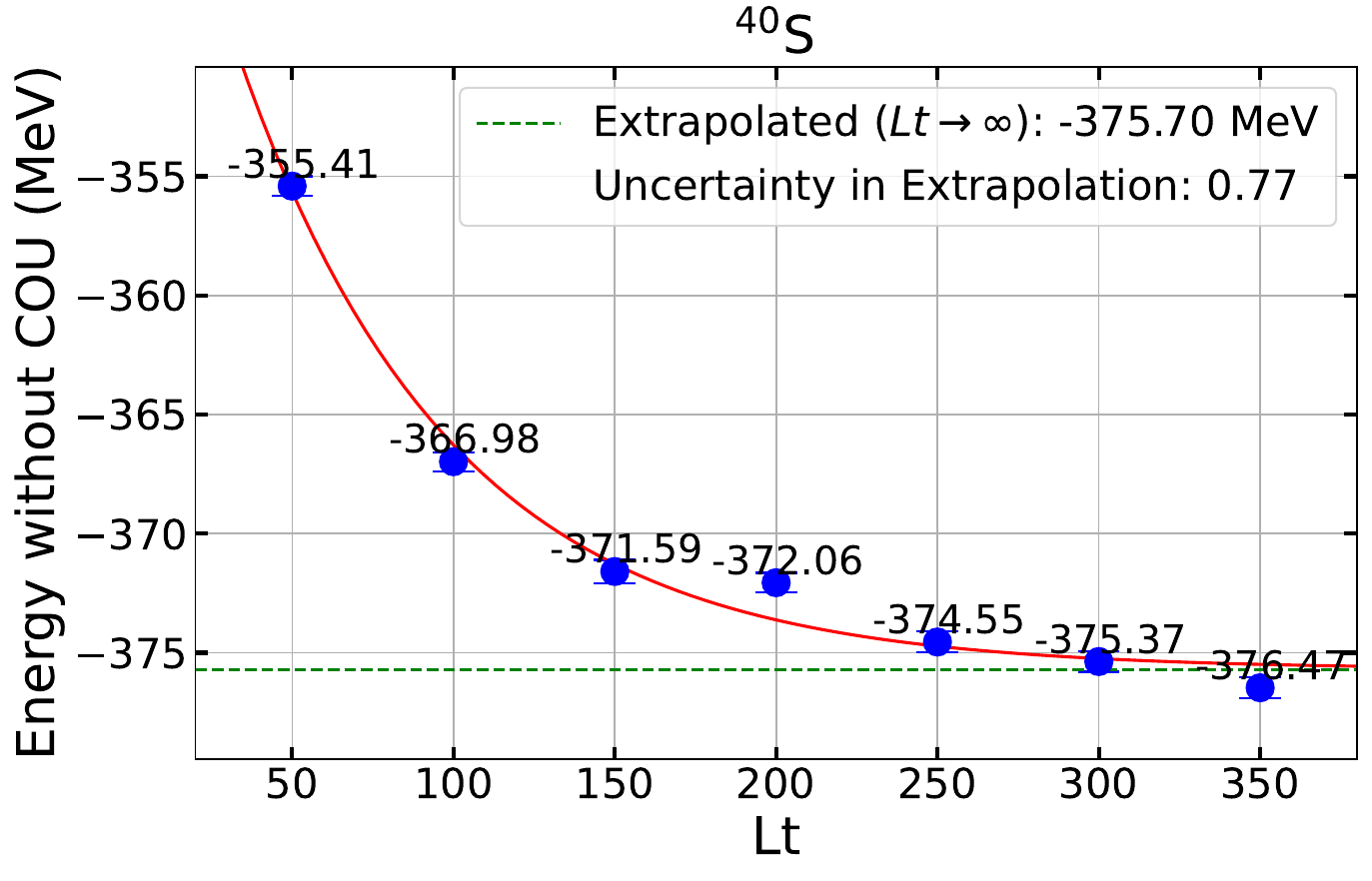}
    \end{minipage}%

    \vspace{0.5cm}
    
    \begin{minipage}{0.46\textwidth}
        \centering
        \includegraphics[width=\textwidth]{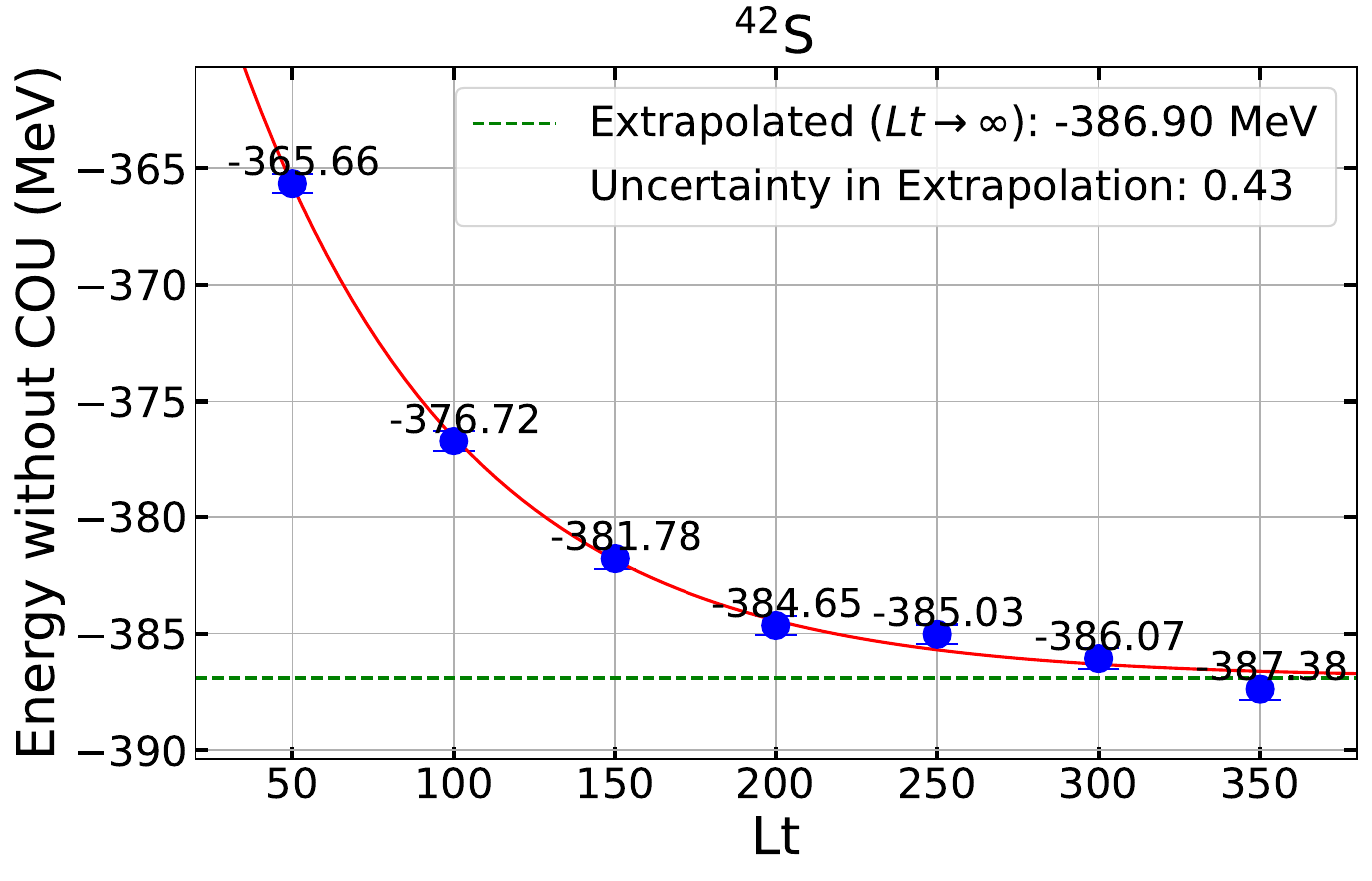}
    \end{minipage}%
    \begin{minipage}{0.46\textwidth}
        \centering
        \includegraphics[width=\textwidth]{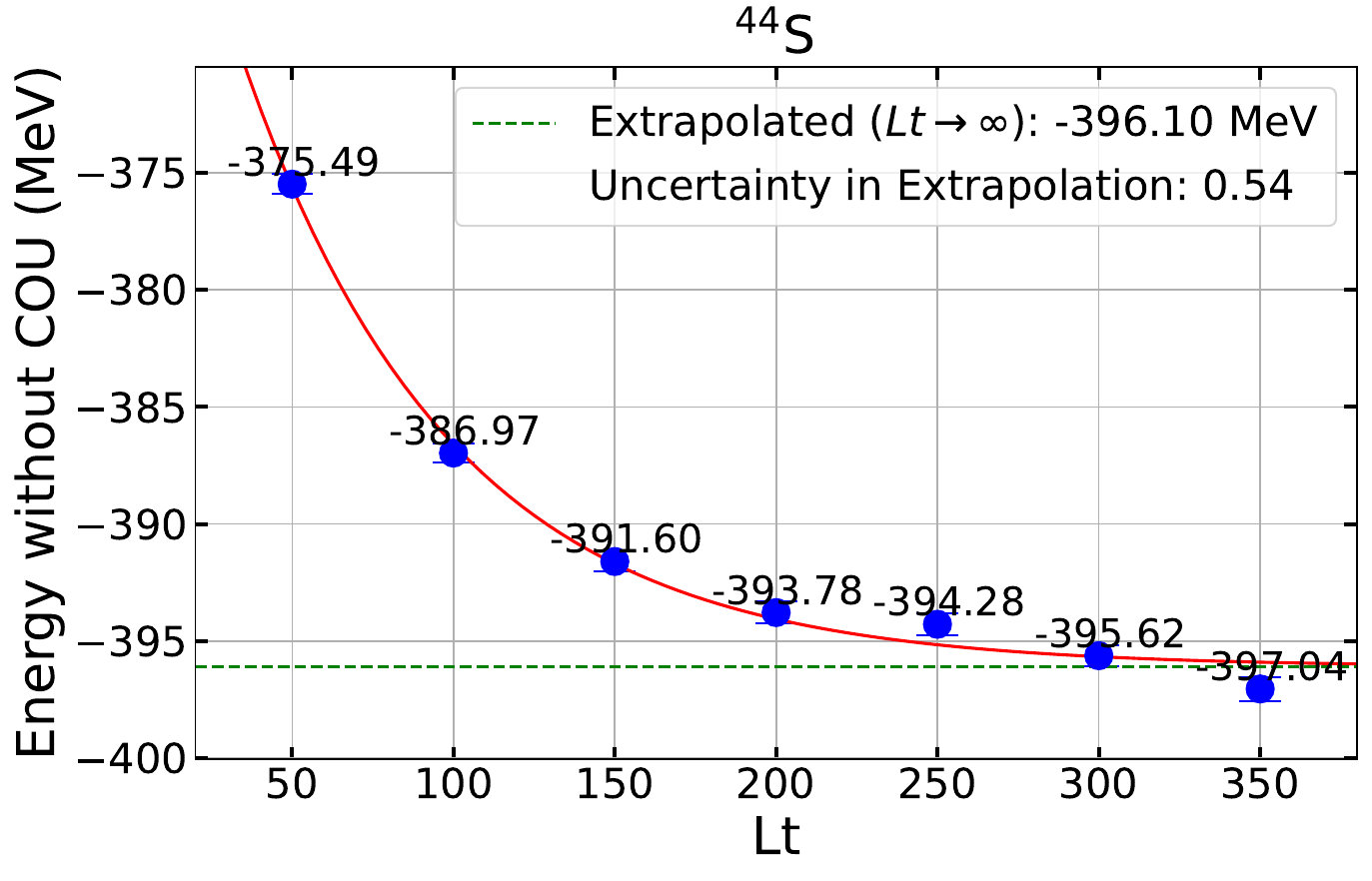}
    \end{minipage}%

\end{figure}

\begin{figure}[H]
    \vspace{0.5cm}
    
    \begin{minipage}{0.46\textwidth}
        \centering
        \includegraphics[width=\textwidth]{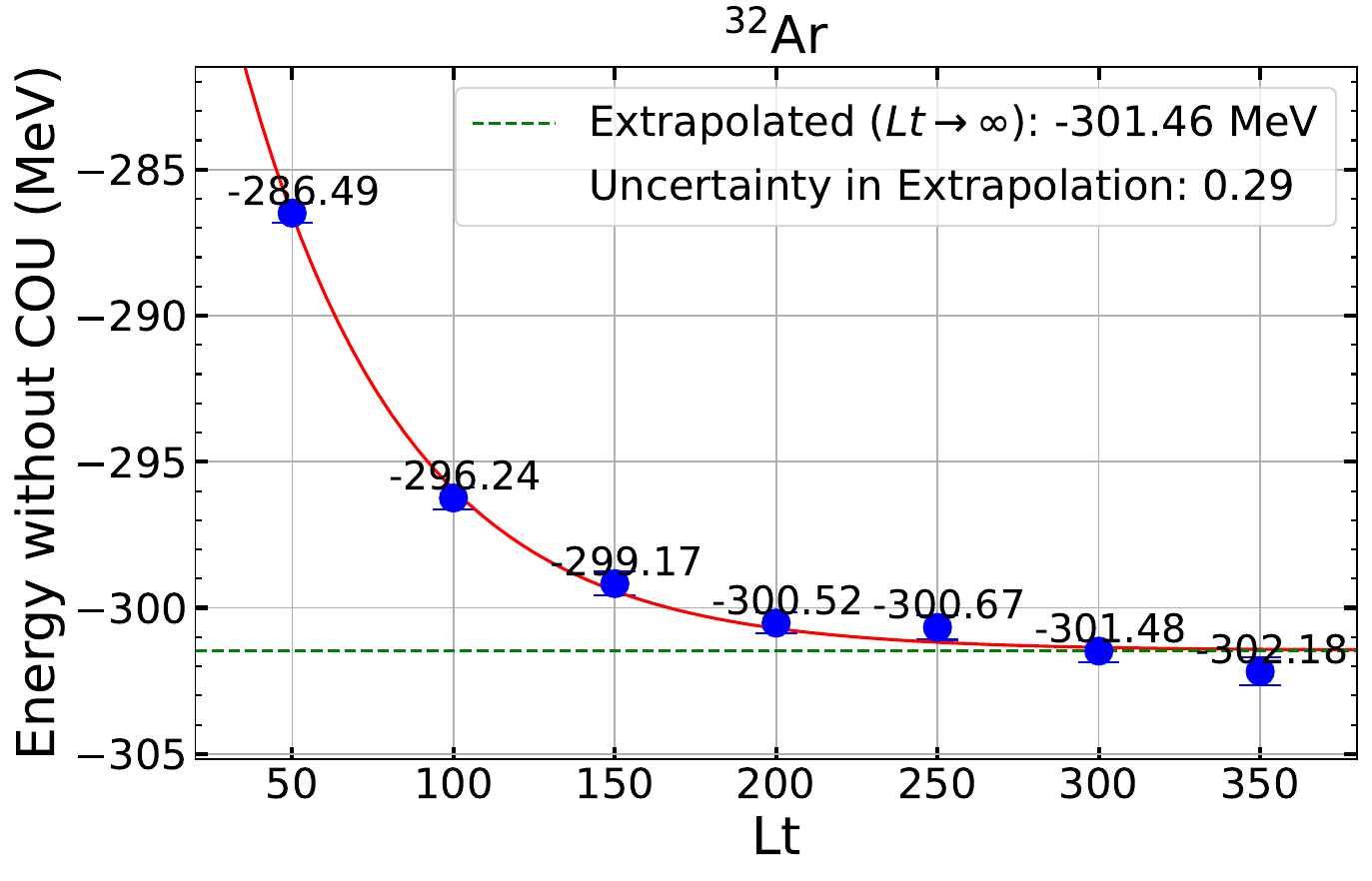}
    \end{minipage}%
    \begin{minipage}{0.46\textwidth}
        \centering
        \includegraphics[width=\textwidth]{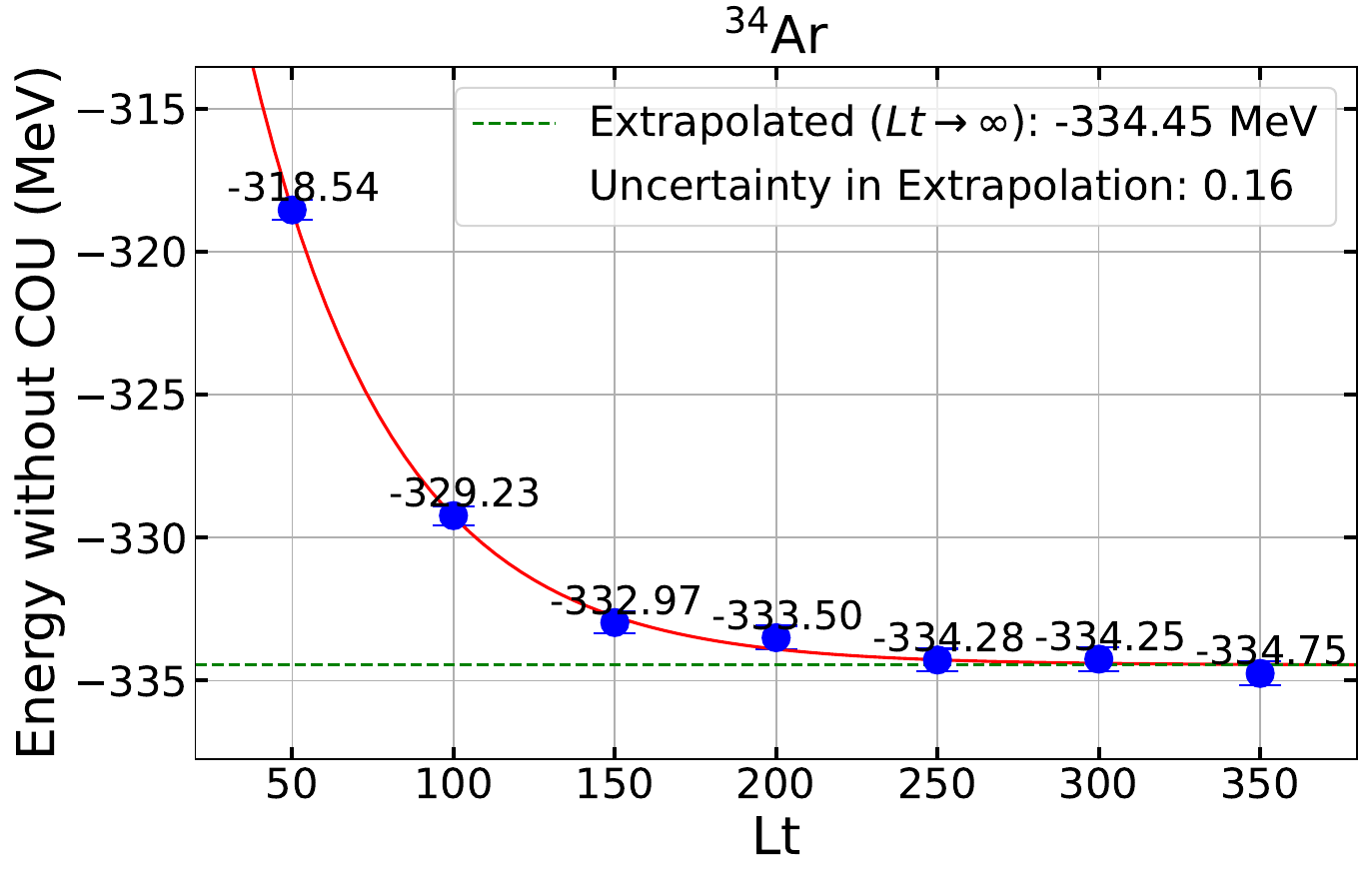}
    \end{minipage}%

    \newpage
    \vspace{0.5cm}
    
    \begin{minipage}{0.46\textwidth}
        \centering
        \includegraphics[width=\textwidth]{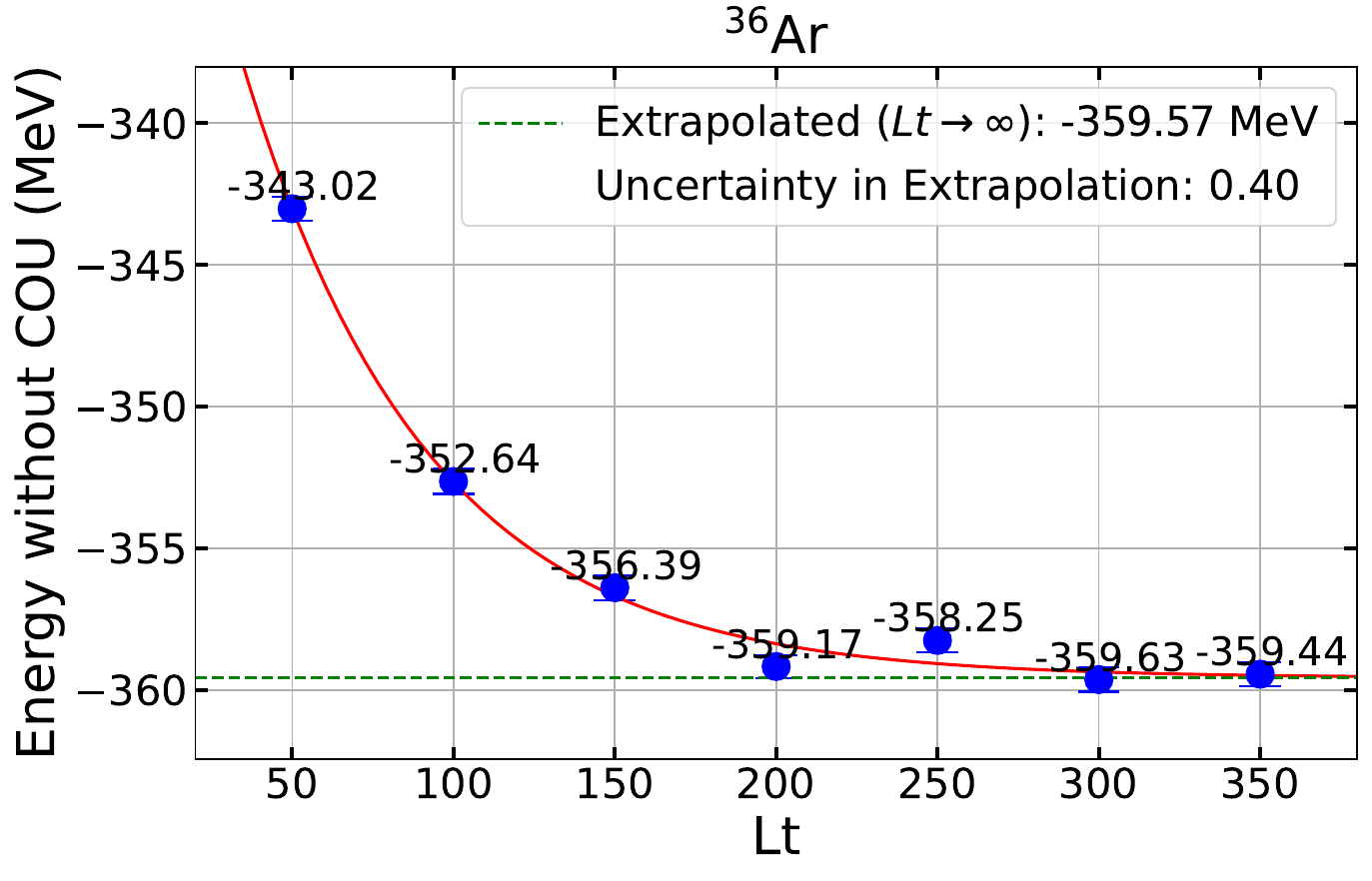}
    \end{minipage}%
    \begin{minipage}{0.46\textwidth}
        \centering
        \includegraphics[width=\textwidth]{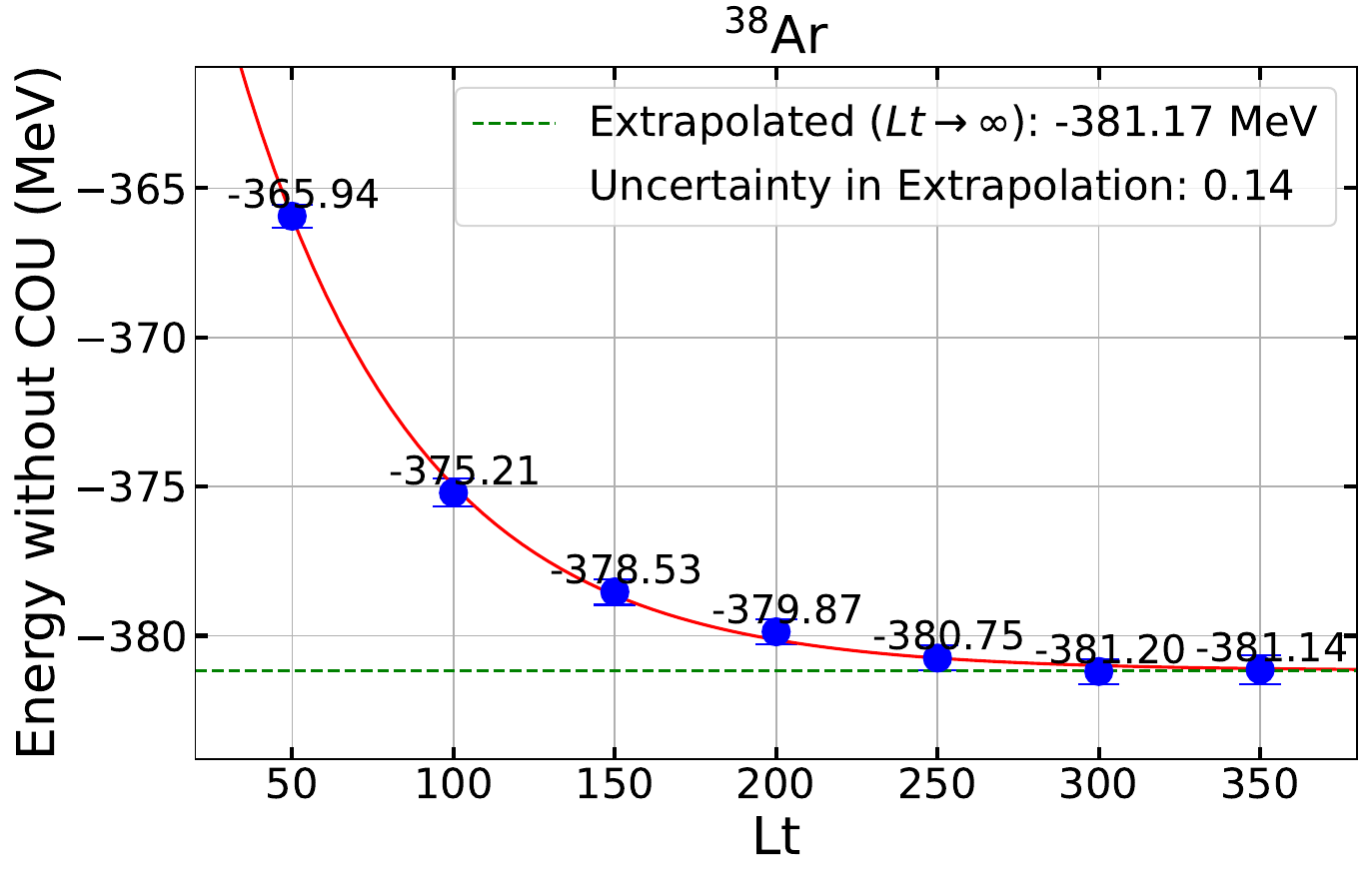}
    \end{minipage}%

    \vspace{0.5cm}

    \begin{minipage}{0.46\textwidth}
        \centering
        \includegraphics[width=\textwidth]{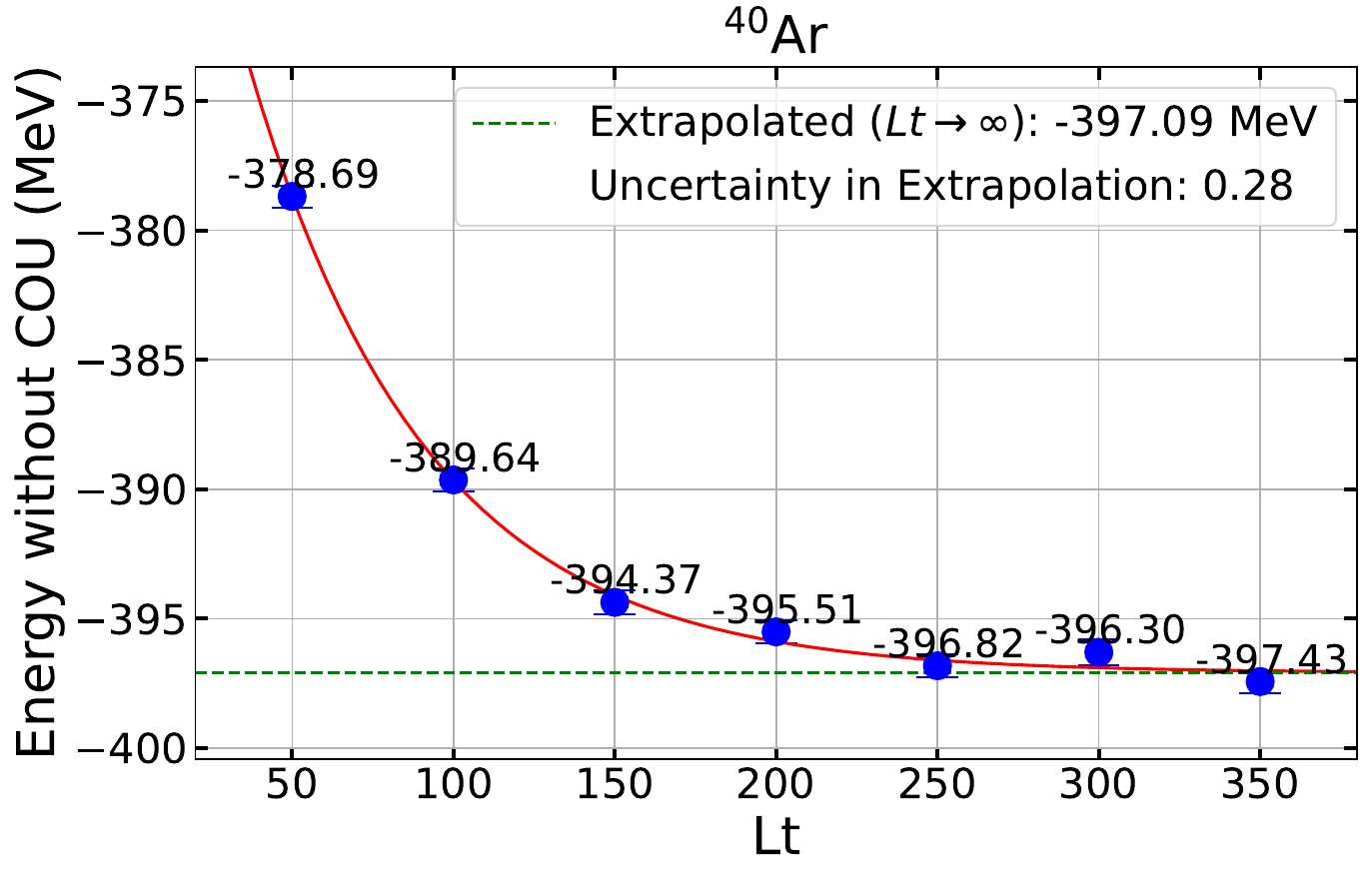}
    \end{minipage}%
    \begin{minipage}{0.46\textwidth}
        \centering
        \includegraphics[width=\textwidth]{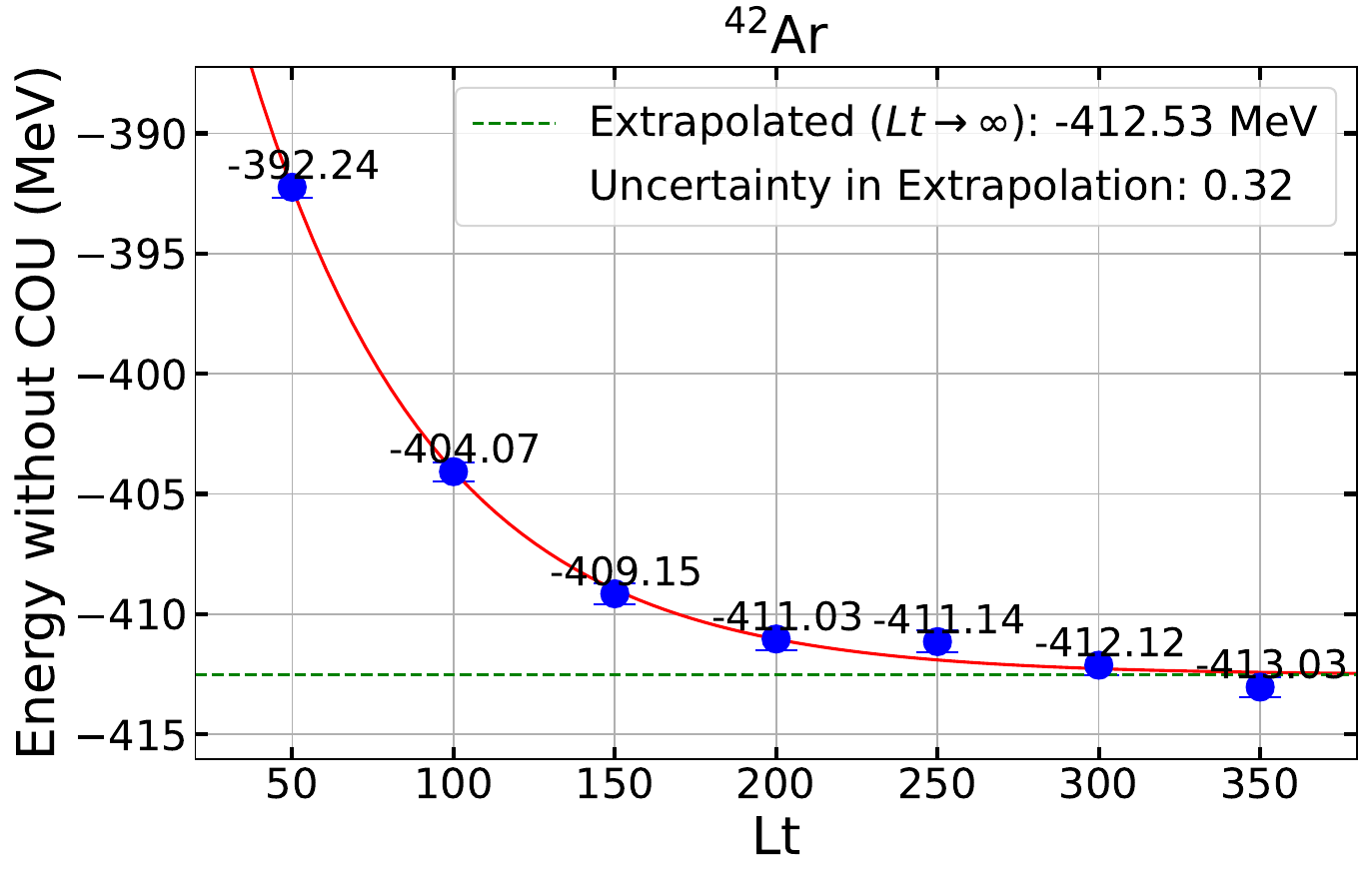}
    \end{minipage}%

    \vspace{0.5cm}
    
    \begin{minipage}{0.46\textwidth}
        \centering
        \includegraphics[width=\textwidth]{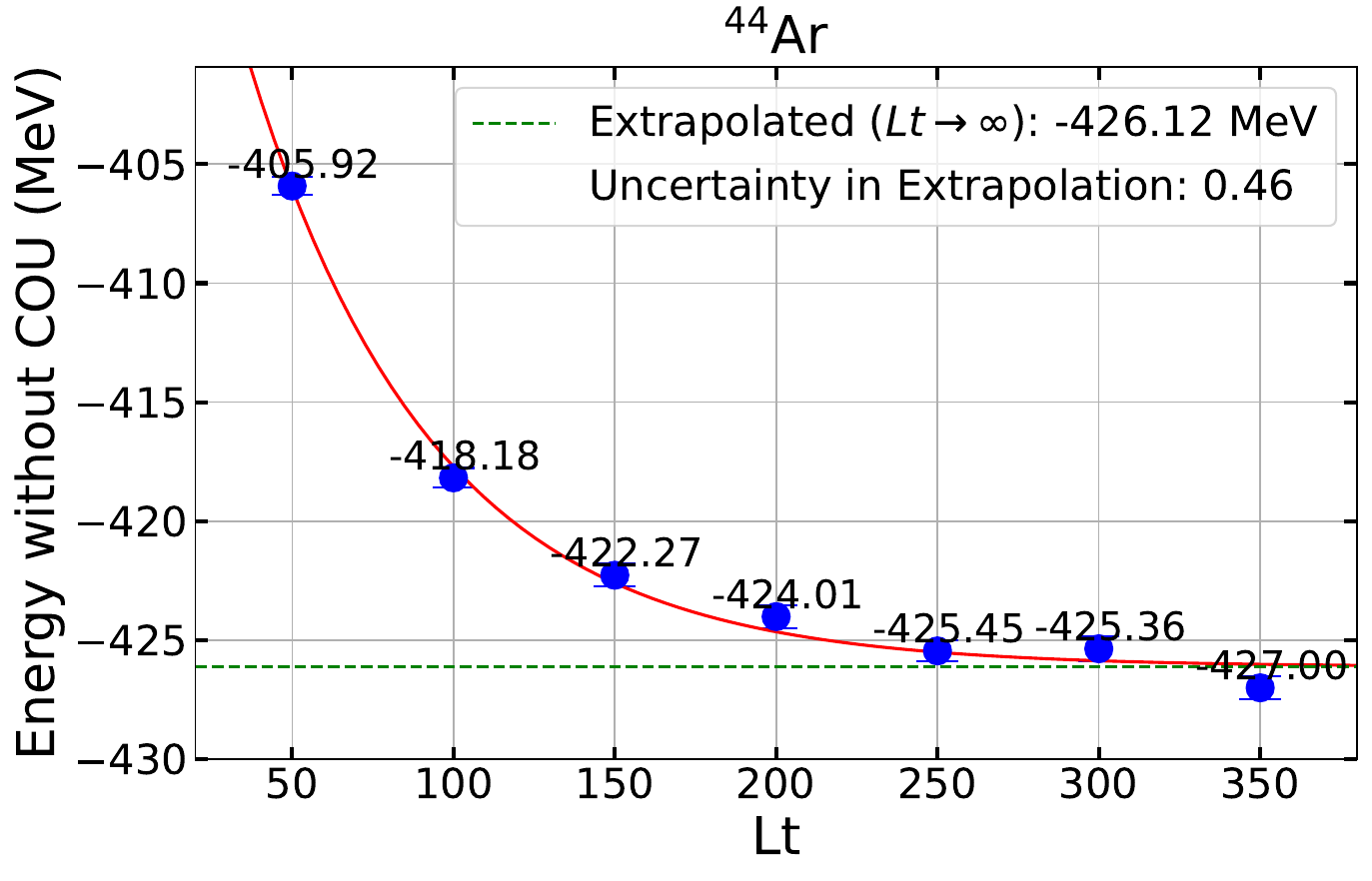}
    \end{minipage}%
    \begin{minipage}{0.46\textwidth}
        \centering
        \includegraphics[width=\textwidth]{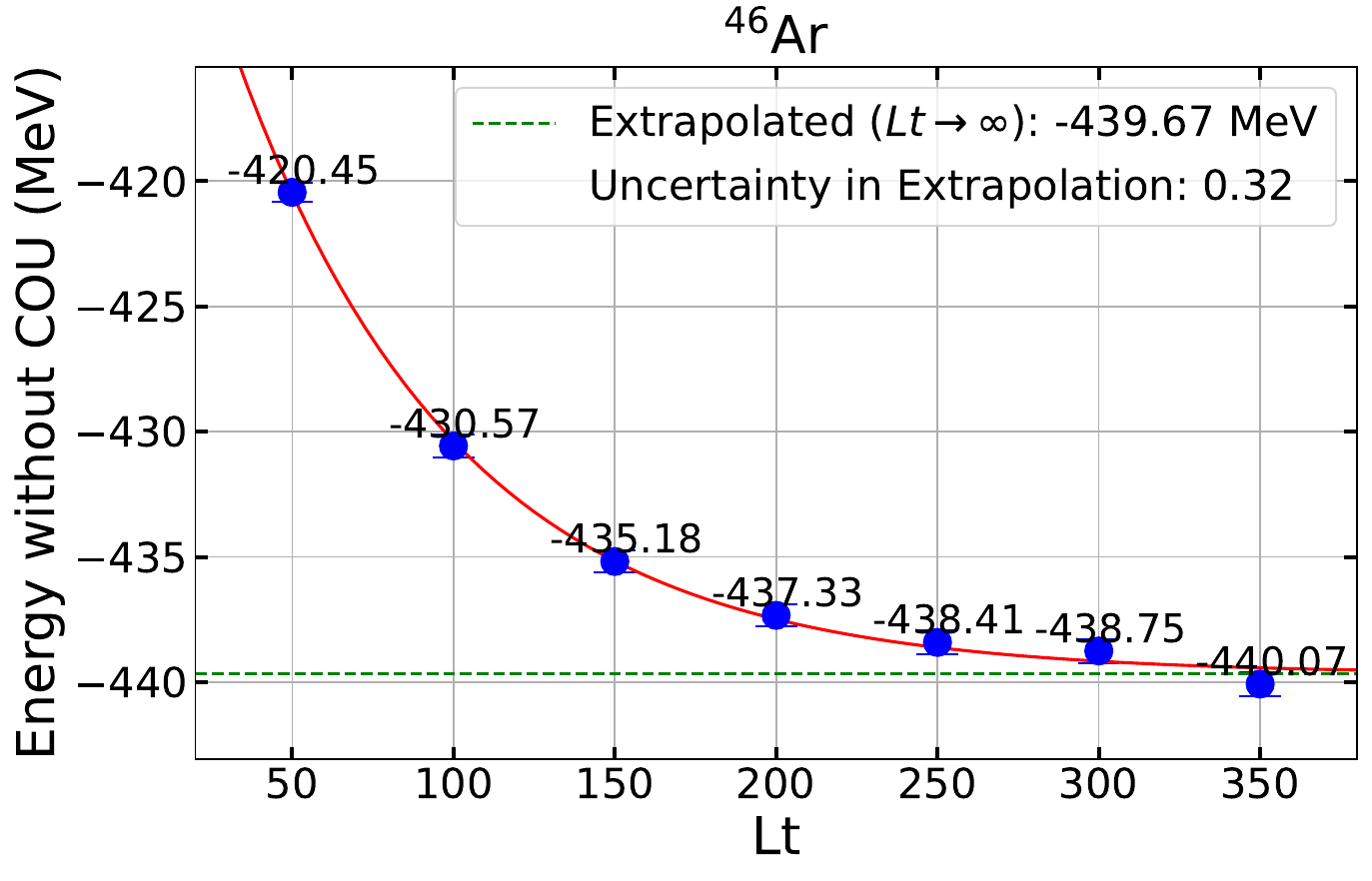}
    \end{minipage}%

\end{figure}

\begin{figure}[H]
    \vspace{0.5cm}

    \begin{minipage}{0.46\textwidth}
        \centering
        \includegraphics[width=\textwidth]{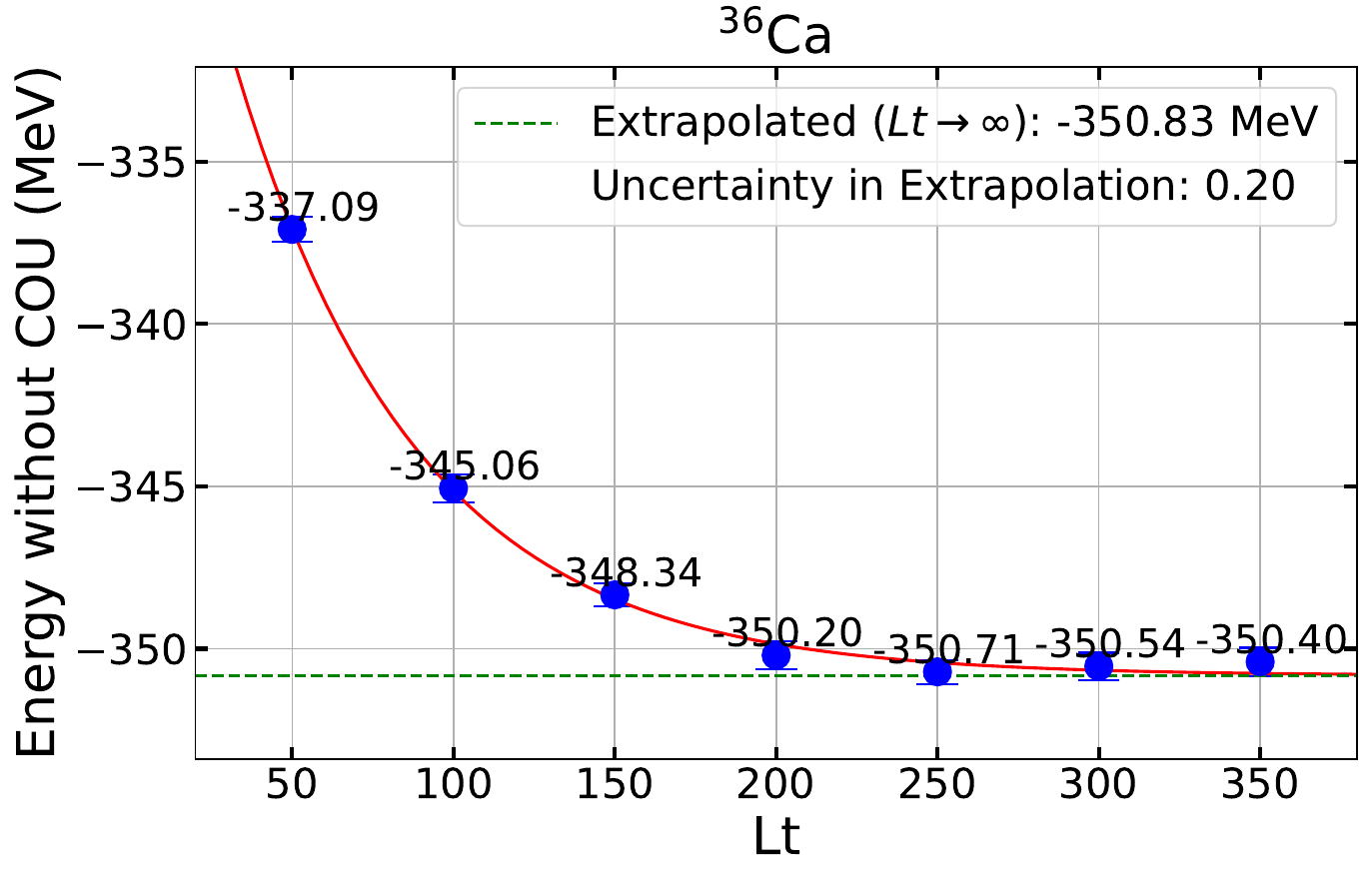}
    \end{minipage}%
    \begin{minipage}{0.46\textwidth}
        \centering
        \includegraphics[width=\textwidth]{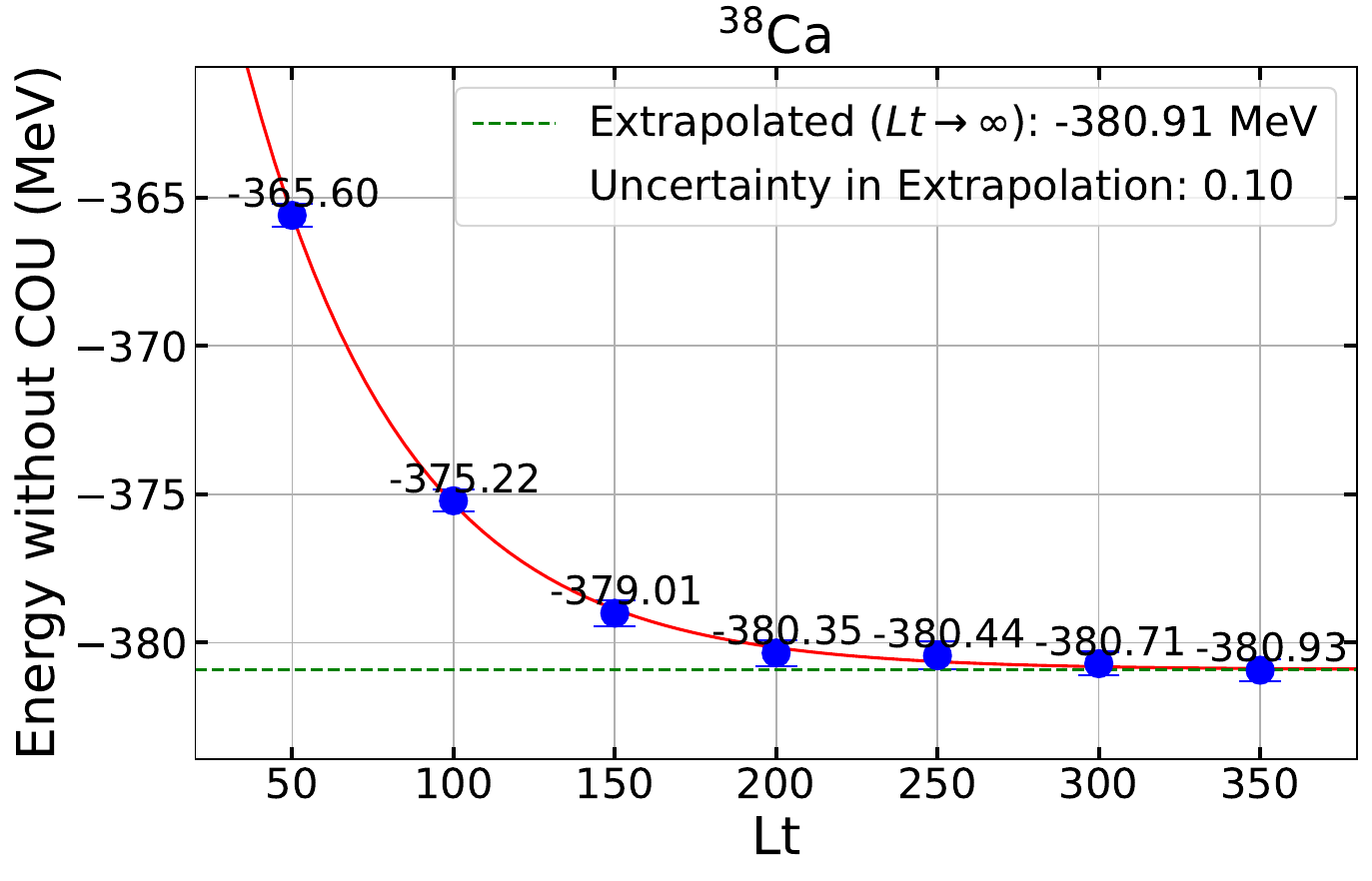}
    \end{minipage}%

    \vspace{0.5cm}
    
    \begin{minipage}{0.46\textwidth}
        \centering
        \includegraphics[width=\textwidth]{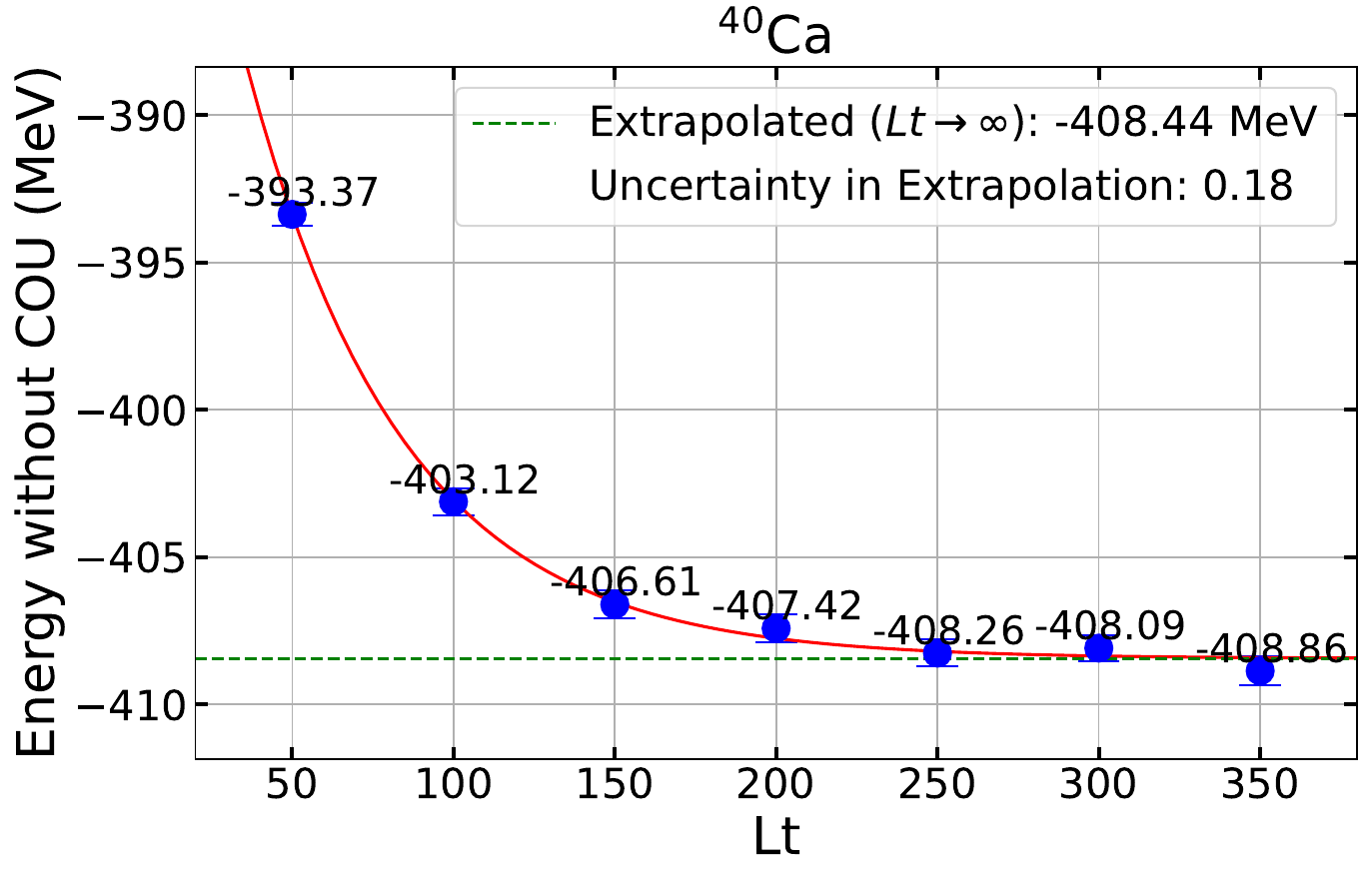}
    \end{minipage}%
    \begin{minipage}{0.46\textwidth}
        \centering
        \includegraphics[width=\textwidth]{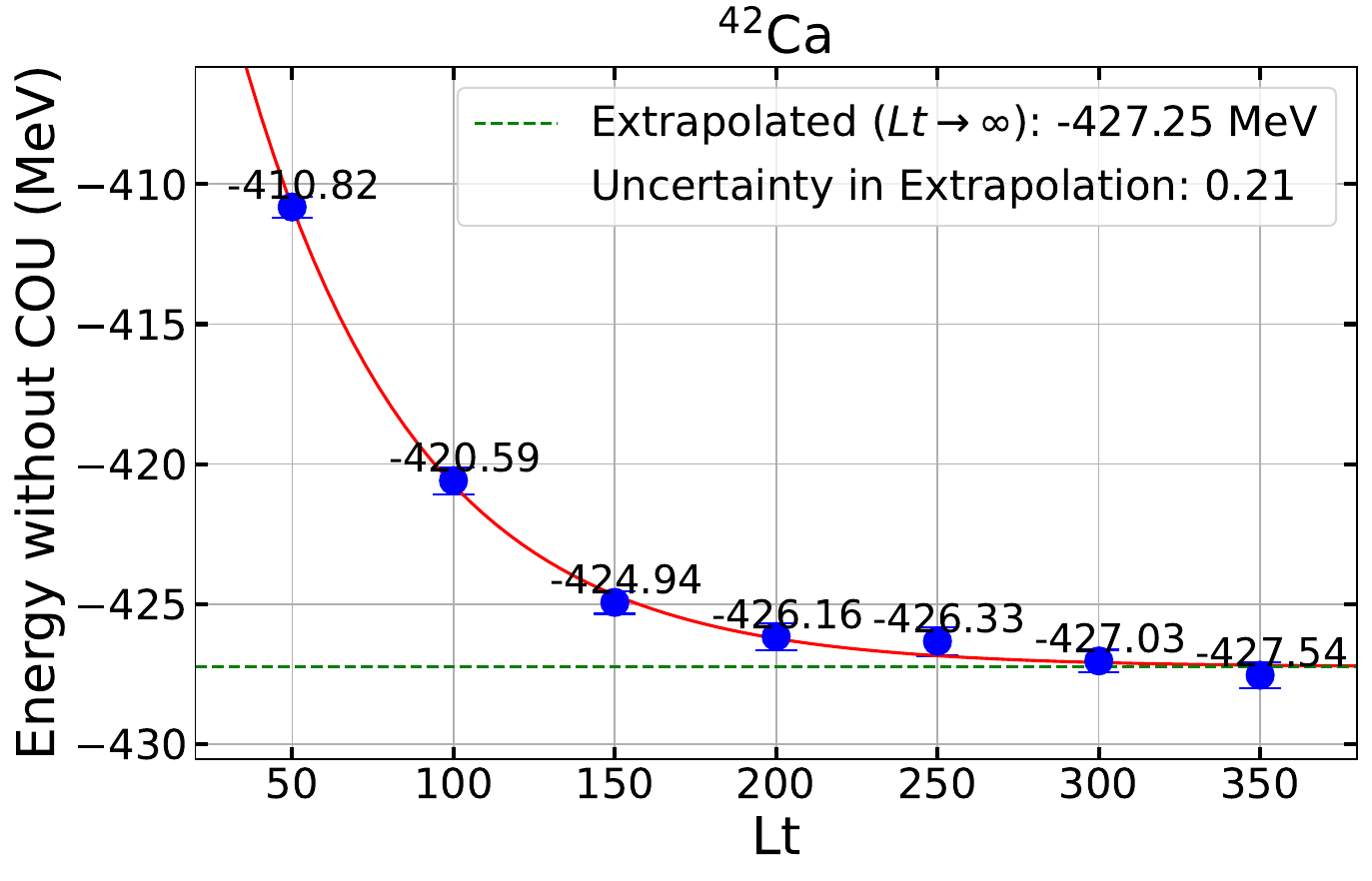}
    \end{minipage}%

    \newpage
    \vspace{0.5cm}

    \begin{minipage}{0.46\textwidth}
        \centering
        \includegraphics[width=\textwidth]{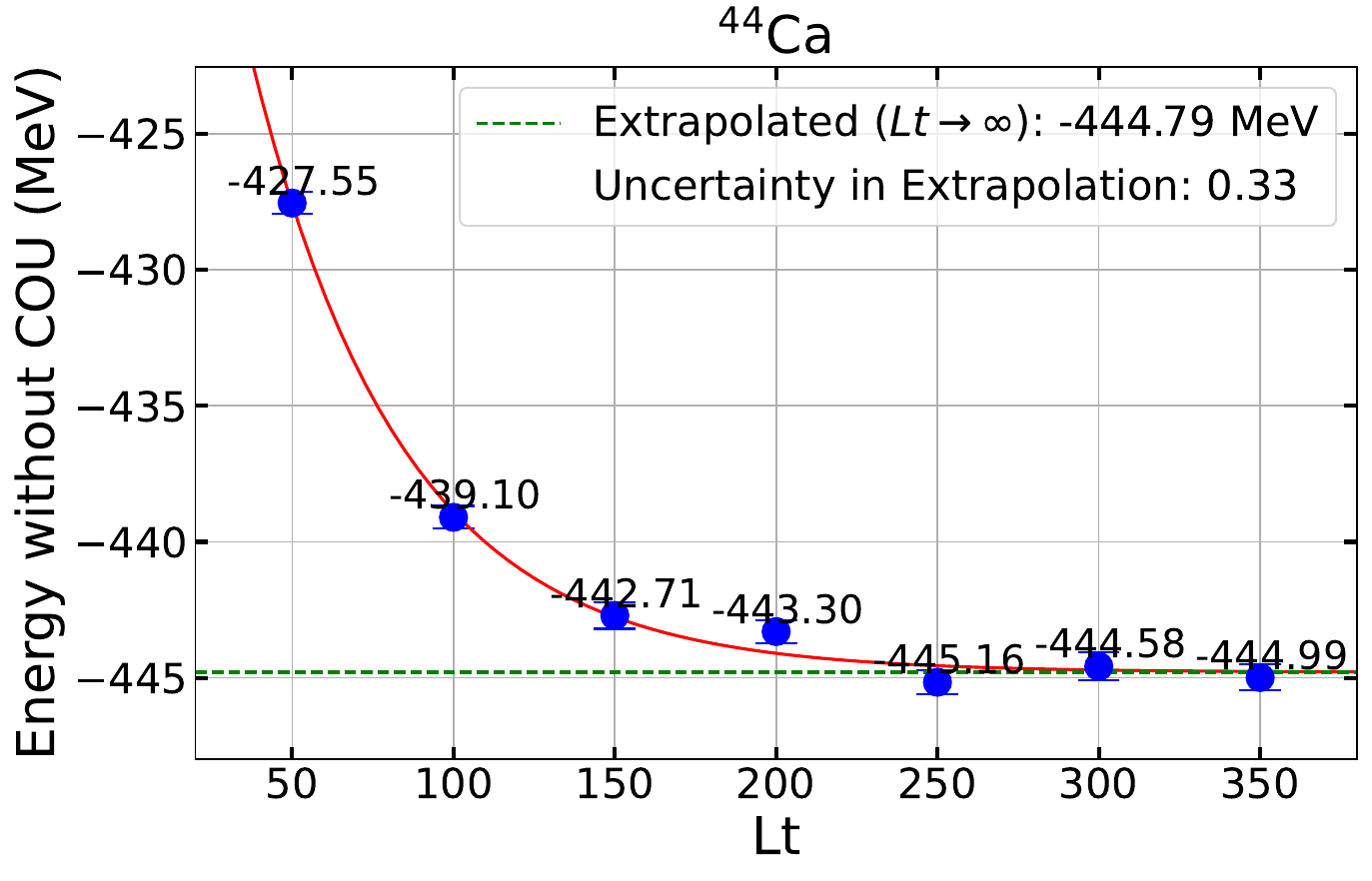}
    \end{minipage}%
    \begin{minipage}{0.46\textwidth}
        \centering
        \includegraphics[width=\textwidth]{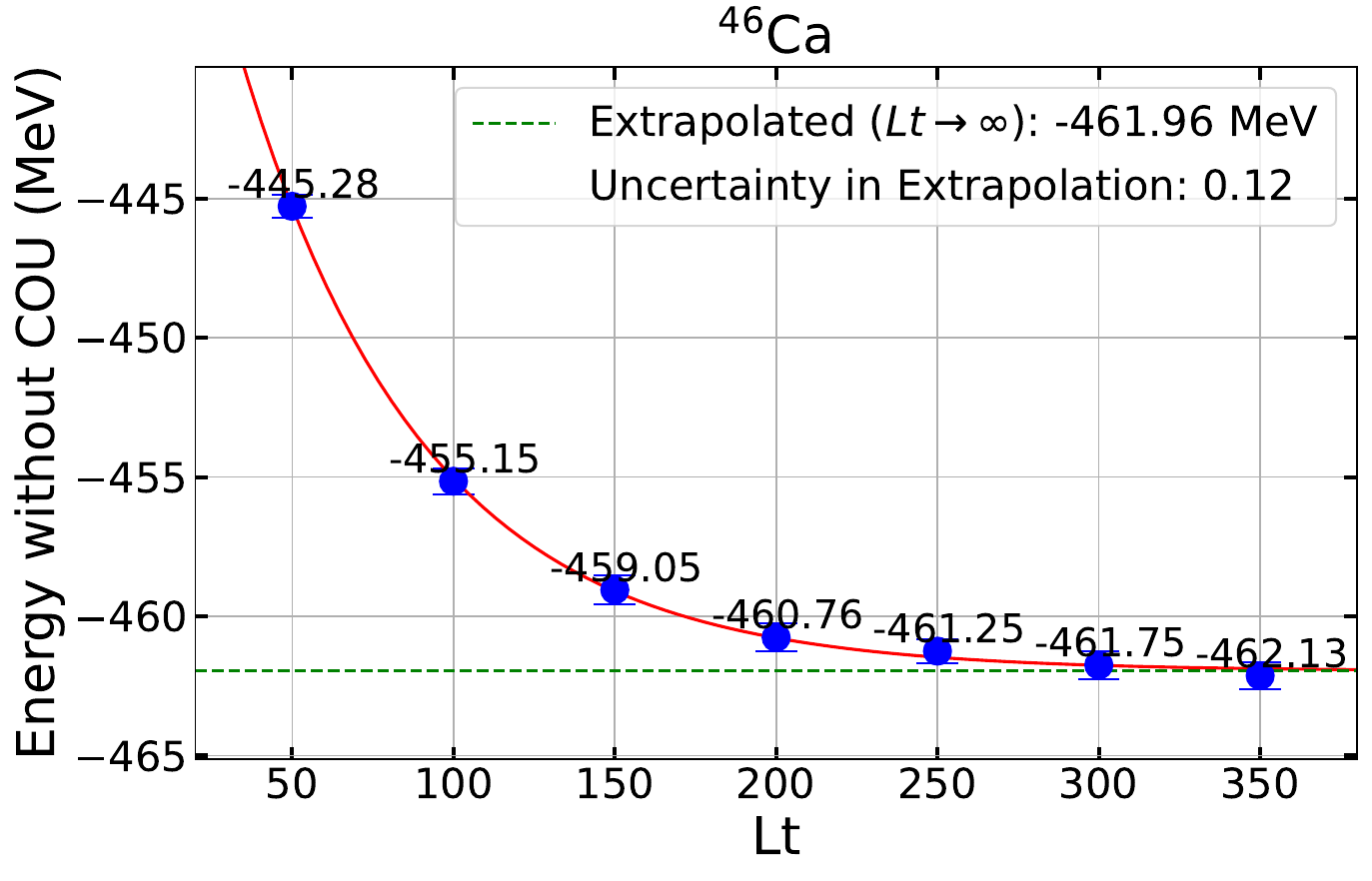}
    \end{minipage}%

    \vspace{0.5cm}
    
    \begin{minipage}{0.46\textwidth}
        \centering
        \includegraphics[width=\textwidth]{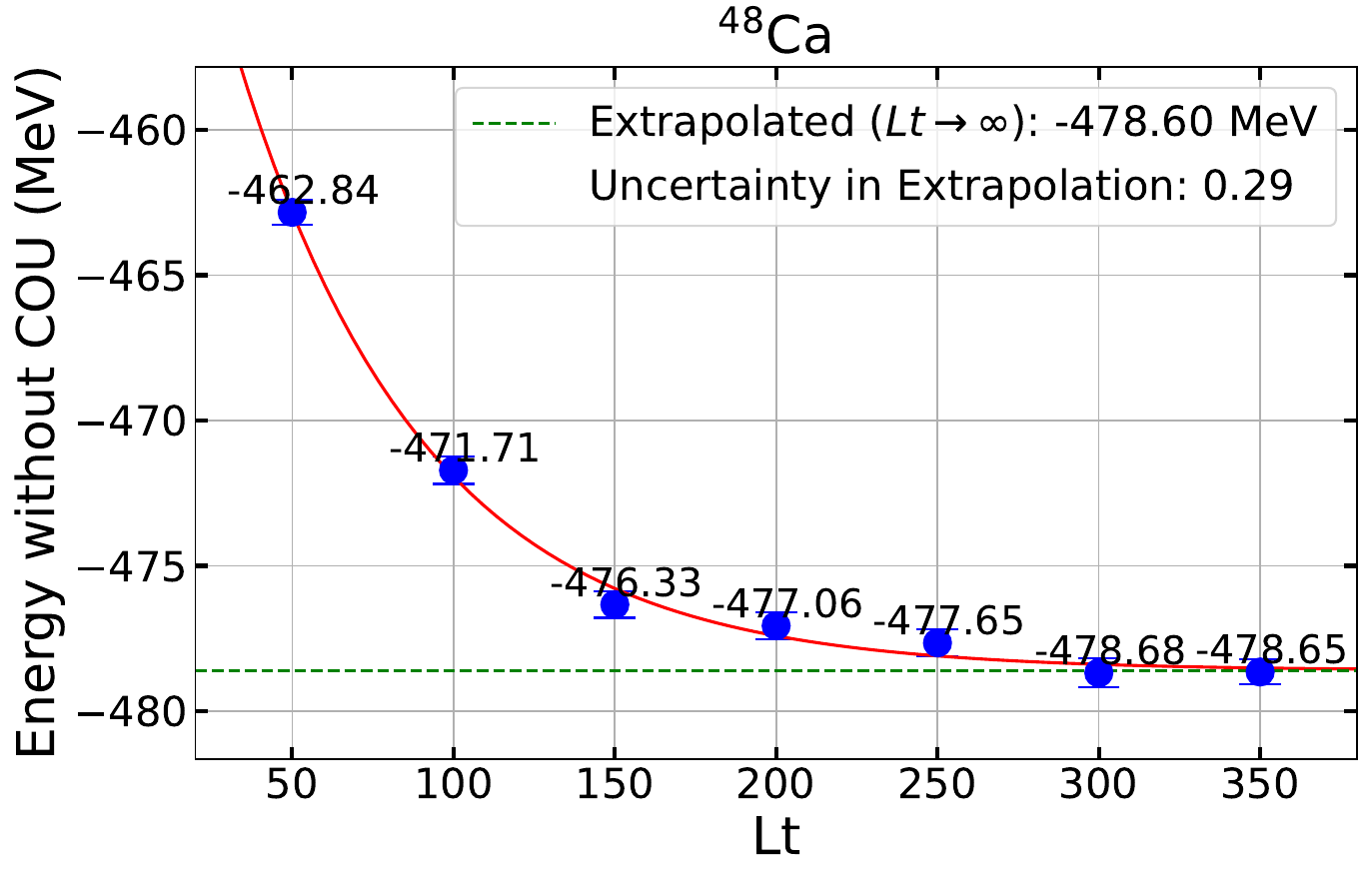}
    \end{minipage}%
    \begin{minipage}{0.46\textwidth}
        \centering
        \includegraphics[width=\textwidth]{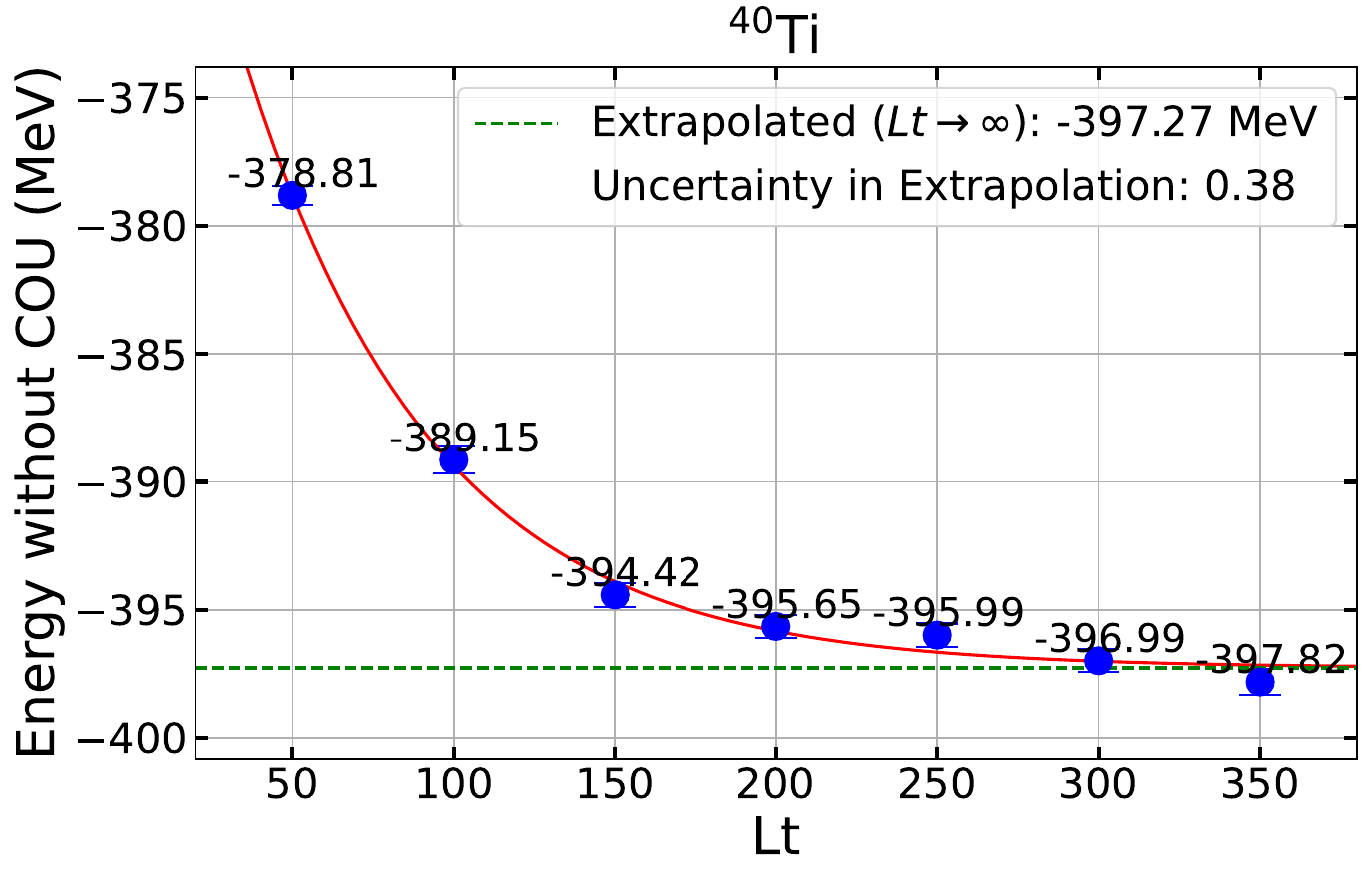}
    \end{minipage}%

\end{figure}

\begin{figure}[H]
    \vspace{0.5cm}

    \begin{minipage}{0.46\textwidth}
        \centering
        \includegraphics[width=\textwidth]{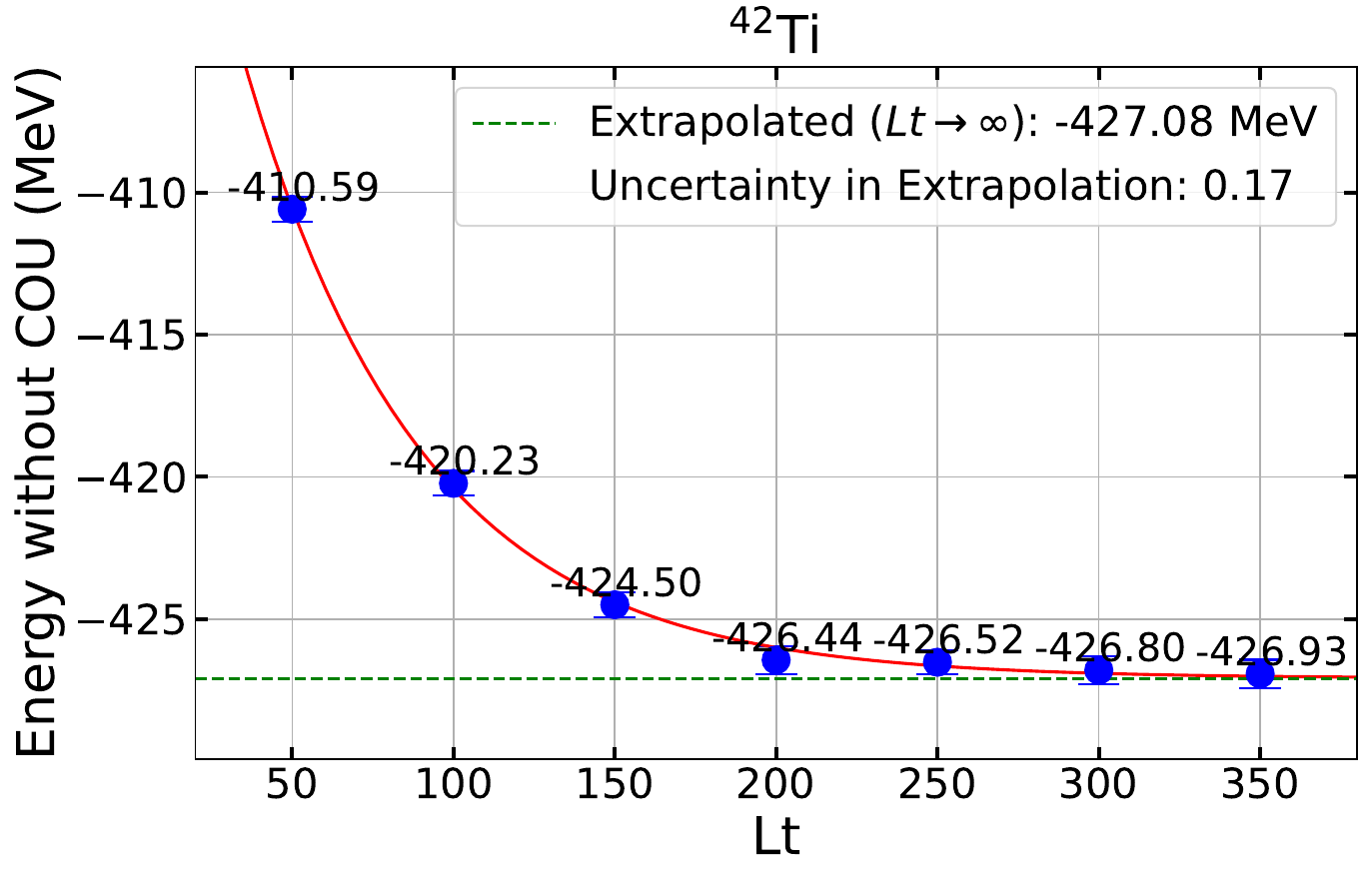}
    \end{minipage}%
    \begin{minipage}{0.46\textwidth}
        \centering
        \includegraphics[width=\textwidth]{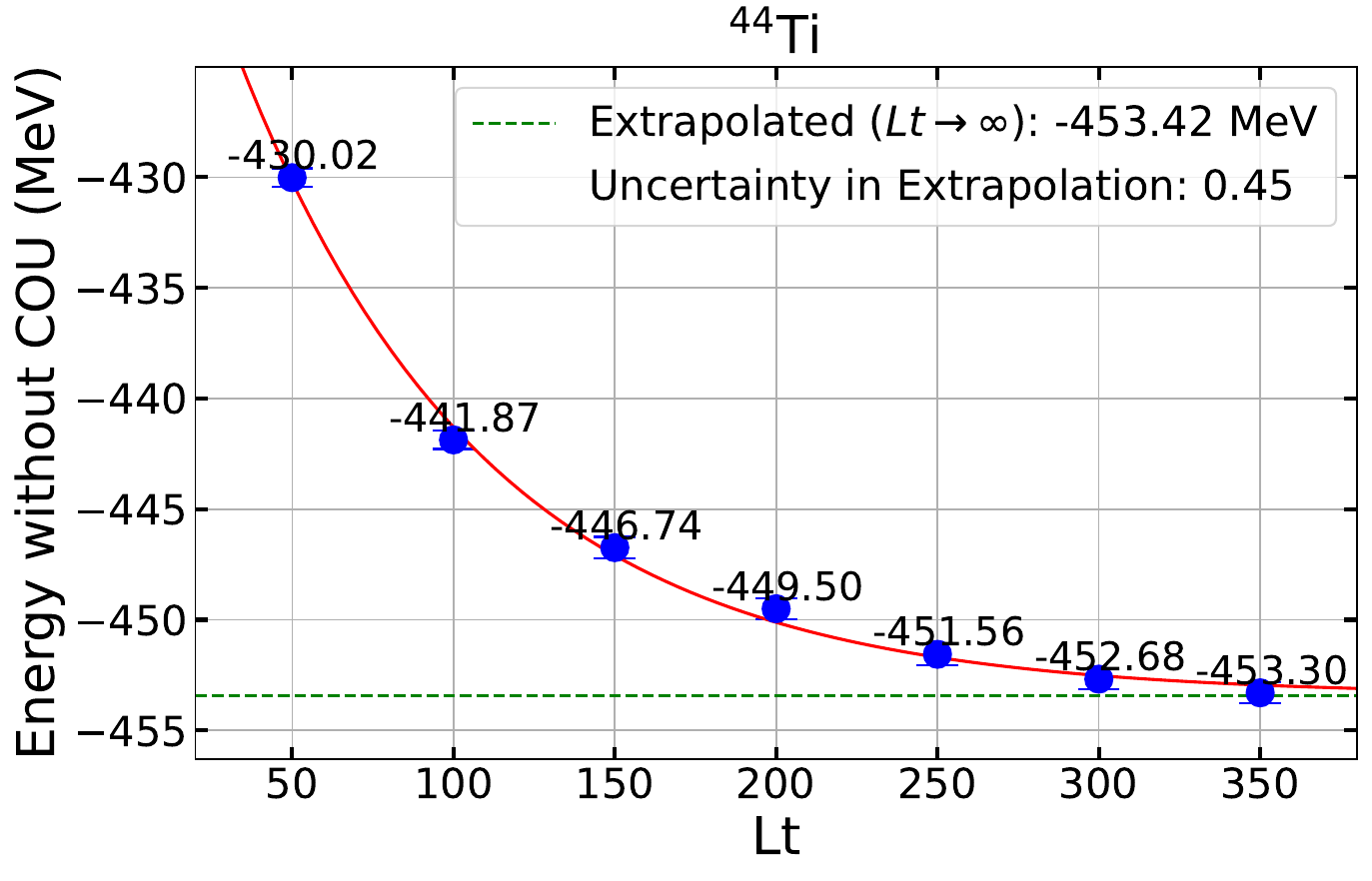}
    \end{minipage}%

    \vspace{0.5cm}
    
    \begin{minipage}{0.46\textwidth}
        \centering
        \includegraphics[width=\textwidth]{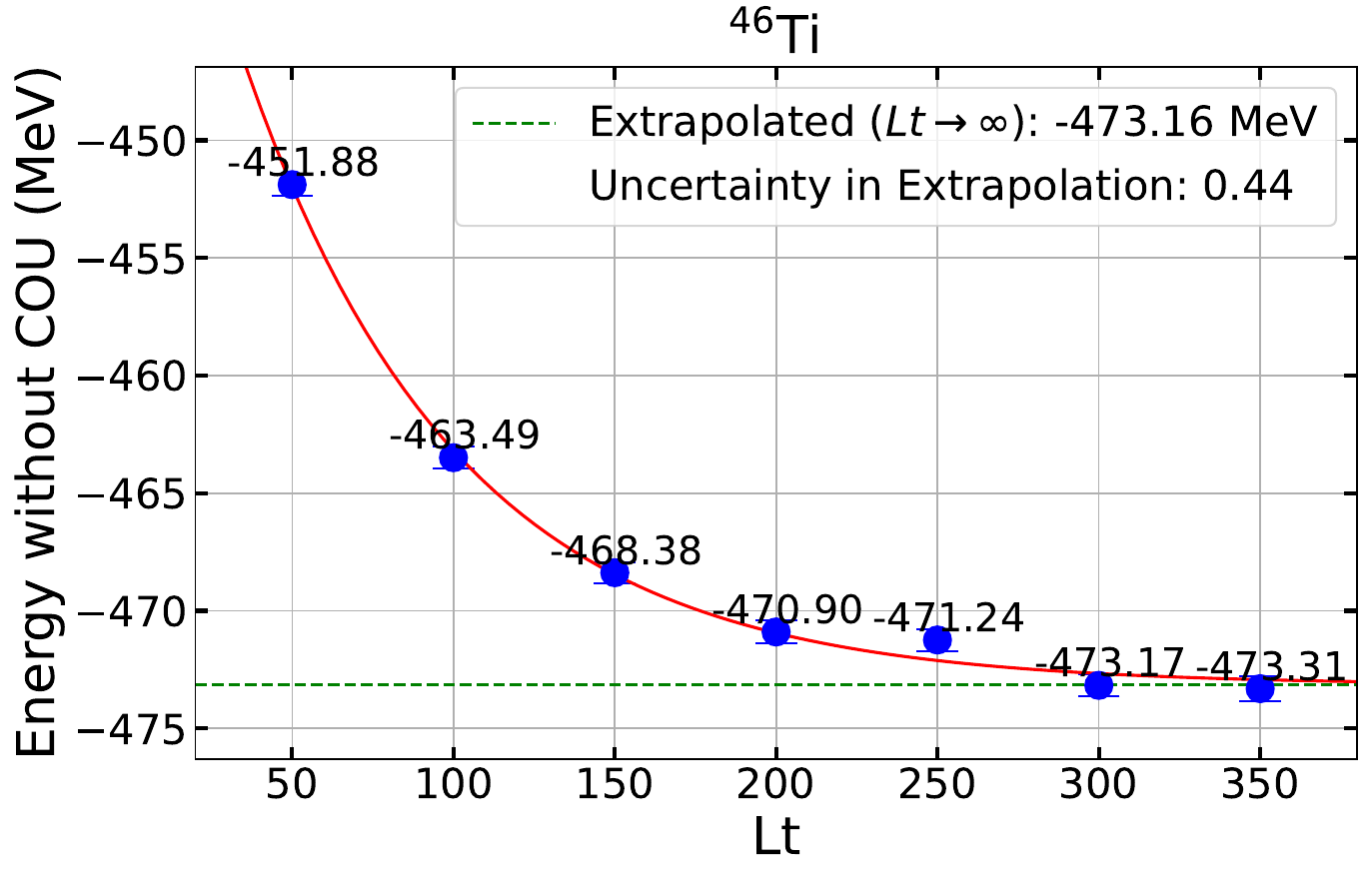}
    \end{minipage}%
    \begin{minipage}{0.46\textwidth}
        \centering
        \includegraphics[width=\textwidth]{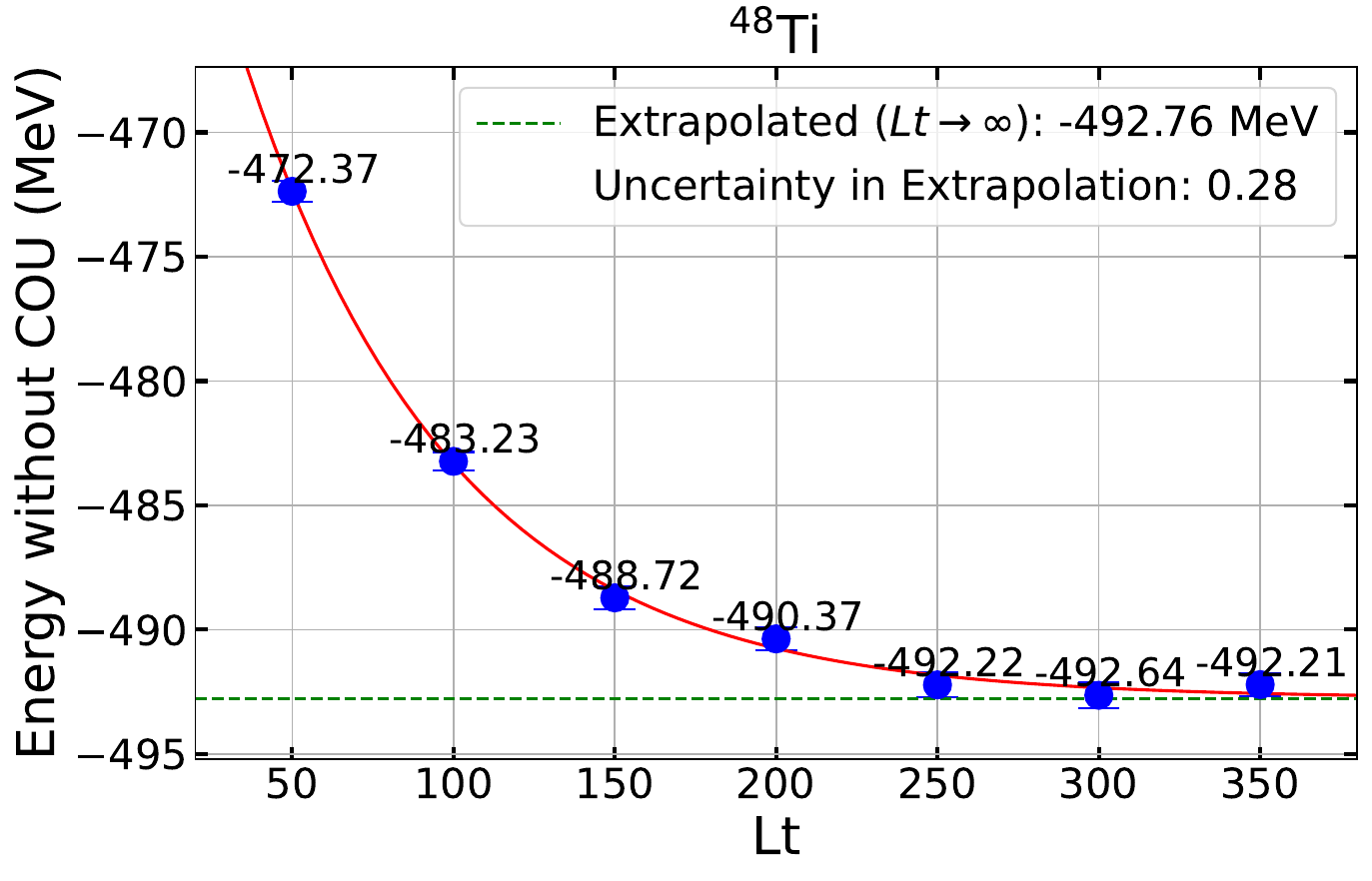}
    \end{minipage}%

    \vspace{0.5cm}

    \begin{minipage}{0.46\textwidth}
        \centering
        \includegraphics[width=\textwidth]{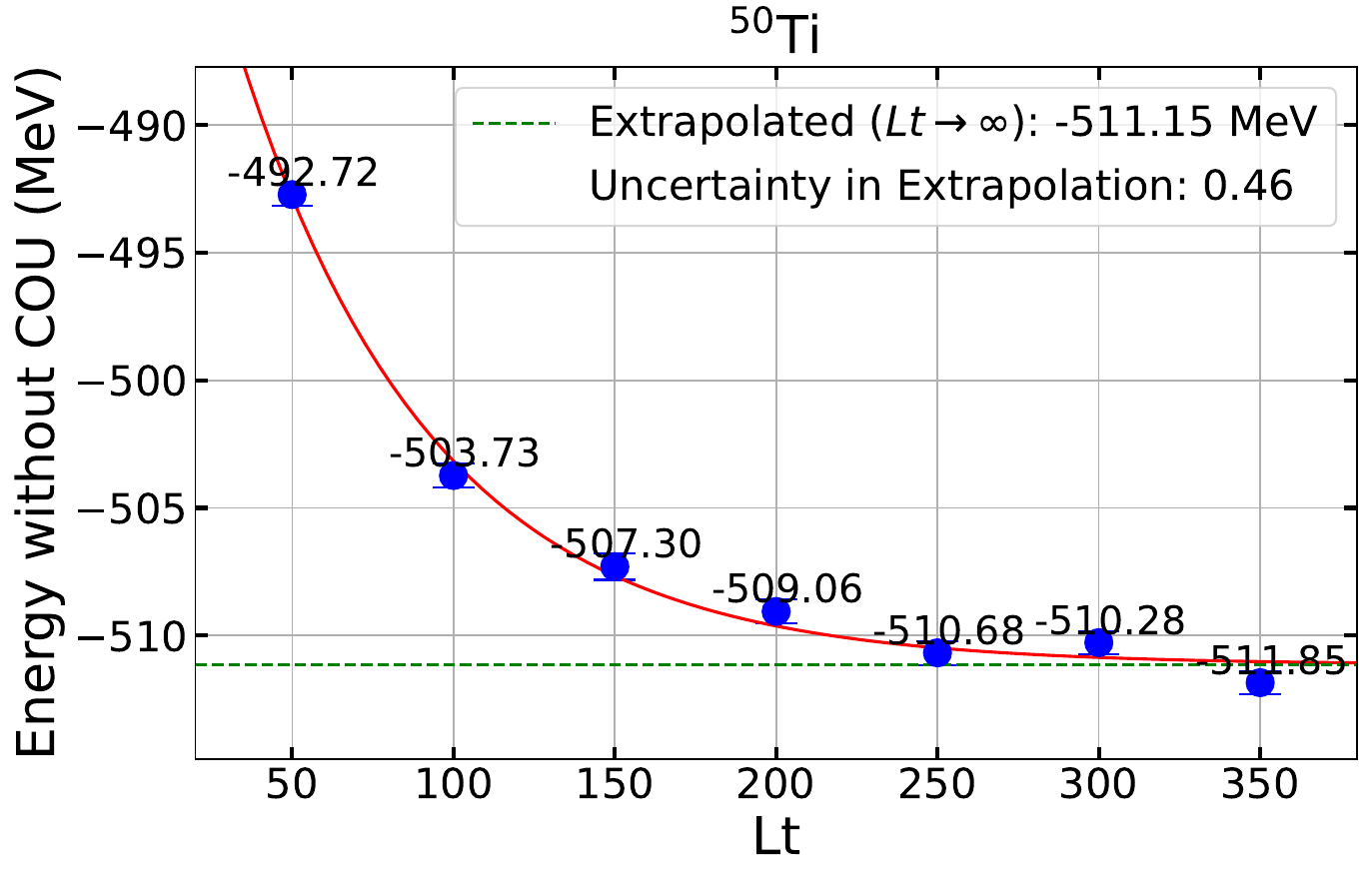}
    \end{minipage}%
    \begin{minipage}{0.46\textwidth}
        \centering
        \includegraphics[width=\textwidth]{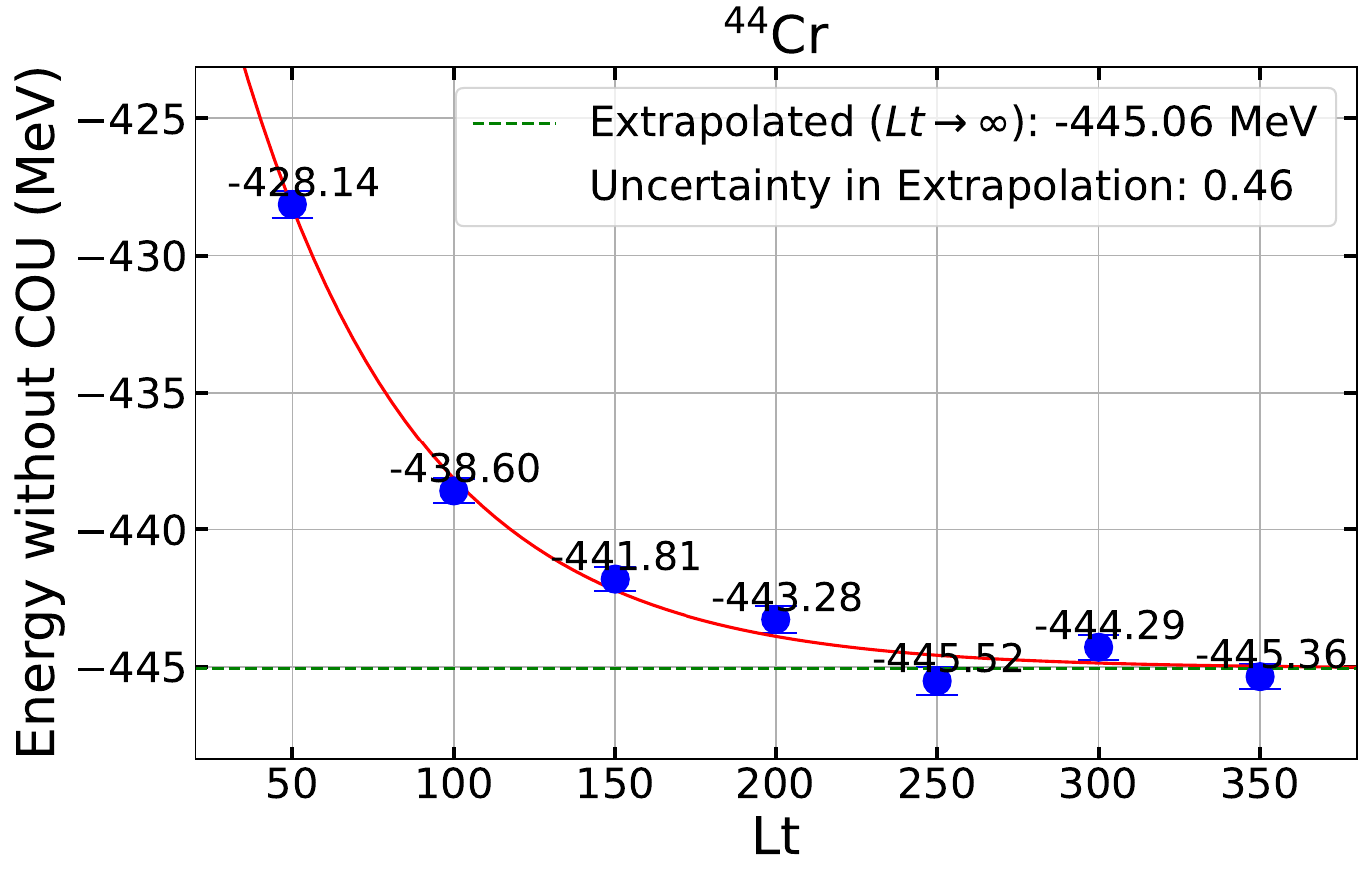}
    \end{minipage}%

    \newpage
    \vspace{0.5cm}
    
    \begin{minipage}{0.46\textwidth}
        \centering
        \includegraphics[width=\textwidth]{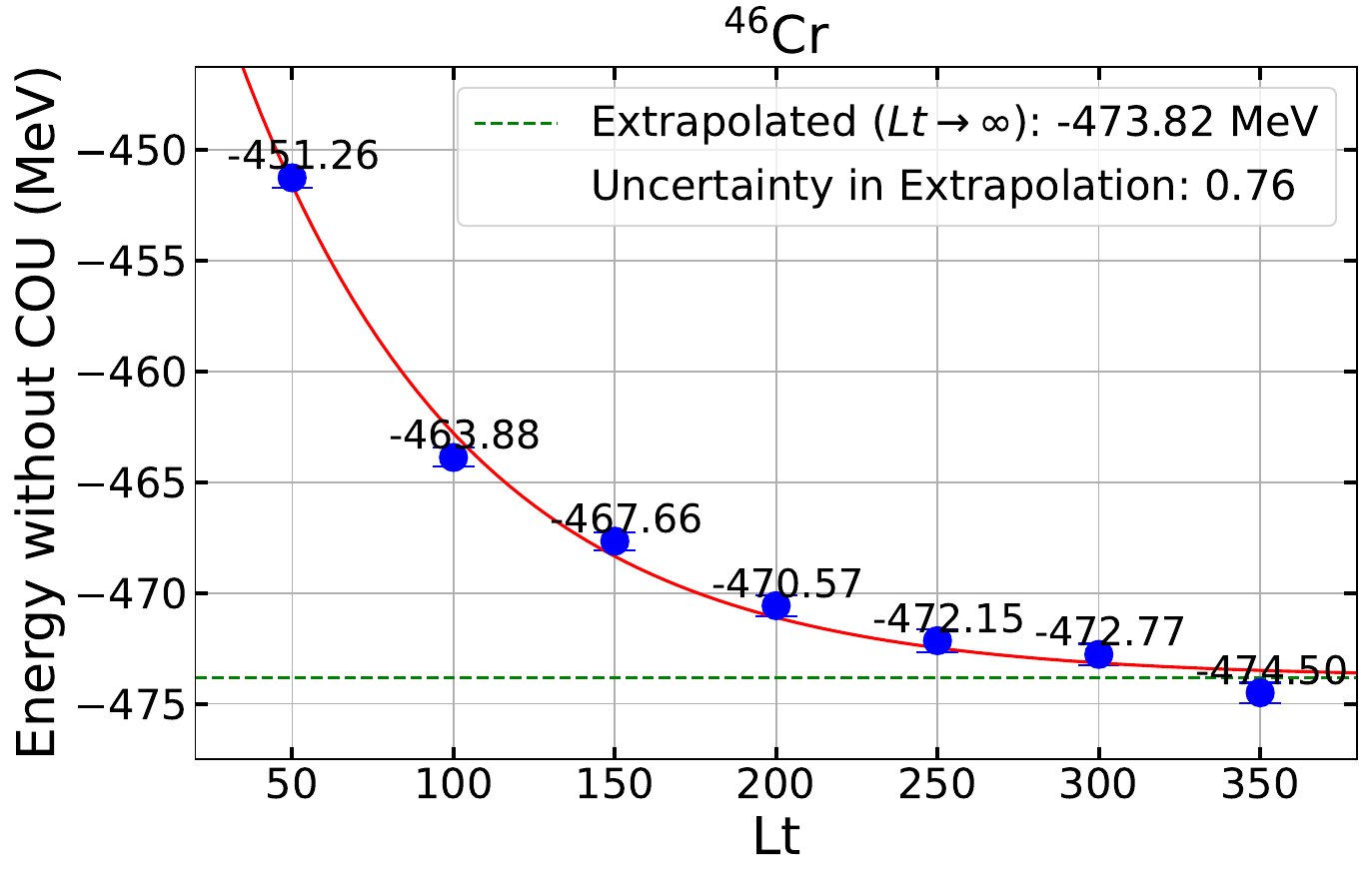}
    \end{minipage}%
    \begin{minipage}{0.46\textwidth}
        \centering
        \includegraphics[width=\textwidth]{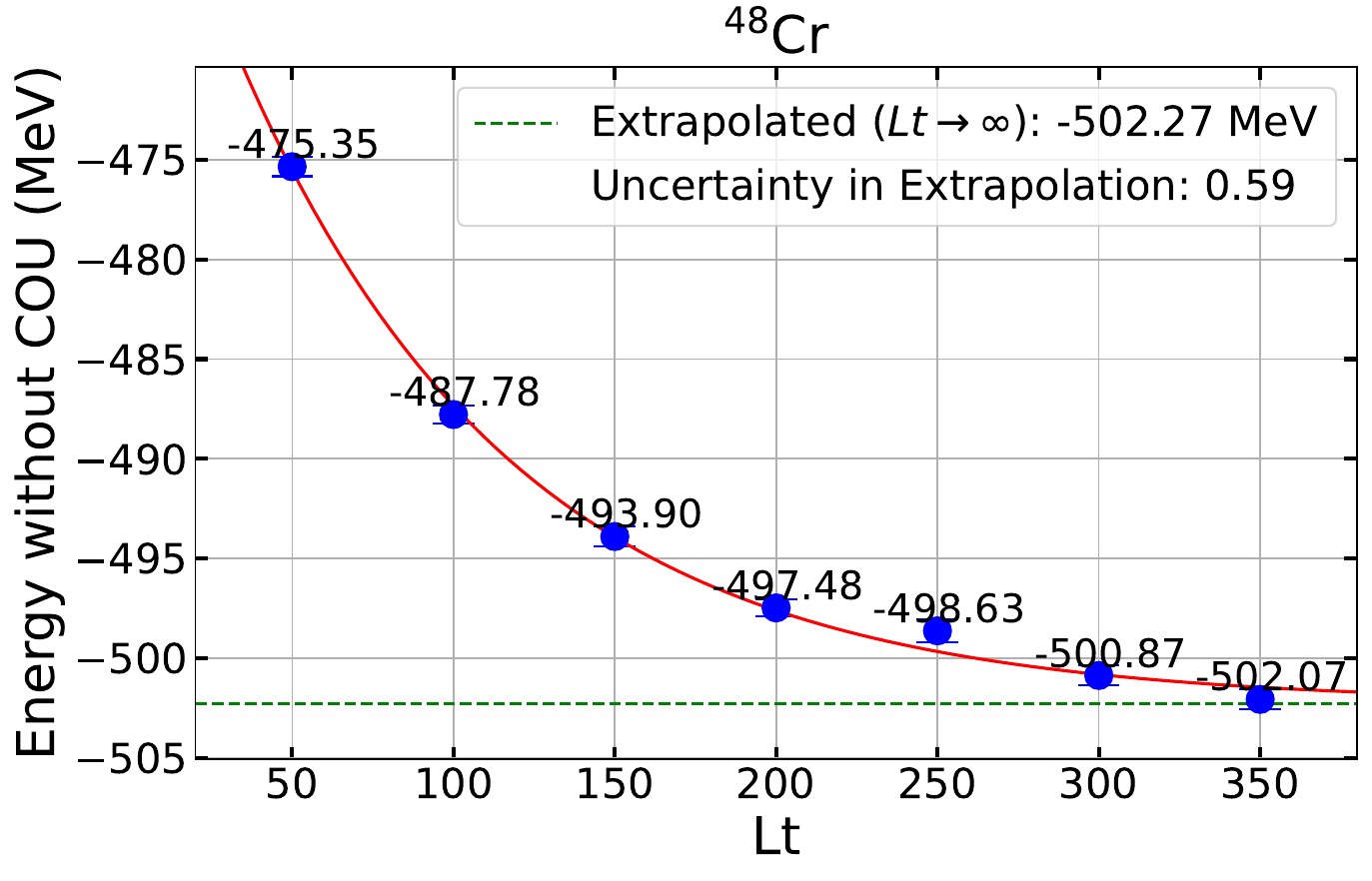}
    \end{minipage}%

\end{figure}

\begin{figure}[H]
    \vspace{0.5cm}
    \begin{minipage}{0.46\textwidth}
        \centering
        \includegraphics[width=\textwidth]{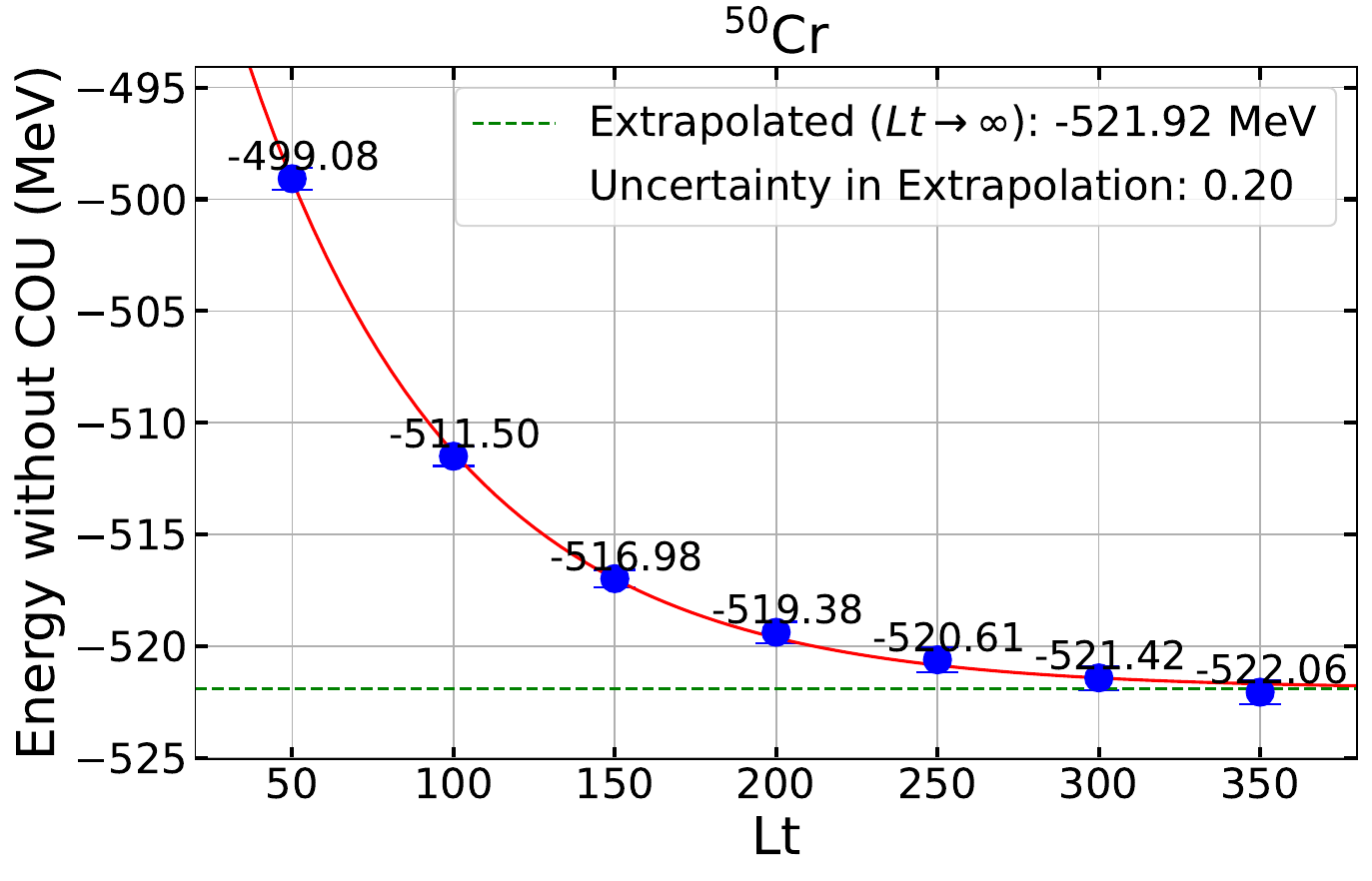}
    \end{minipage}%
    \begin{minipage}{0.46\textwidth}
        \centering
        \includegraphics[width=\textwidth]{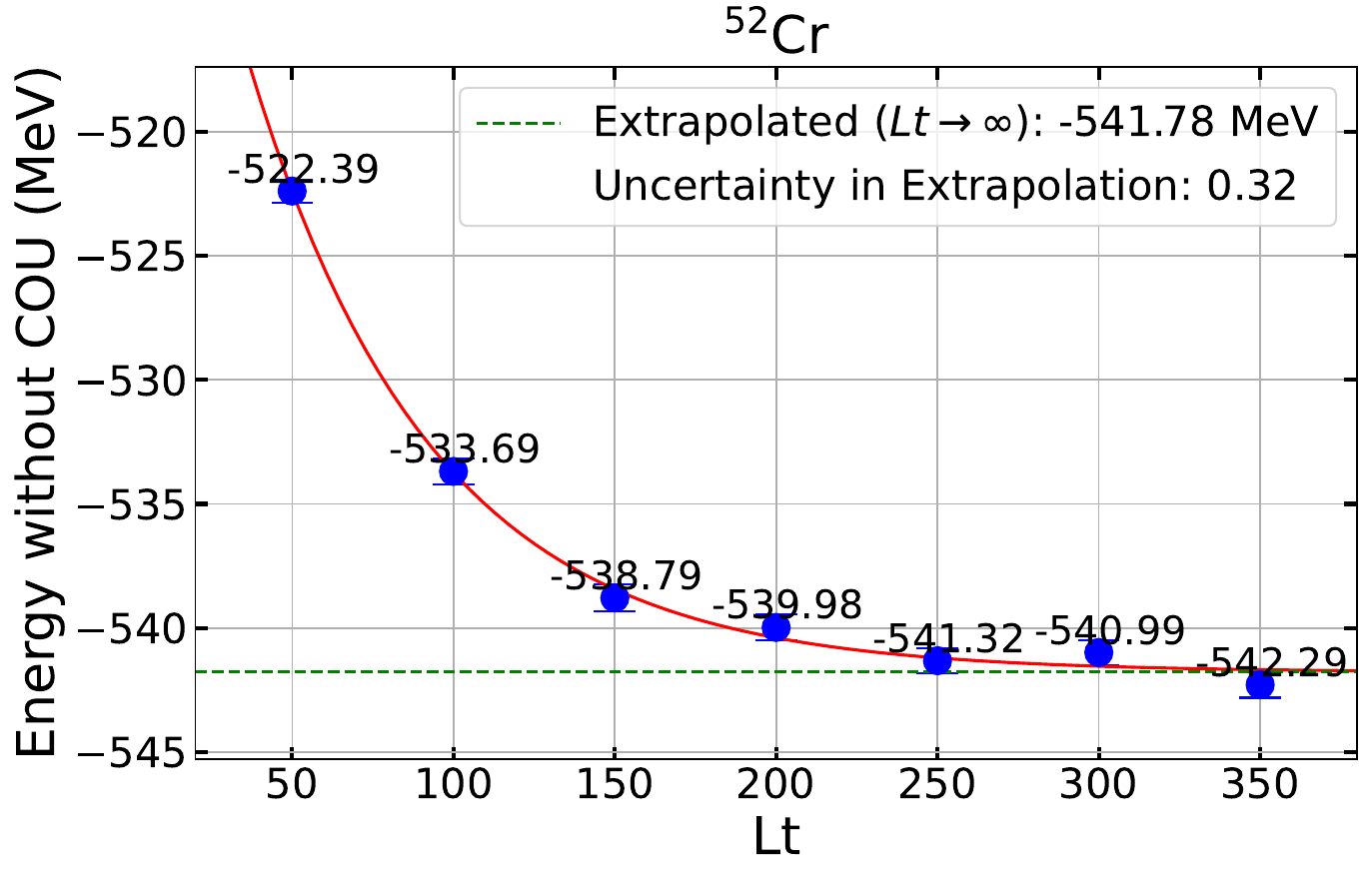}
    \end{minipage}%

    \vspace{0.5cm}
    
    \begin{minipage}{0.46\textwidth}
        \centering
        \includegraphics[width=\textwidth]{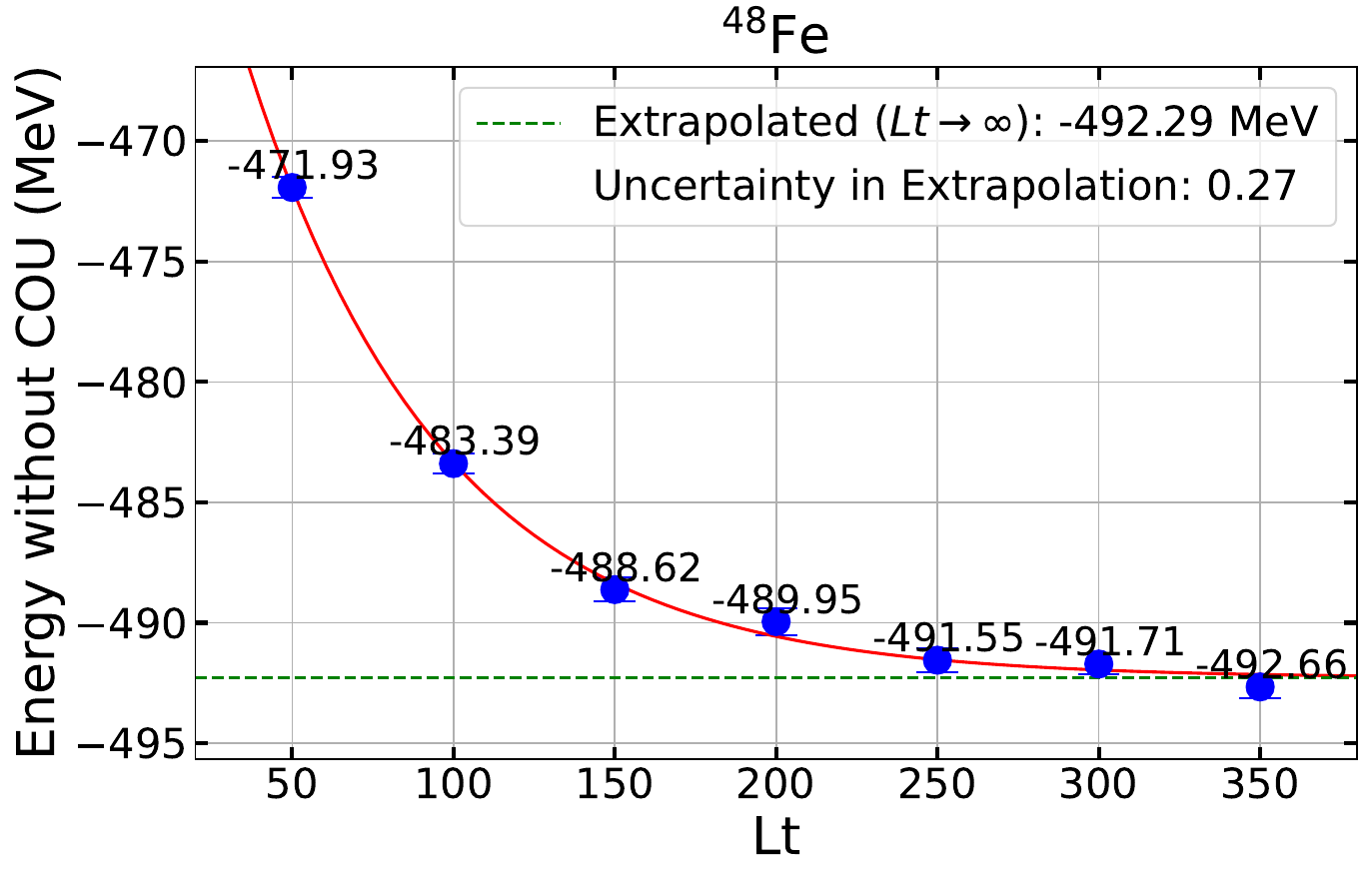}
    \end{minipage}%
    \begin{minipage}{0.46\textwidth}
        \centering
        \includegraphics[width=\textwidth]{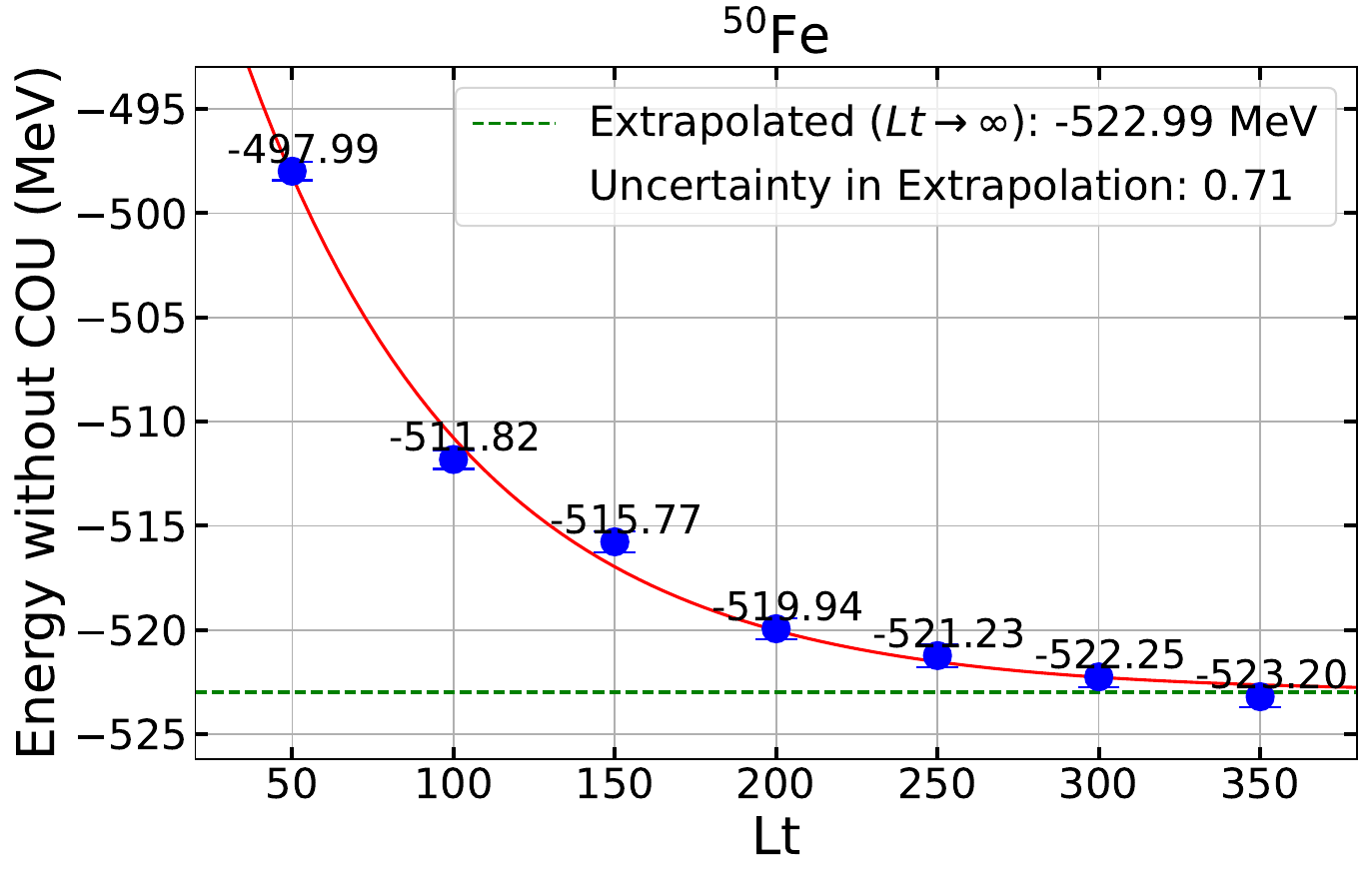}
    \end{minipage}%

    \vspace{0.5cm}
    \begin{minipage}{0.46\textwidth}
        \centering
        \includegraphics[width=\textwidth]{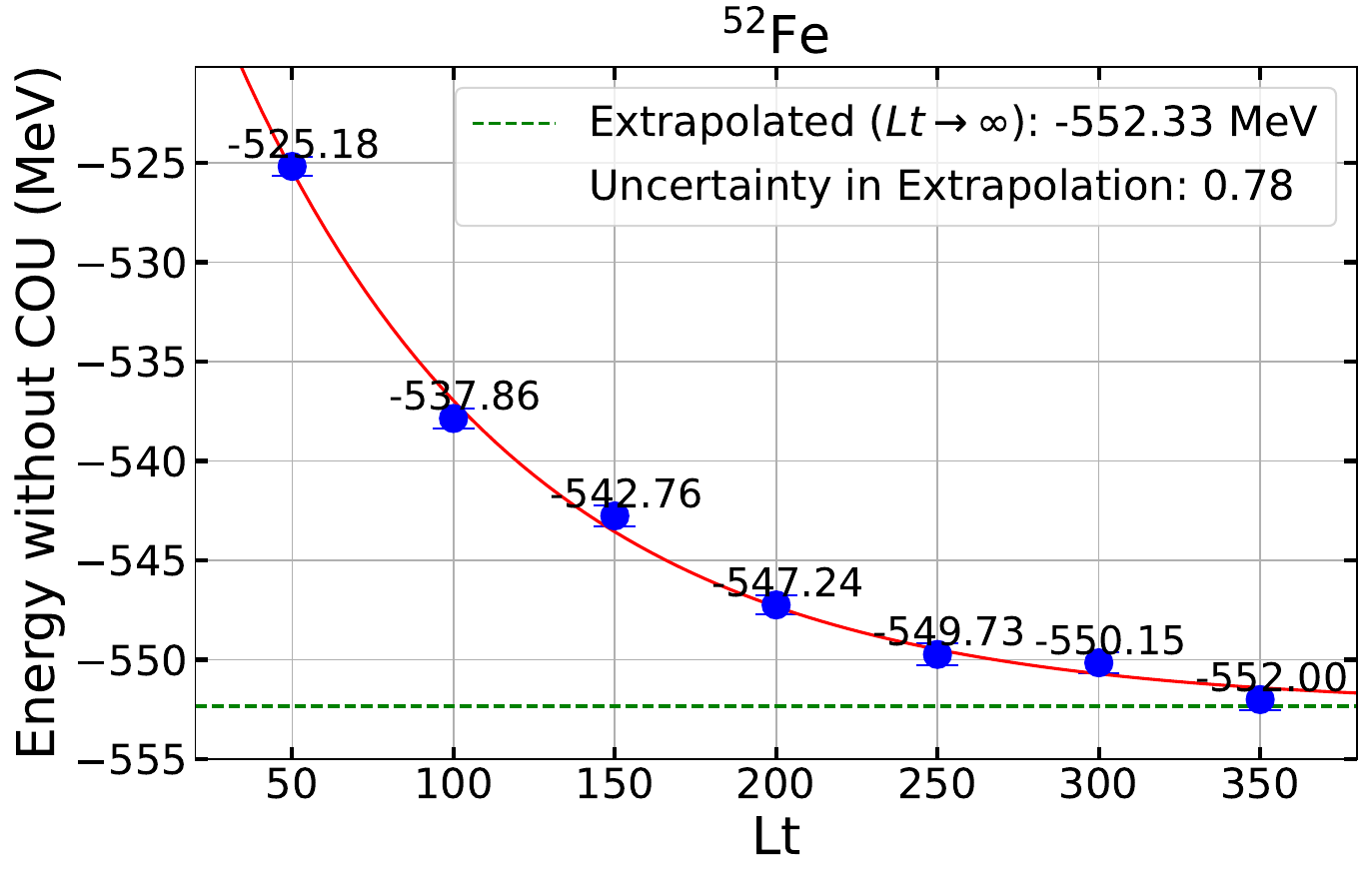}
    \end{minipage}%
    \begin{minipage}{0.46\textwidth}
        \centering
        \includegraphics[width=\textwidth]{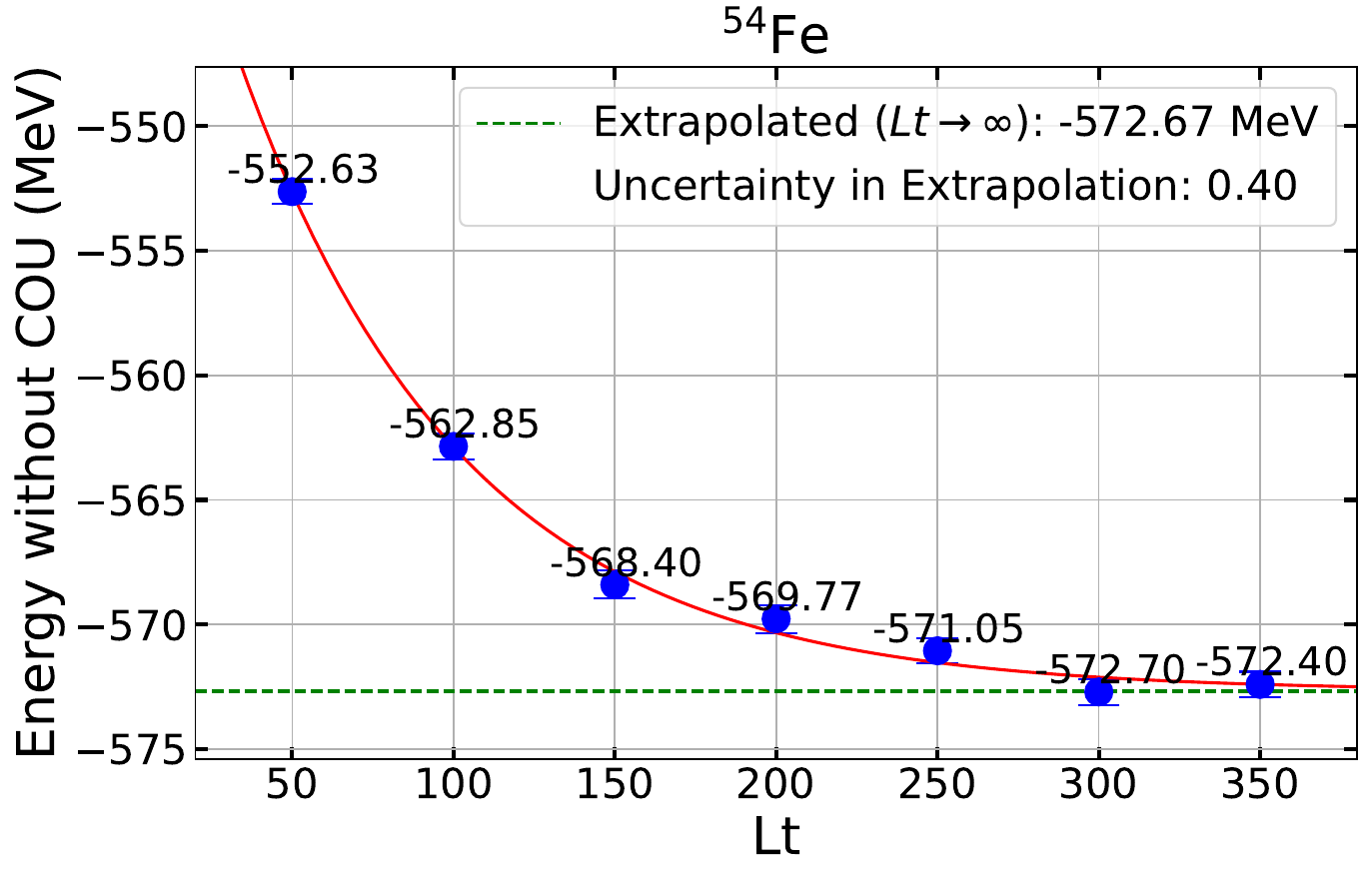}
    \end{minipage}%

    \vspace{0.5cm}
    
    \begin{minipage}{0.46\textwidth}
        \centering
        \includegraphics[width=\textwidth]{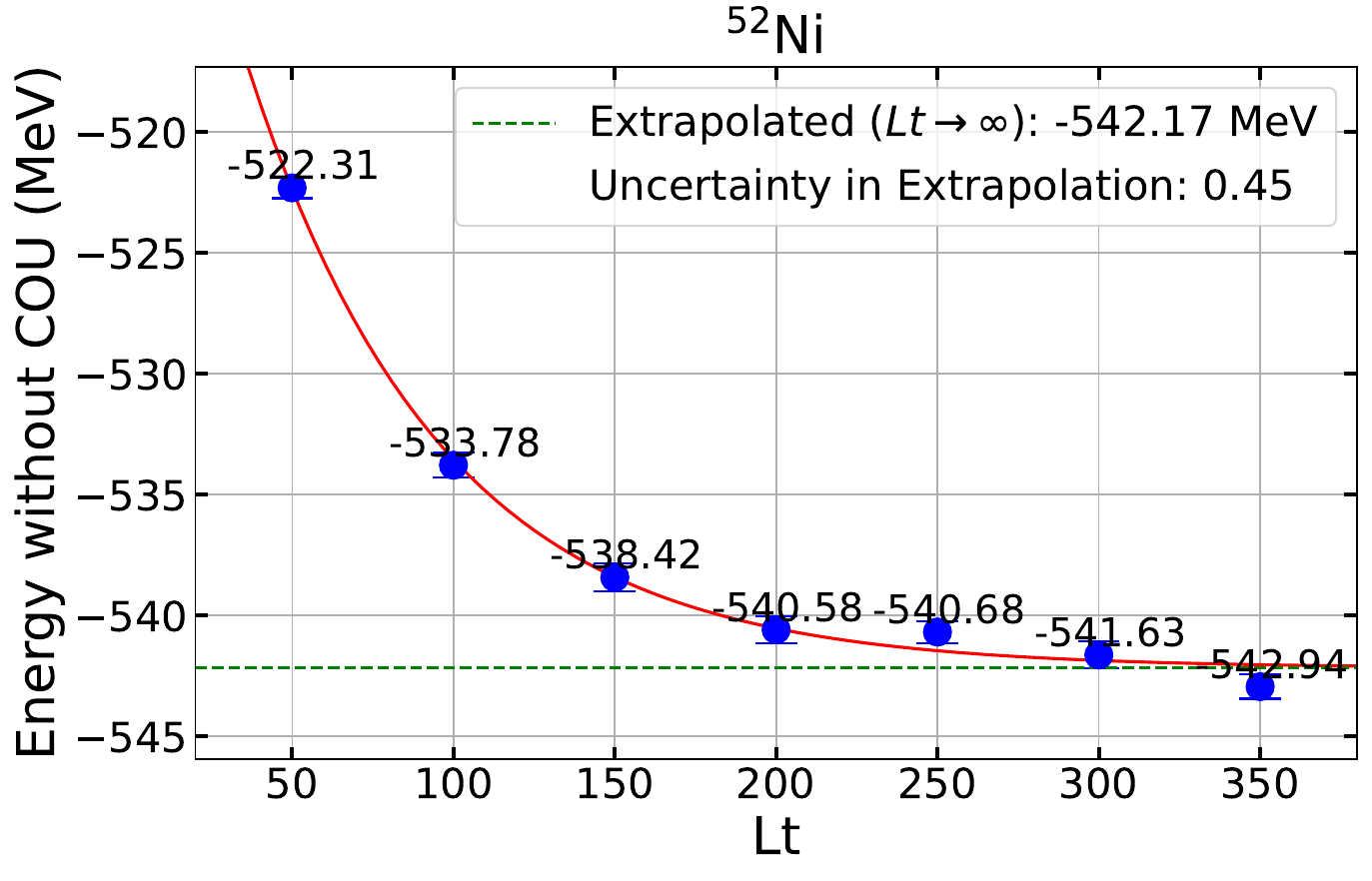}
    \end{minipage}%
    \begin{minipage}{0.46\textwidth}
        \centering
        \includegraphics[width=\textwidth]{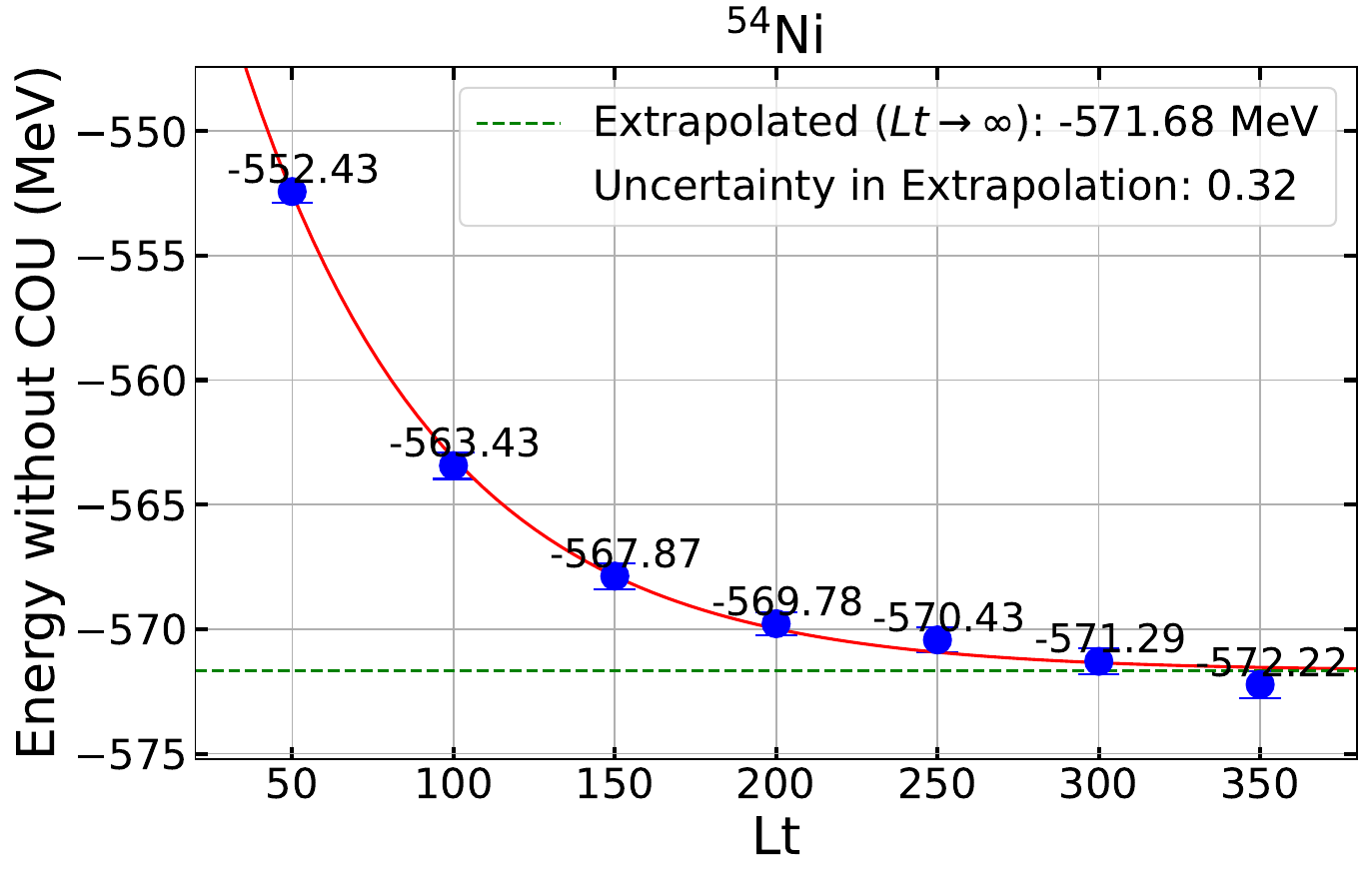}
    \end{minipage}%

\end{figure}
\begin{figure}

    \vspace{0.5cm}
    \begin{minipage}{0.46\textwidth}
        \centering
        \includegraphics[width=\textwidth]{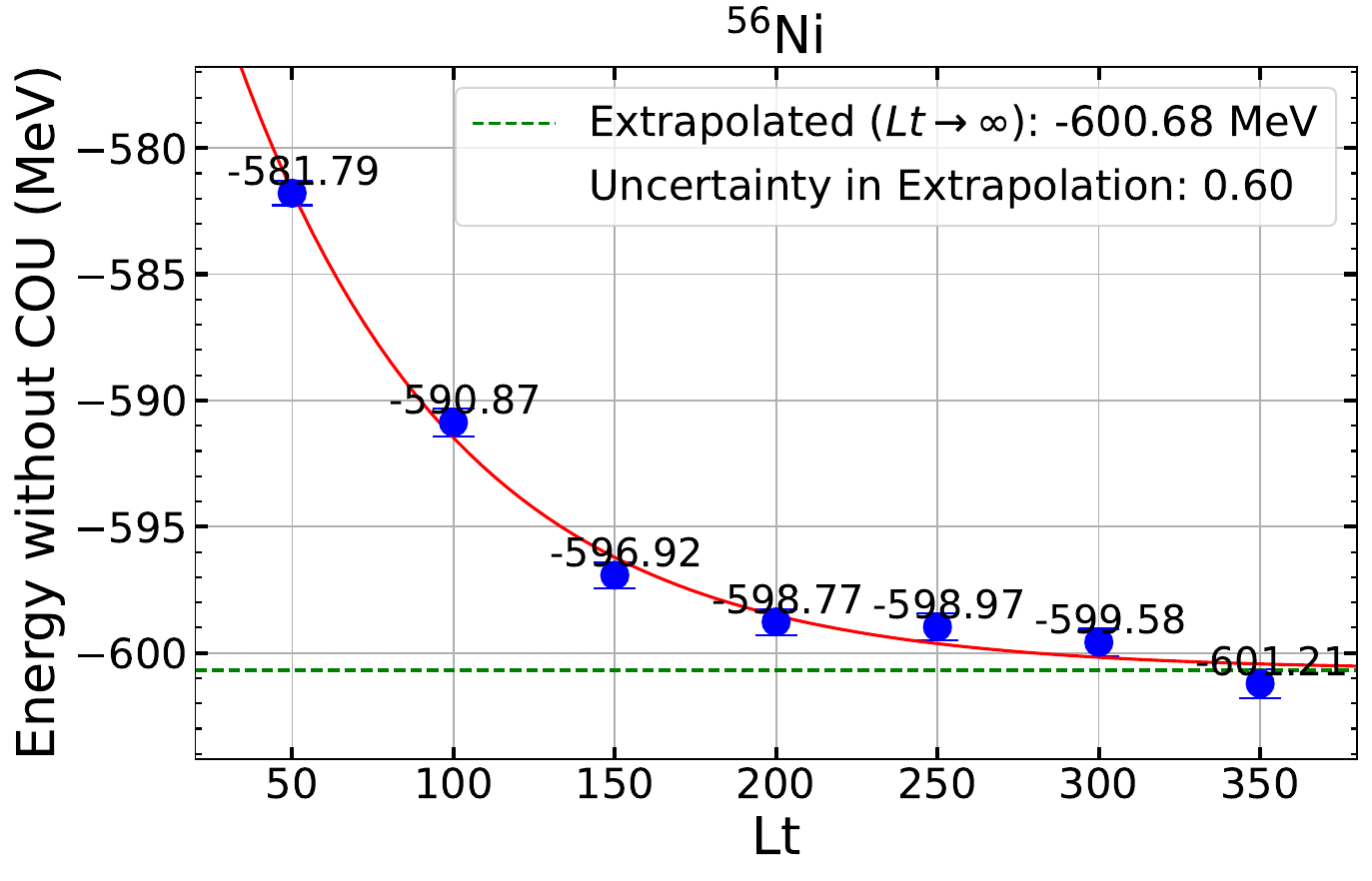}
    \end{minipage}%
    \begin{minipage}{0.46\textwidth}
        \centering
        \includegraphics[width=\textwidth]{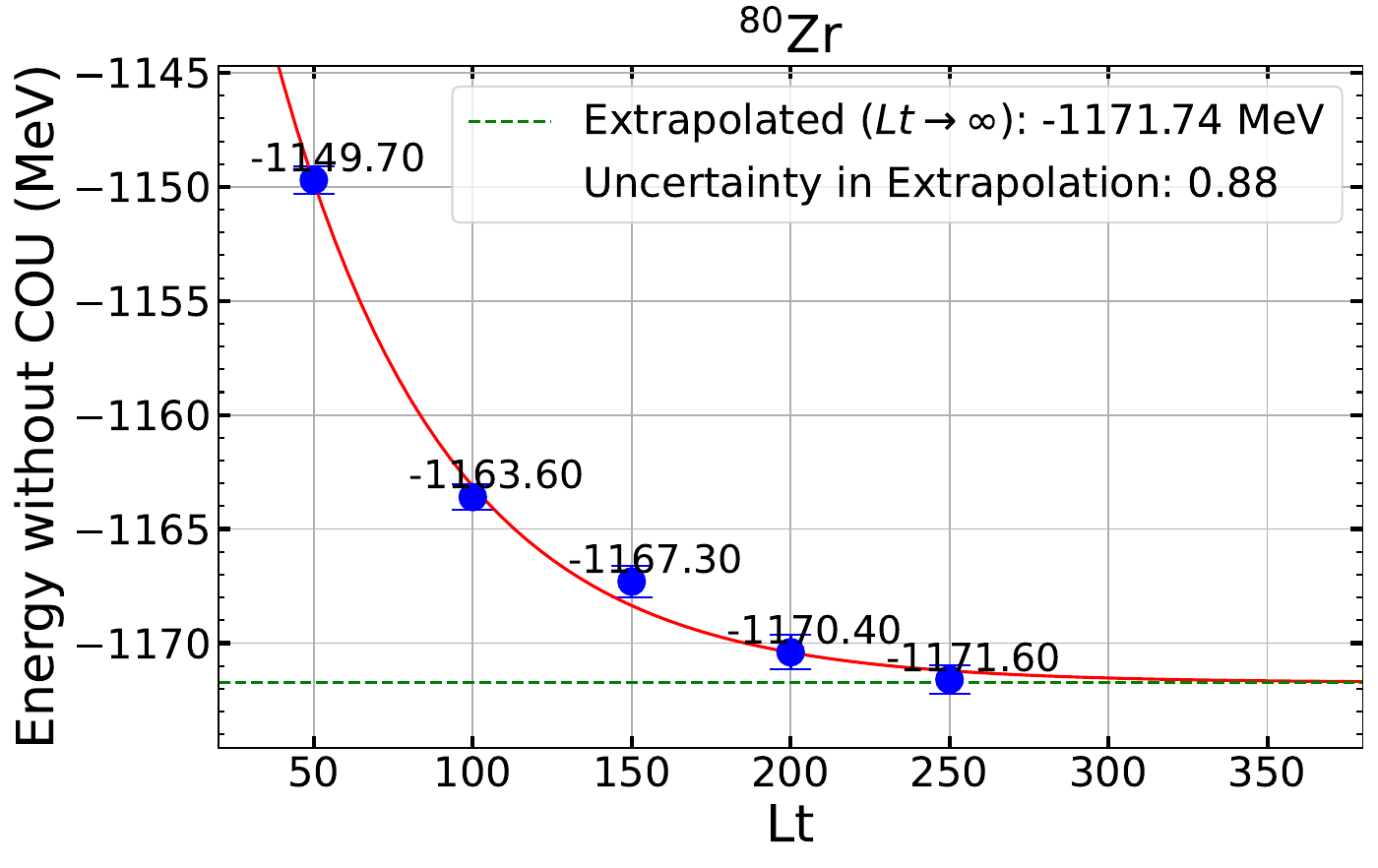}
    \end{minipage}%

    \vspace{0.5cm}
    
    \begin{minipage}{0.46\textwidth}
        \centering
        \includegraphics[width=\textwidth]{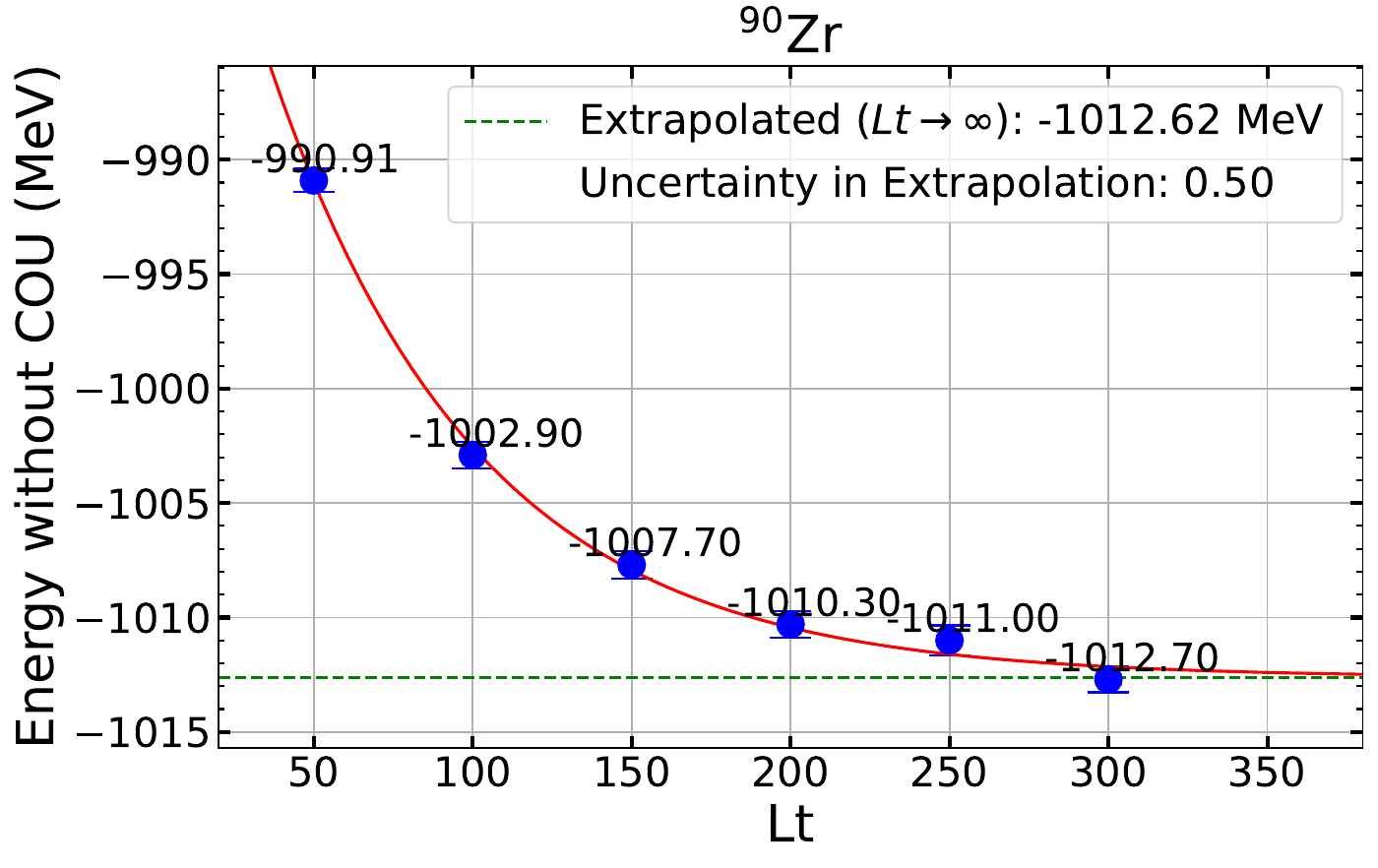}
    \end{minipage}%
    \begin{minipage}{0.46\textwidth}
        \centering
        \includegraphics[width=\textwidth]{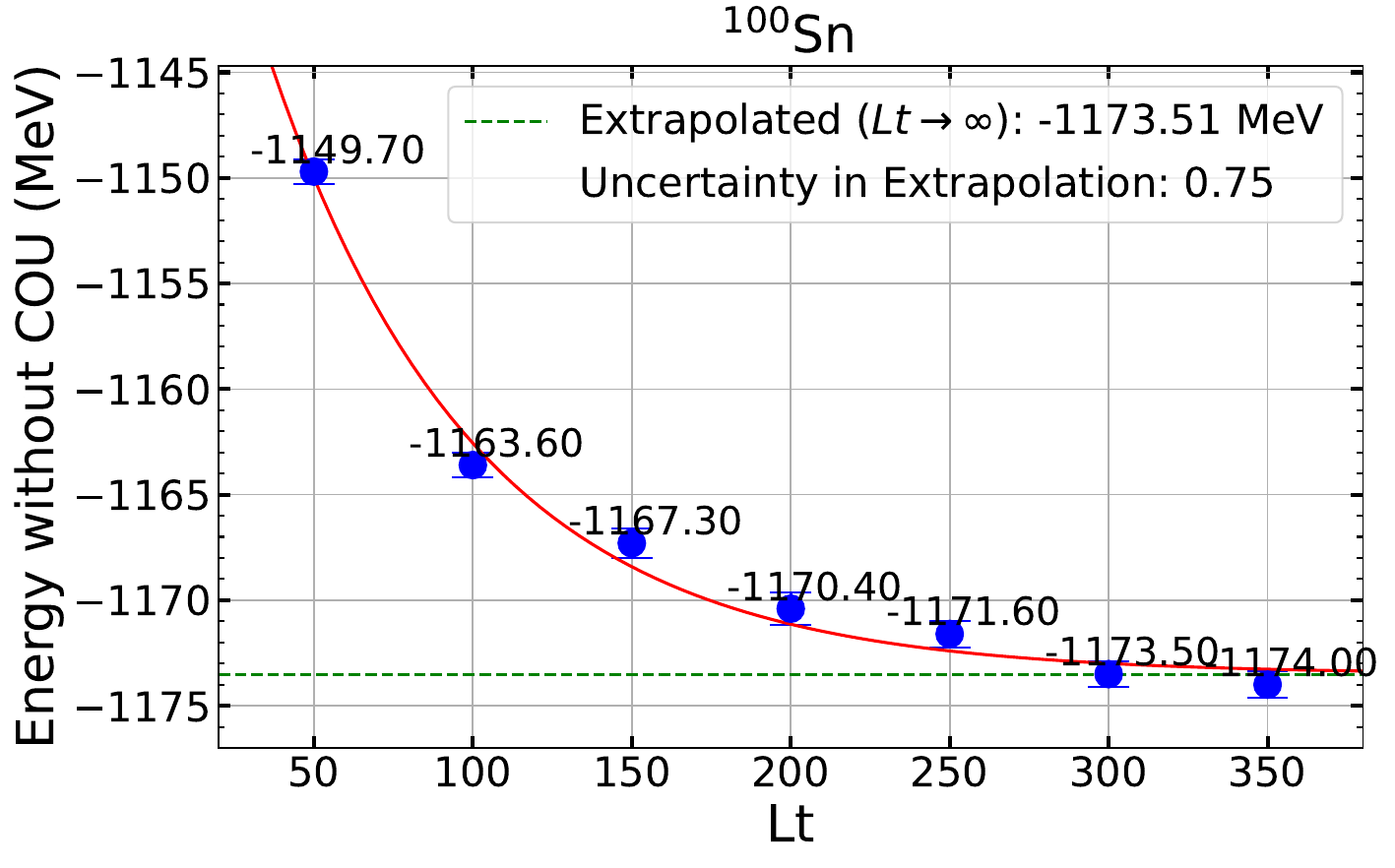}
    \end{minipage}%

\end{figure}

\begin{figure}
    \vspace{0.5cm}
    \begin{minipage}{0.46\textwidth}
        \centering
        \includegraphics[width=\textwidth]{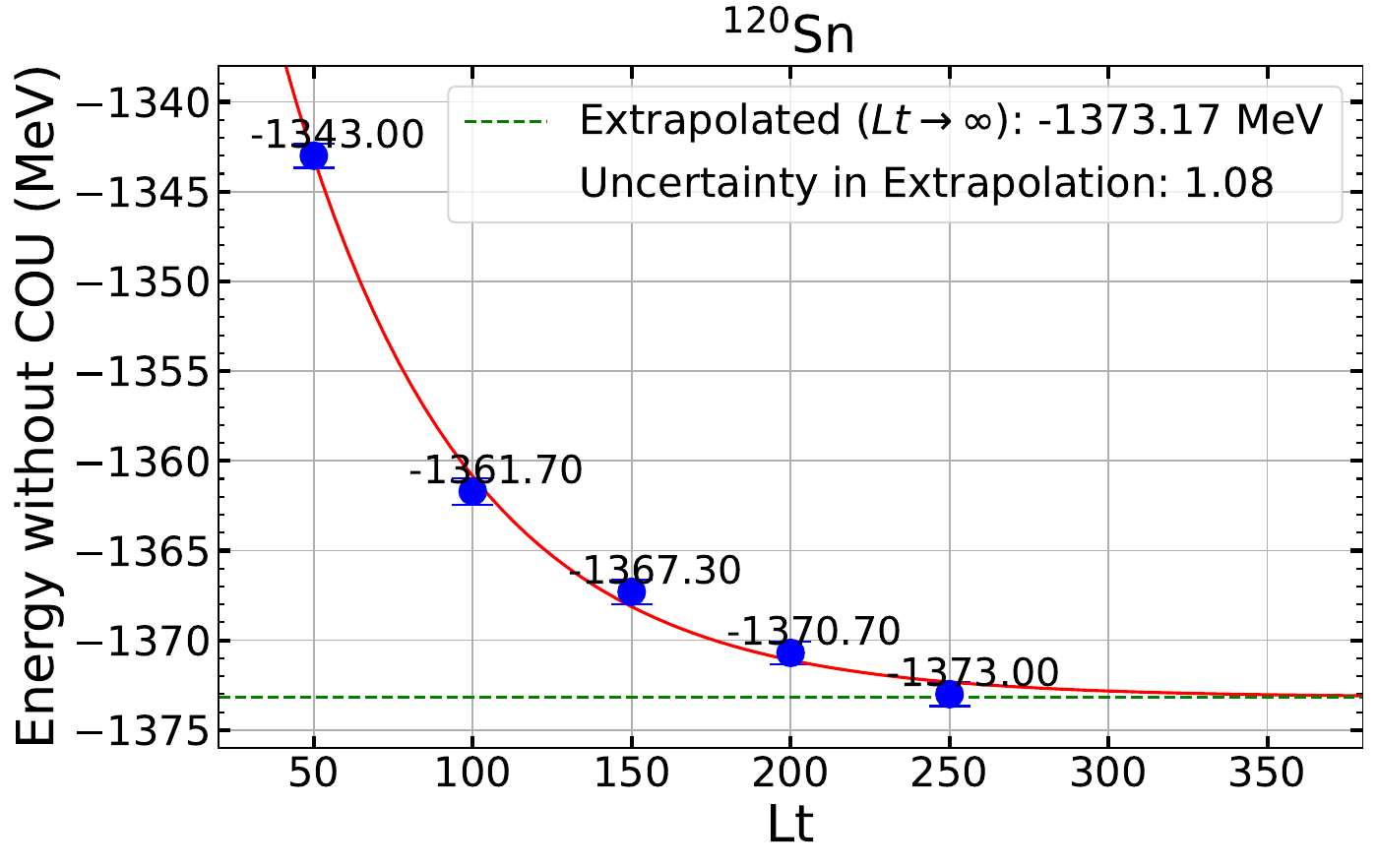}
    \end{minipage}%
    \begin{minipage}{0.46\textwidth}
        \centering
        \includegraphics[width=\textwidth]{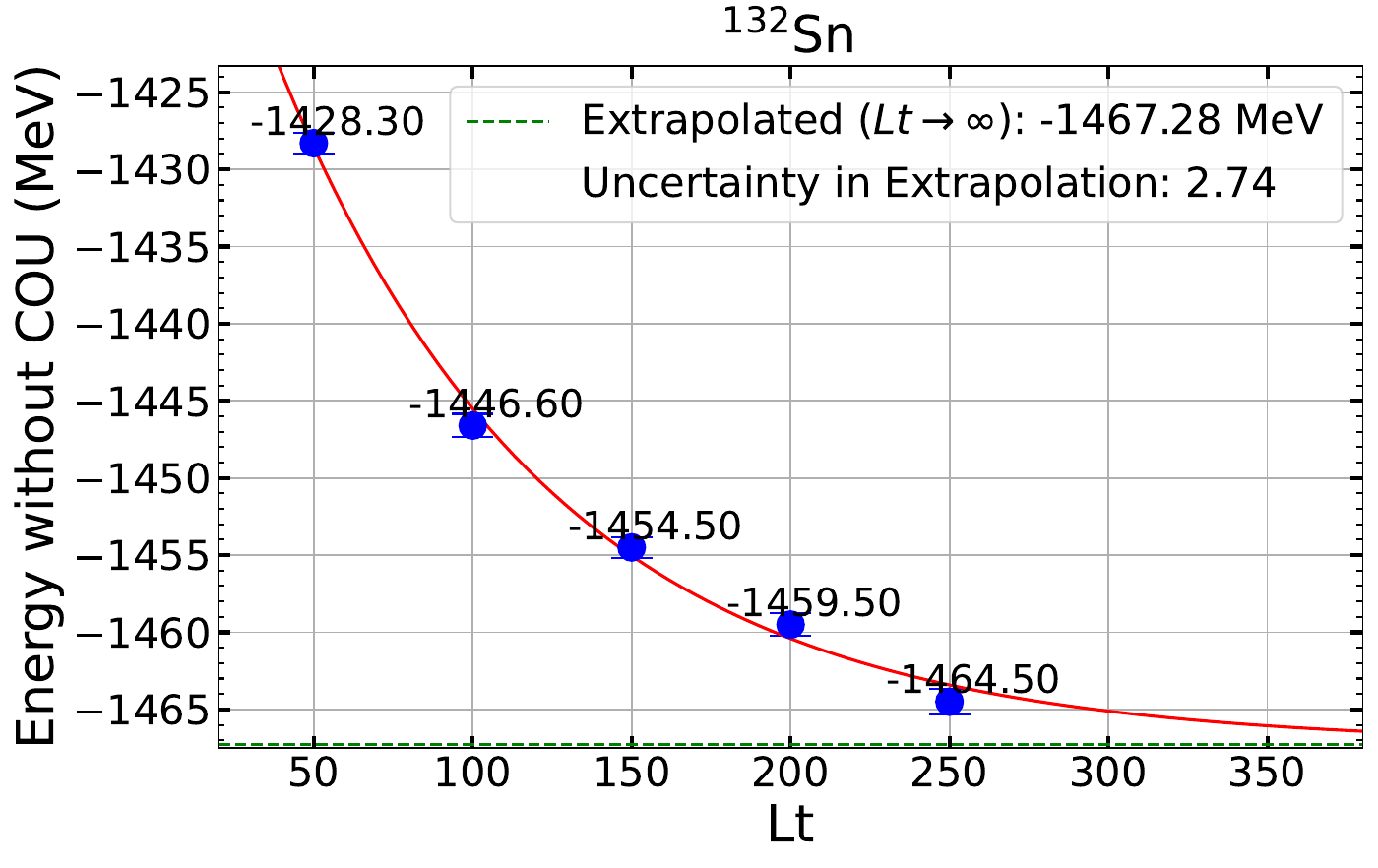}
    \end{minipage}%

\end{figure}

\end{document}